\documentclass[doctor,english,final,pdfdoc,table]{kaist-ucs}

\usepackage{amsmath, amssymb, epsfig, multirow, bbm, mathtools}

\usepackage{microtype}

\usepackage[sort]{natbib}
\usepackage{hyperref}
\usepackage[T1]{fontenc}
\usepackage{wrapfig,lipsum}
\usepackage{graphicx}

\usepackage{soul}
\usepackage{dsfont}
\usepackage{enumerate}
\usepackage{enumitem}
\usepackage{kotex}

\usepackage{amsfonts}
\usepackage{rotating}
\usepackage{dsfont}
\usepackage[Symbol]{upgreek}
\usepackage{lscape}
\usepackage{caption}
\usepackage{balance}
\usepackage{xspace}
\usepackage{float}
\usepackage{wasysym}
\usepackage{makecell}
\usepackage{blindtext}

\usepackage{boldline}
\usepackage{array}
\usepackage{tabu}
\usepackage{csquotes}
\usepackage{wrapfig}
\usepackage{url}
\usepackage{tabulary}
\usepackage{tabularx}
\usepackage{pifont}
\usepackage{bm}

\usepackage[most]{tcolorbox}
\usepackage{tikz}

\usepackage{booktabs} 

\usepackage{subcaption}
\usepackage{tablefootnote, threeparttable}
\usepackage{dashrule}

\newcommand{\highlight}[2]{\colorbox{#1}{\rule[-0.45ex]{-1pt}{0.90em}#2\hspace*{-1pt}}}

\makeatletter
\def\adl@drawiv#1#2#3{%
        \hskip.5\tabcolsep
        \xleaders#3{#2.5\@tempdimb #1{1}#2.5\@tempdimb}%
                #2\z@ plus1fil minus1fil\relax
        \hskip.5\tabcolsep}
\newcommand{\cdashlinelr}[1]{%
  \noalign{\vskip\aboverulesep
           \global\let\@dashdrawstore\adl@draw
           \global\let\adl@draw\adl@drawiv}
  \cdashline{#1}
  \noalign{\global\let\adl@draw\@dashdrawstore
           \vskip\belowrulesep}}
\makeatother

\usepackage{amssymb}
\usepackage{pifont}
\newcommand{\cmark}{\textcolor{PineGreen}{\ding{51}\xspace}}%
\newcommand{\xmark}{\textcolor{red}{\ding{55}\xspace}}%

\renewcommand*\eqref[1]{(\ref{#1})}

\newcommand{\eg}{\emph{e.g.,~}}
\newcommand{\ie}{\emph{i.e.,~}}

\definecolor{Tan}{RGB}{210,180,140}
\definecolor{LightCyan}{rgb}{0.88,1,1}
\definecolor{LightGray}{gray}{0.9}
\definecolor{goldenrod}{rgb}{1.0,0.84,0.3}
\definecolor{shadecolor}{named}{LightGray}

\newcommand{\ourmodelmtl}{CAV2vec\xspace}
\newcommand{\ourmodel}{MoHAVE\xspace}

\usepackage{amsmath,amsfonts,bm}

\def\eqref#1{equation~\ref{#1}}

\def\1{\bm{1}}

\def\rva{{\mathbf{a}}}

\def\rvv{{\mathbf{v}}}

\def\vzero{{\bm{0}}}

\def\vz{{\bm{z}}}

\def\mZ{{\bm{Z}}}

\DeclareMathAlphabet{\mathsfit}{\encodingdefault}{\sfdefault}{m}{sl}
\SetMathAlphabet{\mathsfit}{bold}{\encodingdefault}{\sfdefault}{bx}{n}

\title[english]{Scalable Frameworks for Real-World\\Audio-Visual Speech Recognition}
\title[korean]{현실 환경 오디오-비주얼 음성 인식을\\위한 확장형 프레임워크}

\author[korean] {김}{성 년}
\author[korean2] {김}{성년}    
\author[chinese]{金}{聖 年}
\author[english]{Kim}{Sungnyun}

\advisor[major]{윤 세 영}{Se-Young Yun}{signed}
\advisor[major2]{윤 세 영}{Se-Young Yun}{signed}
\advisorinfo{Professor of Kim Jaechul Graduate School of AI} 

\department{AI}{engineering}{a}

\referee[1]{윤 세 영}
\referee[2]{김 찬 우}
\referee[3]{김 회 린}
\referee[4]{오 태 현}
\referee[5]{정 준 선}

\approvaldate{2025}{12}{5}

\refereedate{2025}{10}{14}

\gradyear{2025}

\begin{document}

   \thesisinfo
    \begin{summary}  
    현실 환경에 존재하는 예측 불가능한 음향 잡음과 시각적 간섭은 오디오-비주얼 음성 인식 시스템의 실용화 및 확장을 가로막는 근본적인 문제이다. 본 학위 논문은 이러한 한계를 극복하고 강건한 확장성을 확보하기 위해, 표현 학습, 아키텍처, 시스템 등의 각 단계에 걸친 체계적이고 계층적인 접근법이 필수적임을 제안한다. 먼저, 표현 학습 단계에서는 다양한 실제 노이즈에 본질적으로 강건한 오디오-비주얼 특징 표현을 학습하여, 별도의 특화된 모듈이 없이도 새로운 환경에 일반화될 수 있는 통합 표현 모델을 구축하는 방법을 연구한다. 아키텍처 단계에서는 입력 데이터의 특성에 따라 계산 자원을 지능적으로 할당하는 프레임워크를 통해, 멀티모달 입력을 적응적이고 신뢰도 높게 사용하면서 모델의 용량을 효율적으로 확장하는 방안을 탐구한다. 마지막으로 시스템 레벨에서는, 대규모 파운데이션 모델과의 모듈식 통합을 통해 음성 인식 시스템의 기능을 확장하고, 파운데이션 모델이 가진 강력한 생성 능력을 활용하여 최종 인식 정확도를 극대화하는 방안을 제시한다. 본 학위 논문은 이 세 가지 계층에 대한 체계적인 해결책을 종합하여, 현실 환경에서도 높은 신뢰도를 갖춘 강건하고 확장 가능한 차세대 오디오-비주얼 음성 인식 시스템을 구축하고자 한다.
    \end{summary}
   
    \begin{Korkeyword}
    오디오-비주얼 음성 인식, 멀티모달 표현 학습, 혼합 전문가 모델, 생성형 오류 수정, 대형 언어 모델
    \end{Korkeyword}

    \begin{abstract}
    The practical deployment of Audio-Visual Speech Recognition (AVSR) systems is fundamentally challenged by significant performance degradation in real-world environments, characterized by unpredictable acoustic noise and visual interference. This dissertation posits that a systematic, hierarchical approach is essential to overcome these challenges, achieving the robust scalability at the representation, architecture, and system levels. At the representation level, we investigate methods for building a unified model that learns audio-visual features inherently robust to diverse real-world corruptions, thereby enabling generalization to new environments without specialized modules. To address architectural scalability, we explore how to efficiently expand model capacity while ensuring the adaptive and reliable use of multimodal inputs, developing a framework that intelligently allocates computational resources based on the input characteristics. Finally, at the system level, we present methods to expand the system's functionality through modular integration with large-scale foundation models, leveraging their powerful cognitive and generative capabilities to maximize final recognition accuracy. By systematically providing solutions at each of these three levels, this dissertation aims to build a next-generation, robust, and scalable AVSR system with high reliability in real-world applications.
    \end{abstract} 
     
    \begin{Engkeyword}
    Audio-Visual Speech Recognition, Multimodal Representation Learning, Mixture-of-Experts, Generative Error Correction, Large Language Models
    \end{Engkeyword}

    \addtocounter{pagemarker}{1}
    \newpage

    \tableofcontents

    \listoftables

    \listoffigures

\chapter{Introduction}

\section{The Challenge of Speech in the Wild}

Automatic Speech Recognition (ASR) has transitioned from a niche technology to a ubiquitous component of modern human-computer interaction, powering applications from virtual assistants to in-car control systems. In acoustically controlled and clean environments, the performance of state-of-the-art ASR systems has achieved, and in some cases surpassed, human-level accuracy~\citep{radford2023robust}. This success, however, is often confined to the laboratory environment. When deployed in the ``real world'', \ie the unconstrained, unpredictable, and acoustically diverse environments of everyday life, the reliability of these systems degrades dramatically. The gap between controlled and in-the-wild conditions remains the foremost obstacle to the universal adoption and trustworthiness of speech technology. This dissertation confronts this challenge directly, positing that true progress requires a fundamental rethinking of how we design and extend ASR systems for robustness and scalability.

The brittleness of conventional ASR systems stems from a multitude of complex, often co-occurring factors that distort the speech signal. These challenges can be broadly categorized into three principal sources of variability: acoustic environment, speaker, and domain specificity.

\paragraph{Acoustic Environment Variability.} The most pervasive challenge is the presence of additive background noise and reverberation. Real-world noise is highly non-stationary and diverse, ranging from the babble of competing speakers in a crowded cafe to the engine hum in a vehicle or the variable acoustics of background music~\citep{gong1995speech}. Such noise can mask crucial phonetic information, leading to a significant increase in recognition error rates. For instance, studies have shown that even moderate levels of background music can substantially reduce recognition accuracy, and the effect is often more severe than that of stationary white noise because of its dynamic spectral and temporal characteristics. Compounding this is reverberation, where sound reflections from surfaces create echoes that blur the temporal boundaries between phonemes, smearing the acoustic features and severely degrading intelligibility for both humans and machines~\citep{ko2017study}. The combined effect of noise and reverberation is often super-additive, presenting a challenge to signal processing and feature extraction front-ends.

\paragraph{Speaker-Related Variability.} Human speech is intrinsically variable. This variability manifests across multiple dimensions, including physiological differences (\eg vocal tract length, age, or gender), speaking style (\eg read speech vs. spontaneous conversation, or emotional state), and speaking rate. Perhaps the most significant challenge in this category is a model's inability to generalize across different accents and dialects. ASR systems trained predominantly on a standard dialect of a language often exhibit substantial performance degradation when confronted with regional or non-native accents. This disparity is not merely a technical issue; it has significant societal implications, leading to technological inequity where systems are less reliable for speakers from marginalized or underrepresented demographic groups~\citep{koenecke2020racial}.

\paragraph{Linguistic and Domain-Specific Challenges.} Beyond acoustic and speaker-level variations, ASR systems also face challenges at the linguistic level. Spontaneous and conversational speech is replete with disfluencies, such as hesitations, repetitions, and false starts, which deviate from the clean, grammatical text on which language models are typically trained. Furthermore, the vocabulary and linguistic constructs used in specialized domains, such as medicine, law, or finance, present a significant out-of-domain problem for general-purpose ASR systems. The presence of specific jargon, acronyms, and named entities not seen during training can lead to catastrophic errors, limiting the utility of these systems in high-stakes professional settings.

\subsection{Imperative for Robustness: Industrial and Research Perspectives}
The necessity of overcoming these challenges is underscored by both compelling industrial demand and fundamental research questions. From the industrial and commercial perspective, robust ASR is a key for the next wave of innovation in human-computer interaction. In the consumer space, the reliability of virtual assistants like Amazon's Alexa, Google Assistant, and Apple's Siri is directly tied to their ability to function in noisy home environments and understand a diverse user base. In the automotive industry, for instance, dependable voice control for navigation and infotainment is a matter of both convenience and safety, as it allows drivers to remain focused on the road. In enterprise and healthcare, the stakes are even higher. Accurate transcription of meetings, legal depositions, and clinical dictations can unlock massive efficiency gains, but only if the technology is trustworthy across a range of acoustic conditions and speakers. The failure to handle real-world variability is not just an inconvenience; it is a direct barrier to market expansion, user adoption, and the realization of significant economic value.

From the research and scientific perspective, the pursuit of robust ASR systems drives progress towards more general and adaptable AI models. The gap between performance on clean benchmarks and messy, real-world data highlights the limitations of current machine learning paradigms, which often struggle with domain shift and generalization to unseen conditions. Solving robustness is a fundamental scientific challenge that pushes the community to develop novel techniques in areas such as self-supervised learning, domain adaptation, and multimodal fusion. Moreover, addressing the biases in ASR performance across different speaker demographics is a critical ethical imperative for the field, ensuring that the benefits of this powerful technology are inclusive for all members of society.

To overcome these profound limitations, this dissertation focuses on a move beyond audio-only paradigms. We push boundaries of Audio-Visual Speech Recognition (AVSR), which enhances robustness by incorporating a second, complementary modality: the visual information from a speaker's lip movements. This visual stream is invariant to acoustic noise and provides crucial cues to disambiguate sounds and recover speech when the audio is corrupted. The powerful synergy between sight and sound in human speech perception, famously illustrated by the McGurk effect~\citep{mcgurk1976hearing}, underscores the profound potential of a multimodal approach to building truly resilient speech recognition systems. However, as we will explore, the effective and scalable fusion of these modalities presents its own set of complex challenges, which forms the core subject of the following chapters.
\section{Research Goal, Scope, and Questions}

\begin{figure}[!t]
    \centering
    \includegraphics[width=\textwidth]{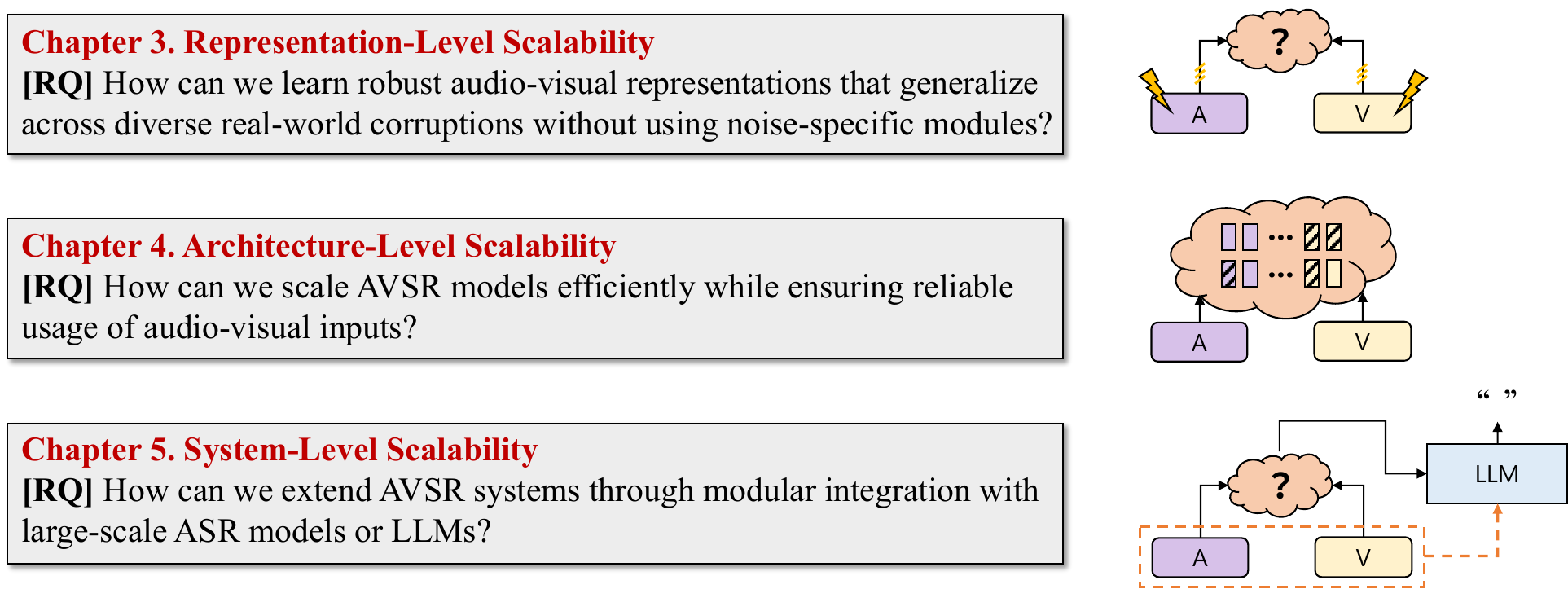}
    \caption[Research questions]{The three levels of scalability in AVSR addressed in this dissertation: representation-level, architecture-level, and system-level scalability. Each part corresponds to a core research question that guides the contributions of this thesis.}
    \label{fig:research_question}
\end{figure}

The overarching goal of this dissertation is to design scalable frameworks for real-world, robust AVSR systems. The central thesis is that achieving true scalability requires a systematic approach that addresses challenges at three distinct and hierarchical levels of the AVSR pipeline: representation level, architecture level, and system level. By developing novel solutions at each of these levels, we can construct AVSR systems that are not only accurate but also adaptable, efficient, and extensible enough for practical and real-world deployment. This research is motivated by three key questions, each corresponding to one level of scalability (see Figure~\ref{fig:research_question}):
\begin{enumerate}

\item \textbf{Representation-Level Scalability:} How can we learn robust audio-visual representations that generalize across diverse, real-world corruptions without resorting to noise-specific modules? The challenge here is to create a foundational model whose learned features are inherently resilient to a wide spectrum of unseen conditions, thus providing a scalable solution that does not require specialized components for every new environment.

\item \textbf{Architecture-Level Scalability:} How can we efficiently scale AVSR model's capacity while ensuring the reliable and adaptive usage of audio-visual inputs? As models grow larger to handle more complex multimodal data, it is crucial to design architectures that can allocate computational resources intelligently based on input characteristics, thereby achieving a scalable balance between performance and efficiency.

\item \textbf{System-Level Scalability:} How can we extend and enhance AVSR systems through the modular integration of large-scale, pre-existing models, such as powerful ASR systems or Large Language Models (LLMs)? This question addresses the need for frameworks that can leverage the rapidly advancing capabilities of foundation models or generative models, allowing AVSR systems to scale in functionality without being rebuilt from scratch.

\end{enumerate}
By systematically answering these three questions, this dissertation constructs a comprehensive roadmap for building the next generation of robust and scalable AVSR technology.

\section{Dissertation Contributions}

This dissertation introduces a series of novel frameworks that directly address the research questions posed in the previous section. The primary contributions are organized according to the three levels of scalability, with each contribution corresponding to a core chapter.
In this dissertation, scalability is considered at three complementary levels: (\textit{i}) representation-level scalability, which focuses on generalizing the representations across diverse corrupted environments; (\textit{ii}) architecture-level scalability, which addresses increasing model capacity without significant computation increase; and (\textit{iii}) system-level scalability, which aims to flexibly integrate external models and modules without retraining the entire AVSR stack.

First, to achieve representation-level scalability, we introduce a self-supervised learning framework, coined as \textbf{CAV2vec} (\underline{C}orrupted \underline{A}udio-\underline{V}isual data \underline{to vec}tors), built on corrupted prediction tasks. Answering our first research question, this method learns to reconstruct clean representations from jointly corrupted audio-visual input representations. By exposing the model to a diverse array of simulated real-world degradations during pretraining, the model develops powerful, disentangled representations that are inherently robust to noise, occlusions, and other diverse distortions. This approach establishes a generalizable and scalable foundation that enhances performance on downstream tasks without relying on specialized noise-specific modules.
Concretely, representation-level scalability here refers to the ability of CAV2vec to generalize across a wide range of audio-visual corruption patterns without requiring extensive domain-specific adaptation or training extra modules, thereby enabling robust AVSR in previously unseen real-world conditions.

Second, to address architecture-level scalability, we propose \textbf{MoHAVE} (\underline{M}ixture \underline{o}f \underline{H}ierarchical \underline{A}udio-\underline{V}isual \underline{E}xperts). This novel architecture directly answers our second research question by providing an effective method for scaling model capacity. Traditional dense models increase in computational cost linearly with size. In contrast, MoHAVE employs a sparse Mixture-of-Experts system, where specialized sub-networks are trained to handle different facets of the data. Our hierarchical gating mechanism dynamically routes inputs only to the most relevant experts or expert groups. This adaptive allocation of computation ensures that model capacity can be efficiently scaled, leading to a system that is more accurate, computationally efficient, and adaptable to varied input characteristics.
In this sense, architecture-level scalability in MoHAVE denotes the ability to grow the effective model capacity by adding more experts and expert groups without a linear increase in compute, since sparse expert routing keeps the per-token computational cost bounded.

Third, to demonstrate system-level scalability, we propose \textbf{DualHyp} (\underline{Dual}-stream \underline{Hyp}otheses), a new paradigm for AVSR that focuses on intelligent generative error correction. In response to our third research question, this framework provides a modular and scalable method for integrating AVSR with powerful LLMs. Instead of relying on a single, potentially flawed hypothesis from a unified ASR or AVSR model, DualHyp generates independent hypotheses from separate, modality-specific audio and visual recognition models. These dual hypotheses are then presented to an LLM, which acts as a compositional reasoner, intelligently integrating the strengths of each modality in the language space to produce a highly accurate transcription. This approach allows the AVSR system to scalably leverage the immense world knowledge and reasoning capabilities of external LLMs, pushing the boundaries of recognition accuracy in challenging conditions.
Here, system-level scalability means that DualHyp can flexibly leverage existing or newly introduced ASR, VSR, and LLM components without retraining the entire AVSR model, in contrast to prior audio-visual LLM approaches that require end-to-end retraining whenever the front-end recognizer or language model is replaced or upgraded.

Together, these three contributions form a cohesive, end-to-end strategy for developing scalable frameworks for real-world audio-visual speech recognition. In turn, we show that our proposed solutions at each level can be effectively combined, leading to state-of-the-art performance on challenging benchmarks.

\begin{figure}[!t]
    \centering
    \includegraphics[width=\textwidth]{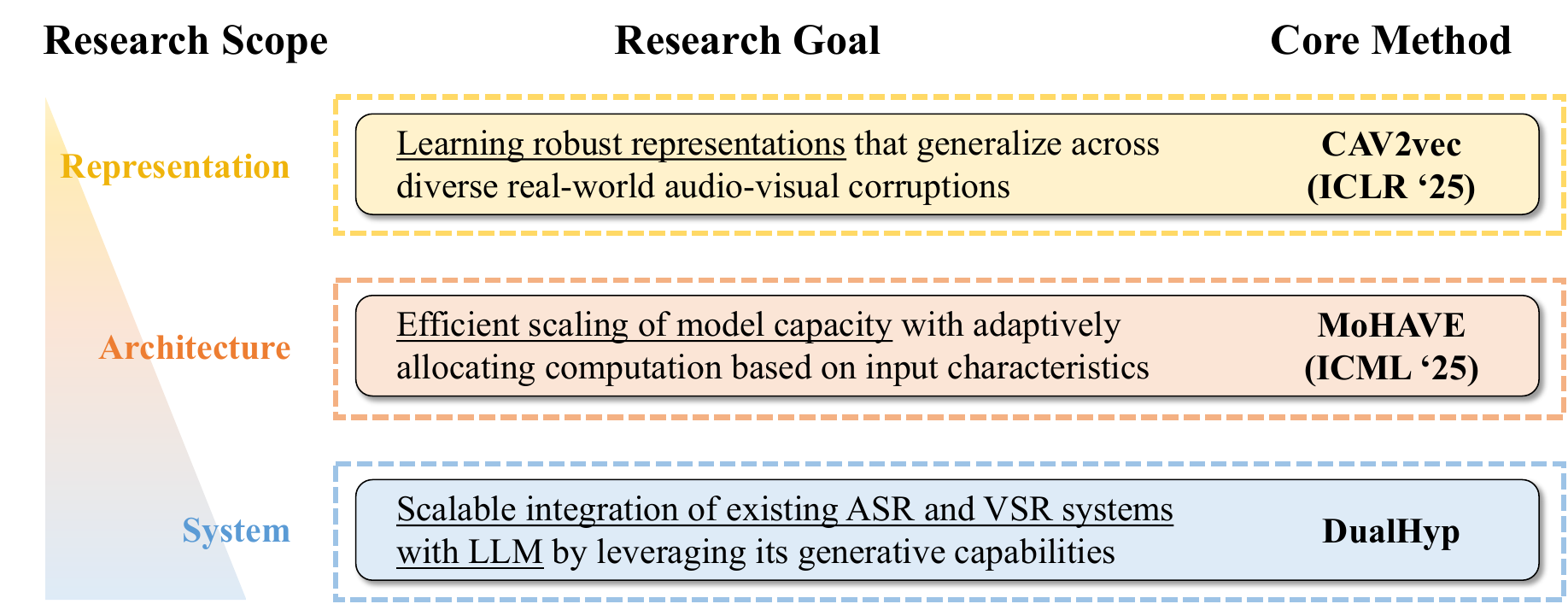}
    \caption[Research goal]{The three main contributions of this dissertation, each addressing a core research question at a different scope of scalability in AVSR.}
    \label{fig:research_contribution}
\end{figure}
\section{Chapter Guide}

The remainder of this dissertation is structured as follows:

\begin{itemize}

\item \textbf{Chapter~\ref{chap:background}} provides background and related work for this dissertation, offering a review of the foundational concepts and prior studies in audio-only, visual-only, and audio-visual speech recognition. It also covers relevant background on representation learning, model architectures, and the recent integration of LLMs for diverse speech processing tasks.

\item \textbf{Chapter~\ref{chap:representation_scalability}} presents CAV2vec, a multi-task corrupted prediction approach for learning robust audio-visual speech representation. This chapter details our proposed self-supervised framework for learning robust multimodal representations from corrupted data, corresponding to our first contribution on the representation-level scalability.

\item \textbf{Chapter~\ref{chap:architecture_scalability}} introduces MoHAVE, a mixture of hierarchical audio-visual experts model designed to enhance the scalability and adaptability of AVSR systems. This chapter discusses the architectural innovations and their impact on performance and efficiency across diverse audio-visual tasks.

\item \textbf{Chapter~\ref{chap:system_scalability}} proposes DualHyp, a novel framework that explores the system-level challenges and suggests solutions for deploying a powerful AVSR system. This chapter explains the methodology of using dual modality-specific hypotheses and reliability-informed prompting to achieve state-of-the-art results, fulfilling our third contribution on system-level scalability.

\item \textbf{Chapter~\ref{chap:conclusion}} summarizes the key findings and contributions of this dissertation. We present that our proposed solutions at each chapter can be effectively combined, and that each strategy leads to improved performance on challenging benchmarks. It concludes with a discussion of the broader implications of this research and outlines promising directions for future investigation in the field of audio-visual speech processing.

\end{itemize}

\chapter{Background and Related Work}
\label{chap:background}

This chapter provides a comprehensive review of the literature that forms the foundation for this dissertation. We start by outlining the evolution and core components of modern speech processing and recognition in Section~\ref{sec:background:speech_processing}.
In Sections~\ref{sec:background:speech_representation}, \ref{sec:background:model_architectures}, and \ref{sec:modular_integration}, we delve into the specific research areas that are directly related to the main contributions of this dissertation, including self-supervised learning of speech representations for robustness, various model architectures for speech processing, and modular integration of speech information with LLM-based foundation models.

\section{Speech Processing and Recognition}
\label{sec:background:speech_processing}

The automatic transcription of human speech by machines has been a central goal of AI research for over half a century. Early systems relied on complex, multi-stage pipelines that involve separate acoustic models, pronunciation lexicons, and language models, often based on Gaussian Mixture Model-Hidden Markov Model (GMM-HMM)~\citep{jelinek1998statistical, jelinek2005continuous}. While foundational, these systems were brittle and have been largely superseded by end-to-end deep learning approaches that directly map speech signals to text. This shift has enabled more robust and scalable solutions, leveraging large-scale human speech data for training.

\subsection{Automatic Speech Recognition}

The advent of deep learning has revolutionized the field of ASR, leading to the dominance of end-to-end models that directly map a sequence of acoustic features to a sequence of text tokens~\citep{graves2014towards, hannun2014deep, chorowski2014end, miao2015eesen, chan2016listen, bahdanau2016end, collobert2016wav2letter, sak2017recurrent}. These models can be broadly categorized by their sequence-to-sequence transduction mechanisms.

First, Connectionist Temporal Classification (CTC)~\citep{graves2006connectionist} based models introduce a special ``blank'' token to resolve alignment issues between variable-length audio inputs and variable-length text outputs. 
This allows the model to be trained directly on input-output pairs without requiring explicit, frame-by-frame alignment data.
To model temporal information, the CTC loss function is often coupled with an RNN model, which performs well in end-to-end speech recognition~\citep{graves2014towards, hannun2014deep}.
Deep Speech 2~\citep{amodei2016deep} is a prominent example of the adoption of this CTC-RNN mechanism.

RNN-Transducer (RNN-T)~\citep{graves2012sequence} model combines the strengths of CTC and RNN-based language models~\citep{cho2014learning}. It processes the audio stream and predicts output tokens conditioned on both the acoustic input and the previously generated text tokens, allowing for powerful online, streaming recognition capabilities. This has become a dominant architecture for on-device and low-latency ASR~\citep{li2019improving, zhang2020transformer}.

Additionally, attention-based encoder-decoder models~\citep{chorowski2015attention} have been proposed. These models, often called listener-encoder-decoder architectures, first encode the entire input audio sequence into a set of high-level representations. A decoder, equipped with an attention mechanism, then selectively focuses on relevant parts of the encoded audio to generate the output text one token at a time. LAS~\citep{chan2016listen} was a pioneering example of this approach.
Also, attention-based encoder-decoder models often share their encoders with a CTC model to facilitate multi-task learning strategy~\citep{kim2017joint, ueno2018acoustic, kim2019attention}.

More recently, the Transformers architecture~\citep{vaswani2017attention}, with its self-attention mechanism, has become the state-of-the-art, demonstrating superior performance in capturing long-range dependencies in speech. The power of large-scale, weakly-supervised training has been showcased by models like OpenAI's Whisper~\citep{radford2023robust} and Meta's SeamlessM4T~\citep{barrault2023seamlessm4t, barrault2023seamless}, which can handle hundreds of languages and multiple tasks~\citep{pratap2024scaling}.

\subsection{Open Challenges in ASR}
Despite significant progress in controlled, academic settings, the performance of ASR systems often degrades substantially when deployed in real-world, \textit{in the wild} scenarios. This gap is primarily due to the vast acoustic and linguistic variability not captured in clean and constrained training corpora.

\begin{itemize}
    \item \textbf{Acoustic Environment}: One of the most significant factors is environmental noise. This includes stationary noise like fans and non-stationary, unpredictable sounds like background chatter, music, or traffic. Another critical challenge is reverberation, where sound reflections from surfaces in a room cause temporal smearing of the speech signal and degrade the intelligibility~\citep{ko2017study}. State-of-the-art ASR systems like Whisper still struggle under these corrupted conditions.
    \item \textbf{Speaker Variability and Multilinguistics}: Human speech is inherently variable. ASR systems struggle with accents and dialects for which they have insufficient training data. Studies have shown significant performance disparities across different demographic groups, highlighting issues of fairness and bias~\citep{koenecke2020racial}. Multilingual ASR is also still far from its best use, since the available training data is often focused on English (or Latin-originated languages), lacking capabilities on low-resource or under-represented languages. Moreover, multilingual systems must handle code-switching~\citep{li2019towards, yue2019end}, where speakers alternate between languages within a single utterance, posing additional challenges for language modeling and acoustic variability.
    \item \textbf{Streaming and Interactive System}: Real-time speech recognition requires low-latency processing and interactive capabilities, which is challenging for traditional ASR models that often rely on a complete single-turn input sequence. Streaming ASR systems must process audio in chunks, making predictions based on partial information. This necessitates the development of new architectures and algorithms capable of maintaining accuracy while operating under these constraints. As the model size grows with Transformers, low-latency processing becomes increasingly difficult, requiring innovative solutions to balance performance and responsiveness~\citep{fang2025llama, jia2025efficient}.
\end{itemize}

\subsection{Multimodal Understanding of Speech}

To address the limitations of audio-only ASR, particularly in noisy environments, research has increasingly turned to multimodal approaches that incorporate complementary sources of information. Visual Speech Recognition (VSR), also known as lip-reading, is the task of recognizing speech solely from visual cues of a speaker's mouth movements~\citep{zhou2014review}. As the visual modality is immune to acoustic noise, VSR offers a powerful signal for robust recognition.
Early works utilized CNNs to predict phonemes or visemes from video frames~\citep{noda2014lipreading, koller2015deep}, and later LSTMs were employed to recognize short words or phrases~\citep{petridis2016deep, wand2016lipreading}.
A landmark model, LipNet~\citep{assael2016lipnet}, first demonstrated the feasibility of end-to-end sentence-level lip-reading. However, VSR is fundamentally challenging due to the ambiguity of visemes, where multiple distinct phonemes correspond to the same visual lip movement.

Audio-Visual Speech Recognition (AVSR) seeks to combine audio and visual information to achieve more robust performance than either modality alone~\citep{galatas2012audio, noda2015audio, tamura2015audio, afouras2018deep, shi2022learning, xu2020discriminative, chen2023leveraging}. This is inspired by human speech perception, as demonstrated by the McGurk effect~\citep{mcgurk1976hearing}, where conflicting audio and visual signals lead to the perception of a third, different sound. The central challenge in AVSR is the fusion of the two modalities. Fusion strategies are typically categorized as early fusion (feature-level), late fusion (decision-level), or hybrid, which involves complex interactions like cross-modal attention at various network layers~\citep{ma2021end}. Recent work continues to explore more sophisticated fusion mechanisms to better handle challenging multimodal data.

\subsection{Datasets and Benchmarks}

The advancement of speech recognition has been propelled by the availability of large-scale public datasets.
For ASR, Switchboard-1 and WSJ corpora~\citep{godfrey1992switchboard, paul1992design} mark the early foundations of human speech datasets. 
LibriSpeech~\citep{panayotov2015librispeech} provides hundreds of hours of clean, read English speech, forming a standard for academic research, and VCTK corpus~\citep{yamagishi2012english} offers multi-speaker data with various accents, useful for speaker adaptation and accent-robustness studies. While LibriSpeech and VCTK remain foundational benchmarks for clean speech, the research community has developed more challenging datasets to better reflect real-world conditions. For instance, CHiME challenge series~\citep{barker2015third} provides speech data recorded in noisy home environments with distant microphones, serving as a key benchmark for noise robustness. 

Beyond English, numerous large-scale corpora have been developed for other languages. The AISHELL corpora~\citep{bu2017aishell, du2018aishell} are prominent open-source resources containing hundreds of hours of Mandarin speech, driving research and development for one of the world's most spoken languages.
Mozilla Common Voice~\citep{ardila2020common} has emerged as an enormous, multilingual, crowdsourced speech corpus, and VoxPopuli~\citep{wang2021voxpopuli} suggests a large-scale corpus sourced from the European Parliament, supporting semi-supervised tasks and non-native English speakers.

In contrast to ASR, there have been much fewer works for AVSR benchmarks. Lip Reading in the Wild (LRW)~\citep{chung2016lip, chung2018learning}, BBC-Oxford Lip Reading Sentences 2 (LRS2)~\citep{son2017lip}, and LRS3-TED~\citep{afouras2018lrs3} datasets, sourced from BBC and TED/TEDx talks, have been the de facto standards for training and evaluating audio-visual models in unconstrained English settings. To push the boundaries towards global applicability, the MuAViC corpus~\citep{anwar2023muavic} was recently introduced as a large-scale, multilingual audio-visual benchmark. It contains transcribed video data from 9 diverse languages, providing a critical resource for developing and evaluating AVSR systems that can generalize across different languages and visual contexts.
Audio-visual datasets are also critical for emotion and dialogue understanding. The IEMOCAP~\citep{busso2008iemocap} and MELD~\citep{poria2019meld} datasets provide multimodal data for emotion recognition, while CREMA-D~\citep{cao2014crema} focuses on the acted emotional speech and song. MultiDialog~\citep{park2024let} is a synthesized multimodal dialogue dataset, containing large-scale conversations and turns within audio, video, and emotion labels.

Across the field, the primary evaluation metric is the Word Error Rate (WER), which measures the number of substitutions, deletions, and insertions required to align the ASR system's predicted transcript with the ground-truth reference.
While WER is the most common metric for English and other space-delimited languages, the Character Error Rate (CER) is often preferred for languages without explicit word boundaries, such as Mandarin or Japanese, as it provides a more consistent measure of performance. Beyond lexical accuracy, the practical utility of an ASR system is often measured by task-specific or performance-based metrics. For streaming applications, such as live captioning or voice assistants, Real-Time Factor (RTF), which is the ratio of processing time to audio duration, is a critical measure of system latency. In goal-oriented dialogue systems, metrics like Intent Accuracy and Slot Filling F1-score are more important than WER~\citep{zhang2016joint}, as they evaluate whether the system correctly understood the user's command, regardless of minor transcription errors.

\section{Speech Representation Learning}
\label{sec:background:speech_representation}

The performance of any speech recognition model is fundamentally dependent on the quality of its input features or representations. While traditional ASR systems relied on hand-crafted features like Mel-Frequency Cepstral Coefficients (MFCCs), modern end-to-end systems have demonstrated the power of learning representations directly from data. This section reviews the paradigm shift towards self-supervised learning (SSL) for creating powerful, generalizable, and robust speech representations from vast quantities of unlabeled data, which directly informs the first major contribution of this dissertation.

\subsection{Self-Supervised Learning from Audio}

The core idea behind SSL is to leverage large unlabeled data by creating a \textit{pretext} task that does not require manual annotations~\citep{chen2020simple}. The model learns to solve this task, and in doing so, produces intermediate representations that are useful for various downstream tasks, such as ASR, after fine-tuning on a much smaller labeled dataset.

\paragraph{Predictive Coding and wav2vec.} An early and influential approach was based on contrastive predictive coding (CPC)~\citep{oord2018representation}. The first wav2vec model operationalized this idea for speech by training a model to distinguish a true future audio segment from a set of negative samples, given a context of past audio (positive samples). The model consists of an encoder network that maps raw audio to feature representations and a context network that summarizes these representations to make predictions about the future~\citep{schneider2019wav2vec}. This demonstrated that rich phonetic and linguistic information could be learned from raw audio without any labels.

\paragraph{Masking, Quantization, and wav2vec 2.0.} The seminal wav2vec 2.0 model~\citep{baevski2020wav2vec} represented a major breakthrough by adapting the masked language modeling paradigm from NLP~\citep{devlin2019bert} to speech. It works by first converting the raw audio waveform into a sequence of latent representations using a convolutional feature encoder. A random portion of these representations is then masked. The model, which uses a Transformer-based context network, is trained on a contrastive task to identify the true quantized representation of the masked timesteps from a set of distractors. By learning to solve this task over massive amounts of unlabeled speech, wav2vec 2.0 produces representations that are highly effective for ASR, achieving state-of-the-art results with as little as ten minutes of labeled fine-tuning data.

\paragraph{Further Advancements.} Building on this success, subsequent models like HuBERT~\citep{hsu2021hubert} or wavLM~\citep{chen2022wavlm} refined the pretraining objective by adopting a ``teacher-student'' approach. It first discovers discrete acoustic units by clustering features from an existing model and then trains a new model to predict the cluster assignments for masked segments. These audio-only SSL models have become the foundation of modern high-performance ASR systems.
Further research has focused on creating more efficient variants of these models~\citep{chang2022distilhubert, wang2022lighthubert, jang2023recycle, jang2024star, lee2022fithubert, peng2023dphubert}, which reduce model size and inference time while maintaining competitive performance, making them suitable for deployment on edge devices.

\begin{figure*}[!t]
    \centering
    \includegraphics[width=.55\linewidth]{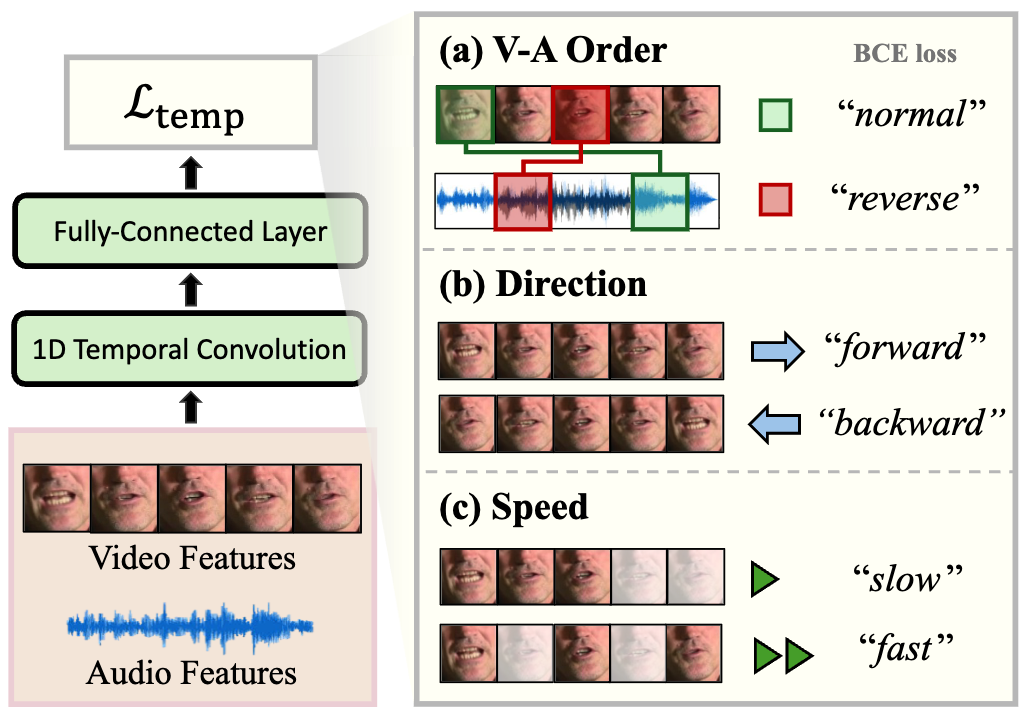}
    \hfill
    \includegraphics[width=.43\linewidth]{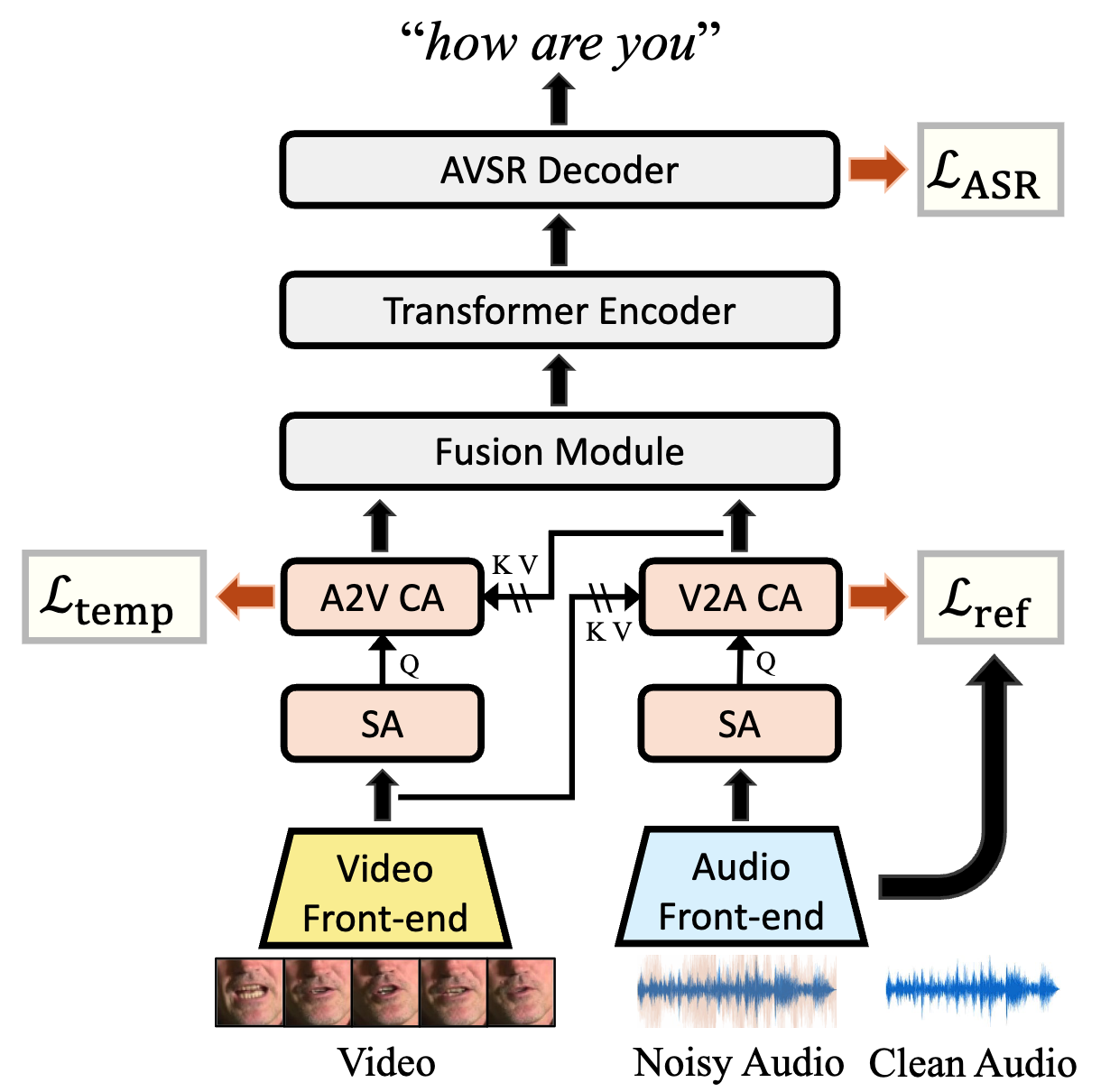}
    \caption[Overview of CMA]{(Left) Temporal dynamics guidance involves predicting (a) the context order considering both video and audio modalities, (b) playback direction, and (c) whether certain frames are skipped or not. Each video temporal predictor consists of 1D convolution and fully-connected layers. (Right) Cross-modal attention (CMA) structure is inserted between the feature extractors and the AVSR encoder. This structure leverages clean video to refine audio, and then learns video temporal dynamics given the refined audio features. Note that the gradient is not backpropagated between the two modalities. Figures adapted from \citet{kim2024learning}.}
    \label{fig:cma}
\end{figure*}

\subsection{Robust and Multimodal Speech Representations}

While audio-only SSL provides a powerful foundation, robustness in real-world settings necessitates representations that can effectively handle noise and leverage complementary modalities. This has motivated the extension of SSL principles to the audio-visual domain.

The central idea is to adapt the masking strategy to multimodal inputs. AV-HuBERT~\citep{shi2022learning}, for example, extends the HuBERT objective by randomly masking either the audio or visual stream (or both) and training the model to predict the discrete audio-visual units from the unmasked parts of the input. This forces the model to learn cross-modal dependencies; for instance, it must learn to reconstruct audio information from visual cues when the audio stream is masked, and vice-versa. Other works, such as MAViL~\citep{huang2023mavil} and CAV-MAE~\citep{gong2023contrastive}, have explored learning by predicting raw audio and visual features from masked inputs, further improving the generalizability of the learned representations.

However, a key limitation of these approaches is that they often treat the video modality as a secondary source of information to be fused with the audio, rather than an equally important signal to be strengthened by its own. Recognizing that the video modality's power lies in its temporal dynamics, \citet{kim2024learning} have designed pretext tasks specifically to make the video encoder a better speech representer (Figure~\ref{fig:cma}). They propose training the video encoder on a series of self-supervised tasks: predicting the correct context order, playback direction, and speed of video frames. Crucially, this training is not done in isolation; cross-modal attention is used to enrich the video features with audio information during this pre-training, and vice versa. This encourages the model to learn temporal dynamics that are semantically linked to speech, rather than just generic motion, resulting in video representations that are far more robust and informative.

These approaches demonstrate a clear trajectory towards more sophisticated pretraining objectives for AVSR. Yet, a key open challenge, directly addressed by the first contribution of this thesis, is to design a pretraining objective that specifically encourages robustness to the unstructured, diverse, and modality-specific corruptions encountered in real-world scenarios, going beyond masking or learning general temporal dynamics from clean signals.

\section{Model Architectures for Speech Processing}
\label{sec:background:model_architectures}

Beyond the quality of representations, the architectural design of a model is paramount to its performance, efficiency, and scalability. This section traces the evolution of architectures for speech processing, from early recurrent models to the current state-of-the-art, and introduces scalable designs that directly motivate the second contribution of this dissertation.

\subsection{From Recurrence to Self-Attention}
Early end-to-end models heavily relied on RNNs, particularly variants like Long Short-Term Memory (LSTM)~\citep{hochreiter1997long} and Gated Recurrent Units (GRUs)~\citep{cho2014properties}, to model the temporal dependencies in speech. These models process sequences step-by-step, maintaining a hidden state that captures information from past timesteps. While effective, their sequential nature makes them difficult to parallelize and can lead to challenges in capturing very long-range dependencies.

The introduction of the Transformers architecture marked a paradigm shift~\citep{vaswani2017attention}. Its core component, the self-attention mechanism, allows the model to weigh the importance of all other tokens in the sequence when processing a given token, regardless of their distance. This global receptive field and high parallelizability made Transformers exceptionally powerful for modeling long sequences, and they quickly became the dominant architecture for ASR encoders or decoders, as seen in \citep{dong2018speech, zeyer2019comparison, karita2019comparative, radford2023robust}.

The impact of such models has spurred significant research not only in their application but also in their reproducibility. For instance, the Open Whisper-style Speech Model (OWSM) project~\citep{peng2023reproducing} is a notable effort to reproduce Whisper-style training using exclusively open-source toolkits~\citep{watanabe2018espnet} and publicly available data, aiming to democratize research and address issues of efficiency, robustness, and bias that are difficult to study with closed models.

\subsection{Hybrid Architectures: Conformer and Beyond}

While Transformers excelled at capturing global context, it was less adept at modeling fine-grained local patterns, for which CNNs are well-suited. This observation led to the development of hybrid architectures that seek to get the best of both worlds.

The Conformer architecture~\citep{gulati2020conformer} explicitly combines self-attention and convolution to effectively model both local and global dependencies in speech signals. A Conformer block serially stacks a multi-head self-attention module and a convolution module within its feed-forward layers. This hybrid design proved highly effective, outperforming both Transformers-only or CNN-only models and setting a new standard for ASR~\citep{guo2021recent}.

The success of the Conformer inspired further research into how to best combine these two complementary operations. The Branchformer architecture proposed a parallel design instead of a serial one. A Branchformer block consists of two parallel branches: one with a multi-head self-attention mechanism and the other with a convolutional spatial gating unit (CSGU) mechanism. The outputs of these two branches are then merged. This parallel structure offers a different inductive bias and has demonstrated strong performance, suggesting that serial stacking is not the only effective combination strategy~\citep{peng2022branchformer}. Concurrent works like Squeezeformer~\citep{kim2022squeezeformer} also explored more efficient hybrid designs by carefully arranging convolutional blocks and reducing the model size at later layers, achieving an improved balance of performance and computational cost. These developments show that the optimal fusion of local and global modeling remains an active and important area of architectural research.

The evolution of these parallel designs culminated in E-Branchformer~\citep{kim2023branchformer}, which further optimized the parallel structure for improved performance and efficiency. The significance of this advanced hybrid architecture was powerfully validated when it was adopted as the backbone for the large-scale OWSM v3.1, v4, and OWSM-CTC projects~\citep{peng2024owsmctc, peng2024owsm, peng2025owsm}. By replacing the original Transformers encoder with E-Branchformer, these models demonstrated that state-of-the-art hybrid designs are not only effective in academic benchmarks but are also scalable and robust to serve as the foundation for massive industry-scale training on over 166K hours of public data in 75 languages.

\subsection{Architectural Scalability: Mixture-of-Experts (MoE)}
As models continue to grow in size to absorb massive datasets, the computational cost of training and inference for these dense, monolithic architectures becomes a major bottleneck. The Mixture-of-Experts (MoE) paradigm offers a compelling solution for scaling model capacity while maintaining a constant computational budget per input.

An MoE layer replaces a standard dense feed-forward network with a set of parallel expert sub-networks and a trainable gating network that learns to sparsely route each input token to a small subset of the experts (typically one or two)~\citep{shazeer2017outrageously}. This allows for a dramatic increase in the total number of parameters in the model, but since only a fraction of these parameters are activated for any given input, the computational cost (FLOPs) remains manageable. This principle has been used to scale language models to over a trillion parameters, as demonstrated by Switch Transformers~\citep{fedus2022switch}, or sharding a larger model into millions of smaller experts~\citep{he2024mixture}.

The success of MoE has inspired its application in the speech domain. SpeechMoE~\citep{you2021speechmoe, you2022speechmoe2}, for example, demonstrated how sparsely-gated MoE layers could be used to build high-capacity, efficient speech recognition models. A variety of works have utilized MoEs for multilingual ASR, showing that the implicit language routers can effectively specialize each task~\citep{kwon2023mole, gaur2021mixture, kumatani2021building, hu2023mixture}. However, designing effective MoE models for the multimodal AVSR domain—where the gating mechanism must learn to route tokens based on complex and potentially conflicting audio-visual signals—remains a significant research challenge. This gap directly motivates the second contribution of this thesis, which explores a hierarchical MoE framework for robust and scalable audio-visual speech recognition.
\section{Modular Integration with Foundation Models}
\label{sec:modular_integration}

The latest paradigm shift in speech processing has moved away from designing isolated, task-specific models and towards the modular integration of specialized speech encoders with large-scale pretrained foundation models, particularly Large Language Models (LLMs)~\citep{devlin2019bert, brown2020language, raffel2020exploring,achiam2023gpt, touvron2023llama, ouyang2022training}. This approach aims to leverage the generative capabilities, world knowledge, and powerful reasoning of LLMs to move beyond simple transcription towards genuine speech context understanding and robust error correction. This trend can be broadly categorized into two main strategies: cascaded systems for post-processing and end-to-end integration with Multimodal Large Language Models (MLLMs)~\citep{team2023gemini, liu2023visual}.

\subsection{Cascaded Systems with LLMs}

The most straightforward and modular approach to combining speech models with LLMs is through a cascaded system~\citep{min2023exploring}. In this two-stage pipeline, a dedicated ASR model first transcribes an input audio signal into a text hypothesis. This text is then fed as input to a standard LLM, which performs a desired downstream task. The primary advantage of this approach is its simplicity and flexibility; any state-of-the-art ASR system can be readily paired with any off-the-shelf LLM without requiring architectural changes or joint training. However, this design is also vulnerable to error propagation, where transcription errors from the ASR model can mislead the LLM in the second stage. Research in this area largely focuses on two goals: refining the intermediate text representation or developing strategies to make the LLM robust to upstream errors.

One major application of the cascaded pipeline is Generative Error Correction (GER), where the LLM's explicit task is to refine the ASR output. The seminal work Hyporadise~\citep{chen2023hyporadise, dighe2024leveraging} established an open baseline for this task, demonstrating that prompting an LLM with ASR hypotheses could significantly enhance the system by correcting grammatical, semantic, and recognition errors. Subsequent works like ClozeGER~\citep{hu2024clozeger} reformulated the task as a cloze problem (\ie fill in the blank with multiple choices) to better guide the correction process. This principle was also extended to the audio-visual domain with models like LipGER~\citep{ghosh2024lipger} and AV-GER~\citep{liu2025avger}, which provide visual features to the LLM to help resolve acoustic ambiguities.

\subsection{End-to-End Integration with Multimodal LLMs}

A more recent and tightly-coupled approach involves using MLLMs that are capable of directly processing raw or encoded speech signals as part of their input sequence. These models aim to create a single, unified system for end-to-end speech understanding and generation~\citep{tang2024extending, yu2024connecting}. SpeechGPT~\citep{zhang2023speechgpt} was an early example, demonstrating how an LLM could be adapted to handle speech inputs for tasks like speech-to-text and spoken dialogue.
WavLLM~\citep{hu2024wavllm}, Qwen-Audio~\citep{chu2023qwen, chu2024qwen2}, and LTU~\citep{gong2024listen} are other notable examples of MLLMs that integrate speech processing capabilities.

More advanced MLLMs like SALMONN~\citep{tang2024salmonn} and video-SALMONN~\citep{sun2024video, sun2025videosalmonno, tang2025video} utilize connectionist modules, such as Q-Former~\citep{li2023blip}, to convert continuous speech representations into a sequence of discrete tokens that are intelligible to the LLM's text embedding space. This allows the LLM to perform zero-shot or few-shot reasoning on spoken instructions. Extending this to the visual domain, recent \textit{omni}-modal models including Gemini~\citep{team2023gemini}, GPT-4o~\citep{hurst2024gpt}, VITA~\citep{fu2024vita}, or Qwen2.5-Omni~\citep{xu2025qwen2} and Qwen3-Omni~\citep{xu2025qwen3} are designed to natively handle interleaved audio, visual, and text inputs, representing the frontier of this research direction. These models hold the potential to perform complex, context-aware audio-visual reasoning tasks far beyond the scope of traditional AVSR.

\section{Chapter Summary}

This chapter has provided a detailed survey of the background and related work essential for contextualizing the contributions of this dissertation. The review began by tracing the trajectory of ASR, from early systems to the current paradigm of end-to-end deep learning models, including CTC, RNN-T, and attention-based encoder-decoders. It established the significant performance gap between controlled and real-world conditions, detailing persistent open challenges such as environmental noise, speaker variability, and interactive system. To address these limitations, the review introduced multimodal approaches, particularly AVSR, as a promising direction for enhancing robustness. We also discussed the widely used datasets and benchmarks for both ASR and AVSR.

Building on this foundation, the chapter then delved into the three core technical areas that align with the research pillars of this thesis. First, we examined speech representation learning, highlighting the transformative impact of self-supervised or unsupervised learning. Foundational models like wav2vec, wav2vec 2.0, and HuBERT were discussed, showing how powerful representations can be learned from unlabeled audio data. This concept was extended to the multimodal domain with models like AV-HuBERT and MAViL, identifying a research gap in designing SSL pretext tasks specifically for robustness against diverse, real-world corruptions.

Second, the review charted the evolution of model architectures, from recurrent networks to the highly effective Transformers and the specialized Conformer or Branchformer, which effectively combines convolution and self-attention for speech. To address the challenge of ever-growing model sizes, the MoE paradigm was introduced as a method for achieving scalable model capacity with efficient computation. The discussion identified the application of MoE to the complex multimodal dynamics of AVSR as a key research frontier.

Third, we explored the emerging trend of modular integration with foundation models, particularly LLMs. Two main strategies were outlined: cascaded systems that use LLMs for post-processing ASR outputs, and end-to-end integrations with MLLMs capable of directly processing audio-visual inputs. The review highlighted the potential of these approaches to leverage the reasoning and world knowledge of LLMs for tasks like generative error correction and spoken language understanding, while also noting challenges such as error propagation in cascaded systems.

Collectively, this review of the literature establishes the technological foundations and identifies the critical research gaps in robust representation learning and scalable architectures that this dissertation aims to address. The subsequent chapters will present novel frameworks that build upon these established concepts to advance the state-of-the-art in real-world AVSR.
\chapter{Representation-Level Scalability of AVSR}
\label{chap:representation_scalability}

\begin{center}
    \vspace{0.5em}
    \begin{tcolorbox}[
        colback=gray!10, 
        colframe=gray!50, 
        boxrule=0.5pt,
        title=Summary: Chapter based on work published at ICLR 2025~\citep{kim2025multitask}, fonttitle=\bfseries,
        width=0.92\linewidth,
        coltitle=black
    ]
    Audio-visual speech recognition (AVSR) incorporates auditory and visual modalities to improve recognition accuracy, particularly in noisy environments where audio-only speech systems are insufficient. While previous research has largely addressed audio disruptions, few studies have dealt with visual corruptions, \eg lip occlusions or blurred videos, which are also detrimental. To address this real-world challenge, we propose \textbf{CAV2vec}, a novel self-supervised speech representation learning framework particularly designed to handle audio-visual joint corruption. CAV2vec employs a self-distillation approach with a corrupted prediction task, where the student model learns to predict clean targets, generated by the teacher model, with corrupted input frames. Specifically, we suggest a \textit{unimodal multi-task learning}, which distills cross-modal knowledge and aligns the corrupted modalities, by predicting clean audio targets with corrupted videos, and clean video targets with corrupted audios. This strategy mitigates the dispersion in the representation space caused by corrupted modalities, leading to more reliable and robust audio-visual fusion. Our experiments on robust AVSR benchmarks demonstrate that the corrupted representation learning method significantly enhances recognition accuracy across generalized environments involving various types of corruption. The code for this chapter is available at \url{https://github.com/sungnyun/cav2vec}.
    \end{tcolorbox}
    \vspace{0.5em}
\end{center}

\section{Multi-Task Corrupted Prediction for Learning Robust Audio-Visual Speech Representation}
\label{sec:ch3_intro}

\begin{figure}[!t]
    \centering
    \begin{minipage}[b]{0.66\textwidth}
        \begin{subfigure}[b]{\textwidth}
            \centering
            \small
            \includegraphics[width=\linewidth]{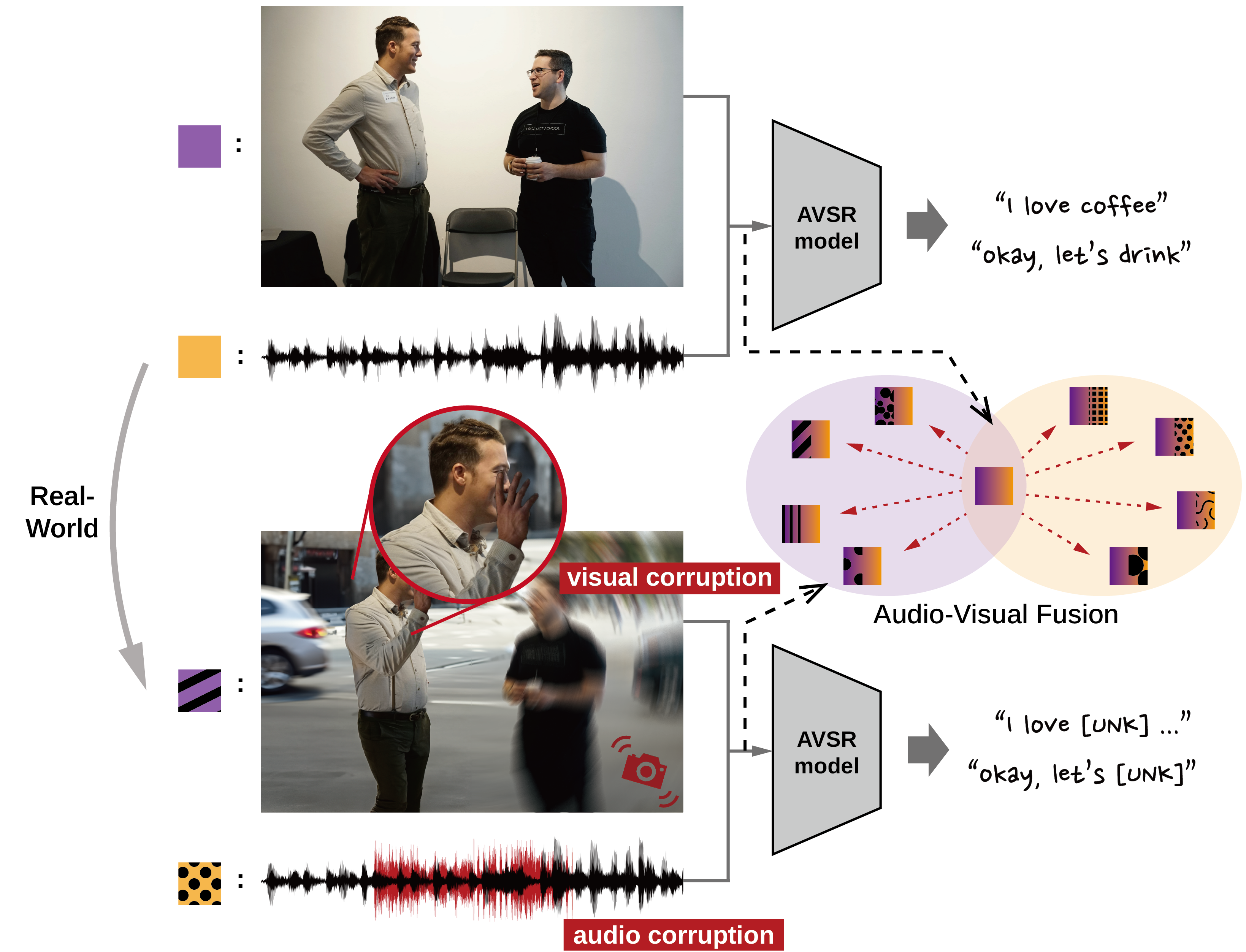}
            \vspace*{-10pt}
            \caption{}
            \label{fig:corrupted_recognition-1}
        \end{subfigure}
    \end{minipage}
    \hfill
    \begin{minipage}[b]{0.32\textwidth}
        \begin{subfigure}[t]{\textwidth}
            \centering
            \includegraphics[width=\textwidth]{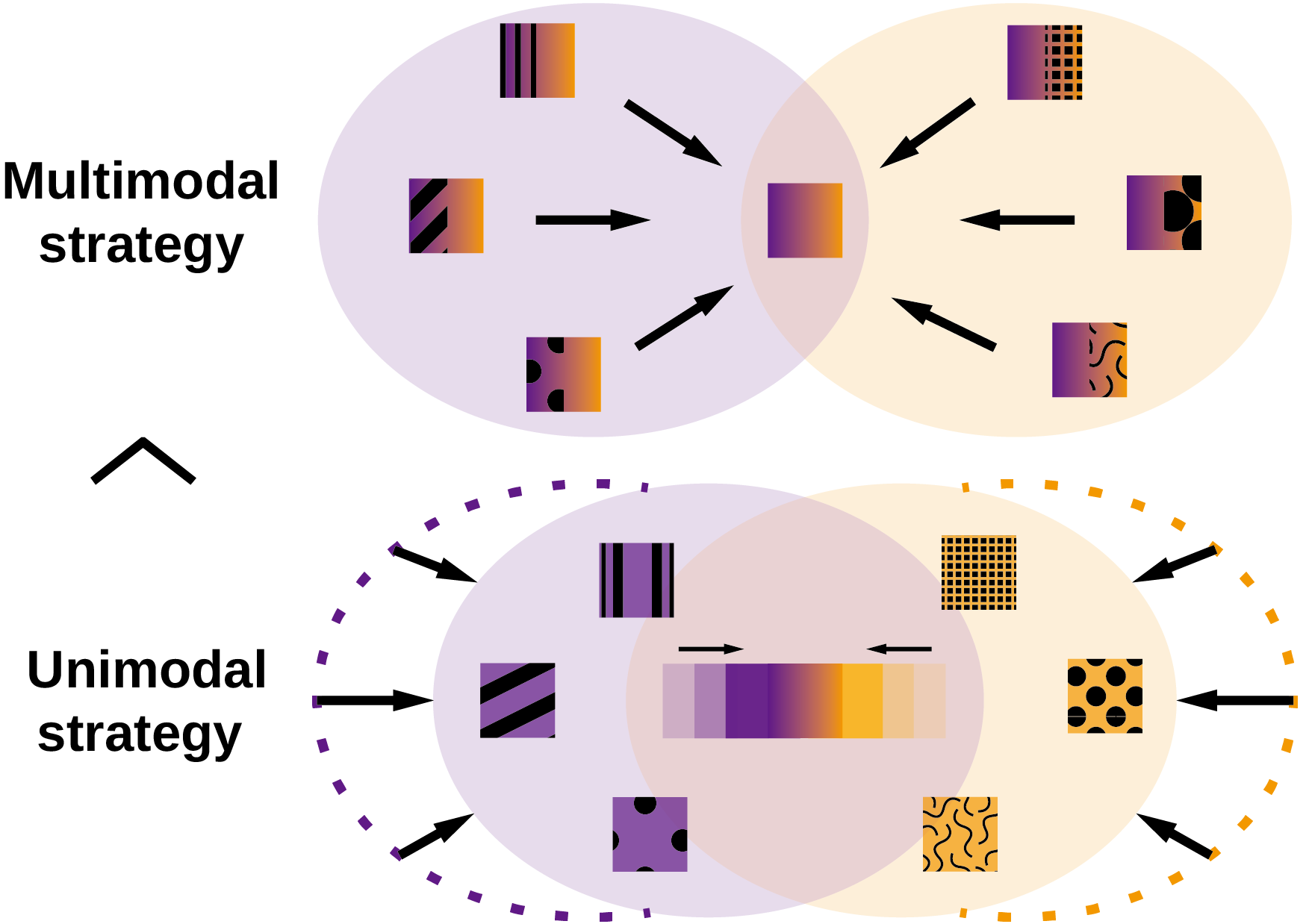}
            \caption{}
            \label{fig:corrupted_recognition-2}
        \end{subfigure}
       
        \begin{subfigure}[b]{\textwidth}
            \centering
            \vspace*{10pt}
            \includegraphics[width=.9\textwidth]{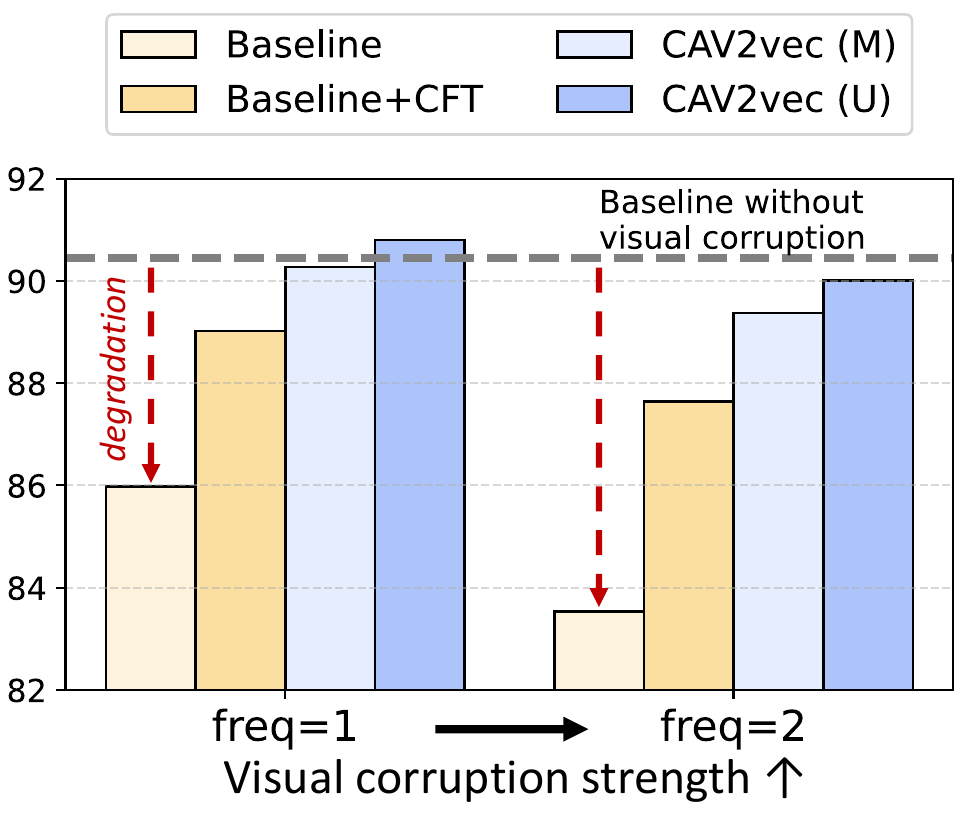}
            \vspace*{-7pt}
            \caption{}
            \label{fig:corrupted_recognition-3}
        \end{subfigure}
    \end{minipage}
    \caption[Overview of corrupted representation learning in CAV2vec for robust AVSR]{(a) Real-world speech recognition challenges. AVSR models suffer from maintaining robust representations under the corrupted environments and fail to recognize utterances. 
    (b) Our corrupted representation learning strategies with multimodal and unimodal corrupted prediction tasks. 
    (c) Speech recognition accuracy (100\,$-$\,WER\,\%), where frequency denotes the number of visual corruption events in a sequence. Our representation learning framework, \ourmodelmtl with a unimodal strategy\,(U), significantly improves robustness compared to the baseline model and even outperforms the multimodal strategy\,(M).
    }
    \label{fig:corrupted_recognition}
    \vspace*{10pt}
\end{figure}

Audio-visual speech recognition (AVSR)~\citep{noda2015audio, afouras2018deep, ma2021end, shi2022learning, hsu2022u, hu2023mir} represents a significant advancement in speech recognition by integrating both auditory and visual modalities to enhance performance. This multimodal integration proves particularly vital in contexts where audio-only speech recognition systems suffer from ambient or background noise, as visual speech information like lip movements significantly improves recognition capabilities~\citep{makino2019recurrent, ren2021learning, chen2023leveraging}. In this sense, previous works on AVSR have primarily focused on overcoming audio disruptions, \eg \citet{xu2020discriminative} training an audio enhancement sub-network, or \citet{shi2022robust} pretraining with noise-augmented audio. Nonetheless, real-world applications often encounter scenarios where video corruption is as critical as audio disturbances. For example, we may consider an outdoor interview where not only is the audio disturbed by ambient noise or traffic sound but visual cues are also intermittently occluded, either when the speaker's hands obstruct the view of their face or a camera is out of focus (see Figure\,\ref{fig:corrupted_recognition-1}). AVSR models often fail to accurately recognize the utterance under these corrupted environments.

Despite the effectiveness of current methods in addressing audio corruptions, there is a lack of solutions for visual corruption, highlighting the necessity for a more robust approach to tackle audio-visual joint corruption in AVSR systems. Recent efforts have addressed the visual corruption using techniques such as scoring modules to assess the reliability of audio and video frames~\citep{hong2023watch}, or generative pipelines to reconstruct occluded face images~\citep{wang2024restoring}. However, these methods often rely on specific architectures or external modules, which limit their applicability.
Building on recent advances in audio-visual self-supervised learning that highlight the efficacy of modality-fusion representations~\citep{shi2022learning, lian2023av, zhang2023self}, we propose \textbf{\ourmodelmtl}, a novel audio-visual speech representation learning method designed to handle jointly \textbf{c}orrupted \textbf{a}udio-\textbf{v}isual data. \ourmodelmtl is trained through a \textit{corrupted prediction task}, where the model learns to predict clean targets from corrupted input sequences. For this, we employ a teacher-student self-distillation framework \citep{caron2021emerging, ruan2023weighted}, which has proven effective in learning contextualized speech representations \citep{baevski2020wav2vec, baevski2022data2vec, shi2022robust, zhu2024multichannel, liu2024dinosr} without requiring architectural changes or additional modules. In this framework, corrupted sequences are fed into the student model, while the self-evolving teacher model generates the clean targets online.

Within the \ourmodelmtl representation learning framework, the corrupted prediction task can be defined in a multimodal or unimodal strategy. 
A multimodal strategy, inspired by the masked prediction task in AV-data2vec~\citep{lian2023av}, involves corrupting both audio and video inputs to generate corrupted multimodal features, with the model learning to predict clean multimodal targets. 
While this approach enhances the robustness of multimodal representations, it is less capable of isolating the effects of corruption on individual modalities, as both inputs and targets contain mixed audio-visual information.
Multimodal fusion is often prone to combining redundant information~\citep{hsu2022u, mai2023multimodal}, where discriminative unimodal information is ignored, thereby the multimodal prediction falls into overfitting. Previous approaches have tackled this by improving cross-modal information, such as \citet{mai2023multimodal} applying late-fusion to filter out noisy information from unimodal features, and \citet{hu2023hearing} learning a viseme-phoneme mapping\,\citep{bear2017phoneme} to restore corrupted phonemes through visemes, but both of them rely on the information bottleneck structure before the fusion.

To address the limitations of multimodal prediction approach, we introduce \ourmodelmtl with a \textit{unimodal multi-task learning strategy} for corrupted prediction tasks, which leverages corrupted unimodal sequences to distill cross-modal knowledge. 
Our unimodal strategy involves predicting clean audio targets with corrupted videos and predicting clean video targets with corrupted audios.
In AVSR, this cross-modal alignment~\citep{ren2021learning, hu2023hearing, hu2023cross} is essential for effectively integrating information from both modalities. 
Our unimodal prediction strategy, as depicted in Figure\,\ref{fig:corrupted_recognition-2}, improves cross-modal alignment by reducing the dispersion in representation caused by the corrupted inputs.
Figure\,\ref{fig:corrupted_recognition-3} describes the AVSR performance under audio-visual jointly corrupted environments. While fine-tuning with corrupted data (CFT) improves robustness to some extent, it remains challenging at higher degrees of corruption. By incorporating the corrupted representation learning before CFT, \ourmodelmtl demonstrates superior performance. \ourmodelmtl with unimodal multi-task learning further enhances the corrupted prediction framework, effectively aligning the corrupted modalities and achieving more robust results.
We summarize our contributions as follows.

\begin{itemize}
    \item We propose a novel audio-visual speech representation learning, \ourmodelmtl, specifically designed for robustness under audio-visual joint corruption within a self-distillation framework. 
    \item \ourmodelmtl conducts a unimodal multi-task learning for corrupted prediction tasks, predicting clean targets from corrupted input sequences. This strategy effectively enhances cross-modal alignment between corrupted audio and video for reliable multimodal fusion.
    \item We establish an AVSR benchmark for generalization so that our setup includes novel types of corruption unseen during training, \eg mouth occlusion by hands or face pixelation along with public noise, allowing for diverse assessment of model robustness to audio-visual joint corruption. \ourmodelmtl demonstrates significant performance improvement on our robust AVSR benchmarks.
\end{itemize}

\section{Related Work}
\label{sec:ch3_related_work}

\subsection{Audio-Visual Speech Recognition Models}
Automatic speech recognition (ASR) methods that transcribe speech audio to text have been extensively studied for years~\citep{schneider2019wav2vec, gulati2020conformer, baevski2020wav2vec, hsu2021hubert, chen2022wavlm, chiu2022self}. However, since audio signals are often disrupted by background noise, multimodality, particularly visual information from speech video, has been incorportaed into the ASR system~\citep{makino2019recurrent, pan2022leveraging, shi2022learning, seo2023avformer, ma2023auto}. In these multimodal approaches, the audio modality captures the acoustic features of speech signal, while the video modality provides visual information about the speaker's face and lip movements. This integration enhances the robustness of the ASR system and improves its performance, especially in noisy audio environments. 

Several studies aimed at aligning and jointly training audio-visual modalities have been developed in an end-to-end learning framework~\citep{dupont2000audio, ma2021end, hong2022visual, burchi2023audio} or self-supervised pretraining approach~\citep{ma2021lira, qu2022lipsound2, shi2022learning, seo2023avformer, zhu2023vatlm} that often utilizes masked modeling of speech representations. Recently, among the multimodal speech recognition models leveraging the self-supervised learning, a self-distillation approach~\citep{lian2023av, haliassos2022jointly, haliassos2024braven, zhang2024es3, liu2024dinosr}, which learns the contextualized representations by distilling the self-evolving teacher's knowledge for masked inputs, has demonstrated superior performances.

\subsection{Learning Robustness for AVSR}
The AVSR studies have focused on developing robust models for various types of noise while simultaneously utilizing audio and visual information. Most of this research has initially focused on addressing noise in the audio modality \citep{shi2022robust,  hu2023hearing, hu2023cross, chen2023leveraging, kim2024learning, ithal2024enhancing}. However, more recent efforts have explored disturbances in visual data, creating robust models by adding background noise, such as Gaussian noise, to the speech videos.
In this line of research, \citet{hong2023watch} have considered that human speech often involves a mouth region being occluded by objects and first applied visual occlusion into speech data. This has inspired further research addressing similar challenges by restoring the occluded images by a generative model~\citep{wang2024restoring}. Additionally, \citet{fu2024boosting} have combined prompt learning with contrastive learning to deal with audio-visual asynchrony, while \citet{zhang2024visual} have examined the issue of a completely missing visual modality, generating the visual hallucination during inference. \citet{li2024unified} demonstrate the effectiveness of leveraging unified cross-modal attention and a synchronization module to encode audio and video sequences in a unified feature space.
In our study, we address real-world challenges of audio-visual joint corruption by using corrupted representation learning, offering a general framework without relying on specific architectures or external modules.

\section{Preliminaries}
\label{sec:ch3_preliminaries}

\subsection{Notations}

Let $A = [a_1, a_2, \dots, a_T]$ be the audio sequence and $V = [v_1, v_2, \dots, v_T]$ be the video sequence over time $T$. 
We define a set of corrupted indices for the audio sequence as $C^a \subseteq \{1, 2, \dots, T\}$ and a set of corrupted indices for the video sequence as $C^v \subseteq \{1, 2, \dots, T\}$.
To model the corruption, we define a family of corruption functions $\Omega$ that can transform or corrupt certain frames of data.
Thus, given some corruption functions $\omega^a, \omega^v \in \Omega$, the corrupted audio sequence $\tilde{A} = [\tilde{a}_1, \tilde{a}_2, \dots, \tilde{a}_T]$ and the corrupted video sequence $\tilde{V} = [\tilde{v}_1, \tilde{v}_2, \dots, \tilde{v}_T]$ are defined as follows.
\begin{equation}
\tilde{a}_t = 
\begin{cases} 
\omega^a(a_t) & \text{if } t \in C^a \\
a_t & \text{else }
\end{cases}
\quad \text{and} \quad
\tilde{v}_t = 
\begin{cases} 
\omega^v(v_t) & \text{if } t \in C^v \\
v_t & \text{else }
\end{cases}
\end{equation}

Raw audio and video data are processed by its respective feature extractor, and the resulting features are concatenated before being input to the multimodal Transformer encoder $f_\theta: \mathbb{R}^{T \times D} \rightarrow \mathbb{R}^{T \times D}$. The output feature sequence for this multimodal input is denoted as $\tilde{\mZ}^{av} = f_\theta(\tilde{A};\tilde{V}) = [\tilde{\vz}^{av}_1, \tilde{\vz}^{av}_2, \dots, \tilde{\vz}^{av}_T]$. If $t \in C^a \cup C^v$, then $\tilde{\vz}^{av}_t$ is considered a corrupted feature representation.

\begin{figure}[!t]
    \centering
    \vspace{-10pt}
    \begin{minipage}[t]{0.65\textwidth}
        \centering
        \vspace{0pt}
        \includegraphics[width=\textwidth]{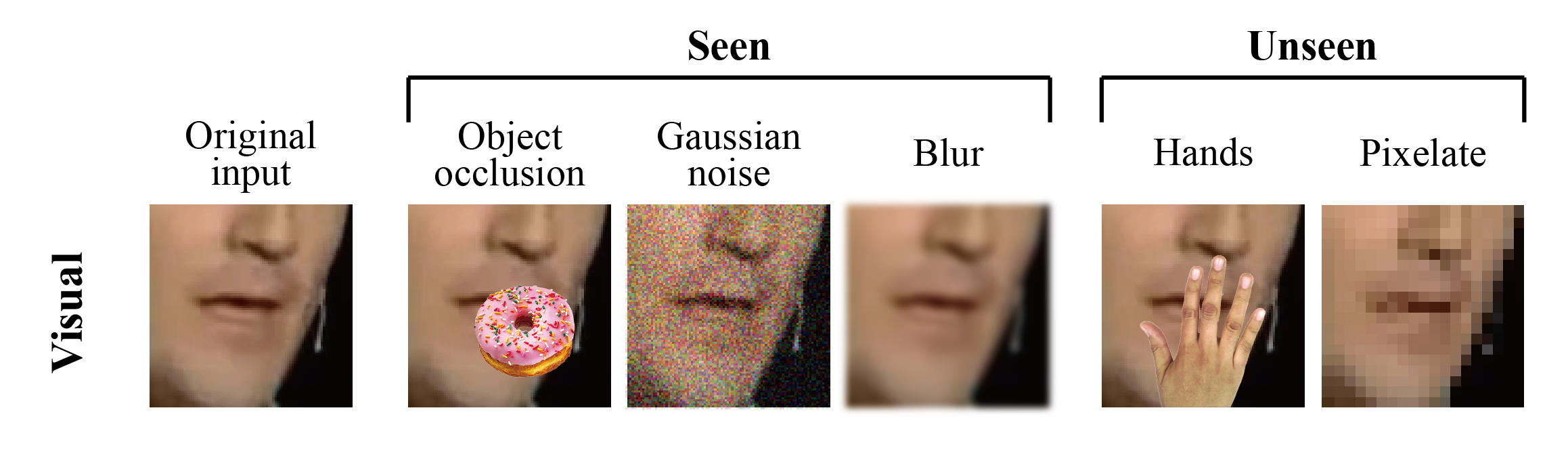}
    \end{minipage}
    \hfill
    \begin{minipage}[t]{0.33\textwidth}
        \centering
        \vspace{10pt}
        \small
        \resizebox{\textwidth}{!}{
        \addtolength{\tabcolsep}{-2pt}
        \begin{tabular}{c|ll}
            \toprule
             & Audio noise (SNR\,=\,$-$10\,dB) & Source \\
            \midrule
            \rotatebox[origin=c]{90}{\makecell[l]{\!\!\!\!\!\!\textbf{Seen}}} & babble, music, natural & MUSAN \\
             & speech (held-out) & LRS3 \\
             \midrule
            \rotatebox[origin=c]{90}{\makecell[l]{\!\!\textbf{Unseen}\!}} & \thead[l]{park, river, cafe, restaurant,\\cafeteria, metro (subway),\\public station, meeting room} & DEMAND \\
            \bottomrule
        \end{tabular}}
    \end{minipage}
    \caption[Audio and visual corruption types used for training and evaluation]{The visual and audio corruption types we use in our training and evaluation phases. Unseen corruption types are only utilized in evaluation to assess the model's generalizability. The speech audio noise from LRS3 is ensured that there is no speaker overlap between train and evaluation sets.}
    \label{fig:corruption_types}
    \vspace*{10pt}
\end{figure}

\subsection{Masked Prediction Task}

\paragraph{Self-distillation framework.}
The self-distillation approach as self-supervised learning has been shown to be highly effective in learning contextualized representations~\citep{baevski2022data2vec, baevski2023efficient, liu2024dinosr} without supervised labels, including in multimodal feature spaces~\citep{zhang2023self, zhang2024es3, zhu2024multichannel}.
In this framework, a reference function $f$, commonly referred to as the teacher model, is updated by an exponential moving average (EMA) of the parameterized student model $f_\theta$ with a decaying parameter $\eta$, $f \leftarrow \eta * f + (1-\eta) * f_\theta$. The student model learns by predicting targets generated online by the teacher. Thus, it enables representation learning without requiring external modules or modifications to the overall model structure.

\paragraph{Masked prediction task loss.}
In AV-data2vec~\citep{lian2023av}, audio and video frames are randomly masked, and a masked prediction task is performed by predicting each masked frame with the target feature. 
The target features are obtained from clean, unmasked data using the teacher model.
Then, the masked prediction loss is defined as:
\begin{equation}
\label{eq:masked_prediction}
    \mathcal{L}_\text{MASK} = \sum_{t \in M^a \cup M^v} \ell \Big( [f_\theta(\texttt{MASK}(A); \texttt{MASK}(V))]_t,~ [f(A; V)]_t \Big)
\end{equation}
where $M^a$ and $M^v$ are the set of indices of masked audio and video frames, respectively. $\ell$ is often used as a mean squared error\,(MSE) loss, and $f(\cdot)$ as the teacher model's average representation---the output sequence averaged over top-$k$ Transformer blocks to establish the target, \ie $f(A;V) = \frac{1}{k}\sum^L_{l=L-k+1} f^l(A; V)$, where $L$ is the number of blocks.

\section{CAV2vec: Unimodal Multi-Task Corrupted Prediction}

\subsection{Visual and Audio Corruption Types}
\label{subsec:corruption_types}

Figure\,\ref{fig:corruption_types} presents the visual and audio corruption types used in this study. We propose a novel evaluation benchmark, introducing corruption types unseen during training, to assess the model's generalizability under diverse and realistic conditions.
During training, visual corruptions include object occlusion, Gaussian noise, and blurring, applied to video frames following~\citet{hong2023watch}.
For object occlusion, we obscure the mouth regions by COCO \citep{lin2014microsoft} object images, as presented in \citet{voo2022delving}. In evaluation, we apply unseen visual corruption types, using the 11k-Hands dataset \citep{afifi201911k} to occlude the mouth regions with diverse hand images. Furthermore, we apply corruption by pixelating the entire frame using the \texttt{opencv} library, with a 3x3 patch interpolation. This allows us to evaluate the model's robustness under conditions where human faces are often pixelated due to privacy or ethical concerns. 

For the audio corruption, we use various types of background noise that are recorded from different sources, at a $-$10\,dB SNR (signal-to-noise ratio). In the training, we apply conventionally used audio noise types~\citep{shi2022robust, hsu2022u}, babble, music, and natural noise sampled from MUSAN~\citep{snyder2015musan}, and speech noise sampled from LRS3~\citep{afouras2018lrs3}, onto the original speech signal. In the evaluation, we introduce new corruption types from the DEMAND dataset~\citep{thiemann2013demand}, which includes real-world indoor and outdoor recordings. Out of 18 categories of recording, we mainly evaluate on 8 relatively noisy environments: park, river, cafe, restaurant, cafeteria, metro, public station, and meeting room.

\subsection{Corrupted Prediction Tasks of CAV2vec}
\label{subsec:corrupted_prediction_task}

\begin{figure}[!t]
    \centering
    \includegraphics[width=1\linewidth]{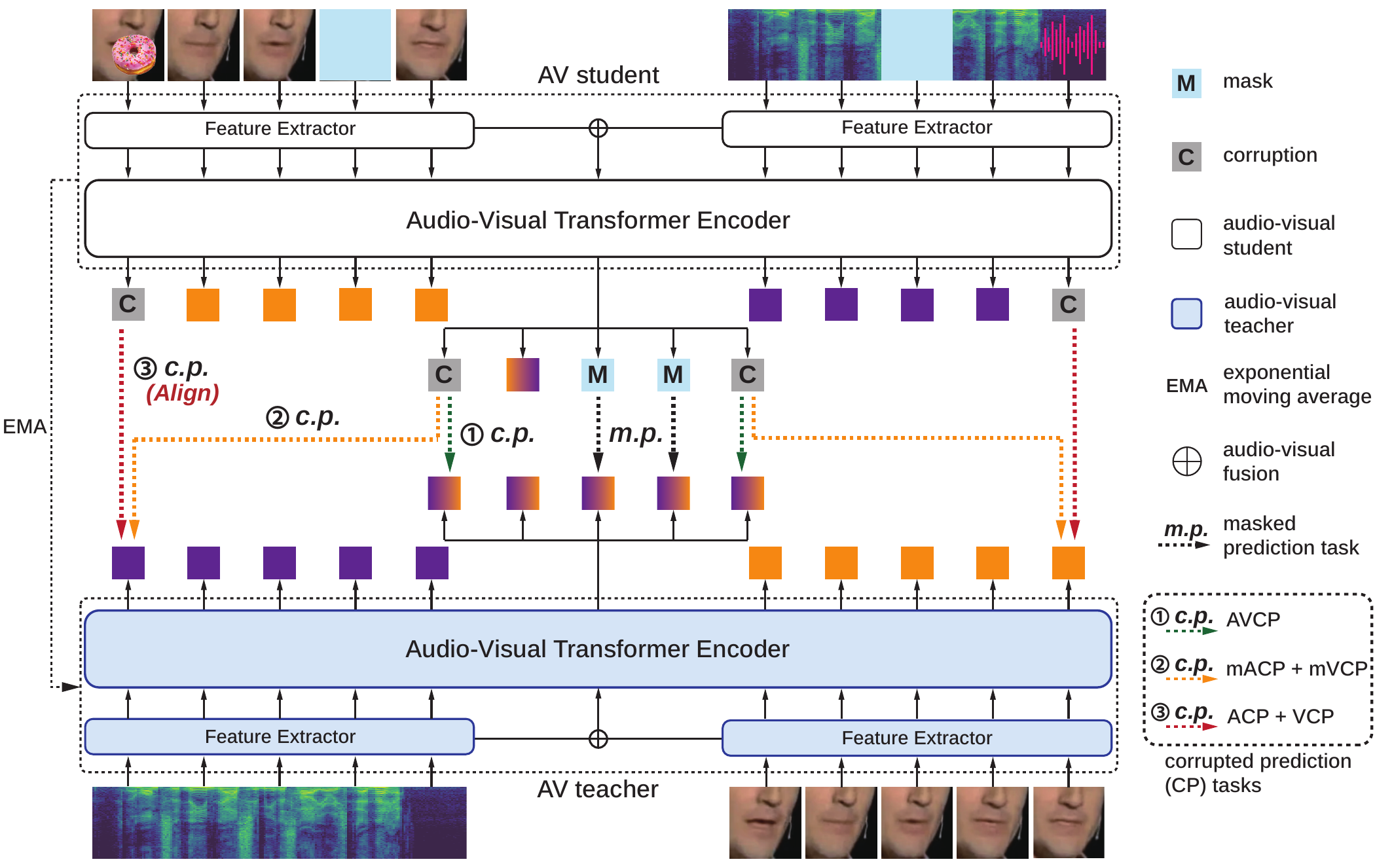}
    \vspace{-15pt}
    \caption[CAV2vec representation learning framework using corrupted prediction tasks]{Overview of our representation learning framework with corrupted prediction tasks. For the corrupted prediction strategies, focusing on the cross-modal alignment through unimodal multi-task learning proves highly effective in gaining multimodal robustness.
    }
    \label{fig:overview}
    \vspace*{10pt}
\end{figure}

We present \textbf{\ourmodelmtl}, a robust representation learning framework to account for corrupted audio-visual sequences. Inspired by the idea of conventional masked prediction strategy~\citep{lian2023av}, \ourmodelmtl is trained through a \textit{corrupted prediction task}. The corrupted prediction task loss is designed to minimize the difference between the student model's output for the corrupted data and the teacher model's output for the uncorrupted data. Figure\,\ref{fig:overview} provides an overview of the corrupted representation learning of \ourmodelmtl.
We apply corruption functions to both video and audio data (see Figure\,\ref{fig:corruption_types}) and perform the corrupted prediction tasks on the corrupted frames, alongside the masked prediction task on the masked frames. 
We suggest different strategies in designing these corrupted prediction tasks, depending on the modality of input and target representations.

\paragraph{Audio-visual corrupted prediction task.}
A straightforward approach is using corrupted multimodal inputs as well as clean multimodal targets (\raisebox{.5pt}{\textcircled{\raisebox{-.9pt} {1}}} in Figure\,\ref{fig:overview}), following the masked prediction strategy.
We name it as audio-visual corrupted prediction (AVCP) task, where the AVCP loss function is analogously defined as:
\begin{equation}
\label{eq:corrupted_prediction}
\mathcal{L}_\text{AVCP} = \sum_{t \in C^a \cup C^v} \ell \Big( \tilde{\vz}^{av}_t,~ [f(A; V)]_t \Big).
\end{equation}
Here, the loss is computed only for the corrupted indices \( t \in C^a \cup C^v \), where \( \tilde{\mZ}^{av} \) represents the student model's predictions for (possibly) corrupted sequences \( \tilde{A} \) and \( \tilde{V} \), and \( f(A; V) \) represents the target for clean sequences \( A \) and \( V \).

\paragraph{Unimodal corrupted prediction tasks.}
While the AVCP task in Eq.\,(\ref{eq:corrupted_prediction}) is effective in learning robustness for multimodal features, the mixed audio-visual information in both inputs and targets makes it hard to isolate corruptions on individual modalities~\citep{mai2023multimodal}.
To address this, we introduce \ourmodelmtl with a \textit{unimodal multi-task learning strategy} for corrupted prediction, leveraging audio-only and video-only sequences (\raisebox{.5pt}{\textcircled{\raisebox{-.9pt} {3}}} in Figure\,\ref{fig:overview}). These unimodal tasks enhance cross-modal alignment, which is essential in AVSR for capturing multimodal correlations~\citep{ren2021learning, hu2023hearing, hu2023cross}, particularly when corruption disrupts the link between two modalities.
We propose the unimodal tasks to distill cross-modal knowledge: audio corrupted prediction\,(ACP) task and visual corrupted prediction\,(VCP) task. Their loss functions are defined as: 
\begin{equation}
    \label{eq:mtl}    
        \mathcal{L}_\text{ACP} = \sum_{t \in C^v} \ell \Big( \tilde{\vz}^{v}_t,~ [f(A;\bm{0})]_t \Big), \quad
        \mathcal{L}_\text{VCP} = \sum_{t \in C^a} \ell \Big( \tilde{\vz}^{a}_t,~ [f(\bm{0};V)]_t \Big)
\end{equation}
where $\tilde{\mZ}^{v} = f_\theta(\bm{0}; \tilde{V})$ and $\tilde{\mZ}^{a} = f_\theta(\tilde{A}; \bm{0})$ denote video-only and audio-only unimodal features, respectively. Thus, the ACP task predicts clean audio targets with corrupted videos, and the VCP task predicts clean video targets with corrupted audios.
To thoroughly examine the impact of each task and the relationship with modality alignment, we also implement multimodal ACP and VCP tasks, called mACP and mVCP, which use multimodal inputs $\tilde{\mZ}^{av}$ and unimodal targets (\raisebox{.5pt}{\textcircled{\raisebox{-.9pt} {2}}} in Figure\,\ref{fig:overview}). These are formulated as $\mathcal{L}_\text{mACP} = \sum_{t \in C^v} \ell (\tilde{\vz}^{av}_t, [f(A;\bm{0})]_t)$ and $\mathcal{L}_\text{mVCP} = \sum_{t \in C^a} \ell (\tilde{\vz}^{av}_t, [f(\bm{0};V)]_t)$, and their effectiveness is investigated in Section\,\ref{sec:ch3_analysis}.

\begin{figure}[!t]
    \centering
    \hfill
    \begin{subfigure}[b]{0.6\textwidth}
        \centering
        \includegraphics[height=3.6cm]{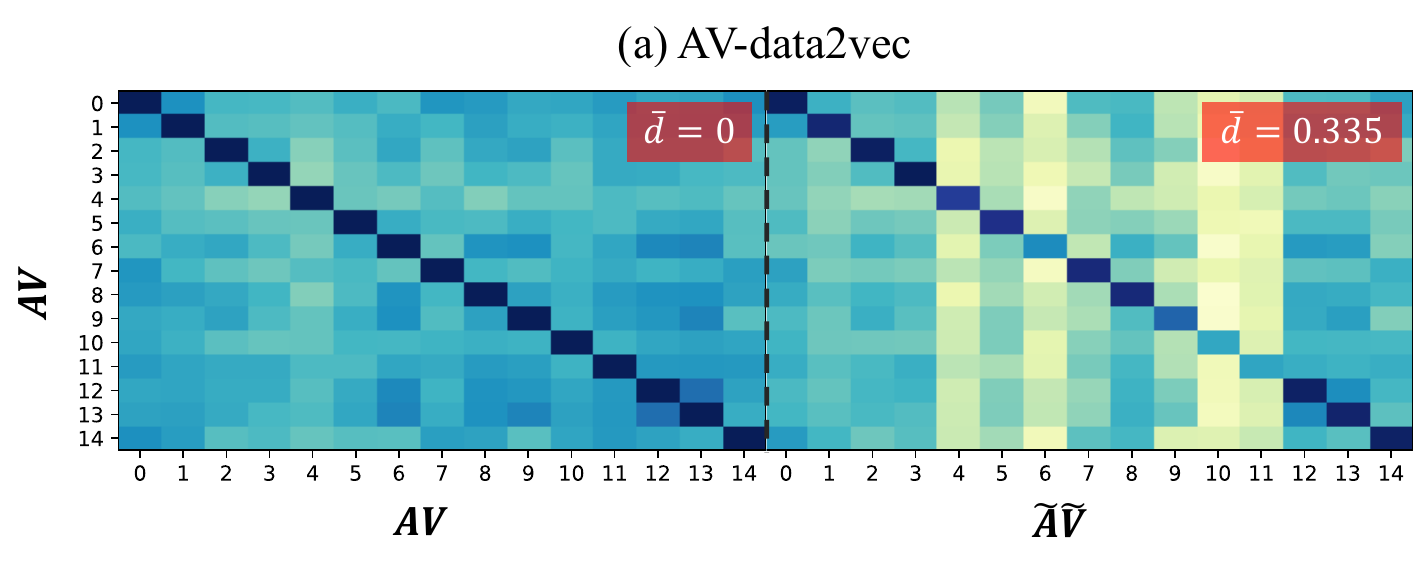}
        \captionlistentry{}
        \label{fig:similarity_matrix_corruption_av2vec}
    \end{subfigure}
    \begin{subfigure}[b]{0.38\textwidth}
        \centering
        \includegraphics[height=3.6cm]{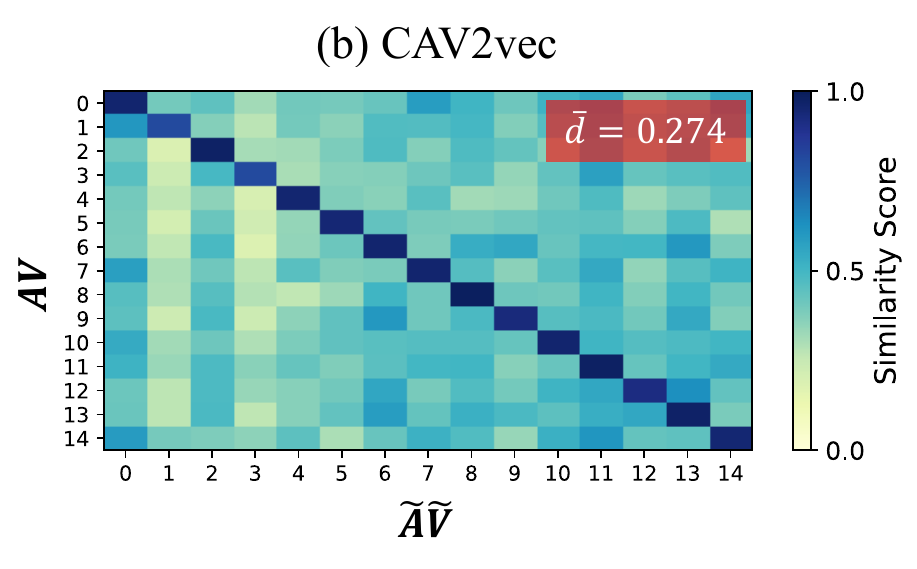}
        \captionlistentry{}
        \label{fig:similarity_matrix_corruption_cav2vec}
    \end{subfigure}
    \hfill
    \vspace{-10pt}
    \caption[L2 distance analysis of feature dispersion for CAV2vec]{Similarity scores measured between audio-visual features of sample sequences.
    Clean sequence representations are compared with corrupted ones from (a) AV-data2vec and (b) \ourmodelmtl. The normalized L2 distance $\overline{d}$ is calculated between the clean and corrupted features per-sample.
    }
    \label{fig:similarity_matrix_corruption}
    \vspace*{10pt}
\end{figure}

\paragraph{Overall multi-task loss of \ourmodelmtl.}
We employ both corrupted prediction in Eq.\,(\ref{eq:mtl}) and masked prediction in Eq.\,(\ref{eq:masked_prediction}), but we do not allow the overlap between masked and corrupted frames to separate the tasks, \ie $(M^a \cup M^v) \cap (C^a \cup C^v) = \emptyset$. We have empirically found this task separation to be effective.
Incorporating the unimodal corrupted prediction tasks with modality dropout \citep{hsu2022u} as well as the masked prediction task for multimodal inputs, the loss function for \ourmodelmtl within multi-task learning is defined as follows:
\begin{equation}
\label{eq:mtl_final}
    \mathcal{L}_\text{CAV2vec} =  \lambda_\text{ACP}\,\mathcal{L}_\text{ACP} + \lambda_\text{VCP}\,\mathcal{L}_\text{VCP} +
    \lambda_\text{MASK}\,\mathcal{L}_\text{MASK} + \lambda_\text{MLM}\,\mathcal{L}_\text{MLM}
\end{equation}
where $\mathcal{L}_\text{MLM}$ is a masked language modeling-style (MLM) loss used in \citet{shi2022learning, zhang2023self}. $\mathcal{L}_\text{MLM}$, which predicts the cluster index of the masked features in a cross-entropy loss form, helps the model converge faster than using $\mathcal{L}_\text{MASK}$ alone~\citep{zhang2023self}.
In our experiments, we set $\lambda_\text{ACP} = \lambda_\text{VCP} = \lambda_\text{MASK} = 1.0$ and $\lambda_\text{MLM} = 2.0$ to match the scales of each loss.

As shown in Figure\,\ref{fig:similarity_matrix_corruption_av2vec}, corruption disrupts audio-visual representations, resulting in reduced similarity scores and increased dispersion of features, which hinders the model’s ability to maintain robust representations. 
In Figure\,\ref{fig:similarity_matrix_corruption_cav2vec}, our corrupted representation learning helps the model encode highly correlated audio-visual representations, restoring high similarity (small distance) between clean and corrupted sequence features and leading to a more compact and resilient representation space.

\section{Experiments and Results}

\subsection{Implementation Details}
\label{subsec:ch3_implementation}

\paragraph{Datasets.}
We train and evaluate our model on LRS3~\citep{afouras2018lrs3}, which contains roughly 433 hours of TED talks from over 5,000 speakers.
Most of our experimental configurations follow \citet{shi2022robust}, including noise augmentation and evaluation protocols. 
Audio noise is extracted from the MUSAN~\citep{snyder2015musan} (\textit{babble}, \textit{music}, \textit{natural}) and LRS3 (\textit{speech}) datasets, partitioned into training, validation, and test sets. This noise is added to audio waveform during both training and evaluation.
The evaluation metric for AVSR is the word error rate (WER, \%). 
For audio feature extraction, we extract 26-dimensional log filter bank features from raw audio at a stride of 10 ms, stacking 4 adjacent frames to achieve a frame rate of 25 fps. Video track is sampled at 25Hz, with a 96$\times$96 region center-cropped on the speaker's mouth. During training, we randomly crop an 88$\times$88 region and apply horizontal flips with probability 50\%.

\paragraph{\ourmodelmtl and baseline models.}
\ourmodelmtl uses 24 Transformer~\citep{vaswani2017attention} block layers for the AVSR encoder and 9 layers for the decoder, based on the AV-HuBERT-\textsc{Large} model~\citep{shi2022learning}.
The visual feature extractor is a modified ResNet-18, while the audio feature extractor is a linear projection layer. The extracted features are concatenated to form fusion audio-visual features, which are input to the Transformer encoder.
We initialize the model using the publicly available checkpoint from \citet{shi2022robust}, the AV-HuBERT encoder pretrained on noise-augmented LRS3~\citep{afouras2018lrs3}\,+\,VoxCeleb2~\citep{chung2018voxceleb2}, and proceed our representation learning phase afterwards.
This training strategy makes the whole process efficient, spanning only 2\% of AV-HuBERT's pretraining cost \citep{shi2022learning}, as well as leveraging high-resource knowledge of VoxCeleb2 (1,326 hours). Our representation learning can thus be considered as an uptraining phase~\citep{ainslie2023gqa}, an additional pretraining step that helps the model adapt to corrupted data before fine-tuning on the supervised speech recognition task.
As in prior self-distillation works~\citep{caron2021emerging, baevski2022data2vec}, we employ single MLP-layer predictors assigned for each task on the student model, which are removed after training.

For baseline models, we compare with (1) V-CAFE~\citep{hong2022visual}, (2) RAVEn~\citep{haliassos2022jointly}, (3) BRAVEn~\citep{haliassos2024braven}, (4) AV-HuBERT~\citep{shi2022learning}, (5) AV-data2vec~\citep{lian2023av}, and (6) AV-RelScore~\citep{hong2023watch}.
Here is a brief summary: 
V-CAFE is an end-to-end supervised model with a relatively small model size, while the other models are pretrained with their own loss functions. 
We re-implemented RelScore on the AV-HuBERT backbone, as the original version based on V-CAFE~\citep{hong2023watch} performs poorly, and further trained the scoring module.
All baseline models are initialized from their respective pretrained checkpoints and then fine-tuned using the same decoder architecture and training configurations as our model.
Below we provide the details on how they have been (re-)implemented.

\textbf{V-CAFE}~\citep{hong2022visual}~~~is an end-to-end supervised model for AVSR, designed to capture the lip movement and generate a noise reduction mask by utilizing visual context. The model is built on a Conformer-Transformer encoder-decoder architecture and employs a joint CTC/Attention loss function \citep{kim2017joint}. V-CAFE incorporates visual context through cross-modal attention to generate a noise reduction mask, where the generated mask is applied to the encoded audio features to mitigate noisy audio representations.

For our experiments, we use the pretrained encoder from the publicly available V-CAFE checkpoint\footnote{\url{https://github.com/ms-dot-k/AVSR}}, which has been trained on the LRS3 dataset. 
To ensure a consistent experimental setup across all baselines and our model, we fine-tune the V-CAFE model by initializing the Transformer decoder and training it for 120,000 steps with a learning rate of $2\times 10^{-3}$.
The encoder remains frozen for the first 48,000 steps, after which the entire model is updated over the remaining 72,000 steps.

\textbf{RAVEn and BRAVEn}~\citep{haliassos2022jointly, haliassos2024braven}~~~are self-supervised learning ASR and VSR models that encode masked inputs and predict contextualized targets generated by momentum encoder teachers. In both models, two unimodal encoders are jointly trained, each serving as a teacher for the cross-modal student encoder. Specifically, the audio student predicts outputs from both audio and video teachers, while the video student predicts only audio targets. BRAVEn is the upgraded version of RAVEn, slightly modifying the self-distillation framework with different hyperparameters to more emphasize ASR than VSR.

We utilize the pretrained encoders of RAVEn and BRAVEn from the public repository\footnote{\url{https://github.com/ahaliassos/raven}}. Both ASR and VSR encoders are loaded, and these models are used to encode the normalized raw audio waveform and video frames, respectively. Although RAVEn and BRAVEn were not originally designed as multimodal models, we follow the approach of \citet{haliassos2024braven} to implement an AVSR framework by fusing the encoded features from each modality. The fusion audio-visual features are then fed into an initialized Transformer decoder. Thus, the modality-fusion MLP layer is also trained with the decoder. We train the model for 120,000 steps with a learning rate of $2\times 10^{-3}$ while the encoder is frozen for the first 96,000 steps. Due to the absence of a pretrained multimodal encoder and the fact that these models were not trained on noise-augmented data, their AVSR performance in corrupted environments is suboptimal despite of their large number of model parameters.

The RAVEn and BRAVEn results in Table\,\ref{tab:main_result} exhibit poor performances than those reported in the original paper. This discrepancy can be attributed to different evaluation settings. Our settings are designed to assess noise-robust audio-visual models under real-world conditions, where the type and extent of modality corruption are unpredictable. These results (WER: RAVEn 2.3\% and BRAVEn 1.8\%) are measured under visual corruption, whereas the published results pertain to clean audio conditions using a standalone ASR model. Although standalone ASR models excel in clean audio environments, their performance significantly degrades under noisy conditions.

BRAVEn has reported low-resource AVSR performance (WER: RAVEn 4.7\% and BRAVEn 4.0\%), but it does not provide results in high-resource settings or under audio-visual corruption. Moreover, the specific hyperparameters used for fine-tuning in conjunction with pretrained ASR and VSR encoders and a decoder have not been detailed. The performance gap might also stem from the larger unlabeled dataset and self-training technique during pretraining and the use of a language model during inference, which were not employed in our experimental setup. Also, while (B)RAVEn used CTC/attention loss for fine-tuning, we only used attention loss to ensure a fair comparison across all models. These variations in decoding approaches can influence outcomes, although they could be orthogonally applied to any models.

\textbf{AV-HuBERT and AV-data2vec}~\citep{shi2022learning, lian2023av}~~~are self-supervised learning models for audio-visual multimodal processing within a single framework, both using masked inputs to predict unmasked targets. They share the same structure for the encoder, with 24 Transformer blocks. The key difference lies in how they generate the targets: AV-HuBERT uses the cluster indices of MFCC\,(mel-frequency cepstral coefficient) features, while AV-data2vec uses the EMA teacher's output sequence. AV-HuBERT updates its targets after each iteration of training using the current model, whereas AV-data2vec generates online targets with a self-evolving teacher. Both methods are effective in learning contextualized audio-visual multimodal features.

Since AV-HuBERT pretrained models are publicly available\footnote{\url{https://github.com/facebookresearch/av_hubert}} but AV-data2vec models are not, we implement AV-data2vec on top of the AV-HuBERT pretrained model, and uptrained it for 60,000 steps in a similar manner to \ourmodelmtl. We use the Adam optimizer with a weight decay of 0.01 and a learning rate of $2\times 10^{-4}$. We fine-tune the decoder for both AV-HuBERT and AV-data2vec models, building on the respective pretrained encoders. The fine-tuning process employs an attention-based sequence-to-sequence cross-entropy loss, with accuracy as the validation metric. The initial learning rate is $10^{-3}$, scheduled with 20,000 warmup steps followed by decaying over the next 40,000 steps. During fine-tuning, the encoder remains frozen for the first 48,000 steps, updating only the decoder. After 48,000 steps, both the encoder and decoder are updated for the final 12,000 steps.

\textbf{AV-RelScore}~\citep{hong2023watch}~~~leverages a reliability scoring module (RelScore) to assess the reliability of each input modality at every frame. RelScore modules are appended after the feature extractor for each modality and consist of three convolutional layers followed by a sigmoid activation, which outputs a scalar value for each frame. The original model is based on the V-CAFE backbone~\citep{hong2022visual}, which has a smaller architecture and subpar performance on large-scale datasets. To ensure a fair comparison with our baselines, we implement RelScore on the AV-HuBERT-\textsc{Large} backbone, leveraging its pretrained knowledge. For training AV-RelScore, we freeze the encoder and feature extractors, training only the RelScore modules and the decoder for 50,000 steps. Afterward, we update the entire model, including the encoder, for an additional 40,000 steps.
Since the RelScore modules introduce additional parameters and are located before the pretrained encoder, AV-RelScore requires more fine-tuning steps until convergence than required by AV-HuBERT.

\subsection{Training and Evaluation}

\paragraph{Training configuration.}
For \ourmodelmtl uptraining, we sample audio noise at 0\,dB SNR and randomly perturb the clean speech signal with a 25\% probability. The encoder is updated for 60K steps, with a maximum of 16K tokens (\ie 640 seconds) per step, which takes 8--10 hours on 4 RTX A6000 GPUs.
Visual corruption is applied to every sequence, randomly corrupting 10--50\% of the sequence length with object occlusion, Gaussian noise, or blurring.
Every clean audio sequence is corrupted by augmenting strong noise of babble, speech, music, or natural, at $-$10\,dB SNR to 30–50\% of the sequence length.

During fine-tuning, the encoder is frozen for the first 48K steps, while the decoder is trained using sequence-to-sequence negative log-likelihood as the AVSR loss function. The entire model is then trained for additional 12K steps.
The visual corruption applied during fine-tuning is identical to that in the uptraining phase, while we randomly corrupt 25\% of the audio signal by an SNR value sampled from a normal distribution with mean 0 and standard deviation 5.
We note that all baseline models follow the same fine-tuning procedure as ours for a fair comparison: initialized from pretrained models and then fine-tuned with audio-visual corrupted inputs.

We perform uptraining using a Transformer-based AV-HuBERT-\textsc{Large} architecture via unimodal corrupted prediction tasks. Both audio and video data are processed at 25 frames per second (fps) and augmented with various corruptions (refer to Figure\,\ref{fig:corruption_types}). Each sample sequence is trimmed to a maximum length of 400 frames during pre-processing. For audio, 25\% of the sequences are applied with noise at SNR\,=\,0\,dB, following the noise augmentation strategy from \citet{shi2022robust}, while the remaining 75\% undergo partial corruption at SNR\,=\,$-$10\,dB. A single chunk within each sequence is corrupted, with the corruption length randomly selected between 30--50\% of the sequence length. The visual modality is corrupted at a frequency of 1, where frequency denotes the number of visual corruption events in the entire sequence. Object occlusion applied to the speaker's lips occurs once, followed by Gaussian noise or blurring, each with a probability of 0.3. The visual corruption length is randomly selected as 10--50\%.

We also apply frame masking for contextualized representation learning. Following the strategy in previous works~\citep{shi2022learning}, 80\% of audio frames are masked, with each mask segment lasting 10 frames, while 30\% of video frames are masked, with each segment lasting 5 frames. The high audio masking ratio helps the model focus on the most relevant information from the audio context. However, we note that masking is applied after corruption, and we do not allow overlap between masked and corrupted frames. Therefore, the effective masking ratio is lower than the initially set probability. 

Additionally, modality dropout is applied to both audio and visual inputs, each with a dropout rate of 0.25. In our implementation of \ourmodelmtl with unimodal multi-task learning, we use audio targets when the audio input is dropped out (video-only input) and video targets when the video input is dropped out (audio-only input). The audio-visual target is always used for the masked prediction task. \ourmodelmtl is optimized using a multi-task loss function that includes ACP\,+\,VCP losses, each weighted at 1.0 ($\lambda_\text{ACP} = \lambda_\text{VCP} = 1.0$). The masked prediction loss coefficient is weighted at 1.0, while the MLM loss is weighted at 2.0 ($\lambda_\text{MASK}=1.0, \lambda_\text{MLM}=2.0$), balancing the scales of self-distillation regression loss and MLM cross-entropy loss. 

We use the Adam optimizer~\citep{kingma2014adam} with a weight decay of 0.01 and a learning rate of $10^{-4}$. The EMA decaying parameter $\eta$ starts from 0.99 and increases up to 0.999 over training. The model is updated for 60,000 steps, using a polynomial decay learning rate scheduler with a warmup phase of the first 5,000 updates. Each model update consumes a batch of 16,000 tokens, equivalent to 640 seconds of audio-visual data.
For fine-tuning, we largely follow AV-HuBERT and apply the same procedure to all baseline models. This involves initializing the 9-layer Transformer decoder and training the AVSR task with a sequence-to-sequence negative log-likelihood loss. For simplicity of fine-tuning, we do not use connectionist temporal classification (CTC) loss~\citep{graves2006connectionist}.

\paragraph{Evaluation details.}
For evaluation, we assess the model's performance under various audio-visual joint corruption scenarios, including corruption types that were not encountered during training. For visual corruption types, object occlusion and Gaussian noise (or blurring) are applied with a frequency of 1 each, consistent with the training phase. For unseen visual corruptions, \ie hands occlusion and pixelated face, we randomly sample a frequency from \{1, 2, 3\}, resulting in an average of two corruption occurrences per sequence, similar to the object occlusion + noise.
The model's AVSR performance is measured using the word error rate (WER) across five SNR values: $\{-10, -5, 0, 5, 10\}$. Audio corruption is introduced using noise from the MUSAN dataset, including babble, music, and natural noise, as well as LRS3 speech noise. To ensure that there is no speaker overlap between the training and test sets, LRS3 speech noise is generated using distinct speakers. Additionally, when evaluating the model with unseen DEMAND noise, we corrupt audio by randomly sampling an SNR value from the range $[-10, 10]$ for each environment category.

\subsection{Robust AVSR Benchmark Results}
\label{sec:ch3_results}

\begin{sidewaystable}
    \centering
    \small
    \caption[CAV2vec performance on corrupted LRS3]{Comparisons of WER\,(\%) with our model and prior works on the LRS3 dataset\,\citep{afouras2018lrs3}. For audio corruption, babble, music, and natural noise are sampled from the MUSAN dataset\,\citep{snyder2015musan}, while speech noise is sampled from the held-out set of LRS3. N-WER averages the results across all four audio noise types and five SNR\,(signal-to-noise ratio) values, while N\,$\ge$\,S denotes noise-dominant scenarios, averaging over \{-10, -5, 0\} SNRs. For visual corruption, we evaluate on (a)\,object occlusion and noise, (b)\,occlusion by hands, and (c)\,pixelated face.}
    \label{tab:main_result}
    \addtolength{\tabcolsep}{0pt}
    \resizebox{\textwidth}{!}{
    \begin{tabular}{c|l|l|cccccc|cccccc|cccccc|cc|c}
    \toprule
    & \multirow{2}{*}{Method} & \multirow{2}{*}{Params} & \multicolumn{6}{c|}{Babble, SNR (dB) $=$} & \multicolumn{6}{c|}{Speech, SNR (dB) $=$} & \multicolumn{6}{c|}{Music + Natural, SNR (dB) $=$} & \multicolumn{2}{c|}{N-WER} & Clean \\
    & & & -10 & -5 & 0 & 5 & 10 & \!\!\textbf{AVG}\!\! & -10 & -5 & 0 & 5 & 10 & \!\!\textbf{AVG}\!\! & -10 & -5 & 0 & 5 & 10 & \!\!\textbf{AVG}\!\! & \textbf{AVG} & N\,$\ge$\,S & $\infty$ \\
    \midrule
    & V-CAFE & 49M & 54.7 & 31.6 & 14.6 & 7.3 & 5.4 & 22.7 & 45.1 & 29.3 & 16.9 & 10.0 & 6.6 & 21.6 & 32.1 & 17.6 & 9.1 & 6.5 & 5.2 & 14.1 & 18.1 & 25.8 & 4.2 \\
    & RAVEn & 673M & 43.5 & 25.4 & 8.5 & 4.0 & 2.8 & 16.9 & 58.9 & 42.3 & 17.9 & 5.1 & 3.1 & 25.5 & 23.3 & 11.3 & 5.4 & 3.3 & 2.5 & 9.2 & 15.2 & 23.0 & 2.3 \\
    & BRAVEn & 673M & 41.1 & 22.2 & 6.1 & 2.5 & 1.8 & 14.7 & 45.6 & 30.4 & 14.9 & 5.2 & 2.2 & 19.7 & 20.3 & 8.3 & 3.5 & 2.1 & 1.8 & 7.2 & 12.2 & 18.7 & 1.8 \\
    & AV-HuBERT & 325M & 30.6 & 14.4 & 5.1 & 2.7 & 2.1 & 11.0 & 7.7 & 4.3 & 3.0 & 2.2 & 2.0 & 3.9 & 10.9 & 5.2 & 3.0 & 2.3 & 1.8 & 4.6 & 6.0 & 8.6 & 1.6 \\
    & AV-data2vec & 325M & 32.1 & 15.1 & 5.3 & 2.5 & 2.0 & 11.4 & 8.3 & 4.9 & 3.1 & 2.2 & 1.9 & 4.1 & 10.9 & 5.5 & 3.0 & 2.1 & 1.8 & 4.6 & 6.2 & 9.0 & \textbf{1.5} \\
    & AV-RelScore & 437M & 30.1 & 14.3 & 5.3 & 2.5 & 1.7 & 10.8 & 6.8 & 4.4 & 2.8 & 2.3 & 2.0 & 3.7 & 10.5 & 5.3 & 2.8 & 2.1 & 1.8 & 4.5 & 5.9 & 8.4 & 1.6 \\
    \cellcolor{white}{\rotatebox[origin=c]{270}{\makecell[r]{\hspace{-85pt}\textbf{(a) Obj.\,+\,Noise}}}} & \cellcolor{blue!10} \textbf{\ourmodelmtl} & \cellcolor{blue!10} 325M & \cellcolor{blue!10} 25.8 & \cellcolor{blue!10} 11.7 & \cellcolor{blue!10} 4.4 & \cellcolor{blue!10} 2.4 & \cellcolor{blue!10} 1.8 & \cellcolor{blue!10} \textbf{9.2} & \cellcolor{blue!10} 5.9 & \cellcolor{blue!10} 3.6 & \cellcolor{blue!10} 2.5 & \cellcolor{blue!10} 2.1 & \cellcolor{blue!10} 1.8 & \cellcolor{blue!10} \textbf{3.2} & \cellcolor{blue!10} 9.6 & \cellcolor{blue!10} 4.3 & \cellcolor{blue!10} 2.6 & \cellcolor{blue!10} 1.8 & \cellcolor{blue!10} 1.7 & \cellcolor{blue!10} \textbf{4.0} & \cellcolor{blue!10} \textbf{5.1} & \cellcolor{blue!10} \textbf{7.2} & \cellcolor{blue!10} \textbf{1.5} \\
    \midrule
    & V-CAFE & 49M & 57.2 & 31.9 & 13.7 & 7.5 & 5.3 & 23.1 & 45.7 & 29.3 & 17.1 & 10.0 & 6.8 & 21.8 & 32.3 & 18.0 & 9.2 & 6.2 & 5.4 & 14.2 & 18.3 & 26.2 & 4.1 \\
    & RAVEn & 673M & 46.0 & 25.8 & 9.0 & 4.2 & 2.8 & 17.6 & 60.6 & 44.0 & 18.9 & 5.5 & 2.9 & 26.4 & 23.7 & 11.4 & 5.1 & 3.4 & 2.7 & 9.3 & 15.6 & 23.7 & 2.2 \\
    & BRAVEn & 673M & 38.3 & 21.6 & 5.7 & 2.5 & 2.0 & 14.0 & 41.8 & 29.0 & 13.9 & 4.6 & 2.4 & 18.4 & 18.5 & 7.5 & 3.3 & 2.2 & 1.9 & 6.7 & 11.4 & 17.4 & 1.7 \\
    & AV-HuBERT & 325M & 32.0 & 15.5 & 5.3 & 2.7 & 2.0 & 11.5 & 8.1 & 4.2 & 3.0 & 2.3 & 2.0 & 3.9 & 11.5 & 5.3 & 2.9 & 2.3 & 1.8 & 4.8 & 6.2 & 9.0 & 1.6 \\
    & AV-data2vec & 325M & 33.2 & 15.6 & 5.7 & 2.6 & 2.0 & 11.8 & 8.1 & 4.7 & 2.8 & 2.3 & 1.9 & 4.0 & 12.3 & 6.0 & 2.9 & 2.2 & 1.9 & 5.0 & 6.5 & 9.4 & \textbf{1.5} \\
    & AV-RelScore & 437M & 31.7 & 14.5 & 5.1 & 2.7 & 2.0 & 11.2 & 7.7 & 4.2 & 2.6 & 2.2 & 1.9 & 3.7 & 11.2 & 5.3 & 3.1 & 2.1 & 1.7 & 4.7 & 6.1 & 8.7 & 1.6 \\
    \cellcolor{white}{\rotatebox[origin=c]{270}{\makecell[r]{\hspace{-85pt}\textbf{(b) Hands Occ.}}}} & \cellcolor{blue!10} \textbf{\ourmodelmtl} & \cellcolor{blue!10} 325M & \cellcolor{blue!10} 26.6 & \cellcolor{blue!10} 12.4 & \cellcolor{blue!10} 4.5 & \cellcolor{blue!10} 2.6 & \cellcolor{blue!10} 1.8 & \cellcolor{blue!10} \textbf{9.6} & \cellcolor{blue!10} 6.2 & \cellcolor{blue!10} 3.6 & \cellcolor{blue!10} 2.6 & \cellcolor{blue!10} 2.2 & \cellcolor{blue!10} 1.7 & \cellcolor{blue!10} \textbf{3.3} & \cellcolor{blue!10} 9.4 & \cellcolor{blue!10} 4.8 & \cellcolor{blue!10} 2.6 & \cellcolor{blue!10} 1.9 & \cellcolor{blue!10} 1.7 & \cellcolor{blue!10} \textbf{4.1} & \cellcolor{blue!10} \textbf{5.2} & \cellcolor{blue!10} \textbf{7.4} & \cellcolor{blue!10} \textbf{1.5} \\
    \midrule
    & V-CAFE & 49M & 55.4 & 31.8 & 13.7 & 7.5 & 5.3 & 22.7 & 44.4 & 28.3 & 16.8 & 10.1 & 7.0 & 21.3 & 31.8 & 17.9 & 9.4 & 6.4 & 5.2 & 14.1 & 18.1 & 25.7 & 4.2 \\
    & RAVEn & 673M & 43.7 & 25.2 & 8.5 & 3.6 & 2.8 & 16.8 & 60.0 & 42.8 & 18.2 & 5.0 & 3.2 & 25.8 & 23.4 & 10.9 & 5.2 & 3.2 & 2.6 & 9.1 & 15.2 & 23.1 & 2.3 \\
    & BRAVEn & 673M & 39.1 & 21.6 & 5.7 & 2.7 & 1.9 & 14.2 & 42.4 & 31.0 & 13.3 & 4.8 & 2.2 & 18.7 & 18.5 & 7.6 & 3.4 & 2.2 & 1.9 & 6.7 & 11.6 & 17.7 & 1.7 \\
    & AV-HuBERT & 325M & 29.8 & 13.6 & 5.0 & 2.6 & 1.9 & 10.6 & 7.4 & 4.1 & 2.7 & 2.1 & 2.0 & 3.7 & 10.9 & 5.1 & 2.8 & 2.2 & 1.9 & 4.6 & 5.8 & 8.3 & 1.6 \\
    & AV-data2vec & 325M & 30.6 & 14.3 & 5.2 & 2.6 & 2.0 & 10.9 & 7.5 & 4.5 & 3.0 & 2.3 & 2.0 & 3.9 & 10.7 & 5.4 & 2.9 & 2.1 & 1.7 & 4.6 & 6.0 & 8.6 & \textbf{1.5} \\
    & AV-RelScore & 437M & 30.1 & 13.1 & 5.2 & 2.6 & 2.0 & 10.6 & 7.3 & 4.3 & 3.0 & 2.2 & 1.9 & 3.7 & 10.3 & 5.0 & 3.1 & 2.1 & 1.8 & 4.5 & 5.8 & 8.3 & \textbf{1.5} \\
    \cellcolor{white}{\rotatebox[origin=c]{270}{\makecell[r]{\hspace{-75pt}\textbf{(c) Pixelate}}}} & \cellcolor{blue!10} \textbf{\ourmodelmtl} & \cellcolor{blue!10} 325M & \cellcolor{blue!10} 26.0 & \cellcolor{blue!10} 12.0 & \cellcolor{blue!10} 4.7 & \cellcolor{blue!10} 2.5 & \cellcolor{blue!10} 1.9 & \cellcolor{blue!10} \textbf{9.4} & \cellcolor{blue!10} 5.8 & \cellcolor{blue!10} 3.6 & \cellcolor{blue!10} 2.5 & \cellcolor{blue!10} 2.3 & \cellcolor{blue!10} 1.7 & \cellcolor{blue!10} \textbf{3.2} & \cellcolor{blue!10} 9.6 & \cellcolor{blue!10} 4.2 & \cellcolor{blue!10} 2.6 & \cellcolor{blue!10} 1.9 & \cellcolor{blue!10} 1.7 & \cellcolor{blue!10} \textbf{4.0} & \cellcolor{blue!10} \textbf{5.1} & \cellcolor{blue!10} \textbf{7.3} & \cellcolor{blue!10} \textbf{1.5} \\
    \bottomrule 
    \end{tabular}
    }
    \vspace*{10pt}
\end{sidewaystable}

\begin{table*}[!t]
    \centering
    \small
    \caption[CAV2vec performance on corrupted LRS3 with DEMAND noise]{Performance comparison on the LRS3 dataset~\citep{afouras2018lrs3} with audio noise sampled from the DEMAND dataset~\citep{thiemann2013demand}. For each noisy environment, WER\,(\%) is measured by randomly sampling the SNR value from the range [$-$10\,dB, 10\,dB].
    }
    \label{tab:demand_result}
    \addtolength{\tabcolsep}{1pt}
    
    \begin{minipage}[t]{\textwidth}
        \centering
        \resizebox{\textwidth}{!}{
        \begin{tabular}{l|ccccccccc}
        \toprule
        Method 
        & PARK & RIVER & CAFE & RESTO & CAFETER & METRO & STATION & MEETING & \textbf{AVG} \\
        \midrule
        BRAVEn & 4.6 & 7.3 & 6.6 & 14.9 & 8.1 & 3.3 & 6.1 & \!\!13.5\!\! & 8.1 \\
        AV-HuBERT & 3.4 & 4.6 & 5.1 & 10.2 & 5.9 & 2.7 & 4.1 & 3.9 & 5.0 \\
        AV-data2vec & 3.4 & 4.5 & 5.1 & 10.3 & 6.2 & 2.7 & 4.1 & 4.4 & 5.1 \\
        AV-RelScore & 3.4 & 4.5 & 5.1 & 9.3 & 5.4 & 2.8 & 3.9 & 3.8 & 4.8 \\
        \rowcolor{blue!10}
        \textbf{\ourmodelmtl} & 2.8 & 4.3 & 4.4 & 8.4 & 5.1 & 2.3 & 3.8 & 3.5 & \textbf{4.3} \\
        \bottomrule
        \multicolumn{10}{c}{\textbf{(a) Object Occlusion + Noise}}
        \end{tabular}
        }
        \vspace{5pt}
    \end{minipage}
    
    \begin{minipage}[t]{\textwidth}
        \centering
        \resizebox{\textwidth}{!}{
        \begin{tabular}{l|ccccccccc}
        \toprule
        Method 
        & PARK & RIVER & CAFE & RESTO & CAFETER & METRO & STATION & MEETING & \textbf{AVG} \\
        \midrule
        BRAVEn & 4.0 & 7.0 & 5.7 & 13.6 & 8.3 & 3.8 & 6.1 & \!\!12.5\!\! & 7.6 \\
        AV-HuBERT & 3.6 & 5.1 & 5.3 & 11.2 & 6.7 & 2.7 & 4.2 & 4.5 & 5.4 \\
        AV-data2vec & 3.3 & 5.0 & 5.8 & 11.9 & 6.5 & 3.2 & 4.1 & 4.4 & 5.5 \\
        AV-RelScore & 3.0 & 5.2 & 5.1 & 10.8 & 6.2 & 2.8 & 3.8 & 4.6 & 5.2 \\
        \rowcolor{blue!10}
        \textbf{\ourmodelmtl} & 3.0 & 4.0 & 4.0 & 8.9 & 5.1 & 2.7 & 3.4 & 3.5 & \textbf{4.3} \\
        \bottomrule
        \multicolumn{10}{c}{\textbf{(b) Hands Occlusion}}
        \end{tabular}
        }
        \vspace{5pt}
    \end{minipage}
    
    \begin{minipage}[t]{\textwidth}
        \centering
        \resizebox{\textwidth}{!}{
        \begin{tabular}{l|ccccccccc}
        \toprule
        Method 
        & PARK & RIVER & CAFE & RESTO & CAFETER & METRO & STATION & MEETING & \textbf{AVG} \\
        \midrule
        BRAVEn & 4.8 & 7.3 & 6.6 & 15.6 & 8.4 & 3.3 & 5.7 & \!\!12.3\!\! & 8.0 \\
        AV-HuBERT & 3.4 & 4.7 & 4.6 & 10.4 & 6.1 & 2.8 & 3.8 & 3.8 & 4.9 \\
        AV-data2vec & 3.4 & 4.7 & 5.0 & 9.3 & 5.5 & 3.0 & 4.1 & 3.9 & 4.9 \\
        AV-RelScore & 3.2 & 4.9 & 4.6 & 10.0 & 6.1 & 2.7 & 3.7 & 3.9 & 4.9 \\
        \rowcolor{blue!10}
        \textbf{\ourmodelmtl} & 3.0 & 3.9 & 4.5 & 8.6 & 4.6 & 2.4 & 3.3 & 3.6 & \textbf{4.2} \\
        \bottomrule
        \multicolumn{10}{c}{\textbf{(c) Pixelated Face}}
        \end{tabular}
        }
        \vspace{5pt}
    \end{minipage}
    
    \vspace*{10pt}
\end{table*}

\paragraph{LRS3 benchmark results with audio-visual joint corruption.}
Through our experiments, we address two questions: \textit{($i$) does representation learning with corrupted prediction task and multi-task learning approach improve the model's robustness to real-world audio-visual corruption?} and \textit{($ii$) does it guarantee generalizability to even unseen types of corruption?}
The evaluation environments in this study are specifically challenging compared to previous works since there exists audio-visual joint corruption.
In Table\,\ref{tab:main_result}, we present the robust AVSR performance of our proposed model, \ourmodelmtl, which is superior to baseline models in diverse conditions.
AV-HuBERT\,\citep{shi2022robust} and AV-data2vec\,\citep{lian2023av} are based on a multimodal encoder and have been pretrained on noise-augmented audio conditions, which result in outperforming RAVEn\,\citep{haliassos2022jointly} and BRAVEn\,\citep{haliassos2024braven} that utilize two separate encoders for ASR and VSR. RAVEn and BRAVEn particularly suffer in severely corrupted environments, \ie SNR\,$\le$\,0\,dB.
While AV-RelScore is specifically designed to address audio-visual joint corruption by assessing the modality reliability\,\citep{hong2023watch} and outperforms other baselines, it entails additional parameters for a scoring module within the encoder, making it hard to incorporate with pretrained models.

\ourmodelmtl consistently demonstrates superior performance across all visual and audio corruption levels. It surpasses all baseline models under the object occlusion and visual noise condition, achieving an average N-WER of 5.1\% (Table~\ref{tab:main_result}(a)), while AV-data2vec and AV-RelScore obtain 6.2\% and 5.9\%, respectively. Furthermore, \ourmodelmtl shows effectiveness in generalizing to unseen types of corruption, with N-WER of 5.2\% and 5.1\% for (b) hands occlusion and (c) pixelated face, respectively, underscoring its practical applicability in real-world scenarios. Occlusion by hands poses a particularly challenging situation, as the obscured region is larger than that of the COCO objects~\citep{lin2014microsoft}, and there is visual similarity between hands and facial tones. While baseline models struggle with such unseen visual corruption type, showing increases in average N-WER (6.5\% for AV-data2vec and 6.1\% for AV-RelScore), \ourmodelmtl maintains robustness, achieving an N-WER of 5.2\%.

In the audio noise-dominant scenarios, characterized by an SNR value less than or equal to 0\,dB (denoted as N\,$\ge$\,S), it is important to fully leverage visual cues to compensate for impaired speech audio signal. In such conditions, corrupted video inputs can be particularly detrimental if the model is not robust to them. \ourmodelmtl, with its unimodal corrupted prediction tasks, effectively learns cross-modal correlations under corruption, mitigating the recognition errors. We also highlight that our model consistently achieves state-of-the-art results even in the signal-dominant conditions, \ie high SNR values, demonstrating its versatility across varying levels of audio corruption. 
In addition, to validate the model's generalizability to real-world audio corruption, Table\,\ref{tab:demand_result} presents the AVSR performance with DEMAND noise that includes more realistic environments.
\ourmodelmtl effectively achieves robust recognition capabilities in environments like indoor stores or outdoor public space.

\paragraph{LRS2 benchmark results.}

\begin{wraptable}{r}{.5\textwidth}
    \centering
    \small
    \vspace*{-15pt}
    \caption[CAV2vec performance on corrupted LRS2]{Comparisons of WER\,(\%) with our model and prior works on the LRS2 dataset.
    We present N-WER, with babble\,(B), speech\,(S), music\,(M), and natural\,(N) noise types, as well as clean WER results. Visual corruption type is used as object occlusion + noise. 
    }
    \label{tab:lrs2_result}
    \addtolength{\tabcolsep}{2pt}
    \resizebox{.5\textwidth}{!}{
    \begin{tabular}{l|cccc|c}
    \toprule
    Method & B & S & M & N & Clean \\
    \midrule
    AV-HuBERT & 11.6 & 5.3 & 6.1 & 6.0 & 3.0 \\
    AV-data2vec & 11.5 & 5.6 & 6.5 & 6.2 & 3.0 \\
    AV-RelScore & 11.1 & 4.8 & 5.9 & 5.5 & 2.9 \\
    \rowcolor{blue!10}
    \textbf{\ourmodelmtl} & \textbf{8.9} & \textbf{4.4} & \textbf{5.1} & \textbf{4.9} & \textbf{2.7} \\
    \bottomrule 
    \end{tabular}
    }
\end{wraptable}

The LRS2 dataset \citep{son2017lip} consists of 224 hours of BBC video recordings, encompassing a wider variety of scenarios than LRS3, including news delivery, panel discussion, and indoor and outdoor interviews. This diversity allows for a more comprehensive evaluation of the model's generalizability. In Table\,\ref{tab:lrs2_result}, \ourmodelmtl demonstrates strong performance on the corrupted LRS2 benchmark, further validating its robustness in real-world conditions.
We note that all models compared are based on the LRS3-pretrained models, with LRS2 used only in the uptraining or fine-tuning phase.

\section{Analysis}
\label{sec:ch3_analysis}

\subsection{Visualization of Modality Gap}

\begin{figure}[!t]
    \centering
    \includegraphics[width=1\linewidth]{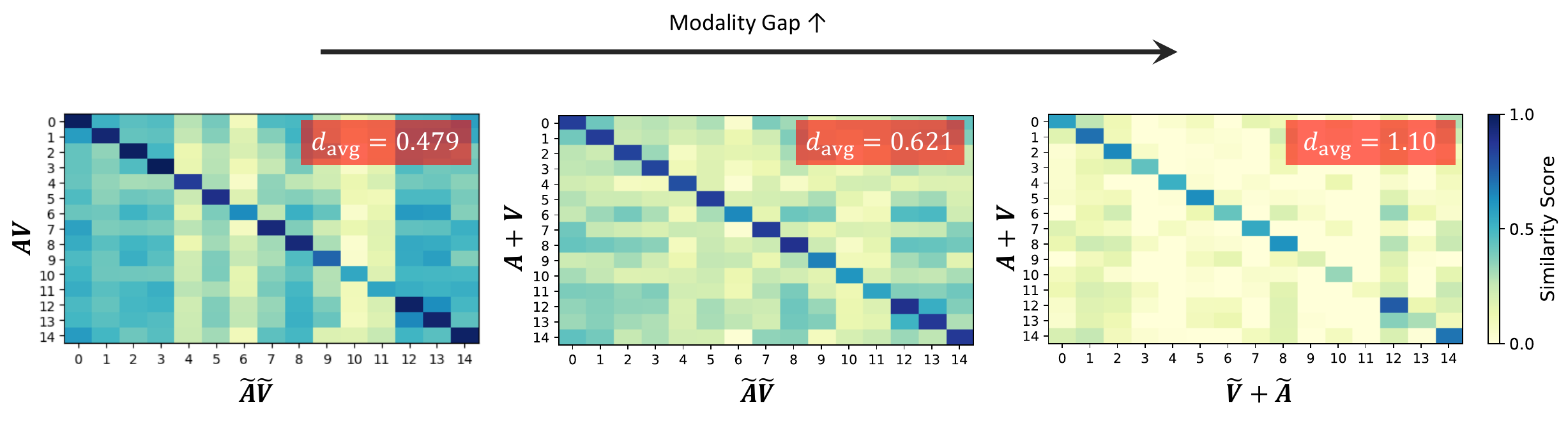}
    \vspace{10pt}
    \includegraphics[width=1\linewidth]{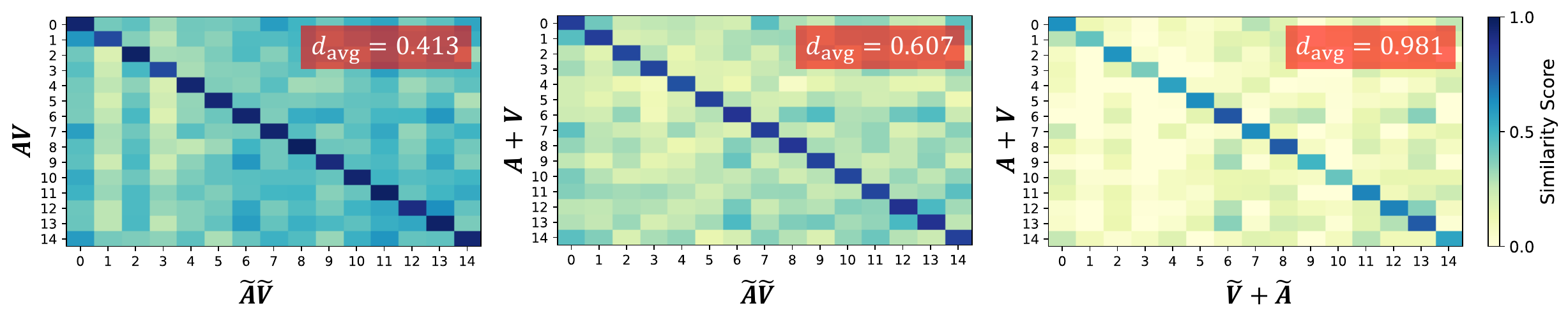}
    \vspace{-30pt}
    \caption[Modality gap analysis across unimodal and multimodal representations]{Visualization of the modality gap of (1st row) AV-data2vec and (2nd row) \ourmodelmtl.
    Each column shows representation comparisons between different modality configurations:
    (1st column) multimodal-to-multimodal comparisons, 
    (2nd column) unimodal-to-multimodal comparisons (averaged across audio-to-multimodal and video-to-multimodal), and
    (3rd column) unimodal-to-unimodal comparisons in a cross-modal manner. $d_\text{avg}$ represents the distance between the average embeddings of each modality.}
    \label{fig:modality_gap_visualization}
    \vspace*{10pt}
\end{figure}

Figure\,\ref{fig:modality_gap_visualization} visualizes the modality gap across different comparisons: multimodal-to-multimodal, unimodal-to-multimodal, and unimodal-to-unimodal. Similar to Figure\,\ref{fig:similarity_matrix_corruption}, we measure the representation similarity scores between sequence features, along with the distance between the average embeddings over 50,000 samples. The diagonal line in the similarity matrix effectively captures the modality gap, as it indicates the distance between different modalities for the same sample. The results demonstrate that the modality gap widens as more unimodal features are introduced, following the intuition that the unimodal-to-unimodal prediction task best improves cross-modal alignment. According to Table\,\ref{tab:ablation_task}, representation learning is enhanced when directly targeting the larger modality gap (\raisebox{.5pt}{\textcircled{\raisebox{-.9pt} {3}}} $>$ \raisebox{.5pt}{\textcircled{\raisebox{-.9pt} {2}}} $>$ \raisebox{.5pt}{\textcircled{\raisebox{-.9pt} {1}}} in Figure\,\ref{fig:overview}). In addition, we present the representation similarity of \ourmodelmtl in the second row of Figure\,\ref{fig:modality_gap_visualization}, observing that the modality gaps are smaller compared to AV-data2vec.

\subsection{Ablation Study for Corrupted Prediction Tasks}

\begin{figure}[!t]
    \centering
    \includegraphics[width=1\linewidth]{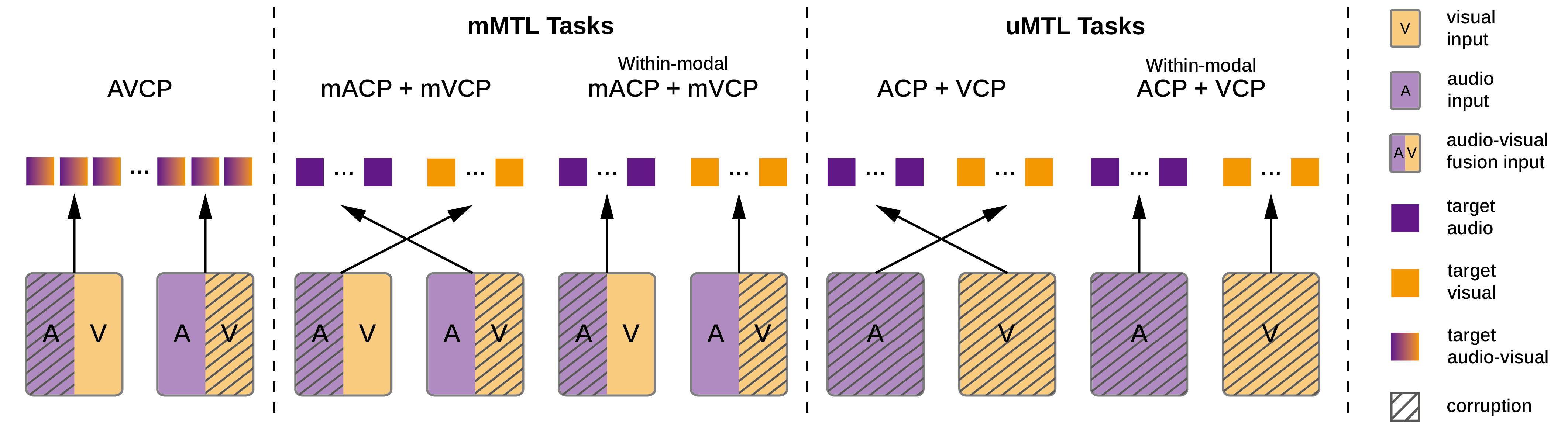}
    \vspace{-20pt}
    \caption[Strategies for corrupted prediction tasks]{Our implemented strategies for corrupted prediction tasks. The AVCP task uses the audio-visual targets. For the multi-task learning\,(MTL) designs that utilize unimodal targets, mACP and mVCP tasks use multimodal inputs (mMTL), while ACP and VCP tasks use unimodal inputs (uMTL). To clearly outline the definitions for each acronym used, these configurations are summarized in Table\,\ref{tab:notations}.}
    \label{fig:ablation_task}
    \vspace*{10pt}
\end{figure}

\begin{table}[!h]
    \centering
    \small
    \caption[Notations for tasks and losses in the CAV2vec framework]{Summary of notations for each task or loss in the CAV2vec framework.}
    \label{tab:notations}
    \resizebox{0.6\textwidth}{!}{
    \addtolength{\tabcolsep}{-1pt}
    \begin{tabular}{llcc}
    \toprule
    Notation & Description & Input modality & Target modality \\
    \midrule
    MP & masked prediction & - & - \\
    CP & corrupted prediction & - & - \\
    \midrule
    AVCP & audio-visual CP & AV & AV \\
    mACP & multimodal audio CP & AV & A \\
    mVCP & multimodal visual CP & AV & V \\
    ACP & unimodal audio CP & V & A \\
    VCP & unimodal visual CP & A & V \\
    \bottomrule
    \end{tabular}
    }
    \vspace{10pt}
\end{table}

\begin{table}[!t]
    \centering
    \small
    \setlength{\fboxsep}{2pt}
    \caption[Ablation study of corrupted prediction tasks in CAV2vec]{Ablation study for the configuration of corrupted prediction tasks. We summarize the AVSR results (average N-WER) across visual corruption types (O: object occlusion + noise, H: hands occlusion, P: pixelate) and audio corruption types (MS: MUSAN and LRS3 noise, DM: DEMAND noise), same evaluation procedures as Tables\,\ref{tab:main_result} and \ref{tab:demand_result}. CRL: corrupted representation learning, mMTL: multimodal multi-task learning, uMTL: unimodal multi-task learning. Refer to Figure\,\ref{fig:ablation_task} for each setting. Note that the masked prediction task is utilized in every setup. 
    \colorbox[HTML]{F7CECD}{Best} and \colorbox[HTML]{D8D8F8}{second best} results.}
    \label{tab:ablation_task}
    \setlength{\fboxsep}{3pt}
    \resizebox{\linewidth}{!}{
    \begin{tabular}{cccl|cc|cc|cc}
    \toprule
    CRL & mMTL & uMTL & Tasks for corruption & O/MS & O/DM & H/MS & H/DM & P/MS & P/DM \\
    \midrule
    \xmark & \xmark & \xmark & - & 6.2 & 5.1 & 6.5 & 5.5 & 6.0 & 4.9 \\
    \midrule
    \cmark & \xmark & \xmark & AVCP & 5.3 & 4.5 & 5.6 & 4.7 & 5.6 & 4.8 \\
    \cmark & \cmark & \xmark & mACP + mVCP & \colorbox[HTML]{F7CECD}{5.1} & \colorbox[HTML]{F7CECD}{4.2} & \colorbox[HTML]{D8D8F8}{5.4} & 4.5 & 5.3 & \colorbox[HTML]{D8D8F8}{4.3} \\
    \cmark & \cmark & \xmark & mACP + mVCP (within-modal) & \colorbox[HTML]{D8D8F8}{5.2} & 4.4 & \colorbox[HTML]{D8D8F8}{5.4} & 4.6 & 5.3 & 4.5 \\
    \cmark & \cmark & \xmark & mACP + mVCP + AVCP & 5.3 & 4.6 & 5.6 & 4.8 & 5.3 & 4.6 \\
    \midrule
    \cmark & \xmark & \cmark & ACP + VCP & \colorbox[HTML]{F7CECD}{5.1} & \colorbox[HTML]{D8D8F8}{4.3} & \colorbox[HTML]{F7CECD}{5.2} & \colorbox[HTML]{F7CECD}{4.3} & \colorbox[HTML]{F7CECD}{5.1} & \colorbox[HTML]{F7CECD}{4.2} \\
    \cmark & \xmark & \cmark & ACP + VCP (within-modal) & \colorbox[HTML]{D8D8F8}{5.2} & 4.6 & \colorbox[HTML]{D8D8F8}{5.4} & 4.7 & \colorbox[HTML]{D8D8F8}{5.2} & 4.4 \\
    {\cmark} & {\xmark} & {\cmark} & {ACP + VCP + AVCP} & \colorbox[HTML]{D8D8F8}{5.2} & 4.4 & 5.6 & 4.6 & 5.3 & 4.6 \\
    \cmark & \cmark & \cmark & mACP + mVCP + ACP + VCP & \colorbox[HTML]{F7CECD}{5.1} & \colorbox[HTML]{D8D8F8}{4.3} & \colorbox[HTML]{F7CECD}{5.2} & \colorbox[HTML]{D8D8F8}{4.4} & \colorbox[HTML]{D8D8F8}{5.2} & \colorbox[HTML]{F7CECD}{4.2} \\
    \bottomrule
    \end{tabular}
    }
    \vspace*{10pt}
\end{table}

In designing our corrupted prediction tasks, we have introduced the AVCP task in Eq.\,(\ref{eq:corrupted_prediction}), along with the unimodal ACP and VCP tasks in Eq.\,(\ref{eq:mtl}). Additionally, we explore the mACP and mVCP tasks for a more comprehensive analysis.
Figure\,\ref{fig:ablation_task} illustrates and compares these tasks, as well as within-modal task designs. The within-modal ACP and VCP tasks are implemented to align the corrupted inputs and targets within the same modality.
Table\,\ref{tab:ablation_task} summarizes the performance results for each task design.

The first observation is that without incorporating the suggested corrupted representation learning, the model struggles to achieve robust speech recognition. Therefore, it is crucial to employ any forms of corrupted prediction task.
When utilizing multimodal inputs, the mACP\,+\,mVCP tasks consistently outperform the AVCP task alone, and even the combination with AVCP. This indicates that distilling knowledge through unimodal targets for the corrupted modality shows effectiveness.
Besides, using unimodal inputs (ACP\,+\,VCP) proves even more effective, as these tasks generally outperform the multimodal input tasks (mACP\,+\,mVCP). 
We also note that within-modal strategies are less effective than cross-modal approaches, as they do not directly target audio-visual correlations.

These findings align with our hypothesis, as depicted in Figure\,\ref{fig:overview}, that directly addressing the cross-modal alignment improves the correlation between the two modalities, which is crucial in generating robust audio-visual features.
Since the masked prediction task is maintained for learning contextualized multimodal representations, the AVCP or mACP\,+\,mVCP tasks become redundant when unimodal tasks are being employed. 
We thus separate the multi-task strategy as leveraging only unimodal inputs for corrupted prediction and multimodal inputs for masked prediction.

\begin{table}[!t]
    \centering
    \small
    \vspace{10pt}
    \caption[Ablation study of corrupted prediction tasks in CAV2vec (full results)]{Ablation study for the configuration of corrupted prediction tasks. We summarize the AVSR results (average N-WER) across visual corruption types (O: object occlusion\,+\,noise, H: hands occlusion, P: pixelate) and audio corruption types (MS: MUSAN and LRS3 noise, DM: DEMAND noise), same evaluation procedures as Tables\,\ref{tab:main_result} and \ref{tab:demand_result}. CRL: corrupted representation learning, mMTL: multimodal multi-task learning, uMTL: unimodal multi-task learning. Refer to Figure\,\ref{fig:ablation_task} for each setting. The masked prediction task is utilized in every setup.
    (w) denotes the within-modal strategies.}
    \label{tab:ablation_task_full}
    \addtolength{\tabcolsep}{-2.5pt}
    \resizebox{\linewidth}{!}{
    \begin{tabular}{cccl|cc|cc|cc}
    \toprule
    CRL & mMTL & uMTL & Tasks for corruption & O/MS & O/DM & H/MS & H/DM & P/MS & P/DM \\
    \midrule
    \xmark & \xmark & \xmark & - & 6.2 & 5.1 & 6.5 & 5.5 & 6.0 & 4.9 \\
    \cmark & \xmark & \xmark & AVCP & 5.3 & 4.5 & 5.6 & 4.7 & 5.6 & 4.8 \\
    \midrule
    \cmark & \cmark & \xmark & mACP + mVCP & \textbf{5.1} & \textbf{4.2} & {5.4} & {4.5} & {5.3} & {4.3} \\
    \cmark & \cmark & \xmark & mACP\,(w) + mVCP\,(w) & 5.2 & 4.4 & {5.4} & {4.6} & {5.3} & {4.5} \\
    \cmark & \cmark & \xmark & mACP + mVCP + AVCP & 5.3 & 4.6 & {5.6} & 4.8 & 5.3 & 4.6  \\
    \midrule
    \rowcolor{blue!10}
    \cmark & \xmark & \cmark & ACP + VCP & \textbf{5.1} & 4.3 & \textbf{5.2} & \textbf{4.3} & \textbf{5.1} & \textbf{4.2} \\
    \cmark & \xmark & \cmark & ACP\,(w) + VCP\,(w) & {5.2} & 4.6 & {5.4} & 4.7 & {5.2} & 4.4 \\
    \cmark & \xmark & \cmark & ACP + VCP + ACP\,(w) & \textbf{5.1} & 4.4 & 5.4 & 4.5 & 5.2 & 4.3 \\
    \cmark & \xmark & \cmark & ACP + VCP + VCP\,(w) & \textbf{5.1} & 4.3 & {5.3} & 4.5 & 5.3 & 4.5 \\ 
    \cmark & \xmark & \cmark & ACP + VCP + ACP\,(w) + VCP\,(w) & \textbf{5.1} & 4.5 & 5.3 & 4.7 & 5.2 & 4.3 \\ 
    \cmark & \xmark & \cmark & ACP + VCP + AVCP & 5.2 & 4.4 & 5.6 & 4.6 & 5.3 & 4.6 \\
    \midrule
    \cmark & \cmark & \cmark & mACP + mVCP + ACP + VCP & \textbf{5.1} & 4.3 & \textbf{5.2} & 4.4 & 5.2 & \textbf{4.2} \\
    \cmark & \cmark & \cmark & mACP\,(w) + mVCP\,(w) + ACP\,(w) + VCP\,(w) & 5.2 & 4.3 & 5.4 & 4.6 & 5.3 & 4.4 \\ 
    \bottomrule
    \end{tabular}
    }
    \vspace{10pt}
\end{table}

In addition to the results presented in Table\,\ref{tab:ablation_task}, we provide complete results for the ablation study on corrupted prediction tasks in Table\,\ref{tab:ablation_task_full}. We have included additional results on within-modal strategies, as \citet{haliassos2022jointly} explored similar training strategies for unimodal encoders. They concluded that, while the VSR student model benefits from using an ASR teacher, the ASR student model requires guidance from both ASR and VSR teachers. This is because the ASR model provides more contextual information, leading to effective training of both encoders.

We can similarly incorporate ACP and VCP tasks in a within-modal manner, which allows us to experiment with an audio-to-audio guidance approach. However, in Table\,\ref{tab:ablation_task_full}, adding the within-modal ACP task has not gained improvement, nor has the within-modal VCP task. 
We find that only using cross-modal ACP\,+\,VCP tasks is sufficient, even surpassing the task configurations with mACP + mVCP tasks.
This is distinct from \citet{haliassos2022jointly} where within-modal strategy is crucial for ASR, while our model is a unified multimodal framework that requires cross-modal knowledge.

\subsection{Sensitivity Study on Corruption Ratios}

\begin{wraptable}{r}{0.55\textwidth}
    \centering
    \small
    \vspace{-10pt}
    \caption[Sensitivity study of corruption ratios in CAV2vec]{Sensitivity study on the corruption ratios in \ourmodelmtl. Object occlusion with Gaussian noise and blurring is used for visual corruption, along with the MUSAN audio corruption, as in Table\,\ref{tab:main_result}\textcolor{red}{a}.
    }
    \label{tab:corruption_ratios}
    \resizebox{0.55\textwidth}{!}{
    \begin{tabular}{cc|c|ccc|c}
    \toprule
    & & & \multicolumn{3}{c|}{N-WER} & \\
    $p_\text{corrupt}^v$ & $p_\text{corrupt}^a$ & $\hat{p}_\text{clean}$ & AVG &  N\,$\ge$\,S &  N\,$<$\,S & Clean \\
    \midrule
    0.1--0.5 & 0.3--0.5 & 0.15 & 5.1 & 7.2 & \textbf{1.9} & \textbf{1.5} \\
    \midrule
    0.1 & 0.3 & 0.22 & 5.3 & 7.4 & 2.0 & 1.6 \\
    0.3 & 0.5 & 0.14 & 5.0 & 7.1 & 2.0 & \textbf{1.5} \\
    0.5 & 0.7 & 0.09 & \textbf{4.7} & \textbf{6.5} & 2.0 & 1.6 \\
    0.7 & 0.9 & 0.04 & \textbf{4.7} & \textbf{6.5} & 2.1 & 1.6 \\
    \bottomrule
    \end{tabular}
    }
\end{wraptable}
We evaluate the model across varying corruption ratios for both audio ($p_\text{corrupt}^a$) and visual ($p_\text{corrupt}^v$) sequences to examine the impact of amount of corruption in learning robustness. As shown in Table\,\ref{tab:corruption_ratios}, higher corruption ratios naturally reduce the empirical proportion of clean inputs ($\hat{p}_\text{clean}$). The result reveals a trade-off between the performance in highly noisy environments (N\,$\ge$\,S) and less corrupted settings (N\,$<$\,S and Clean), depending on the model's exposure to clean or corrupted sequences during training. While higher corruption ratios significantly improve performance in noisy conditions, it is also essential for the AVSR model to maintain reliability in less corrupted scenarios.
To balance this trade-off, we randomly sample the visual corruption ratio from 0.1--0.5 and audio corruption ratio from 0.3--0.5, ensuring a balance between clean and corrupted inputs. 
In terms of masking frames, we follow \citet{shi2022robust, zhang2023self} by using a 0.3 masking ratio for video and 0.8 for audio. For both corruption and masking, higher ratios for audio is necessary as audio holds more critical information in speech. However, we note that masking is applied after corruption, and we do not allow overlap between masked and corrupted frames, which results in effective masking ratios of roughly 0.1 for video and 0.2 for audio. We empirically observed that adjusting the masking ratios has little effect on the \ourmodelmtl's performance.

\subsection{Sensitivity Analysis on Task Loss Coefficients}
\label{appx:loss_coefficients}

\begin{table}[!t]
    \centering
    \small
    \caption[Sensitivity study of task loss coefficients in CAV2vec]{Sensitivity analysis on task loss coefficients, exploring different values for the ACP, VCP, and masked prediction tasks under various noise conditions. We present N-WER for each noise type as well as clean WER results. Visual corruption type is used as object occlusion\,+\,noise.}
    \label{tab:loss_coefficients}
    \resizebox{0.65\textwidth}{!}{
    \begin{tabular}{ccc|cccc|c}
    \toprule
    $\lambda_\text{ACP}$ & $\lambda_\text{VCP}$ & $\lambda_\text{Mask}$ & Babble & Speech & Music & Natural & Clean \\
    \midrule
    0.5 & 0.5 & 1.0 & 9.2 & 3.2 & 4.0 & 4.0 & 1.5 \\
    0.5 & 1.0 & 1.0 & 9.0 & 3.2 & 4.0 & 3.7 & 1.5 \\
    1.0 & 0.5 & 1.0 & 9.0 & 3.1 & 3.9 & 3.9 & 1.4 \\
    \rowcolor{blue!10}
    1.0 & 1.0 & 1.0 & 9.2 & 3.2 & 4.1 & 3.9 & 1.5 \\
    1.0 & 2.0 & 1.0 & 9.3 & 3.2 & 4.2 & 4.1 & 1.6 \\
    2.0 & 1.0 & 1.0 & 9.1 & 3.2 & 3.9 & 3.9 & 1.6 \\
    2.0 & 2.0 & 1.0 & 9.2 & 3.3 & 4.0 & 3.9 & 1.6 \\
    \midrule
    1.0 & 1.0 & 0.5 & 9.1 & 3.1 & 4.1 & 4.0 & 1.6 \\
    \rowcolor{blue!10}
    1.0 & 1.0 & 1.0 & 9.2 & 3.2 & 4.1 & 3.9 & 1.5 \\
    1.0 & 1.0 & 2.0 & 9.4 & 3.2 & 4.1 & 3.9 & 1.6 \\
    \bottomrule
    \end{tabular}
    }
    \vspace*{10pt}
\end{table}

We explore the sensitivity of task loss coefficients for the corrupted prediction and masked prediction tasks, where both tasks employ the self-distillation MSE loss. The loss coefficients for these tasks are initially set to 1.0, matching the scale with that of the MLM-style cross-entropy loss, which is assigned a coefficient of 2.0. Table\,\ref{tab:loss_coefficients} summarizes the result of our experiments varying these loss coefficients within the \ourmodelmtl framework.

Our observation indicates that the model performs consistently well across different coefficient settings. Meanwhile, it is recommended that the masked prediction loss coefficient is not set lower than the coefficients for ACP and VCP tasks. The masked prediction task plays a crucial role in contextualized representation learning.
We also observe that when $\lambda_\text{ACP} = 1.0$ and $\lambda_\text{VCP} = 0.5$, the model outperforms particularly in conditions with low audio noise level. This result may be attributed to the informative nature of the audio target, as similarly reported in \citet{haliassos2024braven}, where stronger audio guidance than visual guidance is utilized for training the ASR model. However, this configuration has resulted in suboptimal performance under more challenging conditions, such as high noise levels or unseen DEMAND noise types.

\subsection{Pretraining from Different Initializations}

In our experiments, we have utilized pretrained AV-HuBERT model to train CAV2vec, since fully training an audio-visual representation learning model from scratch requires substantial resources. For instance, training AV-HuBERT using LRS3\,(433h)\,+\,VoxCeleb2\,(1326h) requires 600K training steps on 64 GPUs, which spans $\sim$4 days to complete \citep{shi2022learning}. Given our limited resources, conducting full pretraining from scratch has not been feasible. Thus, the use of pre-existing weights allowed us to achieve robust performance with minimal additional training cost, demonstrating the efficiency of our method in achieving robustness with fewer resources.

Nevertheless, to investigate the effect of initialization, we compared different pretraining strategies, AV-data2vec and CAV2vec, within our resource constraints. This includes the performance of models with random initialization, pretraining on LRS3\,(433h) for 120K steps with 4 GPUs, using 4 forward passes per update step to compensate for the small batch size.
Table\,\ref{tab:random_init} shows that CAV2vec pretraining with corrupted prediction significantly outperforms AV-data2vec. Furthermore, the performance enhancement gap becomes more pronounced when the models are trained from scratch, implying the potential benefits of our method if more extensive resources were available.

\begin{table}[!t]
    \centering
    \small
    \caption[Analysis of different pretraining initializations]{Comparison between the models pretrained from different initializations.}
    \label{tab:random_init}
    \resizebox{0.9\textwidth}{!}{
    \addtolength{\tabcolsep}{-1pt}
    \begin{tabular}{lll|cc|cc|cc}
    \toprule
    Method & Unlabeled hrs & Init. & O/MS & O/DM & H/MS & H/DM & P/MS & P/DM \\
    \midrule
    AV-data2vec & 433h & random & 10.5 & 8.5 & 11.1 & 9.0 & 10.0 & 8.2 \\
    \rowcolor{blue!10}
    \textbf{CAV2vec} & 433h & random & 8.8 & 7.7 & 9.0 & 7.8 & 8.9 & 7.4 \\
    \midrule
    AV-data2vec & 433h + 1326h & pretrained & 6.2 & 5.1 & 6.5 & 5.5 & 6.0 & 4.9 \\
    \rowcolor{blue!10}
    \textbf{\ourmodelmtl} & 433h + 1326h & pretrained & 5.1 & 4.3 & 5.2 & 4.3 & 5.1 & 4.2 \\
    \bottomrule
    \end{tabular}
    }
    \vspace*{10pt}
\end{table}

\subsection{Comparison between Self-supervised Pretraining with Corrupted Data}

In the framework of CAV2vec, we designed the corrupted prediction tasks to highlight the importance of robust representation learning, and demonstrated the advantages of a unimodal strategy in corrupted representation learning. To dissect the impact of data corruption and CAV2vec’s training by corrupted prediction tasks, other SSL methods could also be trained on corrupted data during the representation learning. We conducted uptraining within the AV-HuBERT and AV-data2vec pretraining frameworks, using the same data corruptions and same number of training steps as CAV2vec. As demonstrated in Table\,\ref{tab:uptrain_corruption}, the incorporation of corrupted prediction tasks in CAV2vec is critical in robust representation learning with corrupted data, highlighting its effectiveness compared to other methods which do not gain large robustness through corrupted data augmentation.

\begin{table}[!t]
    \centering
    \small
    \caption[Analysis of self-supervised frameworks pretrained on corrupted data]{Comparison between different self-supervised pretraining frameworks under corrupted environments. When pretraining (PT) with corrupted data, we use the same training and data configurations across all models.}
    \label{tab:uptrain_corruption}
    \addtolength{\tabcolsep}{-2.5pt}
    \resizebox{.8\linewidth}{!}{
    \begin{tabular}{lc|cc|cc|cc}
    \toprule
    Method & PT w/ corrupted data & O/MS & O/DM & H/MS & H/DM & P/MS & P/DM \\
    \midrule
    AV-HuBERT & \xmark & 6.0 & 5.0 & 6.2 & 5.4 & 5.8 & 4.9 \\
    AV-data2vec & \xmark & 6.2 & 5.1 & 6.5 & 5.5 & 6.0 & 4.9 \\
    \midrule
    AV-HuBERT & \cmark & 5.8 & 4.9 & 5.9 & 5.1 & 5.8 & 4.7 \\
    AV-data2vec & \cmark & 5.8 & 4.9 & 6.0 & 5.0 & 5.9 & 4.8 \\
    \rowcolor{blue!10}
    \textbf{\ourmodelmtl} & \cmark & 5.1 & 4.3 & 5.2 & 4.3 & 5.1 & 4.2 \\
    \bottomrule
    \end{tabular}
    }
    \vspace*{10pt}
\end{table}

\section{Additional Results}
\label{sec:ch3_additional_results}

\subsection{Additional DEMAND Noise Types}
\label{appx:additional_demand}

The original DEMAND dataset~\citep{thiemann2013demand} contains 18 categories of recorded environments. In Table\,\ref{tab:demand_result}, we have presented results for 8 out of these categories, specifically excluding relatively quieter environments. In Table\,\ref{tab:demand_result_full}, we provide full results across all 18 categories, which include the following noise environments: park, river, cafe, restaurant, cafeteria, metro (subway), public station, meeting room, kitchen, living room, washroom, sports field, hallway, office, public square, traffic intersection, bus, and car.
Still, \ourmodelmtl generally outperforms the baselines in the remaining, relatively less noisy environments.

\begin{sidewaystable}
    \centering
    \small
    \caption[CAV2vec performance on corrupted LRS3 with DEMAND noise (full noise results)]{
    Performance comparison on the LRS3 dataset~\citep{afouras2018lrs3} with audio noise sampled from the DEMAND dataset~\citep{thiemann2013demand}. For each noisy environment, WER\,(\%) is measured by randomly sampling the SNR value from the range [$-$10\,dB, 10\,dB].
    }
    \label{tab:demand_result_full}
    \addtolength{\tabcolsep}{-1pt}
    \resizebox{\textwidth}{!}{
    \begin{tabular}{l|cccccccccccccccccc|c}
    \toprule
    Method & \rotatebox[origin=r]{315}{PARK\!\!\!} & \rotatebox[origin=r]{315}{RIVER\!\!\!} & \rotatebox[origin=r]{315}{CAFE\!\!\!} & \rotatebox[origin=r]{315}{RESTO\!\!\!} & \rotatebox[origin=r]{315}{CAFETER\!\!\!} & \rotatebox[origin=r]{315}{METRO\!\!\!} & \rotatebox[origin=r]{315}{STATION\!\!\!} & \rotatebox[origin=r]{315}{MEETING\!\!\!} & \rotatebox[origin=r]{315}{KITCHEN\!\!\!} & \rotatebox[origin=r]{315}{LIVING\!\!\!} & \rotatebox[origin=r]{315}{WASH\!\!\!} & \rotatebox[origin=r]{315}{FIELD\!\!\!} & \rotatebox[origin=r]{315}{HALL\!\!\!} & \rotatebox[origin=r]{315}{OFFICE\!\!\!} & \rotatebox[origin=r]{315}{SQUARE\!\!\!} & \rotatebox[origin=r]{315}{TRAFFIC\!\!\!} & \rotatebox[origin=r]{315}{BUS\!\!\!} & \rotatebox[origin=r]{315}{CAR\!\!\!} & \rotatebox[origin=r]{315}{\textbf{AVG}\!\!\!} \\
    \midrule
    BRAVEn & 4.0 & 7.0 & 5.7 & 13.6 & 8.3 & 3.8 & 6.1 & 12.5 & 2.2 & 3.9 & 1.7 & 1.9 & 2.2 & 1.8 & 2.9 & 3.1 & 2.0 & 1.8 & 4.7 \\
    AV-HuBERT & 3.6 & 5.1 & 5.3 & 11.2 & 6.7 & \textbf{2.7} & 4.2 & 4.5 & 2.2 & 3.2 & 1.6 & \textbf{1.8} & \textbf{1.9} & 1.8 & 2.6 & \textbf{2.5} & 2.0 & 1.6 & 3.6 \\
    AV-data2vec & 3.3 & 5.0 & 5.8 & 11.9 & 6.5 & 3.2 & 4.1 & 4.4 & 2.2 & 3.2 & 1.6 & \textbf{1.8} & 2.2 & 1.8 & {2.5} & 2.7 & \textbf{1.9} & \textbf{1.5} & 3.6 \\
    AV-RelScore & \textbf{3.0} & 5.2 & 5.1 & 10.8 & 6.2 & 2.8 & 3.8 & 4.6 & 2.2 & 3.2 & 1.7 & 1.9 & 2.1 & 1.7 & 2.2 & 3.0 & 2.1 & 1.7 & 3.5 \\
    \rowcolor{blue!10}
    \textbf{\ourmodelmtl} & \textbf{3.0} & \textbf{4.0} & \textbf{4.0} & \textbf{8.9} & \textbf{5.1} & \textbf{2.7} & \textbf{3.4} & \textbf{3.5} & \textbf{1.8} & \textbf{2.8} & \textbf{1.5} & {1.9} & \textbf{1.9} & \textbf{1.6} & \textbf{2.1} & {2.6} & \textbf{1.9} & {1.6} & \textbf{3.0} \\
    \bottomrule
    \multicolumn{20}{c}{\textbf{(a) Hands Occlusion}} \\[5pt]
    \midrule
    BRAVEn & 4.8 & 7.3 & 6.6 & 15.6 & 8.4 & 3.3 & 5.7 & 12.3 & 2.5 & 3.9 & \textbf{1.7} & \textbf{1.7} & 2.2 & 1.8 & 2.8 & 3.2 & 2.1 & 1.7 & 4.9 \\
    AV-HuBERT & 3.4 & 4.7 & 4.6 & 10.4 & 6.1 & 2.8 & 3.8 & 3.8 & 2.1 & 3.0 & \textbf{1.7} & \textbf{1.7} & 1.9 & {1.7} & 2.5 & 2.4 & 1.8 & 1.6 & 3.3 \\
    AV-data2vec & 3.4 & 4.7 & 5.0 & 9.3 & 5.5 & 3.0 & 4.1 & 3.9 & 2.1 & 3.2 & \textbf{1.7} & {1.8} & 2.1 & \textbf{1.6} & 2.3 & 2.6 & 1.8 & \textbf{1.5} & 3.3 \\
    AV-RelScore & 3.2 & 4.9 & 4.6 & 10.0 & 6.1 & 2.7 & 3.7 & 3.9 & \textbf{1.9} & 3.0 & 1.8 & 1.8 & 2.0 & 1.8 & 2.4 & \textbf{2.3} & 1.8 & 1.7 & 3.3 \\
    \rowcolor{blue!10}
    \textbf{\ourmodelmtl} & \textbf{3.0} & \textbf{3.9} & \textbf{4.5} & \textbf{8.6} & \textbf{4.6} & \textbf{2.4} & \textbf{3.3} & \textbf{3.6} & \textbf{1.9} & \textbf{2.6} & \textbf{1.7} & \textbf{1.7} & \textbf{1.8} & {1.7} & \textbf{2.2} & {2.4} & \textbf{1.7} & \textbf{1.5} & \textbf{2.9} \\
    \bottomrule
    \multicolumn{20}{c}{\textbf{(b) Pixelated Face}} \\
    \end{tabular}
    }
    \vspace*{10pt}
\end{sidewaystable}

\begin{sidewaystable}
    \centering
    \small
    \vspace{5pt}
    \caption[CAV2vec performance on corrupted LRS2 (full SNR results)]{Comparisons of WER\,(\%) with our model and prior works on the LRS2 dataset\,\citep{son2017lip}. We follow the experimental setup as in Table\,\ref{tab:main_result}.}
    \label{tab:lrs2_full_result}
    \addtolength{\tabcolsep}{-1pt}
    \renewcommand{\arraystretch}{1.1}
    \resizebox{\textwidth}{!}{
    \begin{tabular}{c|l|l|cccccc|cccccc|cccccc|cc|c}
    \toprule
    & \multirow{2}{*}{Method} & \multirow{2}{*}{Params} & \multicolumn{6}{c|}{Babble, SNR (dB) $=$} & \multicolumn{6}{c|}{Speech, SNR (dB) $=$} & \multicolumn{6}{c|}{Music + Natural, SNR (dB) $=$} & \multicolumn{2}{c|}{N-WER} & Clean \\
    & & & -10 & -5 & 0 & 5 & 10 & \!\!\textbf{AVG}\!\! & -10 & -5 & 0 & 5 & 10 & \!\!\textbf{AVG}\!\! & -10 & -5 & 0 & 5 & 10 & \!\!\textbf{AVG}\!\! & \textbf{AVG} & N\,$\ge$\,S & $\infty$ \\
    \midrule
    & AV-HuBERT & 325M & 28.8 & 15.2 & 6.7 & 3.9 & 3.5 & 11.6 & 9.4 & 5.6 & 4.3 & 3.6 & 3.4 & 5.3 & 11.7 & 6.8 & 4.7 & 3.7 & 3.3 & 6.0 & 7.2 & 9.7 & 3.0 \\
    & AV-data2vec & 325M & 28.1 & 14.8 & 6.8 & 4.3 & 3.7 & 11.5 & 9.6 & 6.2 & 5.0 & 4.0 & 3.5 & 5.6 & 12.3 & 7.2 & 5.1 & 3.8 & 3.4 & 6.4 & 7.5 & 10.0 & 3.0 \\
    & AV-RelScore & 437M & 28.5 & 14.3 & 5.8 & 3.5 & 3.2 & 11.1 & 8.3 & 5.4 & 4.1 & 3.3 & 2.9 & 4.8 & 11.9 & 6.3 & 4.1 & 3.3 & 2.9 & 5.7 & 6.8 & 9.2 & 2.9 \\
    \cellcolor{white}{\rotatebox[origin=c]{270}{\makecell[r]{\hspace{-50pt}\textbf{(a) Object}}}} & \cellcolor{blue!10} \textbf{\ourmodelmtl} & \cellcolor{blue!10} 325M & \cellcolor{blue!10} 21.2 & \cellcolor{blue!10} 10.9 & \cellcolor{blue!10} 5.5 & \cellcolor{blue!10} 3.7 & \cellcolor{blue!10} 3.1 & \cellcolor{blue!10} \textbf{8.9} & \cellcolor{blue!10} 7.1 & \cellcolor{blue!10} 4.9 & \cellcolor{blue!10} 3.6 & \cellcolor{blue!10} 3.2 & \cellcolor{blue!10} 3.2 & \cellcolor{blue!10} \textbf{4.4} & \cellcolor{blue!10} 9.4 & \cellcolor{blue!10} 5.5 & \cellcolor{blue!10} 3.9 & \cellcolor{blue!10} 3.3 & \cellcolor{blue!10} 3.0 & \cellcolor{blue!10} \textbf{5.0} & \cellcolor{blue!10} \textbf{5.8} & \cellcolor{blue!10} \textbf{7.5} & \cellcolor{blue!10} \textbf{2.7} \\
    \midrule
    & AV-HuBERT & 325M & 28.9 & 15.0 & 7.0 & 4.1 & 3.7 & 11.7 & 8.6 & 5.7 & 4.2 & 3.7 & 3.4 & 5.1 & 12.7 & 7.1 & 4.5 & 3.6 & 3.2 & 6.2 & 7.3 & 9.8 & 3.0 \\
    & AV-data2vec & 325M & 30.2 & 15.2 & 7.4 & 4.3 & 3.4 & 12.1 & 9.7 & 5.9 & 4.6 & 3.8 & 3.7 & 5.5 & 13.3 & 7.2 & 4.7 & 3.8 & 3.5 & 6.5 & 7.6 & 10.3 & 3.2 \\
    & AV-RelScore & 437M & 34.3 & 15.7 & 6.4 & 3.7 & 3.1 & 12.6 & 10.0 & 6.2 & 4.0 & 3.5 & 3.0 & 5.3 & 14.2 & 7.2 & 4.5 & 3.4 & 3.0 & 6.5 & 7.7 & 10.7 & \textbf{2.7} \\
    \cellcolor{white}{\rotatebox[origin=c]{270}{\makecell[r]{\hspace{-50pt}\textbf{(b) Hands}}}} & \cellcolor{blue!10} \textbf{\ourmodelmtl} & \cellcolor{blue!10} 325M & \cellcolor{blue!10} 21.2 & \cellcolor{blue!10} 12.2 & \cellcolor{blue!10} 6.4 & \cellcolor{blue!10} 3.7 & \cellcolor{blue!10} 3.2 & \cellcolor{blue!10} \textbf{9.3} & \cellcolor{blue!10} 7.6 & \cellcolor{blue!10} 5.8 & \cellcolor{blue!10} 4.0 & \cellcolor{blue!10} 3.5 & \cellcolor{blue!10} 3.2 & \cellcolor{blue!10} \textbf{4.8} & \cellcolor{blue!10} 10.4 & \cellcolor{blue!10} 6.3 & \cellcolor{blue!10} 4.5 & \cellcolor{blue!10} 3.6 & \cellcolor{blue!10} 3.5 & \cellcolor{blue!10} \textbf{5.7} & \cellcolor{blue!10} \textbf{6.4} & \cellcolor{blue!10} \textbf{8.3} & \cellcolor{blue!10} \textbf{2.7} \\
    \midrule
    & AV-HuBERT & 325M & 28.3 & 14.7 & 6.1 & 4.1 & 3.5 & 11.4 & 9.2 & 5.8 & 4.0 & 3.8 & 3.3 & 5.2 & 12.4 & 7.4 & 4.6 & 3.6 & 3.2 & 6.2 & 7.3 & 9.7 & 2.9 \\
    & AV-data2vec & 325M & 28.9 & 15.3 & 6.9 & 4.6 & 3.6 & 11.9 & 9.6 & 6.4 & 4.9 & 4.0 & 3.4 & 5.6 & 12.1 & 6.7 & 4.8 & 3.8 & 3.4 & 6.1 & 7.4 & 9.9 & 3.1 \\
    & AV-RelScore & 437M & 34.1 & 16.0 & 6.1 & 4.0 & 3.2 & 12.7 & 9.9 & 5.6 & 4.2 & 3.4 & 3.1 & 5.3 & 14.0 & 6.8 & 4.3 & 3.3 & 3.0 & 6.3 & 7.6 & 10.5 & \textbf{2.7} \\
    \cellcolor{white}{\rotatebox[origin=c]{270}{\makecell[r]{\hspace{-50pt}\textbf{(c)\,Pixelate}}}} & \cellcolor{blue!10} \textbf{\ourmodelmtl} & \cellcolor{blue!10} 325M & \cellcolor{blue!10} 22.2 & \cellcolor{blue!10} 12.5 & \cellcolor{blue!10} 6.2 & \cellcolor{blue!10} 4.3 & \cellcolor{blue!10} 3.3 & \cellcolor{blue!10} \textbf{9.7} & \cellcolor{blue!10} 7.9 & \cellcolor{blue!10} 5.6 & \cellcolor{blue!10} 4.4 & \cellcolor{blue!10} 3.5 & \cellcolor{blue!10} 3.2 & \cellcolor{blue!10} \textbf{4.9} & \cellcolor{blue!10} 10.0 & \cellcolor{blue!10} 6.6 & \cellcolor{blue!10} 4.3 & \cellcolor{blue!10} 3.5 & \cellcolor{blue!10} 3.4 & \cellcolor{blue!10} \textbf{5.6} & \cellcolor{blue!10} \textbf{6.4} & \cellcolor{blue!10} \textbf{8.4} & \cellcolor{blue!10} \textbf{2.7} \\
    \bottomrule 
    \end{tabular}
    }
    \vspace*{10pt}
\end{sidewaystable}

\begin{table*}[!t]
    \centering
    \small
    \caption[CAV2vec performance on corrupted LRS2 with DEMAND noise]{Performance comparison on the LRS2 dataset~\citep{son2017lip} with audio noise sampled from the DEMAND dataset~\citep{thiemann2013demand}. We follow the experimental setup as in Table\,\ref{tab:demand_result}.
    }
    \label{tab:lrs2_demand_result}
    \addtolength{\tabcolsep}{1pt}
    
    \begin{minipage}[t]{\textwidth}
        \centering
        \resizebox{\textwidth}{!}{
        \begin{tabular}{l|ccccccccc}
        \toprule
        Method 
        & PARK & RIVER & CAFE & RESTO & CAFETER & METRO & STATION & MEETING & \textbf{AVG} \\
        \midrule
        AV-HuBERT & 4.8 & 5.2 & 5.8 & 11.6 & 7.1 & 4.2 & 5.5 & 6.0 & 6.3 \\
        AV-data2vec & 5.0 & 6.6 & 7.0 & 11.7 & 6.9 & 4.5 & 5.4 & 6.2 & 6.7 \\
        AV-RelScore & 4.7 & 5.9 & 6.1 & 12.2 & 7.5 & 3.8 & 5.0 & 6.5 & 6.5 \\
        \rowcolor{blue!10}
        \textbf{\ourmodelmtl} & 5.1 & 5.7 & 6.1 & 9.4 & 6.8 & 4.0 & 4.9 & 5.4 & \textbf{5.9} \\
        \bottomrule
        \multicolumn{10}{c}{\textbf{(a) Object Occlusion + Noise}}
        \end{tabular}
        }
        \vspace{5pt}
    \end{minipage}
    
    \begin{minipage}[t]{\textwidth}
        \centering
        \resizebox{\textwidth}{!}{
        \begin{tabular}{l|ccccccccc}
        \toprule
        Method 
        & PARK & RIVER & CAFE & RESTO & CAFETER & METRO & STATION & MEETING & \textbf{AVG} \\
        \midrule
        AV-HuBERT & 5.1 & 5.7 & 6.5 & 12.5 & 7.6 & 4.2 & 5.4 & 6.2 & 6.6 \\
        AV-data2vec & 5.2 & 6.0 & 7.7 & 12.4 & 7.6 & 4.4 & 5.9 & 6.3 & 6.9 \\
        AV-RelScore & 5.0 & 5.9 & 6.7 & 12.9 & 8.8 & 4.6 & 5.0 & 6.5 & 6.9 \\
        \rowcolor{blue!10}
        \textbf{\ourmodelmtl} & 4.6 & 5.6 & 6.0 & 9.6 & 6.7 & 3.7 & 5.2 & 5.0 & \textbf{5.8} \\
        \bottomrule
        \multicolumn{10}{c}{\textbf{(b) Hands Occlusion}}
        \end{tabular}
        }
        \vspace{5pt}
    \end{minipage}
    
    \begin{minipage}[t]{\textwidth}
        \centering
        \resizebox{\textwidth}{!}{
        \begin{tabular}{l|ccccccccc}
        \toprule
        Method 
        & PARK & RIVER & CAFE & RESTO & CAFETER & METRO & STATION & MEETING & \textbf{AVG} \\
        \midrule
        AV-HuBERT & 4.6 & 6.5 & 6.4 & 10.1 & 7.4 & 4.1 & 5.0 & 6.3 & 6.3 \\
        AV-data2vec & 5.1 & 5.7 & 6.9 & 10.9 & 7.4 & 4.2 & 5.5 & 6.1 & 6.5 \\
        AV-RelScore & 5.1 & 5.8 & 6.7 & 11.8 & 7.5 & 4.0 & 5.2 & 6.4 & 6.6 \\
        \rowcolor{blue!10}
        \textbf{\ourmodelmtl} & 5.0 & 5.7 & 6.3 & 10.0 & 6.5 & 4.1 & 5.4 & 5.3 & \textbf{6.0} \\
        \bottomrule
        \multicolumn{10}{c}{\textbf{(c) Pixelated Face}}
        \end{tabular}
        }
        \vspace{5pt}
    \end{minipage}
    
    \vspace*{10pt}
\end{table*}

\subsection{Full Results of LRS2 Evaluation}

In Tables\,\ref{tab:lrs2_full_result} and \ref{tab:lrs2_demand_result}, we show the full results of our evaluation on the LRS2 benchmark \citep{son2017lip}, with the same experimental setup as LRS3 evaluation. CAV2vec outperforms the baselines across all audio-visual corruption, effectively demonstrating the model’s robustness under real-world conditions.

\subsection{ASR and VSR Results}

\begin{table}[!ht]
    \centering
    \small
    \caption[CAV2vec performance of unimodal tasks on LRS3]{ASR and VSR task results on the LRS3 benchmark.
    }
    \label{tab:asr_vsr}
    \addtolength{\tabcolsep}{-1pt}
    \resizebox{.8\textwidth}{!}{
    \begin{tabular}{l|ccccc|cccc}
    \toprule
    & \multicolumn{5}{c|}{\textbf{ASR}} & \multicolumn{4}{c}{\textbf{VSR}} \\ 
    Method & Babble & Speech & Music & Natural & Clean & Object & Hands & Pixelate & Clean \\
    \midrule
    AV-HuBERT & 35.8 & 22.6 & 13.9 & 12.8 & 1.6 & 34.9 & 37.2 & 35.6 & 28.7 \\
    AV-RelScore & 36.2 & 22.5 & 15.0 & 13.5 & 1.7 & 34.2 & 37.7 & 36.0 & 28.6 \\
    \rowcolor{blue!10}
    \textbf{\ourmodelmtl} & 35.2 & 20.3 & 13.6 & 12.3 & 1.5 & 33.9 & 36.6 & 35.6 & 27.9 \\
    \bottomrule 
    \end{tabular}
    }
    \vspace*{10pt}
\end{table}

Completely missing modalities are common in real-world scenarios, and evaluating the single-modality performance, \ie ASR and VSR, under corrupted conditions is also essential. In Table\,\ref{tab:asr_vsr}, we evaluated ASR and VSR tasks, comparing three models: AV-HuBERT, AV-RelScore, and CAV2vec. For ASR, we used audio corruptions such as babble, speech, music, and natural noises (N-WER), and for VSR, we used object occlusion with noise, hands occlusion, and pixelated face. We also report results in clean conditions.
The results confirm that our CAV2vec framework robustly works with both unimodal and multimodal data across all conditions, demonstrating its effectiveness in producing robust representations even when modalities are partially or completely corrupted.

\section{Chapter Summary}

In this chapter, we presented \ourmodelmtl, a novel audio-visual representation learning framework designed to address the challenges of audio-visual joint corruption in speech recognition. By employing a corrupted prediction task, \ourmodelmtl enhances multimodal robustness, while incorporating a unimodal multi-task learning strategy to improve cross-modal alignment shows effectiveness. Our experiments on robust AVSR benchmarks including LRS3 and LRS2 demonstrate that \ourmodelmtl significantly outperforms existing baseline models, consistently exhibiting superior performance across a variety of corrupted environments. Particularly in challenging audio noise-dominant scenarios, \ourmodelmtl effectively aligns corrupted modalities, leading to more reliable and robust audio-visual fusion. Additionally, our model showcases strong generalization abilities, achieving state-of-the-art results even in unseen corruption types such as pixelated faces with public audio noise. These findings establish \ourmodelmtl as a robust, adaptable framework for handling corrupted audio-visual data, setting a new benchmark for multimodal speech recognition systems.

\chapter{Architecture-Level Scalability of AVSR}
\label{chap:architecture_scalability}

\begin{center}
    \vspace{0.5em}
    \begin{tcolorbox}[
        colback=gray!10, 
        colframe=gray!50, 
        boxrule=0.5pt,
        title=Summary: Chapter based on work published at ICML 2025~\citep{kim2025mohave}, fonttitle=\bfseries,
        width=0.92\linewidth,
        coltitle=black
    ]
    Audio-visual speech recognition (AVSR) has become critical for enhancing speech recognition in noisy environments by integrating both auditory and visual modalities. However, existing AVSR systems struggle to scale up without compromising computational efficiency. In this chapter, we introduce \textbf{MoHAVE (Mixture of Hierarchical Audio-Visual Experts)}, a novel robust AVSR framework designed to address these scalability constraints. By leveraging a Mixture-of-Experts (MoE) architecture, MoHAVE activates modality-specific expert groups, ensuring dynamic adaptation to various audio-visual inputs with minimal computational overhead. Key contributions of MoHAVE include: (1)\,a sparse MoE framework that efficiently scales AVSR model capacity, (2)\,a hierarchical gating mechanism that dynamically utilizes the expert groups based on input context, enhancing adaptability and robustness, and (3)\,remarkable performance across robust AVSR benchmarks, including LRS3 and MuAViC transcription and translation tasks, setting a new standard for scalable speech recognition systems.
    \end{tcolorbox}
    \vspace{0.5em}
\end{center}

\section{Mixture of Hierarchical Audio-Visual Experts for Robust Speech Recognition}
\label{sec:ch4_intro}

Audio-visual speech recognition (AVSR)~\citep{noda2015audio, afouras2018deep, ma2021end, shi2022learning, hsu2022u, hu2023mir} has emerged as a pivotal technology in enhancing the robustness and accuracy of speech recognition systems, particularly in noisy environments. By integrating auditory and visual modalities, AVSR leverages the complementary information from both speech signals and lip movements~\citep{makino2019recurrent, ren2021learning, chen2023leveraging}, offering significant advantages over audio-only automatic speech recognition\,(ASR) approaches. This multimodal approach is indispensable in situations where auditory data alone is insufficient for reliable recognition.

Despite significant advances, AVSR systems have not kept pace with advancements in model scalability as seen in ASR \citep{radford2023robust} or large language models (LLMs) \citep{kaplan2020scaling, clark2022unified, achiam2023gpt}. Contemporary AVSR models, such as AV-HuBERT \citep{shi2022learning}, AV-data2vec \citep{lian2023av}, and Auto-AVSR \citep{ma2023auto}, generally employ fewer than 0.5B parameters, a stark contrast to large-scale ASR models like Whisper~\citep{radford2023robust} or Seamless~\citep{barrault2023seamlessm4t} which boasts up to 1.6B and 2.3B parameters, respectively. This disparity is not merely a matter of size but reflects a fundamental challenge in AVSR scalability: increasing the model size often disproportionately enhances audio semantic understanding without similarly improving visual processing capabilities~\citep{dai2024study, kim2024learning}. Moreover, the computational complexity and latency of larger models pose challenges for efficient deployment, especially in scenarios where AVSR users often require rapid processing and low latency. These factors make the integration of larger, more computationally intensive models impractical for many real-world applications.

To address the scalability challenges in AVSR systems, we leverage a sparse Mixture-of-Experts (MoE) architecture~\citep{shazeer2017outrageously, fedus2022switch} that activates only a subset of parameters (\ie experts) for efficient scaling of model capacity. Furthermore, recognizing the inherent bias in AVSR systems toward the audio modality, we find it essential to harness the full potential of both audio and video data. One approach is expert group specialization, also known as \textit{hard routing}~\citep{zhu2022uni, li2023pace, lee2025moai}, which assigns specific roles to expert groups and manually activates them based on input types. While effective, this fixed activation strategy lacks adaptability, making it sub-optimal for AVSR where noise conditions and modality reliability vary. A more flexible routing mechanism is needed to dynamically utilize expert groups.

\begin{figure}[!t]
    \centering
    \includegraphics[width=.7\linewidth]{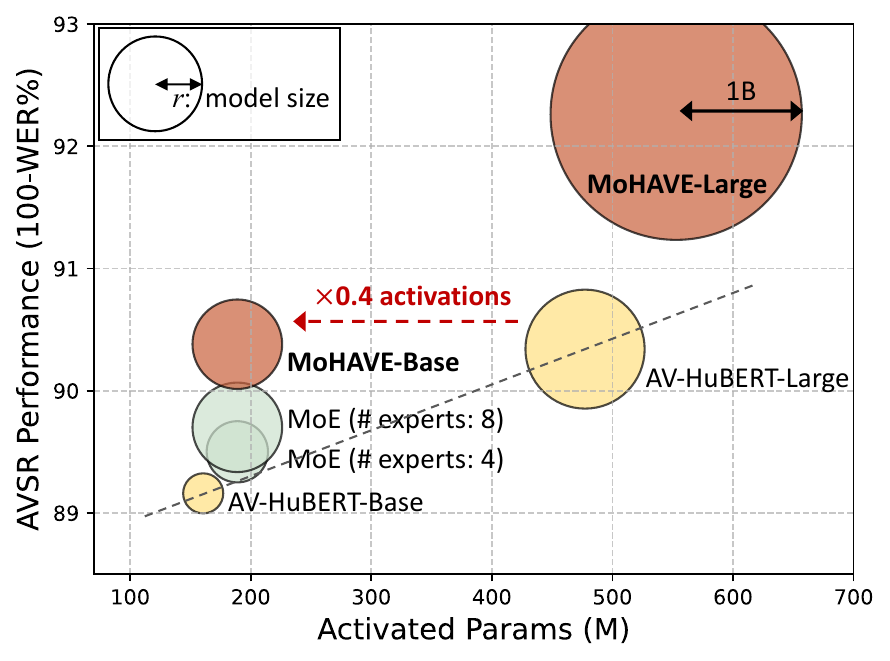}
    \caption[AVSR performance scaling with activated parameters for different model architectures]{Comparison of AVSR models based on standard Transformers (AV-HuBERT, \citealt{shi2022learning}), MoE, and MoHAVE, evaluated under babble noise.
    The MoE structure boosts the model capacity while maintaining the number of activations.
    \ourmodel-\textsc{Base}\,(359M) achieves similar performance to AV-HuBERT-\textsc{Large}\,(477M) while activating only 189M parameters.
    }
    \label{fig:figure1}
    \vspace*{10pt}
\end{figure}

We thus propose a novel MoE framework, \textbf{MoHAVE}\footnote{MoHAVE is pronounced as \textit{Mojave} Desert.} (\underline{M}ixture \underline{o}f \underline{H}ierarchical \underline{A}udio-\underline{V}isual \underline{E}xperts), which employs a hierarchical gating mechanism with two-layer routers.
MoHAVE introduces an inter-modal router that makes decision on utilizing audio and visual expert groups (\S\ref{subsec:hierarchical_gating}). This dynamic routing adapts to input characteristics, specifically trained by our novel load biasing loss\,(\S\ref{subsec:load_biasing_loss}).
MoHAVE achieves state-of-the-art performance\,(\S\ref{subsec:ch4_results}) on the noisy LRS3 benchmark\,\citep{afouras2018lrs3} and in multilingual tasks\,\citep{anwar2023muavic}. 
Empirical analysis shows that MoHAVE adaptively adjusts token distribution based on input context\,(\S\ref{sec:analysis_load}), \eg visual expert group being more actively utilized under high auditory noise.

As shown in Figure\,\ref{fig:figure1}, MoHAVE capitalizes on its increased model capacity to significantly enhance performance while maintaining efficiency. Unlike simple MoE implementations, which yield only modest gains over Transformers, our innovative expert group strategy unlocks substantial advancements in adaptability and robustness.
Our main contributions are outlined as follows:
\begin{itemize}
    \item \textbf{MoE architecture for scaling AVSR systems}: We present MoHAVE, a framework that integrates a sparse MoE architecture to efficiently scale AVSR model capacity and optimally process audio and visual data.
    \item \textbf{Hierarchical gating for adaptive expert utilization}: MoHAVE features a novel hierarchical gating mechanism that dynamically adjusts the usage of audio and visual expert groups based on input context, significantly improving adaptability and robustness.
    \item \textbf{Robust AVSR performance}: Our model showcases substantial improvements across robust AVSR benchmarks including multilingual tasks, delivering high accuracy while maintaining computational overhead.
\end{itemize}
\section{Related Work}
\label{sec:ch4_related_work}

\subsection{Robustness of Audio-Visual Speech Recognition} 

The robustness of AVSR systems has significantly advanced by integrating auditory and visual cues to improve speech recognition, especially in noisy environments. Conventional ASR methods have evolved from relying solely on audio signals \citep{schneider2019wav2vec, gulati2020conformer, baevski2020wav2vec, hsu2021hubert, chen2022wavlm, chiu2022self, radford2023robust} to incorporating visual data from speech videos \citep{makino2019recurrent}.
The multimodal AVSR methods \citep{pan2022leveraging, shi2022learning, seo2023avformer, ma2023auto} have enhanced robustness under audio-corrupted conditions, leveraging visual details like speaker's face or lip movements as well as acoustic features of speech. These advancements have been driven by various approaches, including end-to-end learning frameworks \citep{dupont2000audio, ma2021end, hong2022visual, burchi2023audio} and self-supervised pretraining \citep{ma2021lira, qu2022lipsound2, seo2023avformer, zhu2023vatlm, kim2025multitask}, which focus on audio-visual alignment and the joint training of modalities~\citep{zhang2023self, lian2023av, haliassos2022jointly, haliassos2024braven}.

Furthermore, recent advancements in AVSR highlight the importance of visual understanding alongside audio \citep{dai2024study, kim2024learning}. While initial research primarily targeted audio disturbances \citep{shi2022robust, hu2023hearing, hu2023cross, chen2023leveraging}, latest studies increasingly focus on the visual robustness to address challenges such as real-world audio-visual corruptions~\citep{hong2023watch, wang2024restoring, kim2025multitask} or modality asynchrony~\citep{zhang2024visual, fu2024boosting, li2024unified}. These efforts remark a shift towards a more balanced use of audio and visual modalities. Yet, there has been limited exploration in scaling model capacity or introducing innovative architectural designs, leaving room for further developments in AVSR system that can meticulously balance audio and visual modalities.

\subsection{MoE for Language, Vision, and Speech Models}

Mixture-of-Experts (MoE), first introduced by \citet{jacobs1991adaptive}, is a hybrid structure incorporating multiple sub-models, \ie experts, within a unified framework. The essence of sparsely-gated MoE \citep{shazeer2017outrageously, lepikhin2021gshard, dai2022stablemoe} lies in its routing mechanism where a learned router activates only a subset of experts for processing each token, significantly enhancing computational efficiency. Initially applied within LLMs using Transformer blocks, this structure has enabled unprecedented scalability \citep{fedus2022switch, zoph2022st, jiang2024mixtral, guo2025deepseek} and has been progressively adopted in multimodal models, especially in large vision-language models (LVLMs) \citep{mustafa2022multimodal, lin2024moellava, mckinzie2025mm1}.
Among these multimodal MoEs, \citet{zhu2022uni, shen2023scaling, li2023pace, li2024uni} and \citet{lee2025moai} share the similar philosophy to ours, assigning specific roles to each expert and decoupling them based on distinct modalities or tasks. These models design an expert to focus on specialized segments of input and enhance the targeted processing.

Beyond its applications in LLMs and LVLMs, the MoE framework has also been applied for speech processing \citep{you2021speechmoe, you2022speechmoe2, hu2023mixture, wang2023language}, where it has shown remarkable effectiveness in multilingual and code-switching ASR tasks. In addition, MoE has been employed in audio-visual models \citep{cheng2024mixtures, wu2024robust}, although they primarily focus on general video processing and not specifically on human speech videos. These approaches leverage MoE to model interactions between audio and visual tokens without directly processing multimodal tokens.
Our research advances the application of the MoE framework to AVSR by designing a modality-aware hierarchical gating mechanism, which categorizes experts into audio and visual groups and effectively dispatches multimodal tokens to each expert group. 
This tailored design enhances the adaptability in managing audio-visual speech inputs, which often vary in complexity due to diverse noise conditions.

\section{Preliminaries}
\label{sec:ch4_preliminaries}

\subsection{Sparsely-gated MoE}
\label{subsec:sparse_moe}

\begin{figure}[!t]
    \centering
    \includegraphics[width=.65\linewidth]{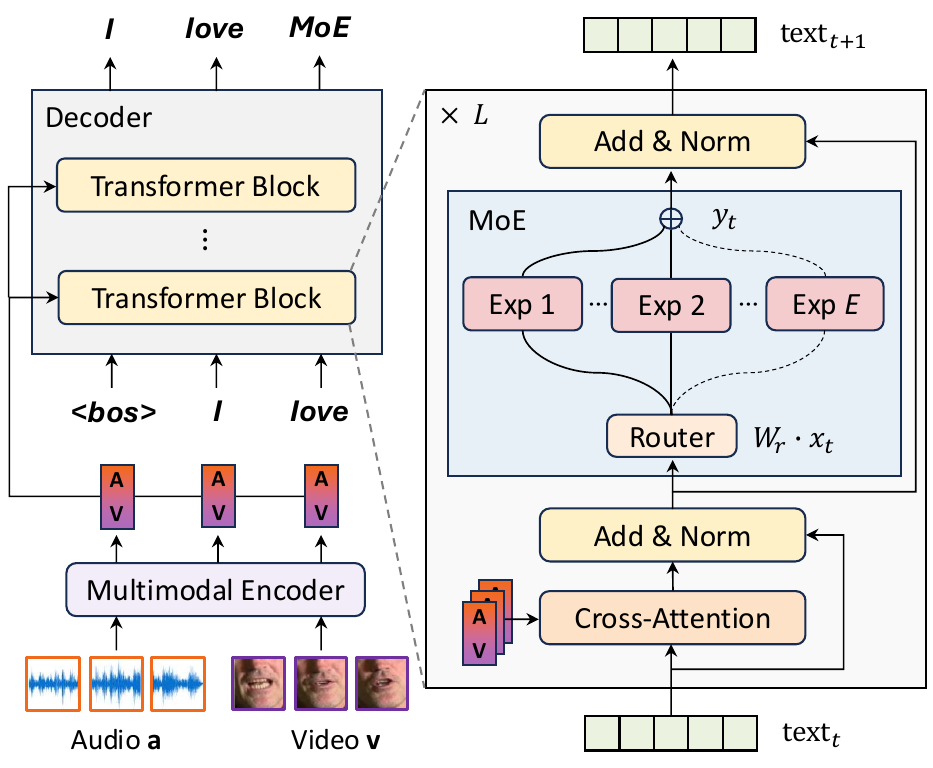}
    \caption[Overview of the sparsely-gated MoE architecture in MoHAVE]{Overview of sparsely-gated MoE for AVSR. A select subset of experts are activated for each token representation ($x_t$).
    }
    \label{fig:mohave_overview}
    \vspace*{10pt}
\end{figure}

In AVSR systems, the multimodal encoder processes a sequence of audio $\rva = [a_1, a_2, \cdots]$ and video $\rvv = [v_1, v_2, \cdots]$ data into combined audio-visual embeddings $\text{Enc}(\rva, \rvv)$. These embeddings are utilized by the decoder to predict subsequent text tokens, where the predicted token is given by $\textsl{text}_{t+1} = \text{Dec}(\text{Enc}(\rva, \rvv), \textsl{text}_t)$. Within the Transformer layer, $x_t$ is the intermediate representation of the token $\textsl{text}_t$, derived by cross-attending to the combined audio-visual embeddings from $\rva$ and $\rvv$ (see Figure\,\ref{fig:mohave_overview}).

The integration of a sparsely-gated MoE framework \citep{shazeer2017outrageously, lepikhin2021gshard} leverages experts $\mathcal{E} = \{E_i\}$ to scale model capacity. Each token representation is routed to a selected subset of these experts through a learned gating mechanism. 
Specifically, the routing function $h(x) = W_r \cdot x$ assigns weights for each token, and the weight for expert $i$ is computed using a softmax function:
\begin{equation}
\label{eq:router_weight}
    p_i(x) = \frac{\exp(h_i(x))}{\sum_{j=1}^{|\mathcal{E}|} \exp(h_j(x))},
\end{equation}
and the output $y$ is the aggregated result of computations from the top-$k$ selected experts:
\begin{equation}
y = \sum_{i \in \text{top}k(\mathcal{E})} \tilde{p}_i(x) E_i(x),
\end{equation}
where $\tilde{p}$ is the normalization of top-$k$ probabilities.
Note that each expert follows the same structure as a feed-forward network\,(FFN) in a Transformer block. Figure\,\ref{fig:mohave_overview} presents the overall MoE architecture and its token routing.

\paragraph{Load balancing.}
To mitigate the load imbalance issue commonly observed in the top-$k$ expert selection strategy, a load balancing loss has been implemented to encourage the balanced token load across all experts. Specifically, we use a differentiable load balancing loss~\citep{fedus2022switch}:
\begin{equation}
    L_B = |\mathcal{E}| \cdot \sum_{i=1}^{|\mathcal{E}|} f_i \cdot P_i,
\end{equation}
where $f_i$ denotes the frequency of expert $i$ being selected as top-1, averaged over all tokens within a batch $\mathcal{B}$,
\begin{equation}
\label{eq:expert_frequency}
    f_i = \frac{1}{T} \sum_{x\in\mathcal{B}} \mathbbm{1} \{\arg\max p(x) = i\}
\end{equation}
and $P_i$ is the average assigned probability for expert $i$,
\begin{equation}
\label{eq:expert_probability}
    P_i = \frac{1}{T} \sum_{x\in\mathcal{B}} p_i(x)
\end{equation}
with $T$ representing the total number of tokens.

An additional router z-loss~\citep{zoph2022st} is employed to stabilize the routing mechanism:
\begin{equation}
    L_Z = \frac{1}{T}\sum_{x\in\mathcal{B}} \Bigg(\log \sum_{i=1}^{|\mathcal{E}|} \exp(h_i(x)) \Bigg)^2.
\end{equation}
This sparse MoE structure ensures that token processing is efficiently managed across multiple experts, utilizing lower compute relative to its expansive capacity.

\begin{figure*}[!t]
    \centering
    \includegraphics[width=\linewidth]{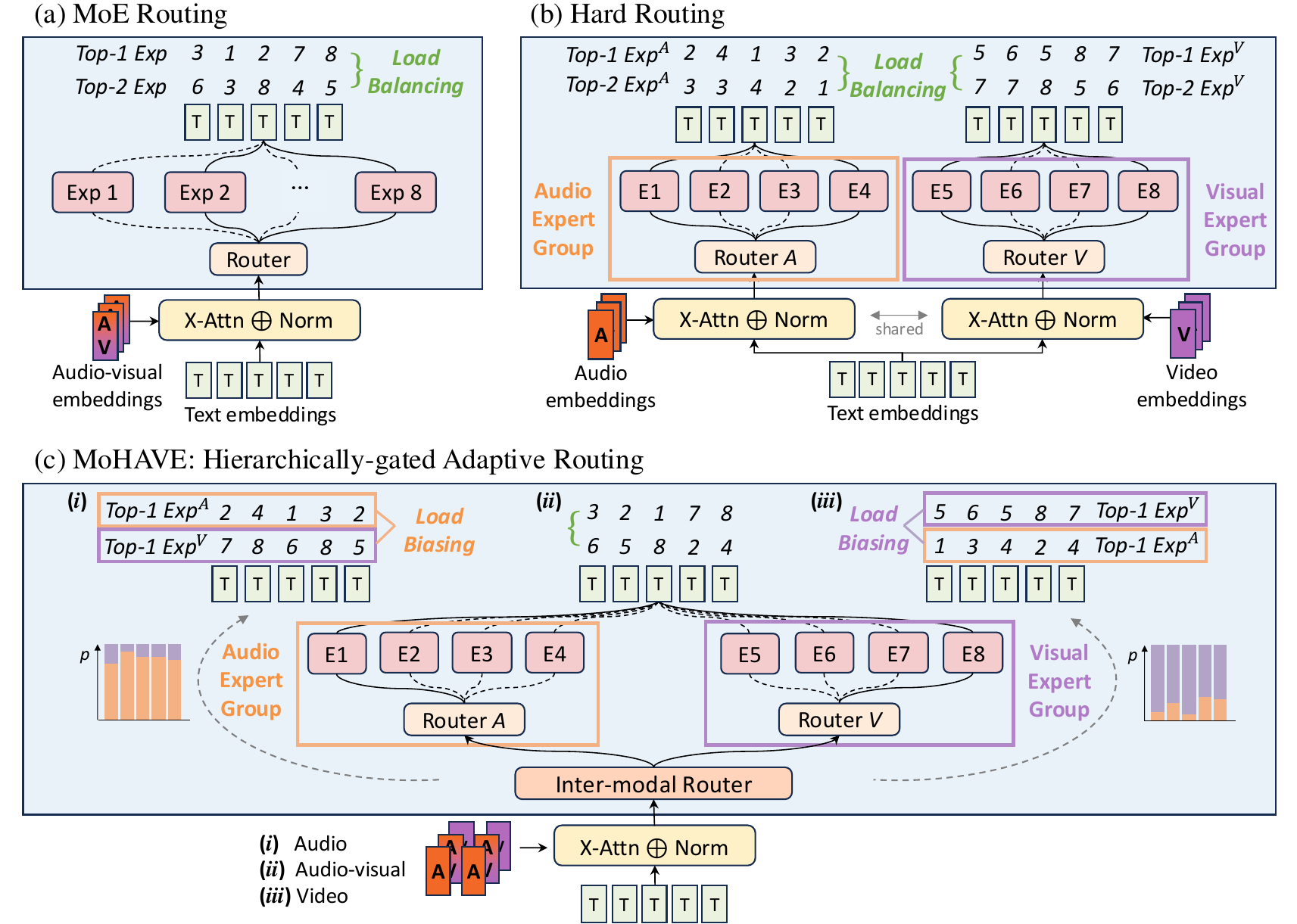}
    \caption[Hierarchically-gated adaptive routing of MoHAVE]{MoE-based routing strategies for AVSR. 
    (a) A conventional MoE approach where a learned router selects the top-2 experts for each token, enforcing the balanced expert load. 
    (b) Experts are explicitly divided into audio and visual groups, with manual activation based on the input modality.
    (c) \ourmodel introduces an inter-modal router that can dynamically assign weights to modality-specific expert groups, followed by intra-modal routers that select the top-1 expert within each group. The inter-modal router is trained by the load biasing loss that guides the expert group specialization.
    }
    \label{fig:routing}
    \vspace*{10pt}
\end{figure*}

\subsection{Expert Group Specialization}
\label{subsec:hard_routing}

To enhance expert management within the AVSR system, a \textit{hard routing} technique can be used for expert group specialization. This approach is inspired by several practices in visual-language MoE models \citep{zhu2022uni, li2023pace, shen2023scaling, lee2025moai} where the role of experts is strictly defined by the input modality, eliminating the need for a trained router.

\paragraph{Hard routing.}
Our hard routing enforces modality-specific activation of expert groups: audio data activate only audio experts, and video data activate only visual experts.
This segregation encourages the independent development of specialized expert groups. As suggested in V/T-MoE \citep{shen2023scaling}, once the group is activated, we use an intra-modal router for the modality-specific experts.

Figure\,\ref{fig:routing}\textcolor{red}{b} visualizes the hard routing mechanism with audio and visual expert groups.
During training, audio or video sequence is randomly dropped, leading to subsets $\mathcal{A}$ and $\mathcal{V}$ within a batch $\mathcal{B}$, consisting of audio-only or video-only sequences, respectively. A token representation $x_t \in \mathcal{A}$ indicates that the cross-attention module processes the input $\textsl{text}_t$ with $\text{Enc}(\rva, \vzero)$---where the visual component is zeroed out---and vice versa for $x_t \in \mathcal{V}$.
For these, we utilize two distinct intra-modal routing networks, $W_r^A$ and $W_r^V$:
\begin{align}
\begin{split}
    h^A(x) &= W_r^A \cdot x \quad \text{for } x \in \mathcal{A}, \\
    h^V(x) &= W_r^V \cdot x \quad \text{for } x \in \mathcal{V}.
\end{split}
\end{align}
These routers calculate the weights $p^{\{A,V\}}(x)$ as in Eq.\,(\ref{eq:router_weight}) within their respective expert group, either $\{E^A\}$ for audio or $\{E^V\}$ for visual. The output for each token is then
\begin{equation}
    y = \begin{cases}
        \sum_{\text{top}k(E^A)} \tilde{p}_i^A(x) E^A_i(x) & \text{if } x \in \mathcal{A}, \\
        \sum_{\text{top}k(E^V)} \tilde{p}_i^V(x) E^V_i(x) & \text{if } x \in \mathcal{V}. \\ 
    \end{cases} 
\end{equation}
For audio-visual sequences, outputs from both groups are averaged, with the top-($k/2$) experts from each group contributing to ensure balanced processing.

\section{MoHAVE: Mixture of Hierarchical Audio-Visual Experts}
\label{sec:mohave}

Despite the benefits of hard routing in specializing expert groups according to decoupled input modality, it lacks the flexibility to autonomously determine the group utilization.
In practice, the optimal balance between audio and visual groups varies depending on ambient conditions such as noise type and intensity (more detailed in Figure\,\ref{fig:expert_load}\textcolor{red}{b}).
To address this limitation and enhance the model’s adaptability, we introduce an adaptive routing mechanism with hierarchical gating \citep{jordan1994hierarchical}, providing a more dynamic approach to manage multimodal inputs.

Our hierarchical model, \ourmodel, features a two-layer routing mechanism: \textit{inter-modal} and \textit{intra-modal} routers, where the inter-modal router learns to assign appropriate weights to each modality-specific group. Figure\,\ref{fig:routing}\textcolor{red}{c} presents the overview of \ourmodel's routing strategy.

\subsection{Hierarchical Gating Structure}
\label{subsec:hierarchical_gating}

The inter-modal router orchestrates the initial token distributions across expert groups. It generates logits through $u(x) = V_r \cdot x$ and determines the dispatch weights for group $i$ with $q_i(x) = \text{softmax}(u(x))_i$. This router dynamically selects the top-$m$ expert groups, and within those, the intra-modal routers select the top-$k$ experts, thus involving $m \times k$ experts in total. For practical efficiency, we set $k=1$ for each group and modify the intra-modal router's probability distribution to a Kronecker delta, $\tilde{p}_{ij} \rightarrow \delta_{j,\text{argmax}(p_i)}$.
The output from this layer integrates these selections:
\begin{align}
  y &= \!\!\!\!\sum_{i \in \text{top}m(G)} \!\!\tilde{q}_i(x)\!\! \sum_{j \in \text{top}k(E_i)} \!\!\tilde{p}_{ij} E_{ij}(x) \\
    &\rightarrow \!\!\!\!\sum_{i \in \text{top}m(G)} \!\!\tilde{q}_i(x)E_{ij}(x), ~~\text{where } j\!=\!\arg\max(p_i) 
\end{align}
where $\tilde{q}$ is the normalization of $q$ across top-$m$ probabilities, $G$ is the number of expert groups, and $E_{ij}(x)$ denotes the output from the $j$-th expert in the $i$-th group.

Focusing on audio-visual applications, we designate two expert groups: audio and visual. Each token $x$, regardless of its modality, is processed by the intra-modal routing networks of both groups, \ie $[h^A(x), h^V(x)] = [W_r^A, W_r^V] \cdot x$. The frequencies $f^{\{A,V\}}$ and probabilities $P^{\{A,V\}}$ for selecting experts are computed in the same manner as Eq.\,(\ref{eq:expert_frequency})--(\ref{eq:expert_probability}) for all $x \in \mathcal{B}$. Thus, the load balancing loss can be computed for both groups:
\begin{equation}
L_B = |E^A| \cdot \sum_{j=1}^{|E^A|} f_j^A \cdot P_j^A + |E^V| \cdot \sum_{j=1}^{|E^V|} f_j^V \cdot P_j^V
\end{equation}
where $f_j^A$ and $f_j^V$ denote the frequencies of token assignments to audio and visual experts, respectively.

\subsection{Group-level Load Biasing Loss}
\label{subsec:load_biasing_loss}

To autonomously manage the expert group loads without manual (de-)activation as hard routing, we introduce a biasing loss that directs the load towards a certain group. This load biasing loss encourages the inter-modal router to assign higher weights to $E^A$ experts for audio sequences and to $E^V$ experts for video sequences.
For audio sequences within a sub-batch $\mathcal{A}$, the frequency and average probability of selecting the $i$-th group is calculated as follows:
\begin{equation}
    g^A_i = \frac{1}{|\mathcal{A}|} \sum_{x\in\mathcal{A}} \mathbbm{1} \{\arg\max q(x) = i\},
\end{equation}
\begin{equation}
    Q^A_i = \frac{1}{|\mathcal{A}|} \sum_{x\in\mathcal{A}} q_i(x).
\end{equation}
Similar calculations for $g^V_i$ and $Q^V_i$ are made for video sequences $x\in\mathcal{V}$. We designate the first group as audio experts and the second group as video experts, then the load biasing loss is defined as:
\begin{equation}
    L_S = L_S^A + L_S^V = (1 - g^A_1 \cdot Q^A_1) + (1 - g^V_2 \cdot Q^V_2).
\end{equation}
Note that $L_S^A$ and $L_S^V$ are only computed over $x\in\mathcal{A}$ and $x\in\mathcal{V}$, respectively. 

For sequences containing both audio and video, we exclude them from the load biasing loss calculation but incorporate them into the load balancing.
Although these tokens are uniformly dispatched on average, the inter-modal router finds the optimal strategy for each token based on its characteristics.
Empirically, we find that \ourmodel learns to assign greater weight to the visual expert group for audio-visual inputs under high auditory noise, and to the audio expert group for less noisy inputs (see \S\ref{sec:analysis_load} for details), demonstrating the model’s adaptability under various noisy conditions.

The overall loss function, combining the cross-entropy\,(CE) for token prediction, is formulated as:
\begin{equation}
L_{tot} = L_{CE} + c_B L_B + c_S L_S + c_Z L_Z.
\end{equation}
Here, $c_B$ and $c_Z$ are set to 1e-2 and 1e-3, respectively, in line with \citep{fedus2022switch, zoph2022st}, and $c_S$ is also set at 1e-2.

\section{Experiments and Results}
\label{sec:ch4_experiments}

\subsection{Implementation Details}
\label{subsec:ch4_implementation}

\paragraph{Datasets.}
For the robust AVSR benchmark, we utilize the LRS3 dataset~\citep{afouras2018lrs3}, which consists of 433 hours of audio-visual speech from 5,000+ speakers. Following the experimental setup of \citet{shi2022robust}, we extract audio noise samples from the MUSAN~\citep{snyder2015musan} dataset, targeting different noise types such as \textit{babble}, \textit{music}, and \textit{natural} noises, along with \textit{speech} noise from LRS3. These noises are randomly augmented into the audio data, corrupting 25\% of the training set with a signal-to-noise ratio (SNR) sampled from $\mathcal{N}(0, 5)$. We measure performance using the word error rate (WER), primarily under noisy conditions with SNRs of \{$-$10, $-$5, 0, 5, 10\}\,dB, specifically N-WER\,\citep{kim2024learning} which highlights the significance of visual cues in noise-corrupted environments.

For multilingual evaluations, the MuAViC dataset \citep{anwar2023muavic} is used, featuring 1,200 hours of audio-visual content from 8,000+ speakers across 9 languages, sourced from LRS3-TED\,\citep{afouras2018lrs3} and mTEDx\,\citep{elizabeth2021multilingual}. We use 8 languages (excluding English) for multilingual AVSR and 6 languages for X-to-English audio-visual speech-to-text translation (AVS2TT) tasks. We assess the models using WER for transcription and the BLEU score\,\citep{papineni2002bleu} for translation.

\paragraph{\ourmodel model description.}

Our \ourmodel framework is developed in two configurations: \textsc{Base} and \textsc{Large}. The \textsc{Base} model consists of 12 Transformer~\citep{vaswani2017attention} encoder layers and 6 decoder layers, while the \textsc{Large} model incorporates 24 encoder layers and 9 decoder layers. Both models’ audio-visual encoders are derived from the AV-HuBERT-\textsc{Base}/-\textsc{Large} models, pretrained on a noise-augmented corpus of LRS3 \citep{afouras2018lrs3} + VoxCeleb2 \citep{chung2018voxceleb2}. The encoder maintains the same structure as AV-HuBERT, while the decoder incorporates MoE layers by replacing every feed-forward network (FFN) layer with expert modules. Each expert in the MoE layers follows the same bottleneck structure with FFN, consisting of two fully-connected layers with an activation function. Our MoE implementation activates top-2 out of 8 experts in every FFN layer within the decoder~\citep{jiang2024mixtral}, while the hierarchical architecture engages the top-1 expert from each audio and visual group.

To facilitate the expert group specialization, load biasing is used with audio or video randomly dropped in 25\% probability. This allows the model to learn modality-aware expert utilization. For expert selection, a router network assigns tokens to a subset of experts, ensuring efficient computation. The router probabilities are always normalized as sum to 1 when computing the output $y$. The routing networks $V_r$ and $W_r$ are parameterized as matrices, with dimensions matching the hidden dimension size by the number of experts.

As summarized in Table~\ref{tab:computation_cost}, the \textsc{Base} model of \ourmodel holds 359M parameters, and the \textsc{Large} configuration contains 1B. Specifically, for \ourmodel-\textsc{Base}, the encoder accounts for 103M parameters, and the decoder 256M, whereas in \textsc{Large}, the encoder holds 325M, and the decoder 681M. Despite its larger model capacity, due to the sparse activation of these parameters, only about half are active during token processing, amounting to 189M for \textsc{Base} and 553M for \textsc{Large} model. This setup ensures computational efficiency which is comparable to the smaller AV-HuBERT counterparts.

For comparison, we evaluate multiple MoE-based AVSR models:
\begin{itemize}
    \item AV-MoE: A simple MoE implementation over AV-HuBERT, activating top-2 out of 4 or 8 experts per token. We follow the same implementation of sparse MoE as \citep{dai2022stablemoe, jiang2024mixtral}.
    \item AV-MoE with Hard Routing: Uses top-2 out of 4 experts for unimodal inputs (audio-only or video-only). For multimodal (audio-visual) inputs, it activates top-1 from each expert group and averages their outputs. This model does not have an explicit router for groups, but within each group, there is an intra-modal router, \ie $W_r^A$ or $W_r^V$.
    \item \ourmodel: Employs top-1 expert per group, with an inter-modal router dynamically adjusting group weight assignments and an intra-modal router uniformly dispatching the tokens to modality-specific experts.
\end{itemize}

\begin{table}[!t]
    \centering
    \small
    \caption[Computational cost and model size comparison of MoHAVE and AV-HuBERT]{Computational cost of AV-HuBERT and \ourmodel in FLOPs, along with their sizes (activated and total parameters).}
    \label{tab:computation_cost}
    \resizebox{.7\columnwidth}{!}{
    \begin{tabular}{l|c|cc}
    \toprule
    \multirow{2}{*}{Model} & Params & Compute & Compute\,/\,FFN \\
     & Act. \& Total & (GFLOPs\,/\,seq) & (MFLOPs\,/\,seq) \\
    \midrule
    AV-HuBERT-\textsc{Base} & 161M \& 161M & 12.1 & 472 \\
    \ourmodel-\textsc{Base} & 189M \& 359M & 14.8 & 921 \\
    AV-HuBERT-\textsc{Large} & 477M \& 477M & 32.2 & 839 \\
    \ourmodel-\textsc{Large} & 553M \& 1.0B & 39.3 & 1,630 \\
    \bottomrule
    \end{tabular}
    }
    \vspace*{10pt}
\end{table}

\subsection{Computation Cost}
\label{subsec:ch4_computation_cost}

Table~\ref{tab:computation_cost} summarizes the parameter sizes and computational costs of \ourmodel. 
To assess actual computation costs when processing inputs, we measure floating point operations per second (FLOPs) using an audio-visual sequence of 500 frames with 50 text tokens. The entire compute cost for AV-HuBERT-\textsc{Base} and \ourmodel-\textsc{Base} are 12.1 GFLOPs and 14.8 GFLOPs, respectively, while for \textsc{Large}, the computes are 32.2 GFLOPs and 39.3 GFLOPs. This indicates a slight increase in FLOPs for \ourmodel, primarily due to the MoE layers replacing FFNs in the decoder. Although the MoE layers require roughly twice the computation cost of standard FFNs (refer to Compute\,/\,FFN), the encoder and attention layers in the decoder remain unchanged. Consequently, the overall computational cost remains comparable to AV-HuBERT counterparts, ensuring scalability without significant computation overhead.

\subsection{Robust AVSR Benchmark Results}
\label{subsec:ch4_results}

\paragraph{Experimental setup.}
We initialize our model using the pretrained checkpoint from \citep{shi2022learning} and fine-tune it on the LRS3 train set for 120K steps. The encoder remains frozen for the first 90K steps, allowing only the AVSR decoder to be trained, after which the entire model is fine-tuned for the remaining 30K steps. Our fine-tuning setup follows the configurations from \citep{shi2022robust}. We employ a sequence-to-sequence negative log-likelihood loss for predicting the next text token, without using connectionist temporal classification (CTC) decoding \citep{watanabe2017hybrid}. The Adam optimizer \citep{kingma2014adam} is used with a learning rate of 5e-4 and a polynomial decay schedule with an initial warmup. Each training step processes 8,000 audio-visual frames, equivalent to 320 seconds of speech data.

For inference, we use beam search with a beam size of 50. The AVSR performance is evaluated using word error rate (WER) across five signal-to-noise ratio (SNR) levels: $\{-10, -5, 0, 5, 10\}$ (lower value means higher noise level). We use audio noise sampled from MUSAN (babble, music, natural) and LRS3 speech noise, ensuring no speaker overlap between training and test sets. 

\begin{table*}[!t]
    \centering
    \small
    \caption[MoHAVE performance on LRS3]{Audio-visual speech recognition performance\,(WER\,\%) on the LRS3 dataset\,\citep{afouras2018lrs3}. The number of parameters for each model includes both encoder and decoder. For evaluation, augmented noise is sampled from the MUSAN dataset\,\citep{snyder2015musan}, while speech noise is sampled from the held-out set of LRS3. N-WER\,\citep{kim2024learning} averages the results across all four noise types and five signal-to-noise ratios, and C-WER indicates the result with a clean audio signal.}
    \label{tab:mohave_main_result}
    \addtolength{\tabcolsep}{-2pt}
    \resizebox{\textwidth}{!}{
    \begin{tabular}{l|cccc|cccc|c|c}
    \toprule
    \multirow{2}{*}{Model} & \multirow{2}{*}{\# Experts} & Specialized & Activated & Total & \multicolumn{4}{c|}{SNR = \{$-$10, $-$5, 0, 5, 10\}} & \multirow{2}{*}{N-WER} & \multirow{2}{*}{C-WER} \\
    & & Groups & Params & Params & babble & speech & music & natural & & \\
    \midrule
    AV-HuBERT-\textsc{Base} & - & - & 161M & 161M & 10.8 & 4.9 & 5.6 & 5.1 & 6.6 & 2.1 \\
    AV-MoE-\textsc{Base} & 4 & \xmark & 189M & 246M & 10.5 & 4.5 & 5.3 & 5.0 & 6.3 & 2.0 \\
    AV-MoE-\textsc{Base} & 8 & \xmark & 189M & 359M & 10.5 & 4.5 & 5.3 & 4.9 & 6.3 & 2.1 \\
    ~~(+) Hard Routing & 8 & \cmark & 189M & 359M & 9.9 & 4.4 & 5.0 & 4.6 & 5.9 & 2.0 \\
    \rowcolor{blue!10}
    \ourmodel-\textsc{Base} \textbf{(ours)} & ~~~2$\times$4\,(H)\!\!\!\! & \cmark & 189M & 359M & \textbf{9.6} & \textbf{4.2} & \textbf{4.7} & \textbf{4.5} & \textbf{5.8} & \textbf{1.8} \\
    \rowcolor{blue!10}
    ~~(--) Load Biasing & ~~~2$\times$4\,(H)\!\!\!\! & \xmark & 189M & 359M & 10.3 & 4.4 & 5.2 & 4.9 & 6.2 & 2.0 \\
    \midrule
    AV-HuBERT-\textsc{Large} & - & - & 477M & 477M & 9.7 & 4.6 & 4.4 & 4.1 & 5.7 & 1.4 \\
    AV-MoE-\textsc{Large} & 4 & \xmark & 553M & 704M & 10.1 & 3.8 & 4.6 & 4.3 & 5.7 & 1.8 \\
    AV-MoE-\textsc{Large} & 8 & \xmark & 553M & 1.0B & 10.1 & 3.8 & 4.6 & 4.2 & 5.7 & 1.8 \\
    ~~(+) Hard Routing & 8 & \cmark & 553M & 1.0B & 8.3 & 3.3 & 4.0 & 3.7 & 4.8 & 1.5 \\
    \rowcolor{blue!10}
    \ourmodel-\textsc{Large} \textbf{(ours)} & ~~~2$\times$4\,(H)\!\!\!\! & \cmark & 553M & 1.0B & \textbf{7.7} & \textbf{3.0} & \textbf{3.7} & \textbf{3.4} & \textbf{4.5} & 1.5 \\
    \rowcolor{blue!10}
    ~~(--) Load Biasing & ~~~2$\times$4\,(H)\!\!\!\! & \xmark & 553M & 1.0B & 9.9 & 3.6 & 4.6 & 4.3 & 5.6 & 1.7 \\
    \bottomrule
    \end{tabular}
    }
    \vspace*{10pt}
\end{table*}

\paragraph{Result.}
Table~\ref{tab:mohave_main_result} presents \ourmodel's robust performance on the AVSR benchmark under diverse noisy conditions, demonstrating exceptional robustness across different noise types and SNR levels: \textbf{N-WER of 5.8\% for \textsc{Base}} and \textbf{4.5\% for \textsc{Large}}. This substantiates the model's potential for effectively scaling AVSR systems without incurring significant computational costs.
The results also reveal that simple MoE implementations (AV-MoE in Table~\ref{tab:mohave_main_result}), despite their larger capacity, fail to achieve remarkable gains. Instead, the key improvement stems from leveraging expert group specialization, as evidenced by the effectiveness of hard routing. By splitting experts into audio and visual groups, MoE is enabled with more targeted and effective processing of multimodal inputs, leading to substantial performance enhancements.
Without our load biasing loss, \ourmodel loses its group specialization capability, comparable to the performance of simple AV-MoEs.

Building upon this expert group strategy, \ourmodel enhances its adaptability through dynamically determining the usage of each group. This adaptive routing approach allows the model to flexibly adjust to varying audio-visual scenarios, contributing to consistent gains in robustness across the benchmark, as detailed in Table~\ref{tab:mohave_main_result}.
An in-depth analysis of this hierarchical gating approach and its impact on token dispatching is discussed in Section~\ref{sec:analysis_load}, underscoring its critical role in advancing MoHAVE’s capabilities in various AVSR environments.

\begin{sidewaystable}
    \centering
    \small
    \caption[MoHAVE performance on LRS3 (full SNR results)]{Audio-visual speech recognition performance\,(WER\,\%) on the LRS3 dataset\,\citep{afouras2018lrs3}. The number of parameters for each model includes both encoder and decoder. For evaluation, augmented noise is sampled from the MUSAN dataset\,\citep{snyder2015musan}, while speech noise is sampled from the held-out set of LRS3. AV-MoE and \ourmodel use 8 experts.}
    \label{tab:lrs3_full}
    \addtolength{\tabcolsep}{0pt}
    \renewcommand{\arraystretch}{1.1}
    \resizebox{\textwidth}{!}{
    \begin{tabular}{l|cccccc|cccccc|cccccc|cccccc}
    \toprule
    \multirow{2}{*}{Model} & \multicolumn{6}{c|}{Babble, SNR (dB) $=$} & \multicolumn{6}{c|}{Speech, SNR (dB) $=$} & \multicolumn{6}{c|}{Music, SNR (dB) $=$} & \multicolumn{6}{c}{Natural, SNR (dB) $=$} \\
    & -10 & -5 & 0 & 5 & 10 & \!\textbf{AVG}\! & -10 & -5 & 0 & 5 & 10 & \!\textbf{AVG}\! & -10 & -5 & 0 & 5 & 10 & \!\textbf{AVG}\! & -10 & -5 & 0 & 5 & 10 & \!\textbf{AVG}\! \\
    \midrule
    AV-HuBERT-\textsc{Base} & 27.6 & 14.1 & 6.1 & 3.8 & 2.7 & 10.8 & 8.6 & 5.8 & 3.9 & 3.3 & 2.8 & 4.9 & 12.2 & 6.4 & 3.8 & 2.8 & 2.5 & 5.6 & 10.9 & 5.4 & 4.0 & 2.8 & 2.3 & 5.1 \\
    AV-MoE-\textsc{Base} & 26.3 & 13.7 & 6.2 & 3.5 & 2.7 & 10.5 & 8.4 & 5.3 & 3.4 & 2.8 & 2.3 & 4.5 & 11.5 & 6.1 & 3.7 & 2.7 & 2.5 & 5.3 & 9.7 & 6.0 & 3.4 & 2.8 & 2.5 & 4.9 \\
    ~~(+) Hard Routing & 25.2 & 13.2 & 5.6 & 3.2 & 2.4 & 9.9 & 8.2 & 5.2 & 3.4 & 2.7 & 2.3 & 4.4 & 11.0 & 5.3 & 3.6 & 2.6 & 2.2 & 5.0 & 9.4 & 5.3 & 3.2 & 2.5 & 2.3 & 4.6 \\
    \rowcolor{blue!10}
    \ourmodel-\textsc{Base} & 25.3 & \textbf{12.2} & \textbf{5.3} & \textbf{2.9} & \textbf{2.3} & \textbf{9.6} & \textbf{7.9} & \textbf{5.1} & \textbf{3.3} & \textbf{2.4} & \textbf{2.3} & \textbf{4.2} & \textbf{10.3} & 5.6 & \textbf{3.3} & \textbf{2.3} & \textbf{2.0} & \textbf{4.7} & 9.7 & \textbf{5.1} & \textbf{3.2} & \textbf{2.4} & \textbf{2.2} & \textbf{4.5} \\
    \rowcolor{blue!10}
    ~~(--) Load Biasing & 26.5 & 13.6 & 5.6 & 3.2 & 2.4 & 10.3 & 8.2 & 5.2 & 3.5 & 2.6 & 2.4 & 4.4 & 11.1 & 6.4 & 3.4 & 2.7 & 2.6 & 5.2 & 10.3 & 5.6 & 3.4 & 2.7 & 2.4 & 4.9 \\
    \midrule
    AV-HuBERT-\textsc{Large} & 27.0 & 12.4 & 4.7 & 2.4 & 1.8 & 9.7 & 11.4 & 4.6 & 2.9 & 2.2 & 1.8 & 4.6 & 10.5 & 4.9 & 2.9 & 2.0 & 1.6 & 4.4 & 9.6 & 4.7 & 2.5 & 2.0 & 1.8 & 4.1 \\
    AV-MoE-\textsc{Large} & 28.1 & 12.5 & 5.0 & 2.7 & 2.1 & 10.1 & 7.9 & 4.0 & 2.9 & 2.4 & 2.0 & 3.8 & 10.4 & 5.4 & 2.9 & 2.3 & 1.9 & 4.6 & 8.9 & 4.8 & 3.1 & 2.0 & 2.0 & 4.2 \\
    ~~(+) Hard Routing & 22.9 & 10.8 & 3.8 & 2.4 & 1.8 & 8.3 & 6.7 & 3.9 & 2.4 & 1.8 & 1.8 & 3.3 & 9.9 & 4.3 & 2.3 & 1.8 & 1.9 & 4.0 & 8.3 & 4.1 & 2.4 & 1.9 & 1.7 & 3.7 \\
    \rowcolor{blue!10}
    \ourmodel-\textsc{Large} & \textbf{21.0} & \textbf{9.8} & 4.1 & \textbf{2.2} & \textbf{1.6} & \textbf{7.7} & \textbf{5.0} & \textbf{3.6} & \textbf{2.3} & 2.0 & 1.9 & \textbf{3.0} & \textbf{8.2} & \textbf{4.0} & 2.6 & \textbf{1.8} & \textbf{1.8} & \textbf{3.7} & \textbf{7.3} & \textbf{3.7} & 2.6 & \textbf{1.9} & \textbf{1.6} & \textbf{3.4} \\
    \rowcolor{blue!10}
    ~~(--) Load Biasing & 27.8 & 12.4 & 4.5 & 2.6 & 2.0 & 9.9 & 6.7 & 4.0 & 3.1 & 2.1 & 1.9 & 3.6 & 10.6 & 5.3 & 3.0 & 2.1 & 1.8 & 4.6 & 9.9 & 4.9 & 2.8 & 2.1 & 2.0 & 4.3 \\
    \bottomrule
    \end{tabular}
    }
    \vspace*{10pt}
\end{sidewaystable}

Since Table~\ref{tab:mohave_main_result} presents SNR-averaged results for each noise type, we provide the full results across all SNR levels in Table~\ref{tab:lrs3_full}.
\ourmodel-\textsc{Large} achieves 5.0\% WER on LRS3 with speech noise at SNR=$-$10, yielding a 56.1\% relative WER improvement over AV-HuBERT-\textsc{Large}, 36.7\% over AV-MoE-\textsc{Large}, and 25.4\% over the hard-routing variant. This indicates that MoHAVE correctly predicts over half of the words that AV-HuBERT misses.

\paragraph{Comparison with state-of-the-art AVSR methods.}

\begin{wraptable}{r}{0.55\textwidth}
    \centering
    \small
    \vspace{-15pt}
    \caption[State-of-the-art comparison on noisy LRS3]{Performance comparison on the noisy LRS3 benchmark with prior works. We present average N-WER, with babble\,(B), speech\,(S), music\,(M), and natural\,(N) noise types. BRAVEn is implemented with separate ASR and VSR encoders, combined and jointly trained with a decoder.}
    \label{tab:sota_comparison}
    \resizebox{\linewidth}{!}{
    \begin{tabular}{lcccc}
    \toprule
    Method & B & S & M\,+\,N & Avg \\
    \midrule
    \multicolumn{4}{l}{\textcolor{gray}{\textit{Joint ASR and VSR encoders}}} \\
    BRAVEn\,\citep{haliassos2024braven} & 13.5 & 15.7 & 7.4 & 11.0 \\
    \rowcolor{blue!10}
    ~~~~+ \ourmodel (\textbf{ours}) & 12.3 & 13.4 & 6.8 & \textbf{9.8} \\
    \midrule
    \multicolumn{4}{l}{\textcolor{gray}{\textit{Noise-augmented AVSR encoder}}} \\
    AV-HuBERT\,\citep{shi2022robust} & 9.7 & 4.6 & 4.2 & 5.7 \\
    ~~~~+ MIR-GAN\,\citep{hu2023mir} & - & - & - & 5.6 \\
    ~~~~+ UniVPM\,\citep{hu2023hearing} & 9.3 & 4.1 & 3.6 & 5.2 \\
    ~~~~+ CMA\,\citep{kim2024learning} & 8.1 & 2.9 & 3.7 & 4.6 \\
    \rowcolor{blue!10}
    ~~~~+ \ourmodel (\textbf{ours}) & 7.7 & 3.0 & 3.6 & \textbf{4.5} \\
    \rowcolor{blue!10}
    ~~~~+ CMA\,+\,\ourmodel (\textbf{ours}) & 7.3 & 2.8 & 3.3 & \textbf{4.2} \\
    \bottomrule
    \end{tabular}
    }
    \vspace*{-10pt}
\end{wraptable}

Table~\ref{tab:sota_comparison} shows how our \ourmodel decoder, when integrated with a range of audio-visual encoders, consistently improves performance compared to existing state-of-the-art methods. While BRAVEn \citep{haliassos2024braven} typically struggles in noisy multimodal scenarios---due to its original design focused on handling unimodal tasks---\ourmodel boosts its accuracy. 

Other recent approaches have advanced by utilizing the noise-augmented AVSR encoder \citep{shi2022learning}, such as additionally learning temporal dynamics with cross-modal attention modules\,(CMA) \citep{kim2024learning}.
MIR-GAN \citep{hu2023mir}, UniVPM \citep{hu2023hearing}, and CMA are all built upon AV-HuBERT-\textsc{Large}, matching the architecture and activated parameter count with the dense (non-MoE) baseline in Table~\ref{tab:mohave_main_result}.
When paired with an AV-HuBERT encoder and trained through the CMA's self-supervised learning, \ourmodel achieves a remarkable performance: \textbf{N-WER of 4.2\%}.

\paragraph{Comparison with previous works.}
Empirical comparisons of \ourmodel with AVMoE~\citep{cheng2024mixtures} and EVA~\citep{wu2024robust} are unfortunately infeasible due to fundamental differences in target tasks and methods. Both AVMoE and EVA primarily address audio captioning for visual contexts (\eg narrating sports game scenes), while our work specifically targets typical AVSR tasks, where both audio and visual inputs directly involve human speech.

Moreover, AVMoE~\citep{cheng2024mixtures} employs a dense MoE; unlike sparse expert structures commonly used in modern LLMs or Transformers, AVMoE’s \textit{MoE} is actually implemented as weighting between unimodal and cross-modal adapters, rather than selecting sparse FFN experts. Specifically, AVMoE uses two entirely separate MoEs for audio encoder and visual encoder, infeasible for processing multimodal tokens. Our approach fundamentally differs by employing a sparse multimodal MoE, dynamically routing tokens based on audio-visual inputs.

Closer to our work is EVA~\citep{wu2024robust}, which simply applies a sparse MoE structure into an audio-visual encoder. Although exact implementation details are unavailable (code and checkpoints unreleased), EVA’s structure aligns closely with our basic MoE implementation which we evaluated as \textit{AV-MoE} in Table~\ref{tab:mohave_main_result}, except ours is in the decoder. As demonstrated in our study (Table~\ref{tab:encoder_mohave}), applying MoE at the encoder-level like EVA falls behind our multimodal decoder approach. Thus, EVA likely cannot achieve comparable robustness or efficiency.

\begin{table*}[!t]
    \centering
    \small
    \addtolength{\tabcolsep}{-2pt}
    \caption[MoHAVE performance on LRS3 with DEMAND noise]{
    Performance comparison on LRS3~\citep{afouras2018lrs3} with audio noise sampled from the DEMAND dataset~\citep{thiemann2013demand}. For each noisy environment, WER\,(\%) is measured by randomly sampling the SNR value from the range [$-$10, 0]\,dB.
    }
    \label{tab:mohave_demand_result_full}
    \resizebox{\textwidth}{!}{
    \begin{tabular}{l|cccccccccccccccccc|c}
    \toprule
    Model & \rotatebox[origin=l]{90}{\!\!\!PARK} & \rotatebox[origin=l]{90}{\!\!\!RIVER} & \rotatebox[origin=l]{90}{\!\!\!CAFE} & \rotatebox[origin=l]{90}{\!\!\!RESTO} & \rotatebox[origin=l]{90}{\!\!\!CAFETER} & \rotatebox[origin=l]{90}{\!\!\!METRO} & \rotatebox[origin=l]{90}{\!\!\!STATION} & \rotatebox[origin=l]{90}{\!\!\!MEETING} & \rotatebox[origin=l]{90}{\!\!\!KITCHEN} & \rotatebox[origin=l]{90}{\!\!\!LIVING} & \rotatebox[origin=l]{90}{\!\!\!WASH} & \rotatebox[origin=l]{90}{\!\!\!FIELD} & \rotatebox[origin=l]{90}{\!\!\!HALL} & \rotatebox[origin=l]{90}{\!\!\!OFFICE} & \rotatebox[origin=l]{90}{\!\!\!SQUARE} & \rotatebox[origin=l]{90}{\!\!\!TRAFFIC} & \rotatebox[origin=l]{90}{\!\!\!BUS} & \rotatebox[origin=l]{90}{\!\!\!CAR} & \rotatebox[origin=l]{90}{\!\!\!\textbf{AVG}} \\
    \midrule
    AV-HuBERT-\textsc{Large} & 3.8 & 6.1 & 6.4 & 13.1 & 8.6 & 3.1 & 4.7 & 5.7 & \textbf{2.2} & 3.6 & \textbf{1.6} & \textbf{1.7} & 2.1 & 1.9 & 2.7 & 3.0 & 1.9 & \textbf{1.6} & 4.1 \\
    \ourmodel-\textsc{Large} & \textbf{3.4} & \textbf{4.4} & \textbf{5.4} & \textbf{11.9} & \textbf{6.4} & \textbf{3.0} & \textbf{4.1} & \textbf{4.5} & \textbf{2.2} & \textbf{3.3} & 1.8 & 1.9 & \textbf{1.9} & \textbf{1.7} & \textbf{2.3} & \textbf{2.5} & \textbf{1.8} & \textbf{1.6} & \textbf{3.6} \\
    \bottomrule
    \end{tabular}
    }
    \vspace*{10pt}
\end{table*}

\paragraph{Evaluation under real-world noise conditions.}
Following the standard practice in robust AVSR works \citep{hong2023watch, kim2024learning}, we have introduced various noise conditions (babble, speech, music, and natural) to evaluate our MoHAVE's robustness and adaptability. To better reflect real-world noise conditions, we conduct further evaluations by augmenting LRS3 with realistic background audio from the DEMAND dataset \citep{thiemann2013demand}, which contains recordings from diverse indoor and outdoor environments, \eg cafeteria or meeting room. 
As summarized in Table~\ref{tab:mohave_demand_result_full}, on this enhanced benchmark, \ourmodel consistently outperforms AV-HuBERT across various real-world settings, achieving an average WER of 3.6\% for 18 environments. These results further confirm MoHAVE’s performance under realistic audio-visual conditions.

\begin{table}[!t]
    \centering
    \small
    \caption[MoHAVE performance on multilingual AVSR and AVS2TT in noisy MuAViC]{Multilingual audio-visual speech task performance, with non-English speech recognition (WER) and X-En speech-to-text translation (BLEU score), in a noisy environment with multilingual babble noise (SNR\,$=$\,0). $^\dagger$Results obtained from \citet{han-etal-2024-xlavs}. $^\ddagger$Re-implemented using the pretrained model from \citet{choi2024av2av}.}
    \label{tab:muavic}
    \addtolength{\tabcolsep}{-2pt}
    \resizebox{\textwidth}{!}{
    \begin{tabular}{llccccccccc}
    \toprule
    && \multicolumn{8}{c}{Source} & \\
    \cmidrule{3-10}
    Model & Pretrain data & Ar & De & El & Es & Fr & It & Pt & Ru & Avg \\
    \midrule
    \multicolumn{11}{c}{\textbf{\textit{Speech Recognition, Test WER $\downarrow$}}} \\
    Whisper large-v2\,\citep{radford2023robust} & \textit{680k hrs, 100+ langs} & 197.9 & 244.4 & 113.3 & 116.3 & 172.3 & 172.4 & 223.6 & 126.2 & 170.8 \\
    u-HuBERT$^\dagger$\,\citep{hsu2022u} & \textit{1.7k hrs, English} & 102.3 & 73.2 & 69.7 & 43.7 & 43.2 & 48.5 & 47.6 & 67.0 & 61.9 \\
    mAV-HuBERT\,\citep{anwar2023muavic} & \textit{1.7k hrs, English} & \textbf{82.2} & 66.9 & 62.2 & 40.7 & 39.0 & 44.3 & 43.1 & 43.1 & 52.7 \\
    XLS-R 300M$^\dagger$\,\citep{babu2022xls} & \textit{1.2k hrs, 9 langs} & 97.3 & 69.8 & 74.8 & 47.6 & 37.1 & 47.9 & 54.4 & 59.8 & 61.1 \\
    XLAVS-R 300M\,\citep{han-etal-2024-xlavs} & \textit{8.3k hrs, 100+ langs} & 91.9 & 53.5 & 49.6 & 28.8 & 29.3 & 32.2 & 32.5 & 46.1 & 45.5 \\
    XLAVS-R 2B\,\citep{han-etal-2024-xlavs} & \textit{1.2k hrs, 9 langs} & 93.5 & 58.5 & 38.6 & 23.9 & 23.5 & 24.6 & 26.1 & 41.0 & 41.2 \\[-1ex]
    \multicolumn{11}{c}{\hdashrule[0.5ex]{1.1\textwidth}{0.5pt}{1.5mm}} \\[-1ex]
    mAV-HuBERT$^\ddagger$ & \textit{7.0k hrs, 100+ langs} & 88.7 & 51.3 & 37.2 & 20.7 & 22.6 & 24.2 & 23.8 & 42.4 & 38.9 \\
    mAV-HuBERT\,+\,Hard Routing & \textit{7.0k hrs, 100+ langs} & 93.4 & 49.3 & 35.7 & 20.3 & 23.6 & 23.4 & 24.1 & 44.7 & 39.3 \\
    \rowcolor{blue!10}
    \ourmodel-\textsc{Large} (\textbf{ours}) & \textit{7.0k hrs, 100+ langs} & 92.9 & \textbf{47.3} & \textbf{35.3} & \textbf{18.7} & \textbf{21.2} & \textbf{21.6} & \textbf{21.9} & \textbf{40.6} & \textbf{37.4} \\
    \midrule
    \multicolumn{11}{c}{\textbf{\textit{X-En Speech-to-Text Translation, Test BLEU $\uparrow$}}} \\
    Whisper large-v2\,\citep{radford2023robust} & \textit{680k hrs, 100+ langs} & - & - & 0.1 & 0.4 & 0.7 & 0.1 & 0.1 & 0.2 & 0.3 \\
    mAV-HuBERT\,\citep{anwar2023muavic} & \textit{1.7k hrs, English} & - & - & 4.2 & 12.8 & 15.0 & 12.5 & 14.8 & 4.6 & 10.7 \\
    XLAVS-R 300M\,\citep{han-etal-2024-xlavs} & \textit{8.3k hrs, 100+ langs} & - & - & 13.2 & 17.4 & 23.8 & 18.7 & 21.8 & 9.4 & 17.4 \\
    XLAVS-R 2B\,\citep{han-etal-2024-xlavs} & \textit{1.2k hrs, 9 langs} & - & - & \textbf{15.7} & 19.2 & 24.6 & 20.1 & 22.3 & \textbf{10.4} & 18.7 \\[-1ex]
    \multicolumn{11}{c}{\hdashrule[0.5ex]{1.1\textwidth}{0.5pt}{1.5mm}} \\[-1ex]
    mAV-HuBERT$^\ddagger$ & \textit{7.0k hrs, 100+ langs} & - & - & 8.9 & 21.5 & 26.5 & 21.2 & 24.2 & 8.8 & 18.5 \\
    mAV-HuBERT\,+\,Hard Routing & \textit{7.0k hrs, 100+ langs} & - & - & 6.7 & 19.9 & 24.7 & 19.6 & 23.0 & 7.2 & 16.8 \\
    \rowcolor{blue!10}
    \ourmodel-\textsc{Large} (\textbf{ours}) & \textit{7.0k hrs, 100+ langs} & - & - & 11.4 & \textbf{22.3} & \textbf{27.1} & \textbf{22.1} & \textbf{25.1} & 9.2 & \textbf{19.5} \\
    \bottomrule
    \end{tabular}
    }
    \vspace*{10pt}
\end{table}

\subsection{Multilingual Audio-Visual Speech Tasks}
\label{subsec:ch4_multilingual}

\paragraph{Experimental setup.}
We evaluate \ourmodel on the MuAViC benchmark \citep{anwar2023muavic} for multilingual AVSR and X-to-English AVS2TT tasks. For multilingual AVSR, the dataset includes 8 non-English languages: Arabic (Ar), German (De), Greek (El), Spanish (Es), French (Fr), Italian (It), Portuguese (Pt), and Russian (Ru), encompassing approximately 700 hours of training data from 3,700 speakers. For X-En AVS2TT, the dataset covers 6 languages: Greek, Spanish, French, Italian, Portuguese, and Russian, where each sample includes audio-visual speech with corresponding English transcriptions.

A single multilingual model is trained for each task, capable of detecting the source language and generating target transcriptions accordingly. The evaluation is conducted on each language separately, as seen in Table\,\ref{tab:muavic}. Using the pretrained multilingual AV-HuBERT from \citep{choi2024av2av}, we fine-tune the model for 120K steps, unfreezing the encoder after 10K steps. Inference is performed with beam size of 5, and the samples with empty ground-truth transcriptions are removed from the evaluation set. 

\paragraph{Result.}
MoEs have demonstrated effectiveness in multilingual speech tasks \citep{hu2023mixture, wang2023language}, as MoE is capable of enabling more diverse routing paths for different language tokens. To evaluate \ourmodel's multilingual capabilities, we train a multilingual model and assess its performance on the MuAViC benchmark~\citep{anwar2023muavic}, evaluating separately for each language.
Following \citet{han-etal-2024-xlavs}, we introduce multilingual babble noise at SNR 0\,dB to 50\% of the input samples during training, where the noise clips are sampled from the MuAViC train set. For inference, we apply beam search with a beam size of 5 and normalize text by punctuation removal and lower-casing before calculating WER. For AVS2TT evaluation, we use SacreBLEU\,\citep{post2018call} with its built-in \textit{13a} tokenizer. To simulate noisy test conditions, we inject babble noise sampled from the MuAViC test set.

Table\,\ref{tab:muavic} summarizes the results, where \ourmodel-\textsc{Large} achieves superior performance in both AVSR and AVS2TT. Whisper \citep{radford2023robust}, a leading multilingual ASR model, is known to perform poorly in noisy setup due to its lack of visual understanding for robustness. While multilingual AV-HuBERT \citep{anwar2023muavic} underperforms the state-of-the-art models like XLAVS-R \citep{han-etal-2024-xlavs}, we have re-implemented it using the pretrained model from \citet{choi2024av2av}, which has been pretrained on a significantly larger dataset including 7,000 hours of speech in 100+ languages. This model outperforms (38.9\% average WER) or remains competitive (18.5\% average BLEU) with much larger XLAVS-R 2B. When integrated with this version, \ourmodel further improves performance, achieving \textbf{37.4\% average WER} and \textbf{19.5\% average BLEU}, setting new benchmarks in almost every language being evaluated.

\begin{table*}[!t]
    \centering
    \small
    \caption[MoHAVE performance on multilingual AVSR and AVS2TT in clean MuAViC]{Multilingual audio-visual speech task performance, with non-English speech recognition (WER) and X-En speech-to-text translation (BLEU score), in a clean environment without auditory noise. $^\dagger$Results obtained from \citet{han-etal-2024-xlavs}. $^\ddagger$Re-implemented using the pretrained model from \citet{choi2024av2av}.}
    \label{tab:muavic_clean}
    \addtolength{\tabcolsep}{-2pt}
    \resizebox{\textwidth}{!}{
    \begin{tabular}{llccccccccc}
    \toprule
    && \multicolumn{8}{c}{Source} & \\
    \cmidrule{3-10}
    Model & Pretrain data & Ar & De & El & Es & Fr & It & Pt & Ru & Avg \\
    \midrule
    \multicolumn{11}{c}{\textbf{\textit{Clean Speech Recognition, Test WER $\downarrow$}}} \\
    Whisper large-v2\,\citep{radford2023robust} & \textit{680k hrs, 100+ langs} & 91.5 & \textbf{24.8} & 25.4 & 12.0 & 12.7 & 13.0 & 15.5 & 31.1 & 28.2 \\
    u-HuBERT$^\dagger$\,\citep{hsu2022u} & \textit{1.7k hrs, English} & 89.3 & 52.1 & 46.4 & 17.3 & 20.5 & 21.2 & 21.9 & 44.4 & 39.1 \\
    mAV-HuBERT\,\citep{anwar2023muavic} & \textit{1.7k hrs, English} & 69.3 & 47.2 & 41.2 & 16.2 & 19.0 & 19.8 & 19.9 & 38.0 & 33.8 \\
    XLS-R 300M$^\dagger$\,\citep{babu2022xls} & \textit{1.2k hrs, 9 langs} & 85.6 & 44.0 & 34.4 & 13.2 & 15.1 & 14.3 & 16.2 & 34.4 & 32.2 \\
    XLAVS-R 300M\,\citep{han-etal-2024-xlavs} & \textit{8.3k hrs, 100+ langs} & 80.0 & 38.0 & 28.1 & 11.7 & 15.3 & 13.8 & 14.4 & 31.2 & 29.1 \\
    XLAVS-R 2B\,\citep{han-etal-2024-xlavs} & \textit{1.2k hrs, 9 langs} & 79.3 & 44.4 & \textbf{19.0} & \textbf{9.1} & \textbf{12.3} & \textbf{10.6} & \textbf{11.2} & \textbf{25.0} & \textbf{26.4} \\
    \multicolumn{11}{c}{\hdashrule[0.5ex]{1.1\textwidth}{0.5pt}{1.5mm}} \\[-1ex]
    mAV-HuBERT$^\ddagger$ & \textit{7.0k hrs, 100+ langs} & \textbf{78.3} & 41.4 & 25.5 & 11.9 & 16.2 & 14.8 & 14.3 & 31.6 & 29.3 \\
    \ourmodel-\textsc{Large} (\textbf{ours}) & \textit{7.0k hrs, 100+ langs} & 85.1 & 38.9 & 25.9 & 11.2 & 14.6 & 14.0 & 13.8 & 30.0 & 29.2 \\
    \midrule
    \multicolumn{11}{c}{\textbf{\textit{Clean X-En Speech-to-Text Translation, Test BLEU $\uparrow$}}} \\
    Whisper large-v2\,\citep{radford2023robust} & \textit{680k hrs, 100+ langs} & - & - & \textbf{24.2} & \textbf{28.9} & \textbf{34.5} & \textbf{29.2} & \textbf{32.6} & \textbf{16.1} & \textbf{29.9} \\
    mAV-HuBERT\,\citep{anwar2023muavic} & \textit{1.7k hrs, English} & - & - & 7.6 & 20.5 & 25.2 & 20.0 & 24.0 & 8.1 & 17.6 \\
    XLAVS-R 300M\,\citep{han-etal-2024-xlavs} & \textit{8.3k hrs, 100+ langs} & - & - & 18.3 & 23.9 & 29.8 & 25.1 & 28.9 & 12.1 & 23.0 \\
    XLAVS-R 2B\,\citep{han-etal-2024-xlavs} & \textit{1.2k hrs, 9 langs} & - & - & 21.6 & 25.1 & 30.6 & 26.6 & 29.9 & 13.9 & 24.6 \\
    \multicolumn{11}{c}{\hdashrule[0.5ex]{1.1\textwidth}{0.5pt}{1.5mm}} \\[-1ex]
    mAV-HuBERT$^\ddagger$ & \textit{7.0k hrs, 100+ langs} & - & - & 11.5 & 24.2 & 29.2 & 23.9 & 28.1 & 10.4 & 21.2 \\
    \ourmodel-\textsc{Large} (\textbf{ours}) & \textit{7.0k hrs, 100+ langs} & - & - & 13.8 & 24.9 & 30.8 & 25.0 & 28.7 & 10.9 & 22.4 \\
    \bottomrule
    \end{tabular}
    }
    \vspace*{10pt}
\end{table*}

\paragraph{Multilingual tasks in clean environments.}
Table\,\ref{tab:muavic_clean} outlines the MuAViC benchmark results in a clean environment, without auditory noise added. The experimental setup remains consistent with Table\,\ref{tab:muavic}, utilizing the same models. Unlike the noisy setting, we observe that \ourmodel does not yield significant performance improvements in clean speech tasks. This is primarily because \ourmodel enhances AVSR under noisy conditions by dynamically adjusting the utilization of audio and visual expert groups. 

Indeed, in clean speech recognition and translation tasks, encoder capacity---particularly when pretrained on large-scale audio data---plays a more crucial role than decoder-specific training methods. In addition, visual information is less essential in noise-free environments, as demonstrated by the strong ASR performance of the Whisper-large-v2 model \citep{radford2023robust}. Even the smaller ASR model, XLS-R 300M \citep{babu2022xls}, surpasses AVSR models such as mAV-HuBERT \citep{anwar2023muavic} or u-HuBERT \citep{hsu2022u} in this setting, underscoring that the advantage of using AVSR models emerges most clearly in robust speech recognition.

\subsection{Number of Activated Experts}

\begin{wraptable}{r}{0.65\textwidth}
    \centering
    \small
    \vspace{-10pt}
    \caption[Sensitivity study of the number of activated experts]{Impact of the number of activated experts on AVSR performance.}
    \label{tab:num_experts}
    \resizebox{0.65\textwidth}{!}{
    \begin{tabular}{c|cccc|c|c}
        \toprule
        ($k^A$, $k^V$) & babble & speech & music & natural & N-WER & C-WER \\
        \midrule
        ($1, 1$) & 9.6 & 4.2 & 4.7 & 4.5 & 5.8 & 1.8 \\
        ($1, 2$) & 9.3 & 4.1 & 4.8 & 4.4 & 5.7 & 1.9 \\
        ($2, 1$) & 10.1 & 4.5 & 5.3 & 4.9 & 6.2 & 2.4 \\
        ($2, 2$) & 11.0 & 5.2 & 5.8 & 5.5 & 6.9 & 3.0 \\
        \bottomrule
    \end{tabular}
    }
    \vspace*{-10pt}
\end{wraptable}

By default, \ourmodel selects one expert from each group---audio and visual---activating a total of two experts per token. This design is to match the compute of standard MoE implementations, which utilizes top-2 out of 8 experts. A natural question arises: \textit{does activating more experts improve performance, or does it simply increase computational costs without substantial gains?}

Table\,\ref{tab:num_experts} presents the results when activating more experts of \ourmodel-\textsc{Base}, where top-$k^A$ experts from the audio group and top-$k^V$ experts from the visual group are selected. Interestingly, increasing the number of audio experts significantly degrades performance, implying that the model might be confused by employing another sub-optimal expert.

In contrast, activating two visual experts while keeping one audio expert improves performance (N-WER of 5.7\%) compared to the default setting of single visual expert. Particularly under the babble noise, WER has decreased from 9.6\% to 9.3\%. 
This suggests that adding an additional visual expert can be beneficial in noisy environments, likely due to the increased robustness from visual information in challenging audio conditions.

\subsection{Unimodal Task Results}

Table\,\ref{tab:mohave_asr_vsr} presents unimodal task results, evaluating model performance on video-only (VSR) sequences and audio-only (ASR) sequences. 
BRAVEn \citep{haliassos2024braven} and Llama-3.1-8B-AVSR \citep{cappellazzo2024large} models achieve the best VSR performance, as these models are specifically pretrained for the VSR task. While using an LLM decoder is highly effective in VSR, since LLMs are able to refine and correct recognition errors, ASR performance is largely determined by the encoder's pretraining strategy as BRAVEn and Whisper encoders.
As an adaptive audio-visual model, \ourmodel does not specialize in unimodal tasks but instead performs robustly in multimodal AVSR. It only exhibits a slight improvement in VSR over AV-HuBERT. These results indicate that unimodal performance is primarily influenced by the effectiveness of the encoder pretraining strategy rather than the MoE-based multimodal approach.

\begin{table}[!h]
    \centering
    \small
    \caption[MoHAVE performance of unimodal tasks on LRS3]{Comparison on the unimodal ASR and VSR task performance.}
    \label{tab:mohave_asr_vsr}
    \begin{tabular}{ll|cc}
    \toprule
    Method & Encoder\,+\,Decoder & V & A \\
    \midrule
    BRAVEn~\citep{haliassos2024braven} & BRAVEn\,+\,Transformer & 26.6 & 1.2 \\
    Llama3.1-8B-AVSR~\citep{cappellazzo2024large} & AV-HuBERT\,+\,LLM & 25.3 & 1.4 \\ 
    Llama3.1-8B-AVSR~\citep{cappellazzo2024large} & Whisper\,+\,LLM & - & 1.1 \\
    \midrule
    AV-HuBERT-\textsc{Large}~\citep{shi2022learning} & AV-HuBERT\,+\,Transformer & 28.6 & 1.4 \\
    \ourmodel-\textsc{Large} (\textbf{ours}) & AV-HuBERT\,+\,MoE & 28.2 & 1.4 \\
    \bottomrule
    \end{tabular}
    \vspace*{10pt}
\end{table}

\begin{table}[!t]
    \centering
    \small
    \caption[Ablation study of the MoHAVE application to the encoder and decoder]{Performance comparison of \ourmodel applied to the encoder, decoder, and both.}
    \label{tab:encoder_mohave}
    \begin{tabular}{l|cccc|c|c}
        \toprule
        Method & babble & speech & music & natural & N-WER & C-WER \\
        \midrule
        Encoder MoHAVE (pretrain only MoE) & 10.5 & 5.0 & 5.6 & 5.0 & 6.5 & 2.4 \\
        Encoder MoHAVE & 10.0 & 4.4 & 5.1 & 4.5 & 6.0 & 1.9 \\
        Encoder + Decoder MoHAVE & 10.1 & 4.7 & 5.3 & 4.7 & 6.2 & 2.0 \\
        \midrule
        Decoder MoHAVE & 9.6 & 4.2 & 4.7 & 4.5 & 5.8 & 1.8 \\
        Decoder MoHAVE (uptrain) & 9.7 & 4.2 & 4.8 & 4.5 & 5.8 & 1.8 \\        
        \bottomrule
    \end{tabular}
    \vspace*{10pt}
\end{table}

\subsection{Variations of MoHAVE Implementations}

\paragraph{\ourmodel in the encoder.}

We have implemented \ourmodel by integrating MoE into the decoder to facilitate text token processing while enhancing multimodal fusion. Since the AVSR decoder incorporates information from both audio and visual modalities along with text tokens, the decoder-based MoHAVE is expected to be the most effective strategy. An alternative approach is to apply MoHAVE within the encoder, by pretraining the encoder using the AV-HuBERT masked prediction strategy \citep{shi2022learning}. For this, we initialize the pretrained encoder (with standard Transformers) and convert the FFN layers into MoE layers by copying the FFN parameters into all the expert modules. Since the \textsc{Base} model consists of 12 encoder layers, we convert 6 of them alternatively to match the number of MoE layers in the decoder \ourmodel. During fine-tuning, all MoE layers in the encoder are also trained following the same procedure.

There are two options for pretraining strategies: (1) pretraining only the MoE layers initialized from the FFN parameters, and (2) pretraining the entire encoder with MoE layers. As shown in Table\,\ref{tab:encoder_mohave}, the latter approach significantly outperforms the former, suggesting that encoder \ourmodel requires full pretraining for effective learning.
However, even with full pretraining, encoder \ourmodel performs inferior to decoder \ourmodel. This is because the encoder only processes audio and visual tokens, whereas the decoder directly integrates audio-visual embeddings with text, finding optimal strategies for text token dispatching that best improves speech recognition. In addition, applying \ourmodel to both the encoder and decoder leads to degraded performance despite the increased computational cost.

\paragraph{Decoder uptraining.}

We also explore a successive training strategy for the decoder, referred to as uptraining \citep{ainslie2023gqa}, where the decoder \ourmodel undergoes additional training after the fine-tuning phase of standard Transformers. However, as seen in Table\,\ref{tab:encoder_mohave}, uptraining does not yield further improvements compared to training from scratch, even after an additional 120K training steps. In fact, we observed the shorter uptraining steps leading to degraded performance. This suggests that training the decoder \ourmodel requires a comprehensive learning phase rather than incremental fine-tuning, as MoE  may fundamentally alter the processing pathways of tokens.

\begin{figure*}[!t]
    \centering
    \includegraphics[width=.9\linewidth]{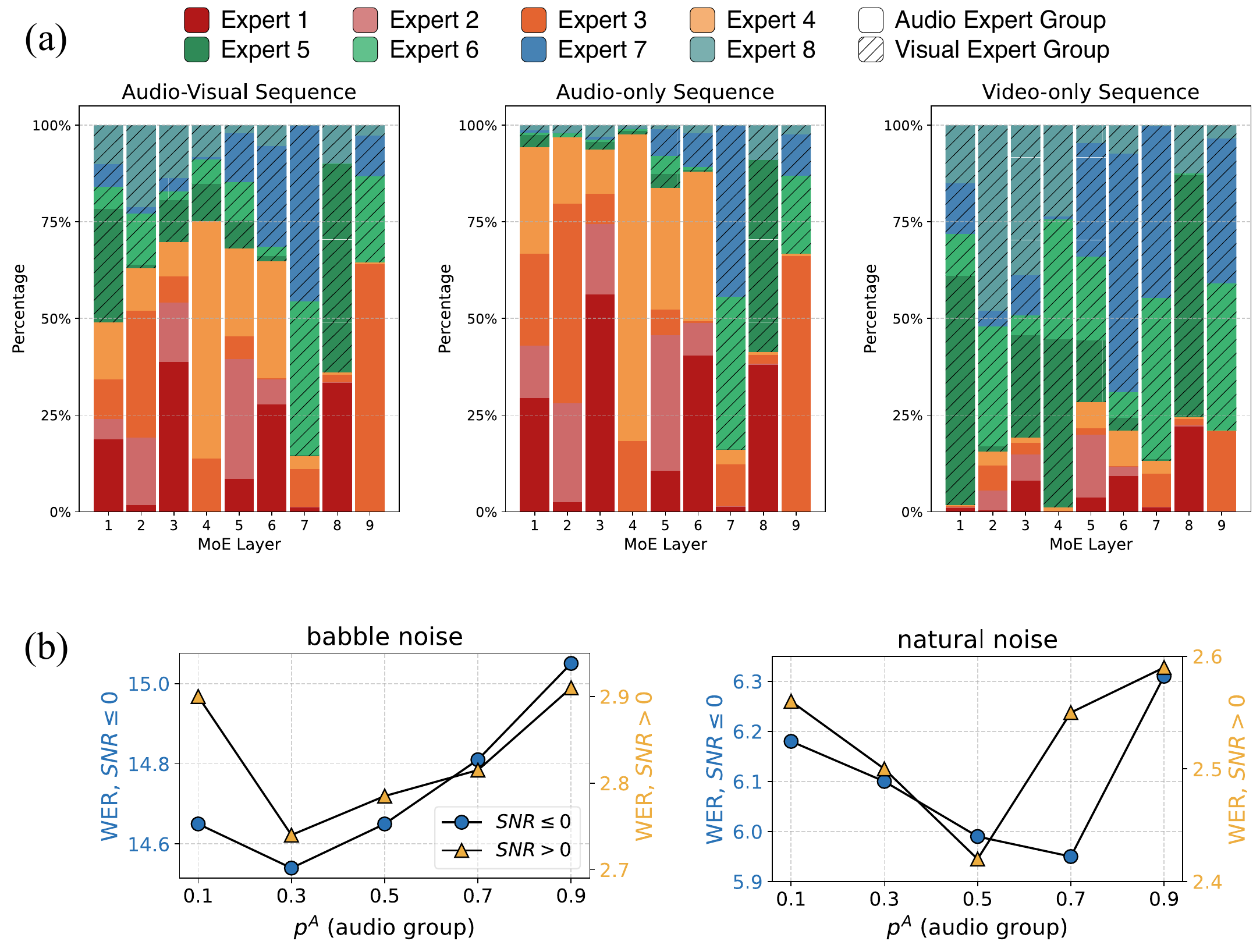}
    \vspace{-5pt}
    \caption[Analysis of expert load for MoHAVE and sensitivity of the hard routing strategy]{(a) Expert load distribution in \ourmodel according to input modalities, with expert selection frequencies weighted by the inter-modal router’s output probability. (b) Performance of the hard routing strategy under different weight assignments to audio expert group. The visual expert group is weighted by $p^V = 1-p^A$.
    }
    \label{fig:expert_load}
    \vspace*{10pt}
\end{figure*}

\begin{figure*}[!t]
    \centering
    \includegraphics[width=.8\linewidth]{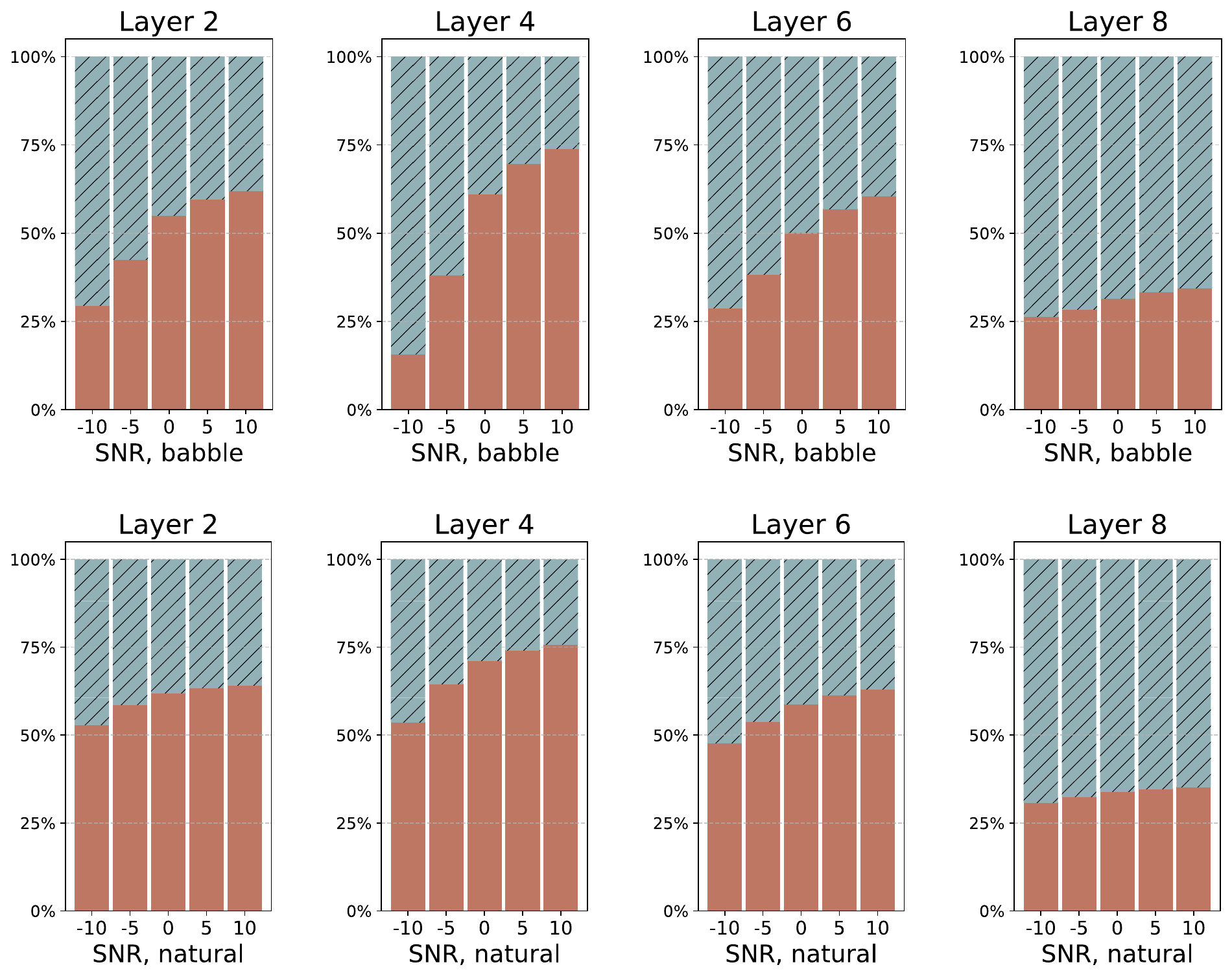}
    \vspace{-5pt}
    \caption[Analysis of expert group load for MoHAVE under audio corruption]{Expert load distribution in \ourmodel for the audio group (solid bars) and visual group (dashed bars) across noisy audio-visual sequences under babble (left) and natural (right) noise. Full layer-wise results are provided in Figure~\ref{fig:appx_group_load}.
    }
    \label{fig:expert_load_noise}
    \vspace*{10pt}
\end{figure*}

\section{Expert and Group Load Analysis}
\label{sec:analysis_load}

\subsection{\ourmodel's Expert Load Distribution}
Figure\,\ref{fig:expert_load}\textcolor{red}{a} illustrates the expert load distribution of \ourmodel according to input types: audio-visual, audio-only, and video-only sequences. For audio-visual inputs, all experts from both the audio and visual groups are selected at similar frequencies, with some layer-dependent variations. In contrast, when processing audio-only sequences, the model predominantly activates the audio expert group, while for video-only sequences, the visual expert group is mainly utilized. This distribution validates the effectiveness of our load biasing loss in guiding the inter-modal router to assign appropriate weights based on input modality.

\subsection{Expert Group Utilization in Noisy AVSR}
To analyze the effectiveness of hierarchical gating in AVSR, we first examine the limitations of hard routing (\S\ref{subsec:hard_routing}) under noisy conditions. Since hard routing relies on manually (de-)activating the audio and visual groups, for audio-visual inputs, we assign a fixed equal weight ($p^A, p^V = 0.5$) to both groups.
However, this equal weighting may not always be optimal in varying environments, such as noise type or intensity.

As shown in Figure\,\ref{fig:expert_load}\textcolor{red}{b}, increasing reliance on the audio group under babble noise degrades performance, with an optimal weight for the audio group being $0.3$.
Unlike babble noise, which confuses the model with multiple overlapping speech signals, natural noise is more distinct from speech, leading to a higher reliance on the audio group ($p^A \ge 0.5$) preferable.
These results indicate that an ideal routing strategy for audio-visual data should be dynamically adjusted.

Figure\,\ref{fig:expert_load_noise} further illustrates \ourmodel's group load distribution across different noise levels. The model adaptively adjusts its reliance between the audio and visual expert groups---under high noise conditions (low SNRs), it shifts more tokens to the visual group, while in cleaner conditions (high SNRs), the audio group is more actively utilized. This behavior also adjusts to noise types, as observed with babble and natural noise, demonstrating the MoHAVE’s adaptability and robustness.
\begin{figure}[!t]
    \centering
    \includegraphics[width=\linewidth]{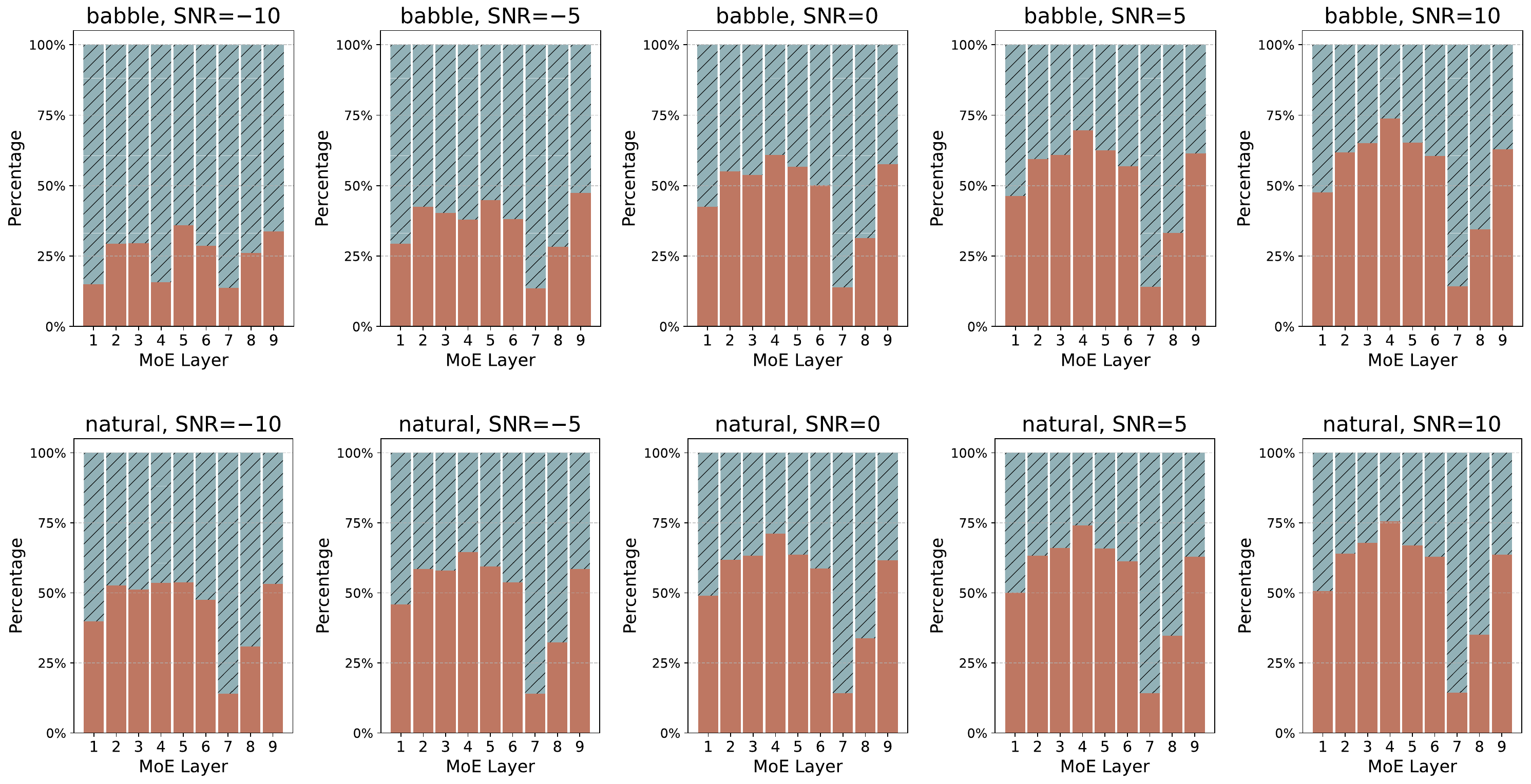}
    \vspace{-20pt}
    \caption[Layer-wise analysis of expert group load for MoHAVE under audio corruption]{Expert load distribution in \ourmodel for the audio group (solid bars) and visual group (dashed bars) across noisy audio-visual sequences under babble (first row) and natural (second row) noise. The frequency of each expert has been weighted by the inter-modal router's output probability.}
    \label{fig:appx_group_load}
    \vspace*{10pt}
\end{figure}

Figure\,\ref{fig:appx_group_load} provides the distribution across all MoE layers, illustrating how \ourmodel dynamically adjusts expert groups based on noise conditions.

\begin{figure}[!t]
    \centering
    \includegraphics[width=\linewidth]{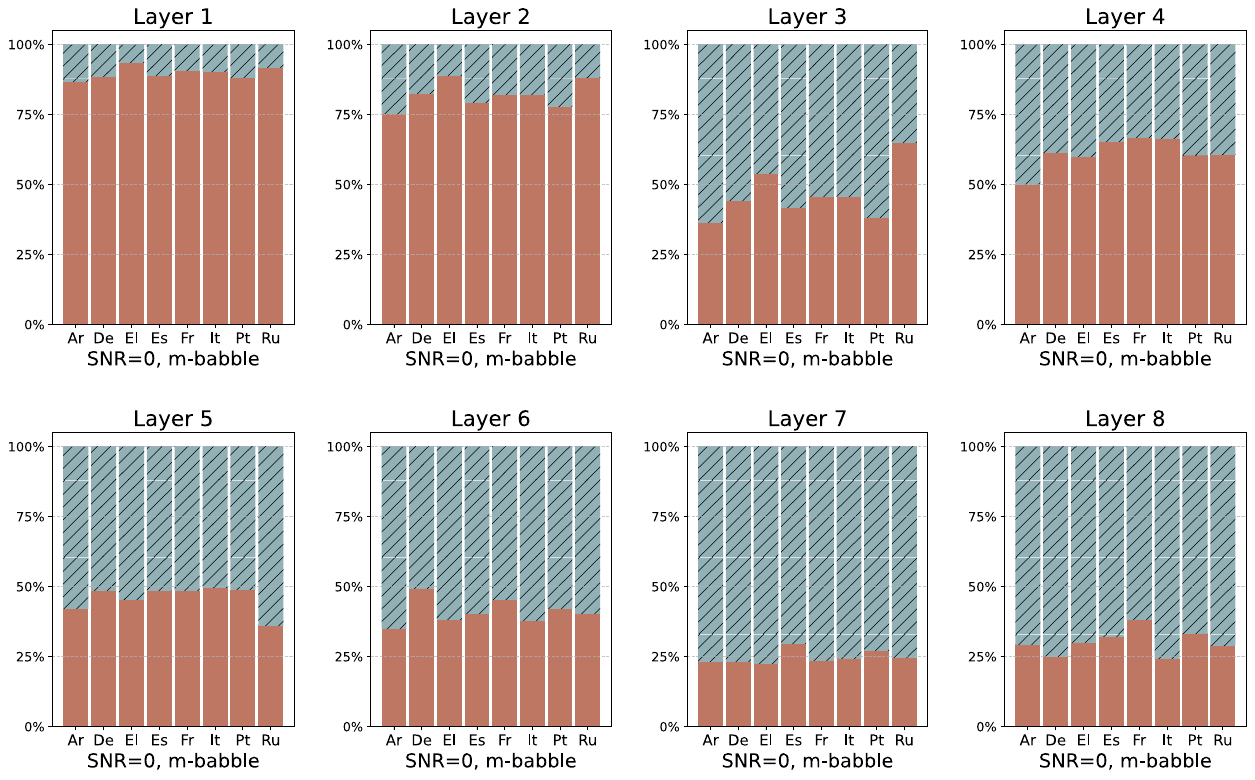}
    \vspace{-20pt}
    \caption[Analysis of expert group load for MoHAVE in multilingual AVSR]{Expert load distribution in multilingual AVSR \ourmodel for the audio group (solid bars) and visual group (dashed bars).}
    \label{fig:mavsr_all}
    \vspace*{10pt}
\end{figure}

\subsection{Language-wise Analysis on Multilingual Tasks}
We additionally provide language-wise analysis on multilingual AVSR in Figure\,\ref{fig:mavsr_all}.
Our analysis indicates language-dependent differences in the expert allocation within \ourmodel. For example, Arabic tokens tend to be routed more frequently toward visual experts, whereas Spanish or French tokens rely more heavily on audio experts. However, we also note that these trends vary by layer. Also, within each expert group, the intra-modal router’s load-balancing ensures a uniform expert utilization across data samples. Thus, there is no explicit language-specific expert selection within groups, consistent with observations found in \citet{zoph2022st}. We suppose that more detailed investigation into expert load distribution across languages and its relation to linguistic/paralinguistic characteristics would serve as valuable future work.

\section{Chapter Summary}

In this chapter, we propose \ourmodel, a hierarchical MoE framework for AVSR, designed to enhance scalability and robustness. By training an inter-modal router that dynamically assigns weights to audio and visual expert groups, \ourmodel enables an adaptive group selection based on input context. Evaluations on robust AVSR benchmarks demonstrate its state-of-the-art performance, with superior noise resilience, further supported by flexible expert load distributions across diverse noisy conditions. This work establishes an adaptive modality-aware MoE paradigm, advancing larger-scale multimodal speech recognition systems.
\chapter{System-Level Scalability of AVSR}
\label{chap:system_scalability}

\begin{center}
    \vspace{0.5em}
    \begin{tcolorbox}[
        colback=gray!10, 
        colframe=gray!50, 
        boxrule=0.5pt,
        title=Summary: Chapter based on preprint work~\citep{kim2025two}, fonttitle=\bfseries,
        width=0.92\linewidth,
        coltitle=black
    ]
    This chapter introduces a new paradigm for generative error correction (GER) framework in audio-visual speech recognition (AVSR) that reasons over modality-specific evidences directly in the language space. Our framework, \textbf{DualHyp}, empowers a large language model (LLM) to compose independent $N$-best hypotheses from separate automatic speech recognition (ASR) and visual speech recognition (VSR) models. To maximize the effectiveness of DualHyp, we further introduce \textbf{RelPrompt}, a noise-aware guidance mechanism that provides modality-grounded prompts to the LLM. RelPrompt offers the temporal reliability of each modality stream, guiding the model to dynamically switch its focus between ASR and VSR hypotheses for an accurate correction. Under various corruption scenarios, our framework attains up to 57.7\% error rate gain on the LRS2 benchmark over standard ASR baseline, contrary to single-stream GER approaches that achieve only 10\% gain. To facilitate research within our DualHyp framework, we release the code and the dataset comprising ASR and VSR hypotheses at \url{https://github.com/sungnyun/dualhyp}.
    \end{tcolorbox}
    \vspace{0.5em}
\end{center}

\section{Two Heads Are Better Than One: Audio-Visual Speech Error Correction with Dual Hypotheses}
\label{sec:ch5_intro}

Recent advancements have introduced GER frameworks that utilize LLMs to refine ASR outputs.
Following the release of $N$-best ASR hypotheses dataset \citep{chen2023hyporadise}, numerous studies demonstrated the efficacy of LLMs in correcting transcriptions based on the hypotheses list \citep{hu2024robustger, hu2024clozeger, mu2024mmger, mu2025mixture}.
These powerful correction frameworks, however, presents a fundamental limitation.
While the performance of the underlying ASR systems is remarkable in controlled environments~\citep{graves2012sequence, zhang2020transformer, chiu2022self, peng2024owsm}, it degrades significantly in noisy real-world conditions where acoustic distortions are prevalent.
To mitigate this challenge, AVSR systems have been developed~\citep{kim2024learning, kim2025mohave, han2024xlavs, shi2022learning, chen2023leveraging}, leveraging complementary visual cues (\eg lip movements) to enhance robustness against noise.

In the realm of AVSR, integrating visual information into GER frameworks remains a nascent area of research.
Existing methods often employ visual adapters~\citep{ghosh2024lipger} or unified AVSR models~\citep{liu2025avger}, both of which process visual data in the feature space.
This feature-level fusion struggles when audio and visual streams are corrupted independently, as noise from one modality can easily contaminate the unified representation \citep{kim2025multitask}.
Moreover, these frameworks heavily rely on a single set of hypotheses generated from one, often error-prone, recognition model.

To address these limitations, we propose \textbf{DualHyp}, the first GER framework that explicitly maintains modality-specific pathways from separate ASR and VSR systems\,(\S\ref{sec:dualhyp}).
LLM intelligently composes these \textbf{dual-stream hypotheses}, leveraging the model's deep contextual understanding in the \textit{language space} rather than forcing the model to interpret complex audio or video embedding subspaces.
Building upon this, we introduce \textbf{RelPrompt}, a \textbf{noise-aware guidance} mechanism that directs the underlying quality of each modality (\S\ref{sec:relprompt}).
Since LLMs for GER primarily operate within the language space, they lack modality-level grounding and may incorrectly prioritize unreliable sources.
To mitigate this, we incorporate reliability predictors to assess the quality of audio and visual streams, which are fed to the LLM to better elicit the compositional capacity of DualHyp.

Our experiments\,(\S\ref{sec:ch5_experiments}) show that this DualHyp approach with RelPrompt significantly outperforms prior single-stream GER frameworks across various audio-visual corruption scenarios.
We also demonstrate its multilingual capabilities as well as improved reasoning with larger LLMs.
Through qualitative analysis\,(\S\ref{sec:ch5_analysis}), we investigate the correction mechanism that makes our framework more effective.
\textit{To facilitate research within the DualHyp framework, we release the dataset comprising ASR and VSR hypotheses.}

\begin{figure*}[!t]
    \centering
    \includegraphics[width=\linewidth]{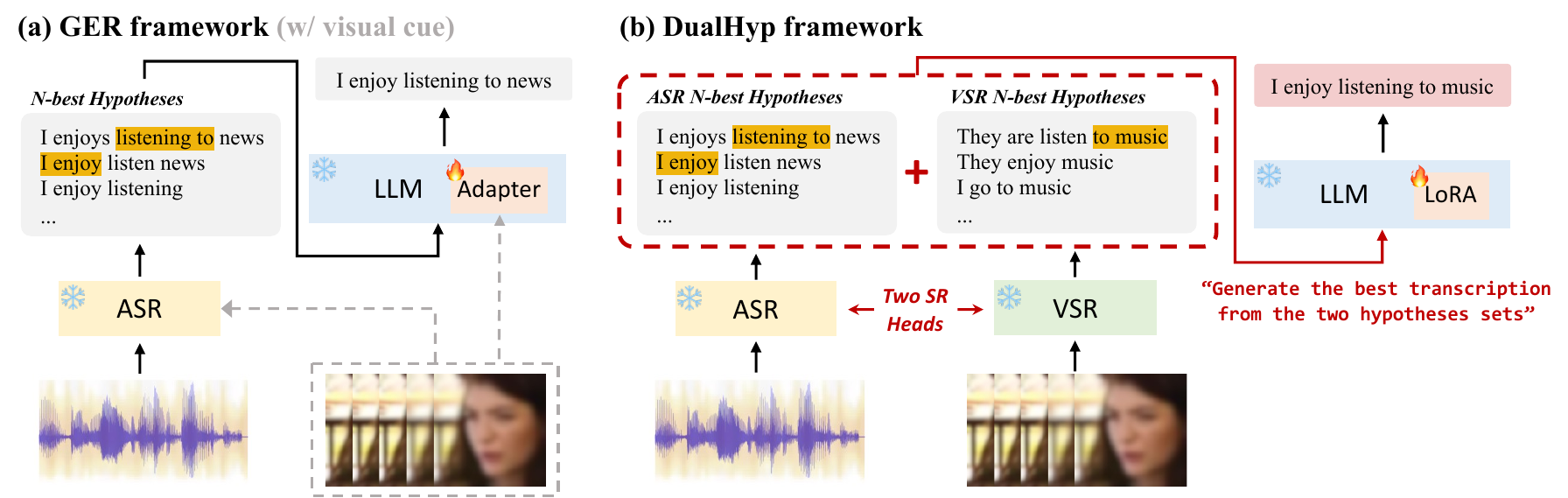}
    \caption[Overview of GER and DualHyp frameworks]{
    (a) Conventional GER frameworks use a single set of ASR hypotheses and (optionally) injects visual features via an adapter or a multimodal encoder. 
    (b) Our DualHyp framework maintains modality separation, using both ASR and VSR heads to generate two distinct sets of textual hypotheses. The LLM performs compositional reasoning on dual hypotheses in the language space to produce a more robust and accurate transcription.
    }
    \label{fig:ch5_overview}
\end{figure*}
\section{Related Work}
\label{sec:ch5_related_work}

\subsection{Generative Error Correction for Speech}
Recently, there has been growing interest in using LLMs for post-hoc correction of speech recognition outputs.
Initial work in GER for ASR demonstrates that LLMs can effectively regenerate transcriptions from $N$-best hypothesis lists~\citep{chen2023hyporadise}.
Subsequent research has refined this paradigm by exploring novel prompting strategies like cloze-style completion~\citep{hu2024clozeger} or by re-injecting acoustic features to better ground the LLM's corrections~\citep{chen2024uadf, radhakrishnan2023whispering, mu2024mmger, mu2025mixture, liu2025denoising}. These foundational works, however, focus exclusively on correcting hypotheses generated from a single, audio-only stream.

\subsection{Modality Fusion in GER for AVSR}
Extending GER to the audio-visual domain presents the central challenge of how to effectively fuse multimodal information. Existing approaches perform this fusion in the feature space, before the final language generation step.
\citet{ghosh2024lipger} involved visual adapters~\citep{houlsby2019parameter, zhang2024llama} to inject lip-reading features directly into the LLM, while \citet{liu2025avger} used dedicated multimodal encoders to create a unified audio-visual representation.
While these methods show promise, their reliance on early, feature-level fusion makes them vulnerable to cross-modal contamination~\citep{hong2022visual}, where corruption in one modality can degrade the quality of the fused representation.

Motivated by prior works highlighting the benefits of modality-specific processing for robustness~\citep{wang2024mlca, kim2025multitask, liu2021audio}, our approach is designed to isolate corruptions specific to each modality before error correction.
In contrast to feature-level fusion methods, we achieve this by deliberately delaying the modality fusion to the generation stage where the LLM operates on independent textual hypotheses from separate ASR and VSR models.

\subsection{End-to-End LLM-based AVSR}
It is important to distinguish our GER framework from an orthogonal line of research that uses LLMs for end-to-end\,(E2E) ASR~\citep{ma2024embarrassingly, yu2024connecting, fathullah2024prompting} and AVSR~\citep{cappellazzo2025large, cappellazzo2025adaptive, cappellazzo2025scaling, yeo2025mms, yeo2024visual}. In that paradigm, encoded audio and visual features serve as direct, multimodal prompts for a single generative model.
While promising, our decoupled approach offers significant advantages in flexibility. 

First, our framework is highly modular and can readily use off-the-shelf ASR systems and LLMs.
This contrasts with monolithic E2E models, which require costly pretraining of the entire system for any component update.
Second, the system can be easily improved by refining text-based prompts. This avoids the inherent complexity of designing and aligning cross-modal prompts, which is a central challenge in E2E systems.

\section{DualHyp Framework}
\label{sec:dualhyp}

\subsection{Uni-modal Generative Error Correction}
\label{subsec:unimodal_ger}

Recent works have successfully employed LLMs for GER~\citep{chen2023hyporadise, hu2024robustger, ghosh2024lipger}, where they aim to refine outputs of a uni-modal ASR system.
Given an input utterance, the ASR model first generates an $N$-best list of candidate transcriptions by beam search decoding, denoted as $\mathcal{H}^{\text{asr}}=\{(h_i^a, s_i^a)\}_{i=1}^N$, where $h_i^a$ is the $i$-th hypothesis and $s_i^a$ is the corresponding log-likelihood score. The LLM takes this hypothesis set as an input and generates a corrected transcription $\hat{y}$ via conditional generation:
\begin{equation}
    \hat{y} = \arg\max_{y} P(y \mid \mathcal{H}^{\text{asr}}; {\theta}_\text{LLM}).
\end{equation}

This approach has proven effective in clean acoustic conditions; however, its performance is fundamentally capped by the quality of the initial ASR hypotheses. When the source audio is severely corrupted by noise such as negative signal-to-noise (SNR) level, the resulting hypotheses are too erroneous to provide useful signal for correction, creating a performance bottleneck.

In contrast, visual information such as lip movements offers a complementary modality that is invariant to acoustic noise.
Visual modality has been shown to be particularly useful in disambiguating homophones or recovering missing segments in noisy environments~\citep{kim2022distinguishing, kim2024learning}.
Motivated by this, we propose to extend GER beyond a single-stream hypothesis by incorporating both audio and visual modalities in a unified framework.

\newpage
\subsection{Oracle Error Analysis of Speech Recognition Systems}
\label{subsec:oracle_error_analysis}

\begin{wraptable}{r}{.55\textwidth}
\centering
\small
\vspace{-10pt}
\caption[Oracle error analysis with different speech recognition heads]{WER\,(\%) analysis with different speech recognition heads, evaluated on noise-augmented LRS2. $o_{nb}$: \emph{$N$-best oracle}, $o_{cp}$: \emph{compositional oracle}. The corruption setting is same as Table~\ref{tab:main_lrs2}\textcolor{Red}{a}, overall results reported.
}
\label{tab:prelim}
\addtolength{\tabcolsep}{0pt}
\resizebox{.55\columnwidth}{!}{
\begin{tabular}{lcccc}
    \toprule
    \textbf{SR Head} & \textbf{Input} & \!\!\textbf{1-best}\!\! & $\pmb{o_{nb}}$ & $\pmb{o_{cp}}$ \\
    \midrule
    Whisper-large-v3 & A & 25.8 & 16.7 & 13.7 \\
    BRAVEn-large & V & 39.7  & 27.8 & 24.6\\
    Auto-AVSR & AV & 24.9 & 16.1 & 13.6 \\
    \midrule
    Whisper\,+\,Auto-AVSR & A\,+\,AV & -- & 7.0 & 4.9 \\
    Whisper\,+\,BRAVEn & A\,+\,V & -- & \textbf{6.4} & \textbf{4.5} \\
    \bottomrule
\end{tabular}
}
\end{wraptable}

To ascertain the potential benefits of incorporating a second modality, we conduct an oracle error analysis of speech recognition systems, in addition to standard 1-best word error rate (WER).
This oracle analysis establishes theoretical lower bounds of ASR and VSR systems in two manners~\citep{chen2023hyporadise}: \textit{N-best oracle} ($o_{nb}$), which selects the single best hypothesis from an $N$-best list, and \textit{compositional oracle} ($o_{cp}$), which constructs an optimal transcript by combining correct words from all $N$-best hypotheses.
Table\,\ref{tab:prelim} summarizes 1-best WERs of three speech recognition heads: Whisper-large-v3 \citep{radford2023robust} for audio-only, BRAVEn-large \citep{haliassos2024braven} for visual-only, and Auto-AVSR \citep{ma2023auto} for audio-visual. Whisper attains 25.8\% WER, while Auto-AVSR is slightly stronger (24.9\%), and BRAVEn is markedly weaker (39.7\%), confirming that VSR alone lags in overall accuracy.

The oracle WER results reveal the limitation of single-stream systems and the compelling potential of dual-stream approaches (A\,+\,AV or A\,+\,V). 
While strong individual models like Whisper and Auto-AVSR perform $o_{cp}$ WERs of 13.7\% and 13.6\%, respectively, combining hypotheses from independent audio and visual heads drastically reduces potential errors: \textbf{Whisper\,(ASR)\,+\,BRAVEn\,(VSR)} plummets to 4.5\%.
This gap indicates that audio and video can provide distinct evidence with highly complementary information.
Consequently, an ideal GER model that can compose across ASR and VSR hypotheses could significantly reduce errors relative to single-stream systems.

\begin{table*}[!t]
\small
\centering
\caption[Error correction examples of DualHyp]{Examples of successful correction via DualHyp framework. The upper hypothesis within each 5-best list has a higher log-likelihood score. The colored highlights trace the origin of word fragments in the final DualHyp output, showing how those are sourced from ASR, VSR, or both, with a word being newly generated by the LLM's internal knowledge. \textbf{\textit{Type 1}} demonstrates the model combining complementary pieces from both modalities, and \textbf{\textit{Type 2}} presents the model identifying and correcting the hypothesis from a more reliable modality.}
\label{tab:dualhyp_case}
\addtolength{\tabcolsep}{-3pt}
\renewcommand{\arraystretch}{1.1}
\setlength{\fboxsep}{1pt} 
\resizebox{\textwidth}{!}{
\begin{tabular}{l|c|l|c}
\toprule
\multicolumn{2}{l|}{\textit{\textbf{Type 1: Multimodal Fragment Composition}}} & \multicolumn{2}{l}{\textit{\textbf{Type 2: Dominant Modality Refinement}}} \\
\midrule
\textbf{Utterance (ASR\,+\,VSR $\rightarrow$ DualHyp)} & \textbf{WER} & \textbf{Utterance (ASR\,+\,VSR $\rightarrow$ DualHyp)} & \textbf{WER} \\
\midrule
\midrule
    \textbf{\underline{ASR 5-best} ($\mathcal{H}^{\text{asr}}$): } & & \textbf{\underline{ASR 5-best} ($\mathcal{H}^{\text{asr}}$):} & \\
    \highlight{yellow!30}{everyone going into the den} has\highlight{Green!10}{ a fresh chance to} talk it around & 35.7 & \texttt{<unk>} & 100.0 \\
    \highlight{yellow!30}{everyone going into the den} is given\highlight{Green!10}{ a fresh chance to} talk it around & 42.9 & thank you & 100.0 \\
    \highlight{yellow!30}{everyone going into the den} gives you\highlight{Green!10}{ a fresh chance to} talk it around & 42.9 & all right & 100.0 \\
    and \highlight{yellow!30}{everyone going into the den} has\highlight{Green!10}{ a fresh chance to} talk around & 35.7 & the president & 100.0 \\
    \highlight{yellow!30}{everyone going into the den} has\highlight{Green!10}{ a fresh chance to} talk to the ground & 42.9 & god bless you & 100.0 \\[5pt]
    \textbf{\underline{VSR 5-best} ($\mathcal{H}^{\text{vsr}}$):} & & \textbf{\underline{VSR 5-best} ($\mathcal{H}^{\text{vsr}}$):} & \\
    but everyone in today \highlight{orange!20}{gets}\highlight{Green!10}{ a fresh chance to}\highlight{purple!15}{ turn things around} & 35.7 & \highlight{yellow!30}{project management}\highlight{orange!20}{ is really}\highlight{Green!10}{ my special} considering & 14.3 \\
    but everyone as i say \highlight{orange!20}{gets}\highlight{Green!10}{ a fresh chance to}\highlight{purple!15}{ turn things around} & 35.7 & \highlight{yellow!30}{project management}\highlight{orange!20}{ is really} by special considering & 28.6 \\
    but everyone on its day \highlight{orange!20}{gets}\highlight{Green!10}{ a fresh chance to}\highlight{purple!15}{ turn things around} & 35.7 & \highlight{yellow!30}{project management}\highlight{orange!20}{ is really}\highlight{Green!10}{ my special}\highlight{purple!15}{ist} theory & 14.3 \\
    but everyone it is a saying \highlight{orange!20}{gets}\highlight{Green!10}{ a fresh chance to}\highlight{purple!15}{ turn things around} & 35.7 & \highlight{yellow!30}{project management} and really\highlight{Green!10}{ my special} considering & 28.6 \\
    but everyone it is the saying \highlight{orange!20}{gets}\highlight{Green!10}{ a fresh chance to}\highlight{purple!15}{ turn things around} & 28.6 & \highlight{yellow!30}{project management}\highlight{orange!20}{ is really}\highlight{Green!10}{ my special} discovery & 28.6 \\
\midrule
\textbf{\underline{DualHyp output} $(\hat{y})$:} & & \textbf{\underline{DualHyp output} $(\hat{y})$:} & \\
\textbf{\highlight{yellow!30}{everyone going into the den}\highlight{orange!20}{ gets}\highlight{Green!10}{ a fresh chance to}\highlight{purple!15}{ turn things around}} & \textbf{14.3} & \textbf{\highlight{yellow!30}{project management}\highlight{orange!20}{ is really}\highlight{Green!10}{ my special}\highlight{purple!15}{ist}\highlight{blue!10}{ area}} & \textbf{0.0} \\
\midrule
\textbf{\underline{Ground-truth:}} & & \textbf{\underline{Ground-truth:}} & \\
but everyone going into the den gets a fresh chance to turn things round & -- & project management is really my specialist area & -- \\
\bottomrule
\end{tabular}
}
\end{table*}

\subsection{DualHyp: Dual-Stream Hypotheses}
\label{subsec:dualhyp}
Existing GER approaches for AVSR either inject visual data into LLMs via adapters~\citep{ghosh2024lipger} or rely on multimodal encoders that perform early fusion of the modalities~\citep{liu2025avger}. Both strategies have notable drawbacks; feature adaptation is insufficient for transferring rich visual cues, whereas early fusion is susceptible to cross-modal interference or modality bias.

Our approach is guided by a different principle, underscored by perceptual phenomena that the premature fusion of conflicting audio-visual signals can distort recognition outcomes~\citep{mcgurk1976hearing}. Inspired by prior work that embeds audio noise into the \textit{language space}~\citep{hu2024robustger}, we suggest that modality-specific information should be explicitly represented in the language space. This allows the LLM to resolve inconsistencies and compose information from both streams without entangling the signals during the upstream feature processing.

Thus, based on our analysis in Section\,\ref{subsec:oracle_error_analysis}, we propose \textbf{DualHyp}, a novel GER framework that explicitly leverages separate hypotheses streams from both audio and video modalities. Instead of relying on a single recognizer, we utilize independent, pretrained ASR and VSR models to process an audio-visual pair.
Each recognizer head generates a distinct $N$-best list:
\[
\mathcal{H}^{\text{asr}}=\{(h_i^{\text{a}}, s_i^{\text{a}})\}_{i=1}^{N}, 
\quad 
\mathcal{H}^{\text{vsr}}=\{(h_j^{\text{v}}, s_j^{\text{v}})\}_{j=1}^{N}.
\]

We then form a combined \emph{dual} hypotheses set, $\mathcal{H}^{\text{dual}}=\mathcal{H}^\text{asr}\cup\mathcal{H}^\text{vsr}$, which preserves the modality-specific information in each hypothesis set. The LLM is conditioned on this enriched set to generate the DualHyp output:
\begin{equation}
\hat{y} = \arg\max_{y} P(y \mid \mathcal{H}^\text{dual}; \theta_\text{LLM}).
\end{equation}

By maintaining separate modality pathways into the language space, this approach avoids the cross-modal contamination issues seen in early-fusion models. It instead enables the LLM to act as an in-context compositional reasoner~\citep{qiu2022evaluating, an2023context}, cross-referencing the audio and visual evidence to resolve ambiguities and reconstruct the intended utterance. Figure~\ref{fig:overview} illustrates the overview of our DualHyp framework, compared to existing GER approaches.

\paragraph{Analysis.}
Table~\ref{tab:dualhyp_case} provides qualitative analysis of DualHyp to show its effectiveness.
We highlight two primary correction mechanisms that exploit the LLM's strengths in the language space. \textit{(Type 1) Multimodal Fragment Composition}, where the model constructs the output by weaving complementary fragments from both ASR and VSR hypotheses. 
\textit{(Type 2) Dominant Modality Refinement}, where the LLM identifies that the ASR hypotheses are inconsistent, discards them, and focuses exclusively on refining the more coherent ones from the dominant VSR stream.
Furthermore, this refinement process retains the LLM's prior knowledge, as it generates a word not present in any source hypothesis, \ie \texttt{area}. We provide more examples, including failure cases, in Section\,\ref{subsec:additional_cases_dualhyp}.

\newpage
\section{Noise-Aware Guidance of DualHyp}
\label{sec:relprompt}

\begin{wrapfigure}[28]{r}{0.55\textwidth}
    \centering
    \vspace{-15pt}
    \begin{tcolorbox}[
        width=.55\textwidth,
        sharp corners=all,
        colback=gray!10,
        boxrule=0.3mm,
        enhanced,
        boxsep=0.1mm,
        left=2mm,
        right=2mm,
    ]
    \small

    \texttt{Below are the best-hypothesis transcribed from ASR and VSR. Revise it using the words which are only included into other-hypotheses, and write the response for the true transcription. Refer to the audio and video masks for reliability.} \\

    \texttt{\#\#\# ASR Best-hypothesis:} $\{h^{\text{a}}_1\}$ \\
    \texttt{\#\#\# ASR Other-hypotheses:}
    $\{h^{\text{a}}_2~||~\cdots~||~h^{\text{a}}_{N}\}$ \\
    \textcolor{red}{\texttt{\#\#\# Audio Mask:~[C][N][N][M][C]} $\cdots$} \\

    \texttt{\#\#\# VSR Best-hypothesis:} $\{h^{\text{v}}_1\}$ \\
    \texttt{\#\#\# VSR Other-hypotheses:}
    $\{h^{\text{v}}_2~||~\cdots~||~h^{\text{v}}_{N}\}$ \\
    \textcolor{red}{\texttt{\#\#\# Video Mask:~[C][C][C][N][N]} $\cdots$} \\

    \texttt{\#\#\# Response:}

    \end{tcolorbox}
    \vspace{-12pt}
    \includegraphics[width=.55\textwidth]{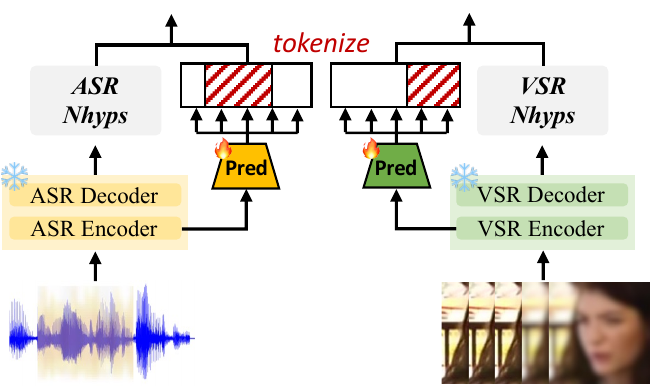}
    \caption[Overview of DualHyp with RelPrompt]{An overview of our DualHyp with RelPrompt. Each predictor uses ASR/VSR encoder features to generate a noise-aware token sequence. These masks accurately guide the LLM to dynamically switch the model's focus between the ASR and VSR hypotheses.}
    \label{fig:overview-2}
    \vspace{-50pt}
\end{wrapfigure}

The DualHyp framework enables an LLM to compose information from separate ASR and VSR hypotheses.
However, since LLM operates purely on these text inputs, the model lacks explicit information about the source signal quality, creating a risk of leveraging unreliable, inaccurate hypotheses \citep{hong2023watch}.
To bridge this gap from an LLM perspective, we introduce \textbf{RelPrompt}, a noise-aware guidance mechanism that explicitly informs the LLM about the temporal reliability of each stream.
RelPrompt is achieved by (1) predicting reliability tokens for each modality using external predictors, which are then (2) provided to the LLM's prompt to serve as temporal guidance.

\subsection{Reliability Mask Prediction}

To generate a compact, time-aligned reliability signal, we segment both the audio and video streams to approximate the duration of a single spoken word.
Grounded in the average native English speaking rate~\citep{yuan2006towards, becker2022horse}, we set the chunk size to 0.4 seconds, \ie 150 wpm.
We process each modality as follows:
\begin{itemize}
    \item \textbf{Audio stream:} The input audio, sampled at 16kHz rate, is divided into chunks of 6,400 samples (16,000 samples/sec × 0.4 sec). 
\end{itemize}

\begin{itemize}
    \item \textbf{Video stream:} The input video, originally at 25 frames per second (fps), is grouped into chunks of 10 frames (25 fps × 0.4 sec).
\end{itemize}

We then employ two lightweight predictors consisting of 1D convolutional neural networks (CNN) that operate on the intermediate features extracted from the ASR and VSR encoders, thus avoiding additional feature extraction.
For each segment, the predictors produce a discrete token $m_i\in\{\texttt{Clean}, \texttt{Noisy}$, \texttt{Mixed}\}, forming a reliability mask that indicates the quality of the source signal to the LLM.
The ground-truth reliability is labeled as \texttt{Clean} if <10\% of its frames are corrupted, \texttt{Noisy} if >60\% of its frames are corrupted, and \texttt{Mixed} otherwise.
Each predictor outputs a sequence of these tokens for its respective modality:
\[
\mathbf{m}^{\text{a}} = (m^{\text{a}}_1,\ldots,m^{\text{a}}_K), \quad
\mathbf{m}^{\text{v}} = (m^{\text{v}}_1,\ldots,m^{\text{v}}_K).
\]

\subsection{Reliability Guidance}

As illustrated in Figure \ref{fig:overview-2}, the reliability token sequences, $\mathbf{m}^\text{a}$ and $\mathbf{m}^\text{v}$ are appended to the dual hypotheses to directly inform the LLM of each modality's temporal reliability.
The entire model is then trained end-to-end, conditioned on both the hypotheses and reliability masks to generate the final transcript:
\begin{equation}
\hat{y}=\arg\max_{y} P\!\left(y \mid \mathcal{H}^{\text{dual}}, \mathbf{m}^{\text{a}}, \mathbf{m}^{\text{v}}; \theta_{\text{LLM}}\right).
\end{equation}

This format allows the LLM to learn the correlation between the reliability tokens and hypotheses quality.
Crucially, this approach avoids the need for explicit word-level alignment, which is infeasible for $N$-best lists with variable lengths and erroneous words \citep{qiu2021learning, gekhman2022red}.
Additionally, the RelPrompt token sequence enhances the interpretability of LLM's reasoning, revealing when the model switches its focus between the ASR and VSR hypotheses.

\section{Experiments and Results}
\label{sec:ch5_experiments}

\subsection{Experimental Setup}

\paragraph{Summary.}
We conduct our experiments on the LRS2 AVSR benchmark~\citep{son2017lip} with the WER metric.
All models are trained and tested under diverse, synthetically corrupted audio-visual conditions, following the protocol of CAV2vec \citep{kim2025multitask}.
Unless specified otherwise, our DualHyp framework is composed of a Whisper-large-v3 \citep{radford2023robust} ASR head, a {BRAVEn-large} \citep{haliassos2024braven} VSR head, and a {TinyLlama} \citep{zhang2024tinyllama} LLM, which we fine-tune using LoRA~\citep{hu2022lora}. 

\paragraph{Release of DualHyp Dataset.}
To facilitate future research within our DualHyp framework, we have publicly released a comprehensive hypotheses dataset. The primary motivation of this dataset construction is to decouple the computationally expensive hypothesis generation step from the LLM fine-tuning process. By providing pre-generated ASR and VSR hypotheses, this dataset will allow researchers to focus directly on developing novel language-space fusion and correction strategies, significantly lowering the barrier to entry.

Our DualHyp hypotheses are mainly created from LRS2~\citep{son2017lip} and LRS3~\citep{afouras2018lrs3} datasets. LRS2 is a benchmark of British English speech from BBC that covers diverse speakers and topics, while LRS3 consists of spoken utterances from TED and TEDx recordings.
For LRS2, our hypotheses dataset covers the standard splits (45,830 training, 1,082 validation, and 1,243 test utterances). For high-resource training (dealt in Section~\ref{subsec:high-resource_training}), we use additional 95,642 utterances, and for the LRS3 experiments (dealt in Section~\ref{app:lrs3_result}), we use 30,775 training utterances. 

For both the ASR head\footnote{\url{https://huggingface.co/openai/whisper-large-v3}} and VSR head\footnote{\url{https://github.com/ahaliassos/raven}}, we generate hypotheses using a beam size of 50. We select the 5-best unique hypotheses from each stream. If fewer than five unique hypotheses are generated, we randomly sample from the existing ones to reach the size 5. This results in a total of 10 hypotheses (5 from ASR, 5 from VSR) that are fed into the LLM for error correction.

Each entry also includes the ground-truth transcription and metadata detailing the specific audio or visual corruption applied. Because different corruption types result in different output hypotheses, we separately save the dataset for each corruption condition. For training, a complete dataset is formed by merging these individual sets and randomly sampling hypotheses.

\paragraph{Implementation Details of DualHyp.}

We fine-tune the LLM using LoRA~\citep{hu2022lora} with a rank of $r=16$. The number of trainable parameters is 4.5M for TinyLlama\footnote{\url{https://huggingface.co/TinyLlama/TinyLlama-1.1B-Chat-v1.0}}, 23.6M for Phi-2\footnote{\url{https://huggingface.co/microsoft/phi-2}}, and 24.3M for Llama-3.2\footnote{\url{https://huggingface.co/meta-llama/Llama-3.2-3B}}. For TinyLlama, we apply LoRA to the attention layers (key, value, query, and projection) only. For the larger Phi-2 and Llama-3.2 models, we apply LoRA to both the attention module and the feed-forward network (FFN) layers to ensure better convergence. All models are trained for 5 epochs with a batch size of 32 and a learning rate of 1e-4.

For the implementation of RelPrompt, our reliability predictors are designed lightweight, with only 1.1M parameters each. The architecture consists of two 1D-convolutional layers followed by average pooling and a final linear classifier to match the segment size. For training these predictors, we create ground-truth labels for each 0.4-second segment based on its constituent frames: a segment is labeled \texttt{[C]} (Clean) if less than 10\% of its frames are corrupted, \texttt{[N]} (Noisy) if more than 60\% of its frames are corrupted, and \texttt{[M]} (Mixed) otherwise.

The full DualHyp\,+\,RelPrompt model is trained for 8 hours on a single NVIDIA A6000 GPU, using a learning rate of 2e-4 for the main LLM (with LoRA) and 1e-4 for the reliability predictors. For all data pre-processing and evaluation, we use the publicly available packages following the LipGER codebase\footnote{\url{https://github.com/Sreyan88/LipGER}}.

\paragraph{Corruption Protocol.}
In our study, all models are trained and evaluated under challenging noisy conditions to assess their robustness in real-world scenarios.
To ensure a robust evaluation, we introduce a diverse set of synthetic corruptions into the LRS2 dataset, following the protocol established by \citet{kim2025multitask}\footnote{\url{https://github.com/sungnyun/cav2vec}}.
\begin{itemize}
    \item Audio corruptions: We augment the audio streams with four types of noise, similar to \citet{shi2022robust}. We use speech noise from the LRS3 dataset~\citep{afouras2018lrs3} and babble, music, and natural sounds from the MUSAN corpus~\citep{snyder2015musan}.
    \item Visual corruptions: We apply four common visual degradation types: object occlusion~\citep{voo2022delving}, hands occlusion, pixelation, and blur~\citep{kim2025multitask}.
\end{itemize}

During training, we randomly apply one of these corruption types to each sample. The duration of the applied corruption is also randomized, with its portion sampled from a Beta distribution ($\alpha,\beta=2.0$) to simulate varying levels of interference. 

For evaluation, we apply background noise to the entire audio sample and corrupt partial video segments to better reflect real-world scenarios. 
For Table~\ref{tab:main_lrs2}\textcolor{red}{a}, audio noise is applied to the entire time duration with SNR randomly sampled from [-10, 10]\,dB, while half of the video segments is occluded with object. 
For Table~\ref{tab:main_lrs2}\textcolor{red}{b}, 0\,dB SNR of speech noise is augmented to the whole audio, while video corruption length is sampled from Beta distribution.

To assess overall performance, we report the average WER from a single comprehensive evaluation run. This run covers all test samples and incorporates a diverse range of noisy conditions to ensure the statistical credibility of our methods.

\paragraph{Baseline Methods.}
For a fair comparison, we train all baseline methods from scratch on the same set of corrupted audio-visual data. We have found that this diverse noise training significantly boosts the performance of all GER-based methods, which establishes a strong set of baselines for our evaluation. Our primary baselines are:
\begin{itemize}
    \item GER \citep{chen2023hyporadise}: The foundational LLM-based error correction framework that operates on \textit{N}-best hypotheses from an ASR model.
    \item RobustGER \citep{hu2024robustger}: An extension of GER designed to improve robustness against noisy audio conditions.
    \item LipGER \citep{ghosh2024lipger}: An audio-visual GER method that incorporates visual features via an adapter but still relies on a single stream of ASR hypotheses for correction.
    \item GER w/ Auto-AVSR \citep{ma2023auto}: A strong baseline we implement by feeding the \textit{N}-best hypotheses from the early-fusion Auto-AVSR model into a standard GER framework.
\end{itemize}

We note that training these single-stream GER baselines presents a significant stability issue when using highly corrupted data. As detailed in our analysis (\S\ref{subsec:snr_to_werr}), the performance of these models is capped by the quality of the initial ASR hypotheses. During training, low-SNR audio produces poor learning signal, which causes the model to learn to over-correct already accurate transcriptions while failing to fix genuinely erroneous ones. We have observed their performance degradation (up to +5\% WER) when trained with the same data as DualHyp. To ensure stable convergence for our baseline comparisons, we therefore construct their training dataset exclusively from audio samples with SNR\,$\ge$\,0\,dB, just as \citet{hu2024robustger, ghosh2024lipger} have constructed their training datasets.

\begin{table*}[!t]
  \centering
  \small
  \caption[DualHyp and RelPrompt performance on corrupted LRS2]{
  WER\% ($\downarrow$) results on the LRS2 test set under joint audio-visual corruption.
  (a) Performance across varying audio noise types, with a fixed visual corruption (50\% segment occluded by an object).
  (b) Performance across varying visual corruption types, with a fixed audio corruption (0\,dB speech noise).
  We also show the relative WER reduction in parentheses compared to the Whisper-large-v3 ASR baseline.
  All the ASR and VSR heads are Whisper-large-v3 and BRAVEn-large, respectively.
  $^\dagger$: We implement a GER model using hypotheses generated from an early-fusion approach, Auto-AVSR~\citep{ma2023auto}, which has been trained on LRS2 with babble noise.
  }
  \label{tab:main_lrs2}
  \resizebox{\textwidth}{!}{
  \begin{tabular}{lc|llll|l}
    \toprule
    \textbf{Method} & \textbf{Input} & \textbf{Babble (B)} & \textbf{Speech (S)} & \textbf{Music (M)} & \textbf{Natural (N)} & \textbf{Overall (O)} \\
    \midrule
    \color{gray}\textit{ASR oracle $o_{nb}$\,/\,$o_{cp}$} & \color{gray}A & \color{gray}30.9 / 26.9 & \color{gray}19.4 / 14.1 & \color{gray}8.0 / 6.6 & \color{gray}8.5 / 7.4 & \color{gray}16.7 / 13.7 \\
    \color{gray}{\textit{ASR\,+\,VSR oracle $o_{nb}$\,/\,$o_{cp}$}} & \color{gray}A\,+\,V & \color{gray}11.7 / 8.8 & \color{gray}6.6 / 4.4 & \color{gray}3.5 / 2.2 & \color{gray}3.6 / 2.7 & \color{gray}6.4 / 4.5 \\
    \midrule
    Whisper-large-v3 
    \citep{radford2023robust} & A & 40.0 & 36.5 & 12.7 & 14.2 & 25.8 \\
    BRAVEn-large \citep{haliassos2024braven} & V & - & - & - & - & 39.7$_{(+53.9\%)}$ \\
    GER~\citep{chen2023hyporadise} & A & 39.3$_{(-1.8\%)}$ & 34.4$_{(-5.8\%)}$ & 11.5$_{(-9.4\%)}$ & 13.2$_{(-7.0\%)}$ & 24.6$_{(-4.7\%)}$ \\
    RobustGER~\citep{hu2024robustger} & A & 39.3$_{(-1.8\%)}$ & 33.8$_{(-7.4\%)}$ & 11.7$_{(-7.9\%)}$ & 13.1$_{(-7.7\%)}$ & 24.5$_{(-5.0\%)}$ \\
    LipGER~\citep{ghosh2024lipger} & AV & 39.3$_{(-1.8\%)}$ & 34.2$_{(-6.3\%)}$ & 12.0$_{(-5.5\%)}$ & 13.4$_{(-5.6\%)}$ & 24.7$_{(-4.3\%)}$ \\
    GER w/ Auto-AVSR$^\dagger$ & AV & \textbf{18.9}$_{(-52.8\%)}$ & 39.0$_{(+6.8\%)}$ & 17.4$_{(+37.0\%)}$ & 18.1$_{(+27.5\%)}$ & 23.3$_{(-9.7\%)}$ \\
    \midrule
    \textbf{DualHyp ({ours})} & A\,+\,V & 21.6$_{(-46.0\%)}$ & 17.9$_{(-51.0\%)}$ & 8.1$_{(-36.2\%)}$ & 9.3$_{(-34.5\%)}$ & 14.2$_{(-45.0\%)}$ \\
    \textbf{+\,RelPrompt ({ours})} & A\,+\,V & 20.4$_{(-49.0\%)}$ & \textbf{16.0}$_{(-56.2\%)}$ & \textbf{8.0}$_{(-37.0\%)}$ & \textbf{8.2}$_{(-42.3\%)}$ & \textbf{13.2}$_{(-48.8\%)}$ \\
    \bottomrule
    \addlinespace[2pt]
    \multicolumn{7}{c}{\textbf{(a) Audio: random noise [-10, 10]\,dB, Video: 50\% segment occluded with object}} \\
    \addlinespace[8pt]
    \toprule
    \textbf{Method} & \textbf{Input} & \textbf{Object} & \textbf{Hands} & \textbf{Pixelate} & \textbf{Blur} & \textbf{Overall} \\
    \midrule
    \color{gray}\textit{ASR oracle $o_{nb}$\,/\,$o_{cp}$} & \color{gray}A & \color{gray}- & \color{gray}- & \color{gray}- & \color{gray}- & \color{gray}10.9 / 6.6 \\
    \color{gray}{\textit{ASR\,+\,VSR oracle $o_{nb}$\,/\,$o_{cp}$}} & \color{gray}A\,+\,V & \color{gray}4.7 / 2.8 & \color{gray}4.5 / 2.6 & \color{gray}4.8 / 2.7 & \color{gray}4.1 / 2.4 & \color{gray}4.5 / 2.6 \\
    \midrule
    Whisper-large-v3 \citep{radford2023robust} & A & 26.7 & 26.7 & 26.7 & 26.7 & 26.7 \\
    BRAVEn-large \citep{haliassos2024braven} & V & 39.7$_{(+48.7\%)}$ & 35.1$_{(+31.5\%)}$ & 39.4$_{(+47.6\%)}$ & 31.7$_{(+18.7\%)}$ & 36.5$_{(+36.7\%)}$ \\
    GER~\citep{chen2023hyporadise} & A & - & - & - & - & 23.9$_{(-10.5\%)}$ \\
    RobustGER~\citep{hu2024robustger} & A & - & - & - & - & 24.9$_{(-6.7\%)}$ \\
    LipGER~\citep{ghosh2024lipger} & AV & 24.2$_{(-9.4\%)}$ & 24.3$_{(-9.0\%)}$ & 24.3$_{(-9.0\%)}$ & 24.1$_{(-9.7\%)}$ & 24.3$_{(-9.0\%)}$ \\
    GER w/ Auto-AVSR$^\dagger$ & AV & 29.5$_{(+10.5\%)}$ & 26.6$_{(-0.4\%)}$ & 29.1$_{(+9.0\%)}$ & 23.5$_{(-12.0\%)}$ & 27.2$_{(+1.9\%)}$ \\
    \midrule
    \textbf{DualHyp ({ours})} & A\,+\,V & {12.0}$_{(-55.1\%)}$ & 11.8$_{(-55.7\%)}$ & 12.7$_{(-52.6\%)}$ & 11.1$_{(-58.4\%)}$ & 11.9$_{(-55.4\%)}$ \\
    \textbf{+\,RelPrompt ({ours})} & A\,+\,V & \textbf{11.9}$_{(-55.4\%)}$ & \textbf{11.0}$_{(-58.8\%)}$ & \textbf{11.9}$_{(-55.4\%)}$ & \textbf{10.2}$_{(-61.8\%)}$ & \textbf{11.3}$_{(-57.7\%)}$ \\
    \bottomrule
    \addlinespace[2pt]
    \multicolumn{7}{c}{\textbf{(b) Audio: speech noise 0\,dB, Video: random segment corrupted}}
  \end{tabular}
  }
\end{table*}

\subsection{LRS2 Benchmark Results}
Table\,\ref{tab:main_lrs2} presents the benchmark results, where we isolate modality-specific robustness by either varying audio noise against fixed visual corruption (Table\,\ref{tab:main_lrs2}\textcolor{Red}{a}) or varying visual corruption against fixed audio noise (Table\,\ref{tab:main_lrs2}\textcolor{Red}{b}).
Our proposed \textbf{DualHyp\,+\,RelPrompt} achieves the lowest \textbf{overall WER of 13.2\%} under audio variability and \textbf{11.3\%} under visual variability, representing a relative improvement of 48.8\% and 57.7\% compared to the ASR baseline, Whisper-large-v3. 
This confirms our core hypothesis that LLMs can perform robust compositional reasoning when provided with separate ASR and VSR hypotheses.

In contrast, all baseline methods show clear limitations.
ASR-only models like GER \citep{chen2023hyporadise} or RobustGER \citep{hu2024robustger} are fundamentally capped by the input audio quality and struggle under low SNRs (also refer to \S\ref{subsec:snr_to_werr}).
Audio-visual approach like LipGER \citep{ghosh2024lipger} fails to improve over the standard GER framework, which shows that injecting video via additional adapter is insufficient for LLM to fully exploit the visual modality while harming its stability due to cross-modal gap~\citep{zhang2024llamaadapter, li2023prompting, gao2023llama}.
Similarly, GER w/ Auto-AVSR \citep{ma2023auto} exhibits strong but narrow performance, excelling only on the babble noise that the model's AVSR head is specifically trained on, failing to generalize to other conditions (see \S\ref{subsec:comparison_avsr} for further analysis).

The success of our DualHyp with RelPrompt approach stems from two aspects: (1) a text-level late fusion strategy and (2) the ability to dynamically leverage the more reliable modality.
Our late fusion provides the LLM with rich, modality-specific evidence in a unified text format that is readily processed by the LLM.
The isolation of modalities also ensures that corruption in one stream does not contaminate the other. 
Then, RelPrompt dynamically leverages the more reliable stream, utilizing visual hypotheses when audio quality is low and falling back on audio hypotheses when the visual stream is degraded.
Notably, this superior performance is achieved even though our VSR model is substantially weaker than ASR, suggesting that our framework's potential is scalable as more powerful VSR models emerge.

\paragraph{Clean Audio or Video Inputs.}
\begin{wraptable}[9]{r}{0.5\textwidth}
    \centering
    \small
    \vspace{-15pt}
    \caption[DualHyp and RelPrompt performance on clean LRS2]{Performance under different modality conditions on LRS2, with clean audio or video (X$^c$) and noisy audio or video (X$^n$), $\text{X}\in\{\text{A}, \text{C}\}$.}
    \label{tab:lrs2_clean}
    \vspace{-5pt}
    \renewcommand{\arraystretch}{0.95}
    \addtolength{\tabcolsep}{2.2pt}
    \resizebox{0.5\textwidth}{!}{
    \begin{tabular}{lc|ccc}
        \toprule
        \textbf{Method} & \textbf{Input} & A$^c$V$^c$ & A$^c$V$^n$ & A$^n$V$^c$ \\
        \midrule
        Whisper-large-v3 & A & 3.8 & 3.8 & 25.8 \\
        BRAVEn-large & V & 26.9 & 36.5 & 26.9  \\
        GER & A & 2.6 & 2.6 & 24.6 \\
        \midrule
        \textbf{DualHyp} & A\,+\,V & \textbf{1.9} & 2.1 & 11.5 \\
        \textbf{+\,RelPrompt} & A\,+\,V & \textbf{1.9} & \textbf{2.0} & \textbf{9.9} \\
        \bottomrule
    \end{tabular}
    }
    \vspace{-20pt}
\end{wraptable}
Even in the clean audio settings (Table~\ref{tab:lrs2_clean}), our DualHyp methods achieve the lowest WER, showing they effectively capitalize on the high-quality audio stream. In the noisy-audio/clean-video setting, while GER is severely hampered by corrupted audio (24.6\%), RelPrompt leverages clean visual hypotheses to dramatically improve to 9.9\%. The gap between DualHyp (11.5\%) and its reliability-guided version demonstrates that dynamically detecting clean signal (in this case video) and giving the LLM explicit hints about which to trust is effective.

\begin{table}[!t]
    \centering
    \small
    \caption[DualHyp and RelPrompt performance with larger LLMs]{The corruption strategy follows Table\,\ref{tab:main_lrs2}\textcolor{Red}{a}, where \textbf{B}, \textbf{S}, \textbf{M}, and \textbf{N} represent each noise type with the overall result (\textbf{O}).}
    \label{tab:larger_llm}
    \addtolength{\tabcolsep}{1.5pt}
    \resizebox{.65\columnwidth}{!}{
    \begin{tabular}{ll|cccc|c}
    \toprule
    \textbf{Method} & \textbf{LLM (Params.)} & \textbf{B} & \textbf{S} & \textbf{M} & \textbf{N} & \textbf{O} \\
    \midrule
    \multirow{3}{*}{GER} & TinyLlama (1.1B) & 39.3 & 34.4 & 11.5 & 13.2 & 24.6 \\
    & Phi-2 (2.7B) & 39.0 & 33.7 & 11.9 & 13.0 & 24.4 \\
    & Llama-3.2 (3.2B) & 38.9 & 34.1 & 11.6 & 12.9 & 24.4 \\
    \midrule
    \multirow{3}{*}{\textbf{DualHyp}} & TinyLlama (1.1B) & 21.6 & 17.9 & 8.1 & 9.3 & 14.2 \\
    & Phi-2 (2.7B) & 21.6 & 19.0 & 7.8 & 8.7 & 14.3 \\
    & Llama-3.2 (3.2B) & 20.4 & 16.0 & \textbf{7.2} & \textbf{8.1} & 12.9 \\    
    \midrule
    \multirow{3}{*}{\textbf{\thead[l]{DualHyp\\+\,RelPrompt}\!\!}} & TinyLlama (1.1B) & 20.4 & 16.0 & 8.0 & 8.2 & 13.2 \\
    & Phi-2 (2.7B) & 21.1 & 18.2 & 8.0 & 8.5 & 14.0 \\
    & Llama-3.2 (3.2B) & \textbf{19.6} & \textbf{14.1} & 7.4 & 8.2 & \textbf{12.3} \\
    \bottomrule
    \end{tabular}
    }
\end{table}

\subsection{Larger LLMs}
\label{subsec:larger_llms}

We investigate the impact of LLM scale by evaluating our methods with three different models: TinyLlama~\citep{zhang2024tinyllama}, Phi-2~\citep{javaheripi2023phi2}, and Llama-3.2~\citep{meta_llama3_2_2024}. The results in Table~\ref{tab:larger_llm} show that the benefits of a larger LLM are most pronounced within our proposed framework. For the GER baseline, scaling the LLM yields only marginal gains, indicating that its performance is limited by the quality of the single-stream input hypotheses. In contrast, our models benefit more significantly from larger LLM's capacity. The effect is greatest for DualHyp + RelPrompt, which achieves the best overall WER of 12.4\% with the Llama-3.2-3B model. This suggests that by providing a richer and more complex input, our framework creates a more sophisticated reasoning task that can effectively leverage the capabilities of LLMs.

\begin{table*}[!t]
    \centering
    \small
    \caption[DualHyp performance on noisy MuAViC]{WER (\%) comparison with multilingual babble noise (SNR\,=\,0\,dB) on the MuAViC dataset.
    Subscript values of mAV-HuBERT indicate the relative WER increase compared to Whisper-large-v3.}
    \label{tab:muavic_full}
    \addtolength{\tabcolsep}{-3pt}
    \resizebox{\textwidth}{!}{
    \begin{tabular}{lc|llllllll}
    \toprule
    \textbf{Method} & \textbf{Input} & \textbf{Ar} & \textbf{De} & \textbf{El} & \textbf{Es} & \textbf{Fr} & \textbf{It} & \textbf{Pt} & \textbf{Ru} \\
    \midrule
    Whisper-large-v3 & A & \textbf{91.7} & \textbf{55.7} & \textbf{54.4} & 49.6 & \textbf{46.8} & 52.3 & 52.7 & \textbf{50.9} \\
    mAV-HuBERT & V & 102.0$_{(+11\%)}$ & 96.6$_{(+74\%)}$ & 87.1$_{(+60\%)}$ & 70.5$_{(+42\%)}$ & 81.7$_{(+75\%)}$ & 73.7$_{(+41\%)}$ & 74.1$_{(+41\%)}$ & 80.9$_{(+59\%)}$ \\
    GER & A & 96.9 & 56.2 & 57.7 & 50.6 & 47.8 & 58.5 & 52.3 & 54.1 \\
    \textbf{DualHyp (ours)} & A\,+\,V & 106.8 & 100.4 & 77.3 & \textbf{47.3} & 47.9 & \textbf{47.2} & \textbf{49.0} & 58.9 \\
    \bottomrule
    \end{tabular}
    }
\end{table*}

\subsection{Multilingual AVSR}
\label{subsec:multilingual_avsr}

To evaluate our framework in a multilingual context, we conduct experiments on the MuAViC dataset~\citep{anwar2023muavic} with adding multilingual babble noise at SNR 0\,dB~\citep{kim2025mohave}.
While the Whisper ASR head remains the same as in prior experiments, a VSR head is fine-tuned from mAV-HuBERT~\citep{kim2024efficient} for each language, due to the absence of strong multilingual VSR system.
Llama-3.2-3B is employed for the multilingual reasoning.
In Table~\ref{tab:muavic_full}, our framework outperforms both Whisper and GER in three of the four languages.
However, this performance gain can be limited when VSR performance is severely degraded, as observed in the French case.
We thus anticipate that the performance gains of our methodology will become even more significant as more powerful multilingual VSR models emerge.

We observe that standard GER shows limited effectiveness across all languages, suggesting inherent difficulty of error correction in non-English contexts, even when the LLM itself is multilingual.
Our DualHyp framework is designed to aid this reasoning by providing the LLM with more comprehensive evidence from both ASR and VSR streams, achieving performance improvements in three languages.

However, our results reveal that a large disparity between the ASR and VSR quality can exacerbate the LLM's inherent weakness in multilingual reasoning.
While for the languages where DualHyp succeeds, the VSR head maintains a relatively consistent performance gap around 40\% higher than the ASR baseline, for the languages where DualHyp underperforms (\eg Greek), this gap widens significantly to over 60\%.
Meanwhile, for Arabic, the hypotheses from both modalities are of exceptionally poor quality (>90\% WER), leaving the LLM with no useful source to compose.

\subsection{High-Resource Training}
\label{subsec:high-resource_training}
\begin{table*}[!t]
    \centering
    \small
    \caption[Effect of high-resource training]{The effect of dataset integration (+\,LRS3) and high-resource (+\,HR) training.}
    \label{tab:high_resource}
    \addtolength{\tabcolsep}{-3pt}
    \resizebox{\textwidth}{!}{
    \begin{tabular}{lcc|ccccc|ccccc}
        \toprule
        &&& \multicolumn{5}{c|}{\textbf{Object occlusion (50\%)}} & \multicolumn{5}{c}{\textbf{Speech noise (SNR = 0\,dB)}} \\
        \textbf{Method} & \textbf{+\,LRS3} & \textbf{+\,HR} & \textbf{Babble} & \textbf{Speech} & \textbf{Music} & \textbf{Natural} & \textbf{Overall} & \textbf{Object} & \textbf{Hands} & \textbf{Pixelate} & \textbf{Blur} & \textbf{Overall} \\
        \midrule
        \multirow{3}{*}{GER} & \xmark & \xmark & 39.3 & 34.4 & 11.5 & 13.2 & 24.6 & - & - & - & - & 23.9 \\
         & \cmark & \xmark & 39.1 & 34.3 & 11.7 & 12.8 & 24.5 & - & - & - & - & 23.9 \\
         & \xmark & \cmark & 39.2 & 34.1 & 11.7 & 13.1 & 24.5 & - & - & - & - & 25.6 \\
        \midrule
        \multirow{3}{*}{\textbf{DualHyp}} & \xmark & \xmark & 21.6 & 17.9 & 8.1 & 9.3 & 14.2 & 12.0 & 11.8 & 12.7 & 11.1 & 11.9 \\
         & \cmark & \xmark & 21.0 & 17.9 & 7.8 & 8.7 & 13.9 & 12.7 & 11.4 & 12.3 & 11.2 & 11.9 \\
         & \xmark & \cmark & 21.9 & 17.7 & \textbf{7.7} & \textbf{8.0} & 13.8 & 12.1 & 11.0 & 11.5 & 10.5 & 11.3 \\
         \midrule
        \multirow{3}{*}{\thead[l]{\textbf{DualHyp}\\\textbf{+\,RelPrompt}}} & \xmark & \xmark & 20.4 & 16.0 & 8.0 & 8.2 & 13.2 & 11.9 & 11.0 & 11.9 & 10.2 & 11.3 \\
         & \cmark & \xmark & \textbf{19.5} & 15.4 & 7.8 & 8.5 & \textbf{12.8} & 11.7 & 11.0 & 11.9 & \textbf{9.6} & 11.1 \\
         & \xmark & \cmark & 20.1 & \textbf{15.1} & \textbf{7.7} & 8.3 & \textbf{12.8} & \textbf{10.5} & \textbf{9.4} & \textbf{10.9} & 9.7 & \textbf{10.1} \\
        \bottomrule
    \end{tabular}
    }
\end{table*}
We investigate how our framework scales with additional training data by augmenting the main LRS2 training set (29 hours) with either the larger LRS3 dataset (59 hours) or a high-resource LRS2 pretraining set (HR, 195 hours). The results in Table~\ref{tab:high_resource} show that GER fails to benefit from more data, showing no to adverse impact on the performance. In contrast, our DualHyp frameworks consistently improve with larger training sets. The best performance is achieved by DualHyp\,+\,RelPrompt when trained with the high-resource data, reaching 12.8\% overall WER on audio corruptions and 10.1\% overall WER on visual corruptions. This indicates that while the bottleneck of single-stream GER is not readily resolved by scaling data, our compositional framework has the capacity to effectively leverage more data to enhance its robustness.

\subsection{LRS3 Results}
\label{app:lrs3_result}

\begin{table*}[!t]
  \centering
  \small
  \caption[DualHyp and RelPrompt performance on corrupted LRS3]{
  WER\% ($\downarrow$) results on the LRS3 test set under joint audio-visual corruption. (a) Performance across varying audio noise types, with a fixed visual corruption (50\% segment occluded by an object). (b) Performance across varying visual corruption types, with a fixed audio corruption (0\,dB speech noise). We also show the relative WER reduction in parentheses compared to the Whisper-large-v3 ASR baseline. All the ASR and VSR heads are Whisper-large-v3 and BRAVEn-large, respectively. $^\dagger$: We implement a GER model using hypotheses generated from an early-fusion approach, Auto-AVSR~\citep{ma2023auto}, which has been trained on LRS3 with babble noise.
  }
  \label{tab:main_lrs3}
  \resizebox{\textwidth}{!}{
  \begin{tabular}{lc|llll|l}
    \toprule
    \textbf{Method} & \textbf{Input} & \textbf{Babble (B)} & \textbf{Speech (S)} & \textbf{Music (M)} & \textbf{Natural (N)} & \textbf{Overall (O)} \\
    \midrule
    \color{gray}\textit{ASR oracle $o_{nb}$\,/\,$o_{cp}$} & \color{gray}A & \color{gray}26.9 / 23.2 & \color{gray}19.6 / 13.5 & \color{gray}4.5 / 3.7 & \color{gray}5.0 / 4.6 & \color{gray}14.0 / 11.2 \\
    \color{gray}{\textit{ASR\,+\,VSR oracle $o_{nb}$\,/\,$o_{cp}$}} & \color{gray}A\,+\,V & \color{gray}7.9 / 5.8 & \color{gray}5.6 / 3.5 & \color{gray}1.6 / 1.0 & \color{gray}1.9 / 1.3 & \color{gray}4.2 / 2.9 \\
    \midrule
    Whisper-large-v3 & A & 32.6 & 34.5 & 7.3 & 7.8 & 20.6 \\
    BRAVEn-large & V & - & - & - & - & 31.9$_{(+54.9\%)}$ \\
    GER~\citep{chen2023hyporadise} & A & 32.4$_{(-0.6\%)}$ & 35.4$_{(+2.6\%)}$ & 7.6$_{(+4.1\%)}$ & 8.0$_{(+2.6\%)}$ & 20.9$_{(+1.5\%)}$ \\
    RobustGER~\citep{hu2024robustger} & A & 32.5$_{(-0.3\%)}$ & 36.0$_{(+4.3\%)}$ & 7.7$_{(+5.5\%)}$ & 8.1$_{(+3.8\%)}$ & 21.1$_{(+2.4\%)}$ \\
    LipGER~\citep{ghosh2024lipger} & AV & 32.4$_{(-0.6\%)}$ & 34.4$_{(-0.3\%)}$ & 7.6$_{(+4.1\%)}$ & 8.1$_{(+3.8\%)}$ & 20.6$_{(-0.0\%)}$ \\
    GER w/ Auto-AVSR$^\dagger$ & AV & 17.9$_{(-45.1\%)}$ & 45.6$_{(+32.2\%)}$ & 14.2$_{(+94.5\%)}$ & 11.0$_{(+41.0\%)}$ & 22.2$_{(+7.8\%)}$ \\
    \midrule
    \textbf{DualHyp (ours)} & A\,+\,V & 16.3$_{(-50.0\%)}$ & 18.2$_{(-47.2\%)}$ & \textbf{5.6}$_{(-23.3\%)}$ & {5.5}$_{(-29.5\%)}$ & 11.4$_{(-44.7\%)}$ \\
    \textbf{+\,RelPrompt (ours)} & A\,+\,V & \textbf{14.9}$_{(-54.3\%)}$ & \textbf{16.2}$_{(-53.0\%)}$ & {5.7}$_{(-21.9\%)}$ & \textbf{5.1}$_{(-34.6\%)}$ & \textbf{10.5}$_{(-49.0\%)}$ \\
    \bottomrule
    \addlinespace[2pt]
    \multicolumn{7}{c}{\textbf{(a) Audio: random noise [-10, 10]\,dB, Video: 50\% segment occluded with object}} \\
    \addlinespace[8pt]
    \toprule
    \textbf{Method} & \textbf{Input} & \textbf{Object} & \textbf{Hands} & \textbf{Pixelate} & \textbf{Blur} & \textbf{Overall} \\
    \midrule
    \color{gray}\textit{ASR oracle $o_{nb}$\,/\,$o_{cp}$} & \color{gray}A & \color{gray}- & \color{gray}- & \color{gray}- & \color{gray}- & \color{gray}8.3 / 5.3 \\
    \color{gray}{\textit{ASR\,+\,VSR oracle $o_{nb}$\,/\,$o_{cp}$}} & \color{gray}A\,+\,V & \color{gray}3.2 / 1.8 & \color{gray}3.0 / 1.6 & \color{gray}3.0 / 1.5 & \color{gray}2.5 / 1.4 & \color{gray}2.9 / 1.6 \\
    \midrule
    Whisper-large-v3 & A & 23.8 & 23.8 & 23.8 & 23.8 & 23.8 \\
    BRAVEn-large & V & 31.9$_{(+34.0\%)}$ & 30.8$_{(+29.4\%)}$ & 29.5$_{(+23.9\%)}$ & 23.8$_{(-0.0\%)}$ & 29.0$_{(+21.8\%)}$ \\
    GER~\citep{chen2023hyporadise} & A & - & - & - & - & 26.0$_{(+9.2\%)}$ \\
    RobustGER~\citep{hu2024robustger} & A & - & - & - & - & 27.1$_{(+13.9\%)}$ \\
    LipGER~\citep{ghosh2024lipger} & AV & 26.2$_{(+10.1\%)}$ & 26.0$_{(+9.2\%)}$ & 25.9$_{(+8.8\%)}$ & 26.0$_{(+9.2\%)}$ & 26.0$_{(+9.2\%)}$ \\
    GER w/ Auto-AVSR$^\dagger$ & AV & 47.5$_{(+99.6\%)}$ & 44.2$_{(+85.7\%)}$ & 42.0$_{(+76.5\%)}$ & 38.2$_{(+60.5\%)}$ & 43.0$_{(+80.7\%)}$ \\
    \midrule
    \textbf{DualHyp (ours)} & A\,+\,V & {12.2}$_{(-48.7\%)}$ & 10.9$_{(-54.2\%)}$ & 10.8$_{(-54.6\%)}$ & {9.6}$_{(-59.7\%)}$ & 10.9$_{(-54.2\%)}$ \\
    \textbf{+\,RelPrompt (ours)} & A\,+\,V & \textbf{11.0}$_{(-53.8\%)}$ & \textbf{10.5}$_{(-55.9\%)}$ & \textbf{10.1}$_{(-57.6\%)}$ & \textbf{8.8}$_{(-63.0\%)}$ & \textbf{10.1}$_{(-57.6\%)}$ \\
    \bottomrule
    \addlinespace[2pt]
    \multicolumn{7}{c}{\textbf{(b) Audio: speech noise 0\,dB, Video: random segment corrupted}}
  \end{tabular}
  }
\end{table*}

Similar to Table~\ref{tab:main_lrs2}, Table~\ref{tab:main_lrs3} presents additional results on the LRS3 dataset~\citep{afouras2018lrs3} to demonstrate the generalizability of our findings. A key difference from the LRS2 experiments is that our VSR head, BRAVEn-large, has also been fine-tuned on LRS3, making it a much stronger, in-domain model supporting the ASR stream. This serves to amplify the benefits of our dual-stream approach.

As shown in Table~\ref{tab:main_lrs3}, our DualHyp\,+\,RelPrompt framework achieves an overall WER of 10.5\% on audio corruptions and 10.1\% on visual corruptions. The performance gap between our method and GER w/ Auto-AVSR is even larger than on LRS2 (also refer to Table~\ref{tab:comparison_avsr}), confirming that as the quality of the independent VSR head improves, the advantage of our language-space fusion becomes more pronounced.
We also observe that on LRS3, the ASR hypotheses, while coherent, are often homogeneous and contain similar errors across the $N$-best list. Our DualHyp approach is particularly effective in this case, as the independent VSR hypotheses provide the diversity to break out of the ASR's error patterns.
\newpage
\section{Analysis}
\label{sec:ch5_analysis}

\subsection{Reliability Mask Prediction}
\begin{wraptable}{r}{0.55\columnwidth}
    \centering
    \small
    \vspace{-10pt}
    \caption[Performance of reliability mask predictors]{Performance (\%) of the reliability mask predictors with randomly corrupted audio and video segments. The metrics evaluate the classification of segments as \texttt{noisy}, which includes the \texttt{mixed} category.}
    \label{tab:mask_acc}
    \addtolength{\tabcolsep}{1pt}
    \resizebox{.55\columnwidth}{!}{
    \begin{tabular}{c|cccc|c}
        \toprule
        \textbf{SNR} & \textbf{Acc.} & \!\!\textbf{Precision}\!\! & \textbf{Recall} & \textbf{F1} & \textbf{WER} \\
        \midrule
        $-$10\,dB & 84.7 & 95.3 & 87.8 & 91.4 & 25.8 \\
        $-$5\,dB & 83.9 & 95.0 & 87.1 & 90.9 & 17.8 \\
        0\,dB & 82.2 & 94.4 & 85.5 & 89.7 & 7.2 \\
        5\,dB & 79.6 & 93.0 & 82.8 & 87.6 & 3.4 \\
        10\,dB & 76.2 & 90.9 & 78.2 & 84.1 & 2.5 \\
        \bottomrule
    \end{tabular}
    }
\end{wraptable}

In Table~\ref{tab:mask_acc}, our evaluation of the reliability predictors reveals two key strengths.
First, the predictor shows consistently high precision (>90\%), which ensures that its \texttt{noisy} flags are highly trustworthy and prevents the main model from incorrectly discarding clean data.
Second, the recall naturally decreases as the SNR increases.
This is a desirable behavior, as the predictor conservatively labels mildly corrupted audio segments as \texttt{clean}, allowing the model to continue exploiting the useful signal.

\begin{table}[!h]
    \centering
    \small
    \vspace{10pt}
    \caption[Analysis of hypotheses generation heads on LRS2 and LRS3]{WER (\%) comparison of different hypotheses from single-stream (GER) and dual-stream (DualHyp) generation heads.
    Note that the AVSR head is trained on LRS2 with babble noise~\citep{ma2023auto}, unlike the ASR and VSR heads.
    The corruption strategy follows Table\,\ref{tab:main_lrs2}\textcolor{Red}{a} and Table\,\ref{tab:main_lrs3}\textcolor{Red}{a}.}
    \label{tab:comparison_avsr}
    \addtolength{\tabcolsep}{1pt}
    \resizebox{\textwidth}{!}{
    \begin{tabular}{lcc|cccc|c|cccc|c}
        \toprule
        & & & \multicolumn{5}{c|}{\textbf{LRS2}} & \multicolumn{5}{c}{\textbf{LRS3
        }} \\
        \midrule
        \textbf{Method} & \!\textbf{Input}\! & \!\!\textbf{\# hyps}\!\! & \textbf{B} & \textbf{S} & \textbf{M} & \textbf{N} & \textbf{O} & \textbf{B} & \textbf{S} & \textbf{M} & \textbf{N} & \textbf{O} \\
        \midrule
        \multirow{3}{*}{GER} & A & 5 & 39.3 & 34.4 & 11.5 & 13.2 & 24.6 & 32.4 & 35.4 & 7.6 & 8.0 & 20.9 \\
         & AV & 5 & 18.9 & 39.0 & 17.4 & 18.1 & 23.3 & 17.9 & 45.6 & 14.2 & 11.0 & 22.2 \\
         & AV & 10 & 18.2 & 38.1 & 16.7 & 17.6 & 22.6 & 18.3 & 44.8 & 14.1 & 10.6 & 21.9 \\
        \midrule
        \multirow{2}{*}{\textbf{DualHyp}} & \!\!A\,+\,AV\!\! & 10 & {17.1} & 26.7 & \textbf{7.2} & {8.4} & 14.8 & 19.3 & 34.8 & 6.0 & {5.4} & 16.4 \\
        & A\,+\,V & 10 & 21.6 & {17.9} & 8.1 & 9.3 & {14.2} & {16.3} & {18.2} & \textbf{5.6} & 5.5 & {11.4} \\
        \midrule
        \textbf{DualHyp} & \!\!A\,+\,AV\!\! & 10 & \textbf{15.4} & 25.9 & 7.3 & 8.8 & 14.3 & 19.0 & 32.9 & 6.2 & 6.9 & 16.3 \\
        \textbf{+\,RelPrompt} & A\,+\,V & 10 & 20.4 & \textbf{16.0} & 8.0 & \textbf{8.2} & \textbf{13.2} & \textbf{14.9} & \textbf{16.2} & {5.7} & \textbf{5.1} & \textbf{10.5} \\
        \bottomrule
    \end{tabular}
    }
\end{table}

\subsection{Comparison with an AVSR Head}
\label{subsec:comparison_avsr}

Our analysis in Table~\ref{tab:comparison_avsr} highlights two key findings regarding hypothesis generation.
First, modality diversity of hypotheses is more crucial than sheer quantity.
Simply increasing the number of hypotheses for the single-stream GER (5\,$\rightarrow$\,10 AV hypotheses) yields only a marginal gain for overall performance (23.3\%\,$\rightarrow$\,22.6\%), compared to DualHyp using 5-best hypotheses from each distinct modality (23.3\%\,$\rightarrow$\,14.2\%).

Second, while AVSR hypotheses might seem viable alternatives to VSR, they remain overly dependent on the audio modality.
This is particularly evident under the speech noise condition, where the visual stream is crucial for disambiguating target utterance from interfering speech.
In this scenario, DualHyp (A\,+\,AV) struggles (26.7\% WER), as the early fusion of AVSR embeddings makes visual information rely on the corrupted audio.
Instead, DualHyp (A\,+\,V) leverages the audio-independent VSR stream to achieve 17.9\%, demonstrating the superiority of using disentangled hypotheses.
These findings are further supported by our LRS3 experiments.

\begin{figure}[!t]
    \centering
    \includegraphics[width=.48\linewidth]{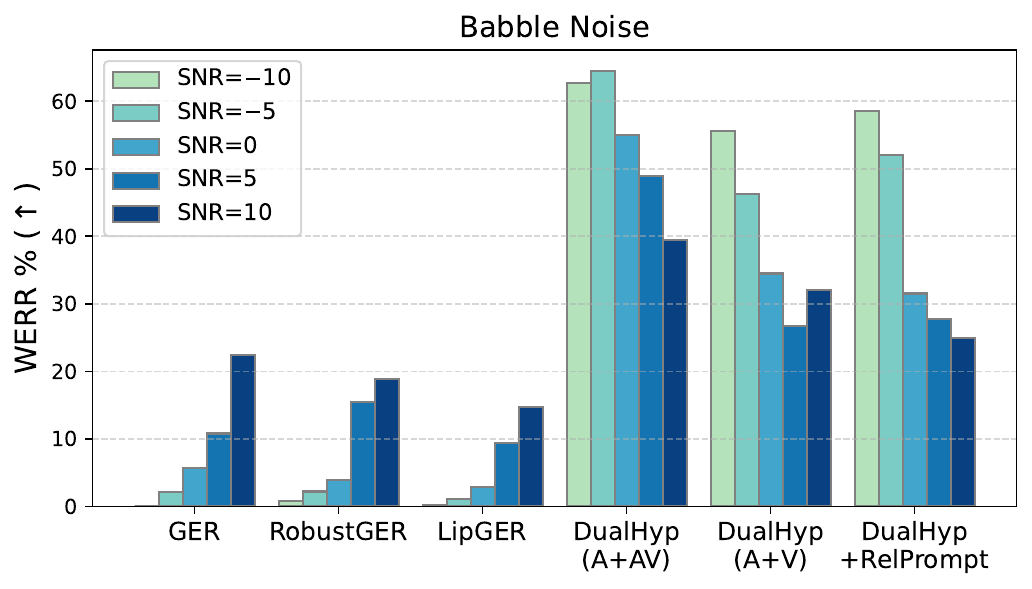}
    \hfill
    \includegraphics[width=.48\linewidth]{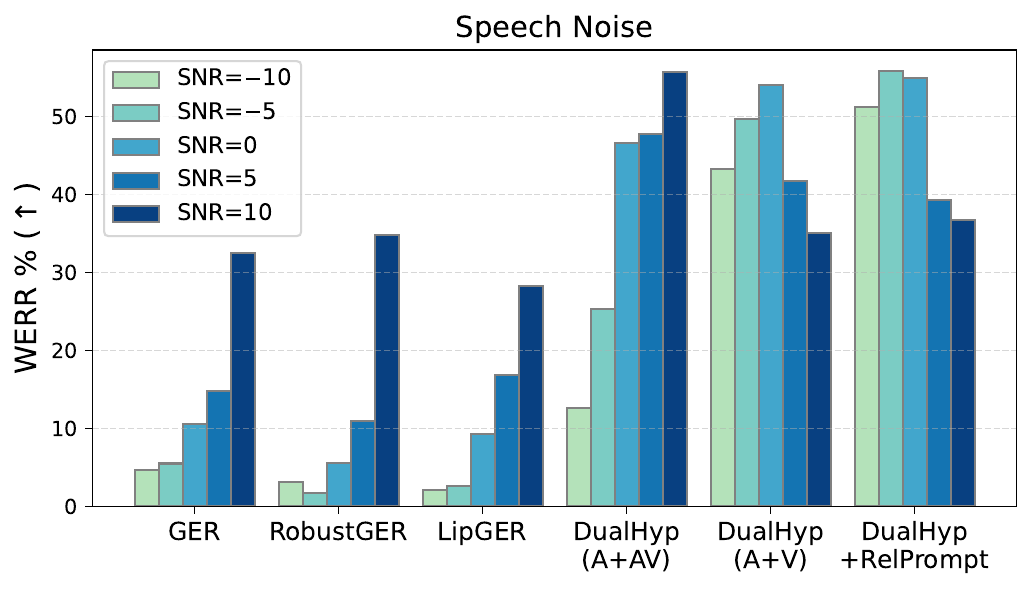}

    \includegraphics[width=.48\linewidth]{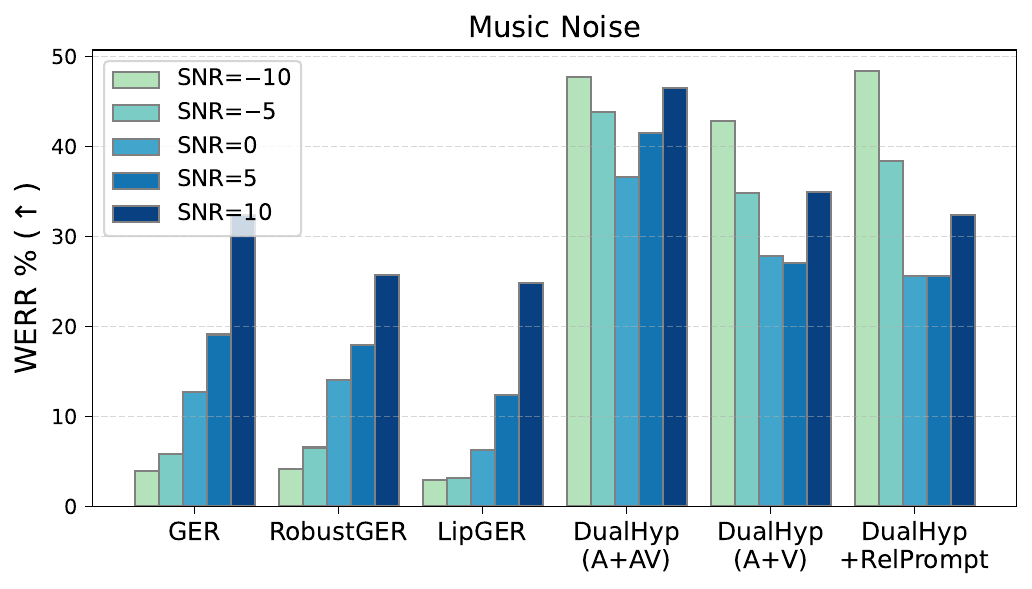}
    \hfill
    \includegraphics[width=.48\linewidth]{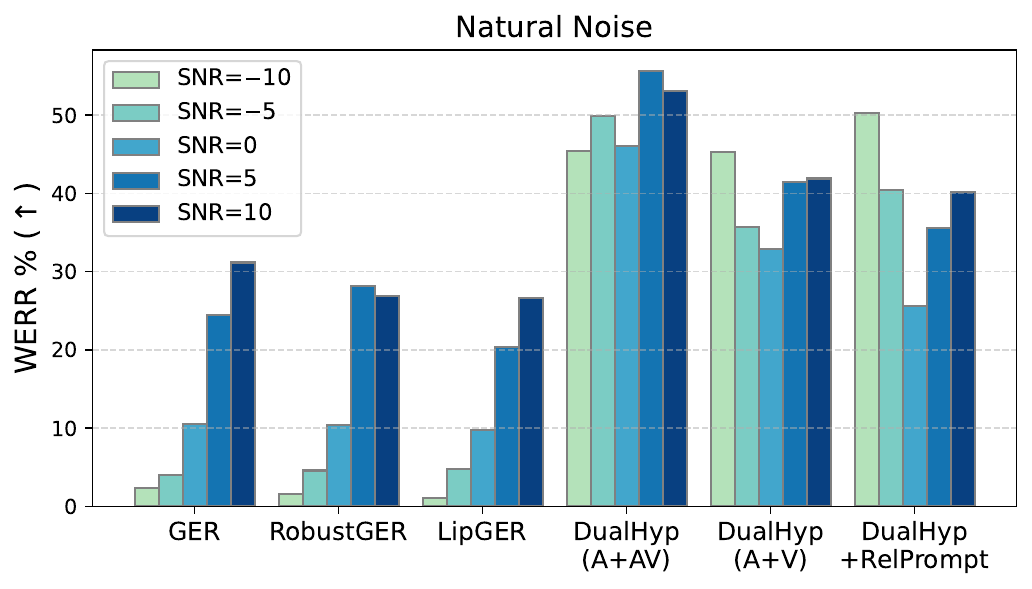}
    \caption[WERR analysis at different SNRs]{Word error rate reduction (WERR) at different audio SNRs, under diverse types of noise. Higher WERR indicates greater improvement over the Whisper ASR baseline. The experimental setup is identical to Table~\ref{tab:main_lrs2}\textcolor{Red}{a}.
    }
    \label{fig:snr_analysis}
\end{figure}

\subsection{SNR-wise WER Improvement}
\label{subsec:snr_to_werr}

Figure~\ref{fig:snr_analysis} reveals opposing trends in WER reduction (WERR, \citet{liu2025avger}) between single-stream and dual-stream methods.
For single-stream methods, WERR increases with better audio quality, as their effectiveness is limited to refining an already decent ASR output.
In contrast, our dual-stream framework maintains a high WERR even at very low SNRs by leveraging VSR hypotheses.
Furthermore, the addition of RelPrompt consistently boosts performance, with the most significant gains observed in low-SNR scenarios.
This confirms that by effectively utilizing the reliability information about corruption provided by RelPrompt, our framework can substantially reduce errors precisely when the audio is most challenging.

The DualHyp (A\,+\,AV) variant also illustrates this principle; it achieves a high WERR on familiar babble noise but does not show such strong correction capabilities on speech noise, especially at low SNRs.
This demonstrates a key limitation of the early-fusion AVSR head: since it is affected by the audio corruption, it may fail to provide a truly independent and useful signal for error correction.
In contrast, our DualHyp frameworks demonstrate superior robustness by maintaining high WERR even at very low SNRs, effectively leveraging the visual stream when the audio is most corrupted.

\begin{figure*}[!t]
    \centering
    \includegraphics[width=\linewidth]{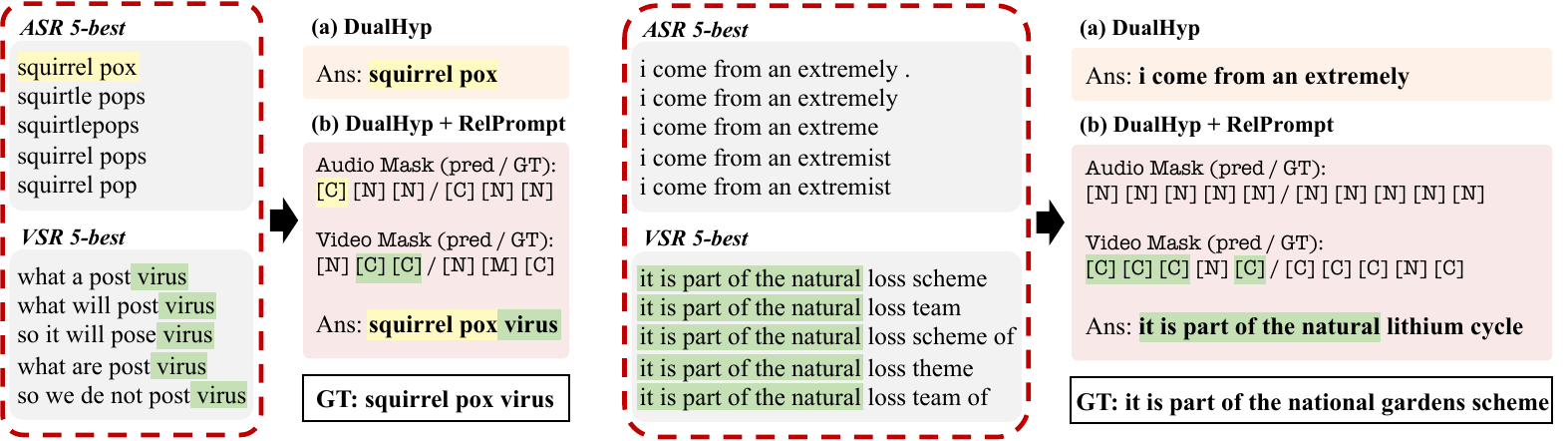}
    \caption[Qualitative analysis of RelPrompt]{Qualitative analysis comparing RelPrompt to the DualHyp baseline. RelPrompt uses reliability tokens (\ie masks) to explicitly inform the input signal quality, correctly guiding the use of ASR and VSR hypotheses.
    }
    \label{fig:qual_analysis}
\end{figure*}

\begin{table*}[!t]
\centering
\small
\vspace{-20pt}
\caption[Success cases of DualHyp]{Successful examples of the GER process using DualHyp. Highlights illustrate how the final output is assembled from partial information scattered across the ASR (audio) and VSR (video) 5-best hypothesis lists. These cases illustrate two primary successful patterns: multi-modal fragment composition and dominant modality refinement.}
\label{tab:successful_cases}
\vspace{-5pt}
\renewcommand{\arraystretch}{0.8}
\setlength{\fboxsep}{1pt} 
\resizebox{.95\textwidth}{!}{
\begin{tabularx}{\linewidth}{l|X|c}
\toprule
\textbf{Method} & \textbf{Utterance} & \textbf{WER (\%)} \\
\midrule
\midrule
\multicolumn{3}{c}{\textit{\textbf{Type 1: Multimodal Fragment Synthesis}}} \\
\midrule
\multirow{5}{*}{ASR 5-best} &
    \highlight{yellow!30}{which upset some}one in the next day & 71.4  \\
    & \highlight{yellow!30}{which} i am saying some of you may or may not understand & 128.6 \\
    & \highlight{yellow!30}{which upset some}one who knew not what to do & 100.0 \\
    & \highlight{yellow!30}{which} i am saying is a lot easier than it is today & 157.1 \\
    & \highlight{yellow!30}{which upset some}one you know what i mean & 85.7 \\
    \midrule
\multirow{5}{*}{VSR 5-best} &
    we jumped at \highlight{orange!20}{some of our female}\highlight{Green!10}{ residents} & 42.9 \\
    & we jumped at \highlight{orange!20}{some of our female} races & 57.1 \\
    & we jumps at \highlight{orange!20}{some of our female}\highlight{Green!10}{ residents} & 42.9 \\
    & we jump set \highlight{orange!20}{some of our female}\highlight{Green!10}{ residents} & 42.9 \\
    & we jumped at \highlight{orange!20}{some of our female} reasons & 57.1 \\
\midrule
DualHyp output &
\textbf{\highlight{yellow!30}{which upset some}\highlight{orange!20}{ of our female}\highlight{Green!10}{ residents}} & \textbf{0.0} \\
\midrule
Ground-truth & which upset some of our female residents & -- \\
\midrule
\midrule
\multirow{5}{*}{ASR 5-best} &
    \highlight{yellow!30}{so} what are the dangers of\highlight{purple!15}{ relying on}\highlight{blue!10}{ this information} & 62.5 \\
    & \highlight{yellow!30}{so} what are the dangers of\highlight{purple!15}{ relying on} disinformation & 87.5 \\
    & \highlight{yellow!30}{so} my other thing is\highlight{Green!10}{ just}\highlight{purple!15}{ relying on}\highlight{blue!10}{ this information} & 50.0 \\
    & \highlight{yellow!30}{so} what are the dangers\highlight{purple!15}{ relying on}\highlight{blue!10}{ this information} & 50.0 \\
    & \highlight{yellow!30}{so} my other thing is\highlight{purple!15}{ relying on}\highlight{blue!10}{ this information} & 50.0 \\
\midrule
\multirow{5}{*}{VSR 5-best} &
    \highlight{yellow!30}{so}\highlight{orange!20}{ rather than}\highlight{Green!10}{ just} regarding all\highlight{blue!10}{ this information} & 25.0 \\
    & \highlight{yellow!30}{so}\highlight{orange!20}{ rather than} regarding all\highlight{blue!10}{ this information} & 37.5 \\
    & \highlight{yellow!30}{so}\highlight{orange!20}{ rather than} to regard all\highlight{blue!10}{ this information} & 37.5 \\
    & \highlight{yellow!30}{so}\highlight{orange!20}{ rather than}\highlight{Green!10}{ just} regarding\highlight{blue!10}{ this information} & 25.0 \\
    & \highlight{yellow!30}{so}\highlight{orange!20}{ rather than} argue\highlight{blue!10}{ this information} & 37.5 \\
\midrule
DualHyp output &
\textbf{\highlight{yellow!30}{so}\highlight{orange!20}{ rather than}\highlight{Green!10}{ just}\highlight{purple!15}{ relying on}\highlight{blue!10}{ this information}} & \textbf{0.0} \\
\midrule
Ground-truth & so rather than just relying on this information & -- \\
\midrule
\midrule
\multicolumn{3}{c}{\textit{\textbf{Type 2: Dominant Modality Refinement}}} \\
\midrule
\multirow{5}{*}{ASR 5-best} &
    \highlight{yellow!30}{the armed forces} were & 33.3 \\
    & \highlight{yellow!30}{the armed forces} go & 33.3 \\
    & and\highlight{yellow!30}{ the armed forces} were & 66.7 \\
    & and\highlight{yellow!30}{ the armed forces} go & 66.7 \\
    & in\highlight{yellow!30}{ the armed forces} but & 66.7 \\
\midrule
\multirow{5}{*}{VSR 5-best} &
    i feel disco & 100.0 \\
    & helpful disco & 100.0 \\
    & i fell this & 100.0 \\
    & time for disco & 100.0 \\
    & helpful to this & 100.0 \\
\midrule
DualHyp output & \textbf{\highlight{yellow!30}{the armed forces}} & \textbf{0.0} \\
\midrule
Ground-truth & the armed forces & -- \\
\midrule
\midrule
\multirow{5}{*}{ASR 5-best} &
    is \highlight{yellow!30}{already}\highlight{orange!20}{ in}\highlight{Green!10}{ the}\highlight{purple!15}{ states} & 25.0 \\
    & \highlight{orange!20}{in}\highlight{Green!10}{ the} united\highlight{purple!15}{ states} & 50.0 \\
    & of\highlight{Green!10}{ the} united\highlight{purple!15}{ states} & 75.0 \\
    & from the rest of\highlight{Green!10}{ the} united\highlight{purple!15}{ states} & 100.0 \\
    & \highlight{orange!20}{in} one of\highlight{Green!10}{ the} other\highlight{purple!15}{ states} & 75.0 \\
\midrule
\multirow{5}{*}{VSR 5-best} &
    \highlight{yellow!30}{already} understand & 75.0 \\
    & \highlight{yellow!30}{already} understanding & 75.0 \\
    & \highlight{yellow!30}{already} understands & 75.0 \\
    & \highlight{yellow!30}{already} understanding that & 100.0 \\
    & we are writing these things & 125.0 \\
\midrule
DualHyp output &
\textbf{\highlight{yellow!30}{already}\highlight{orange!20}{ in}\highlight{Green!10}{ the}\highlight{purple!15}{ states}} & \textbf{0.0} \\
\midrule
Ground-truth & already in the states & -- \\
\bottomrule
\end{tabularx}
}
\end{table*}
\begin{table*}[!t]
\centering
\small
\vspace{-20pt}
\caption[Failure cases of DualHyp]{Failure examples of the GER process using DualHyp. Green highlights illustrate the correct words from ground-truth, whereas red highlights illustrate wrong words from inference. These cases illustrate two primary error patterns: over-reliance on plausible but inaccurate hypotheses and hallucination based on semantic association.}
\label{tab:failure_cases}
\vspace{-5pt}
\renewcommand{\arraystretch}{0.8}
\setlength{\fboxsep}{1pt} 
\resizebox{.95\textwidth}{!}{
\begin{tabularx}{\linewidth}{l|X|c}
\toprule
\textbf{Method} & \textbf{Utterance} & \textbf{WER (\%)} \\
\midrule
\midrule
\multicolumn{3}{c}{\textit{\textbf{Type 1: Over-reliance on Plausible but Inaccurate Hypotheses}}} \\
\midrule
\multirow{5}{*}{ASR 5-best} &
    what could be\highlight{Green!10}{ in the world} & 75.0 \\
    & to be\highlight{Green!10}{ in the world} & 50.0 \\
    & i can not believe\highlight{Green!10}{ the world} & 100.0 \\
    & it should be\highlight{Green!10}{ in the world} & 75.0 \\
    & good to be\highlight{Green!10}{ in the world} & 75.0 \\
\midrule
\multirow{5}{*}{VSR 5-best} &
    \highlight{red!10}{what we do}\highlight{red!30}{ is} & 100.0 \\
    & \highlight{red!10}{when we do} this & 125.0 \\
    & \highlight{red!10}{what we do}\highlight{red!20}{ here}\highlight{red!30}{ is} & 100.0 \\
    & \highlight{red!10}{what we do} with this & 125.0 \\
    & what we are doing\highlight{red!30}{ is} & 100.0 \\
\midrule
DualHyp output &
\textbf{\highlight{red!10}{what we do}\highlight{red!20}{ here}\highlight{red!30}{ is}} & \textbf{125.0} \\
\midrule
Ground-truth & \textbf{probably\highlight{Green!10}{ in the world}} & -- \\
\midrule
\midrule
\multirow{5}{*}{ASR 5-best} &
    that is what i am talking about & 100.0 \\
    & i\highlight{Green!20}{ can not} believe it & 100.0 \\
    & \highlight{Green!10}{i just}\highlight{Green!20}{ can not} believe it & 42.9 \\
    & \highlight{Green!10}{i} am \highlight{Green!10}{just} going to come with you & 85.7 \\
    & that is what i am saying & 114.3 \\
\midrule
\multirow{5}{*}{VSR 5-best} &
    \highlight{Green!10}{i just}\highlight{red!10}{ asked them}\highlight{Green!30}{ to win} & 42.9 \\
    & \highlight{Green!10}{i just}\highlight{red!10}{ asked} him \highlight{Green!30}{ to win} & 42.9 \\
    & \highlight{Green!10}{i just}\highlight{red!10}{ asked them}\highlight{red!20}{ to wait} & 57.1 \\
    & \highlight{Green!10}{i just}\highlight{red!10}{ asked} him\highlight{red!20}{ to wait} & 57.1 \\
    & \highlight{Green!10}{i just}\highlight{red!10}{ ask them}\highlight{Green!30}{ to win} & 42.9 \\
\midrule
DualHyp output &
\textbf{\highlight{Green!10}{i just}\highlight{red!10}{ asked them}\highlight{red!20}{ to wait}} & \textbf{57.1} \\
\midrule
Ground-truth & \textbf{\highlight{Green!10}{i just}\highlight{Green!20}{ can not} seem\highlight{Green!30}{ to win}} & -- \\
\midrule
\midrule
\multicolumn{3}{c}{\textit{\textbf{Type 2: Hallucination and Semantic Association Errors}}} \\
\midrule
\multirow{5}{*}{ASR 5-best} &
    no no no & 100.0 \\
    & no no very good & 166.7 \\
    & love love love & 100.0 \\
    & no no & 100.0 \\
    & i love \highlight{Green!10}{november} & 66.7 \\
\midrule
\multirow{5}{*}{VSR 5-best} &
    it goes \highlight{Green!10}{november} & 66.7 \\
    & \highlight{Green!10}{november} & 66.7 \\
    & it is on \highlight{Green!10}{november} & 66.7 \\
    & it is not \highlight{Green!10}{november} & 66.7 \\
    & it is called \highlight{Green!10}{november} & 66.7 \\
\midrule
DualHyp output &
\textbf{\highlight{Green!10}{november}\highlight{red!10}{ and december}} & \textbf{100.0} \\
\midrule
Ground-truth & \textbf{end of \highlight{Green!10}{november}} & -- \\
\midrule
\midrule
\multirow{5}{*}{ASR 5-best} &
    \highlight{Green!10}{this is the best} bathroom downtown & 50.0 \\
    & \highlight{Green!10}{this is the best} bathroom town\highlight{Green!10}{ in town} & 25.0 \\
    & \highlight{Green!10}{this is the best} bathroom\highlight{Green!10}{ in town} & 25.0 \\
    & \highlight{Green!10}{this is the best} bath\highlight{Green!10}{ hotel in town} & 12.5 \\
    & \highlight{Green!10}{this is the best} basketball town\highlight{Green!10}{ in town} & 50.0 \\
\midrule
\multirow{5}{*}{VSR 5-best} &
    \highlight{Green!10}{this is the best} bad\highlight{Green!10}{ hotel in town} & 12.5 \\
    & \highlight{Green!10}{this is the best} band\highlight{Green!10}{ hotel in town} & 12.5 \\
    & \highlight{Green!10}{this is the best bat hotel in town} & 0.0 \\
    & \highlight{Green!10}{this is the best} pat\highlight{Green!10}{ hotel in town} & 12.5 \\
    & \highlight{Green!10}{this is the best} baton\highlight{Green!10}{ hotel in town} & 12.5 \\
\midrule
DualHyp output &
\textbf{\highlight{Green!10}{this is the best}\highlight{red!10}{ bistro}\highlight{Green!10}{ in town}} & \textbf{25.0} \\
\midrule
Ground-truth & \textbf{\highlight{Green!10}{this is the best bat hotel in town}} & -- \\
\bottomrule
\end{tabularx}
}
\end{table*}

\begin{table*}[!t]
\centering
\small
\vspace{-20pt}
\caption[Qualitative analysis of DualHyp with RelPrompt]{Qualitative examples of successful corrections by DualHyp with RelPrompt. These cases show how RelPrompt improves upon the baseline DualHyp by leveraging the predicted reliability masks (\texttt{pred}) to trust or discard certain parts of hypotheses from the ASR and VSR streams. Ground-truth masks (\texttt{GT}) are also shown for comparison.}
\label{tab:additional_qual_analysis}
\vspace{-5pt}
\renewcommand{\arraystretch}{0.8}
\setlength{\fboxsep}{1pt} 
\resizebox{.95\textwidth}{!}{
\begin{tabularx}{\linewidth}{l|X|c}
\toprule
\textbf{Method} & \textbf{Utterance} & \textbf{WER (\%)} \\
\midrule
\midrule
\multirow{5}{*}{ASR 5-best} &
    to your baby of this year when she asked & 100.0  \\
    & to your baby of this year when she asks & 100.0 \\
    & \highlight{yellow!30}{there was} your baby of this year when she asked & 100.0 \\
    & there is your baby of this year when she asked & 88.9 \\
    & to your baby of this year when she asked & 100.0 \\
    \midrule
\multirow{5}{*}{VSR 5-best} &
    \highlight{yellow!30}{there was}\highlight{orange!20}{ no} air \highlight{Green!10}{so there was no sound} & 22.2 \\
    & \highlight{yellow!30}{there was}\highlight{orange!20}{ no} hit \highlight{Green!10}{so there was no sound} & 33.3 \\
    & \highlight{yellow!30}{there was}\highlight{orange!20}{ no} heat \highlight{Green!10}{so there was no sound} & 33.3 \\
    & \highlight{yellow!30}{there was}\highlight{orange!20}{ no} heart there was no sound & 44.4 \\
    & \highlight{yellow!30}{there was}\highlight{orange!20}{ no} it \highlight{Green!10}{so there was no sound} & 33.3 \\
\midrule
DualHyp output & \textbf{\highlight{yellow!30}{there was} your baby of this year when she asked} & \textbf{100.0} \\
\midrule
\multirow{5}{*}{RelPrompt output} & \texttt{Audio Mask (pred / GT):} & \\
& \texttt{[N][N][N][N][N][N] / [N][N][N][N][N][N]} & \\
& \texttt{Video Mask (pred / GT):} & \\
& \texttt{[C][M][N][N][M][C] / [C][M][N][N][M][C]} \\
& \textbf{\highlight{yellow!30}{there was}\highlight{orange!20}{ no} heat \highlight{Green!10}{so there was no sound}} & \textbf{33.3} \\
\midrule
Ground-truth & there is no air so there is no sound & -- \\
\midrule
\midrule
\multirow{5}{*}{ASR 5-best} &
    it is the same & 80.0  \\
    & at the same time . & 100.0 \\
    & at the same time & 100.0 \\
    & which again opens\highlight{orange!20}{ the elements} & 60.0 \\
    & it is the same . & 100.0 \\
    \midrule
\multirow{5}{*}{VSR 5-best} &
    and it opens your eyes & 100.0 \\
    & \highlight{yellow!30}{it opens to} the enemies & 60.0 \\
    & \highlight{yellow!30}{it opens to} the animation & 60.0 \\
    & and \highlight{yellow!30}{it opens to} the enemies & 80.0 \\
    & and \highlight{yellow!30}{it opens to} the animation & 80.0 \\
\midrule
DualHyp output & \textbf{\highlight{yellow!30}{it opens to} the east} & \textbf{60.0} \\
\midrule
\multirow{5}{*}{RelPrompt output} & \texttt{Audio Mask (pred / GT):} & \\
& \texttt{[N][N][N][C][C] / [M][N][N][C][C]} & \\
& \texttt{Video Mask (pred / GT):} & \\
& \texttt{[C][N][N][N][C] / [C][M][N][N][C]} \\
& \textbf{\highlight{yellow!30}{it opens to}\highlight{orange!20}{ the elements}} & \textbf{40.0} \\
\midrule
Ground-truth & again open to the elements & -- \\
\midrule
\midrule
\multirow{5}{*}{ASR 5-best} &
    \highlight{yellow!30}{like one hundreds of one thousands of people}\highlight{orange!20}{ do every year} & 0.0  \\
    & like one hundreds or one thousands of people\highlight{orange!20}{ do every year} & 9.1 \\
    & one hundreds of one thousands of people do every year & 9.1 \\
    & \highlight{yellow!30}{like one hundreds of one thousands of people}\highlight{orange!20}{ do every year} . & 9.1 \\
    & like one hundreds and one thousands of people\highlight{orange!20}{ do every year} & 9.1 \\
    \midrule
\multirow{5}{*}{VSR 5-best} &
    \highlight{yellow!30}{like one hundreds of one thousands of people} or so every & 27.3 \\
    & \highlight{yellow!30}{like one hundreds of one thousands of people} or so every year & 18.2 \\
    & \highlight{yellow!30}{like one hundreds of one thousands of people} or so often & 27.3 \\
    & \highlight{yellow!30}{like one hundreds of one thousands of people} do every & 9.1 \\
    & \highlight{yellow!30}{like one hundreds of one thousands of people} or so whoever & 27.3 \\
\midrule
DualHyp output &
\textbf{like one hundreds or one thousands of people\highlight{orange!20}{  do every year}} & \textbf{9.1} \\
\midrule
\multirow{5}{*}{RelPrompt output} & \texttt{Audio Mask (pred / GT):} & \\
& \texttt{[C][N][N][N][N][C][C] / [C][N][N][N][N][C][C]} & \\
& \texttt{Video Mask (pred / GT):} & \\
& \texttt{[C][C][C][C][N][N][C] / [C][C][C][N][N][N][N]} \\
& \textbf{\highlight{yellow!30}{like one hundreds of one thousands of people}\highlight{orange!20}{ do every year}} & \textbf{0.0} \\
\midrule
Ground-truth & like one hundreds of one thousands of people do every year & -- \\
\bottomrule
\end{tabularx}
}
\end{table*}

\subsection{Qualitative Analysis}

Our qualitative analysis in Figure~\ref{fig:qual_analysis} illustrates how RelPrompt corrects failures of the baseline DualHyp framework by providing explicit reliability signals.
(\textit{Left}): RelPrompt uses clean video tokens \texttt{[C]} as a cue to trust the last part of the VSR hypotheses, allowing it to recover the word (\texttt{virus}) which the baseline has missed.
(\textit{Right}): The ASR system is presented with fluent but entirely incorrect hypotheses.
By referencing the consistently noisy audio tokens \texttt{[N]}, the LLM correctly identifies the ASR stream as unreliable and pivots to the more accurate VSR candidates.
In contrast, without the RelPrompt mechanism, the model lacks any modality-level grounding and produces a completely incorrect output.
These cases demonstrate that by providing explicit reliability tokens, RelPrompt empowers the LLM to act as an intelligent controller, grounding its compositional reasoning in the predicted quality of the source signals.

\subsection{Additional Cases of DualHyp}
\label{subsec:additional_cases_dualhyp}

\paragraph{Success Case.}
As a supplement to the cases presented in Table~\ref{tab:dualhyp_case}, Table~\ref{tab:successful_cases} provides further qualitative examples that illustrate the successful mechanisms of our DualHyp framework.
These successes can be categorized into two main patterns.

The first pattern, \textit{Multimodal Fragment Composition}, involves the model's ability to recover correct transcriptions by leveraging complementary fragments from both ASR and VSR hypotheses.
This can be seen when the framework fuses the beginning of an ASR hypothesis and the end of a VSR hypothesis as in the first case (\ie combining \texttt{which upset some...} from ASR and \texttt{...female residents} from VSR), or vice versa as in the second case (\ie \texttt{so rather than...} and \texttt{relying on...}).
The compositional fusion of correct sub-sequences from noisy ASR and VSR inputs highlights the DualHyp's robustness by leveraging a generative correction ability of LLM.

The second pattern is \textit{Dominant Modality Refinement}, where the model identifies and grounds the prediction in the more reliable modality, even when that modality's best hypothesis is not perfect.
This is evident in \texttt{the armed forces} and \texttt{already in the states} cases, where the model primarily refines ASR's strong-but-flawed hypotheses while disregarding the less plausible VSR candidates.
These cases highlight that providing the LLM with separate, modality-specific hypotheses is a more effective correction strategy than relying on a single or early-fused representation, as it allows the model to reason over distinct evidences.

\paragraph{Failure Case.}
Following the successful cases, we also present and analyze several typical failures, which often occur when both modalities provide highly ambiguous information.
As illustrated in Table~\ref{tab:failure_cases}, these failures can be categorized into two primary patterns.

The first failure pattern is \textit{Over-reliance on Plausible but Inaccurate Hypotheses}, where LLMs are misled by a semantically incorrect candidate from one modality.
In the \texttt{probably in the world} example, the LLM disregards the partially correct ASR hypotheses and instead adopts the coherent but entirely wrong VSR hypothesis.
Second example shows that the model favors the plausible but incorrect verb \texttt{to wait} from VSR candidates, although the correct verb \texttt{to win} is also present in the other VSR hypotheses. 
These cases show that the ambiguity between hypotheses can make the LLM confuse and incorrectly prioritize a plausible but wrong candidate.
The over-reliance issue is a common drawback in all GER frameworks but can be mitigated to some extent by leveraging our RelPrompt, as shown in Figure~\ref{fig:qual_analysis}.

The second failure pattern involves \textit{Hallucination and Semantic Association Errors}, where the LLM generates words that are not present in any of the provided hypotheses.
This often occurs when the model is biased towards a specific keyword and generates a semantically related but incorrect term, as seen in the \texttt{end of november} example where it generates \texttt{december} out of nowhere.
In the last case, the model's strong prior knowledge can override direct evidence, misinterpreting \texttt{bat hotel} as \texttt{bistro}.
This reveals the fundamental duality of leveraging the LLM's internal knowledge for GER, where context-aware corrections produce not only useful generative revisions but also factually incorrect hallucinations, suggesting a potential direction for future research on controlling this mechanism.

\paragraph{RelPrompt.}
Table~\ref{tab:additional_qual_analysis} provides the qualitative examples that demonstrate how RelPrompt successfully corrects errors for the cases where baseline DualHyp framework fails. In the first example (\texttt{there is no air...}), DualHyp is misled by entirely incorrect ASR hypotheses (\texttt{your baby of...asked}). RelPrompt, in contrast, uses its predicted audio reliability tokens (all \texttt{[N]}) and rather clean video reliability tokens (\texttt{[C]}) to correctly identify the audio stream as unreliable, allowing it to pivot to the more accurate VSR hypotheses for a much better result. 

In the second and third examples, the reliability masks guide the model to capitalize on the structure from the cleaner VSR stream at the beginning of the utterance, while correctly extracting a more accurate key phrase (\ie \texttt{the elements} and \texttt{do every year}) from the ASR stream to form the ending.
The baseline DualHyp method, lacking this guidance, is confused by the conflicting signals and produces errors by incorporating some flawed hypotheses.
These cases demonstrate how the explicit reliability signals empower the model to intelligently arbitrate between hypotheses at a sub-sentence level, composing the final output from the most reliable fragments of each modality.

\section{Chapter Summary}

In this chapter, we introduced DualHyp, a novel GER framework for AVSR that deliberately delays modality fusion to the language space, where an LLM performs compositional reasoning on independent hypotheses from ASR and VSR models. We further enhanced this with RelPrompt, a noise-aware guidance mechanism that guides the LLM with explicit, time-aligned reliability signals for each modality. The experiments showed that our new framework significantly outperforms single-stream GER approaches, highlighting a flexible paradigm that leverages modular integration.

\paragraph{Limitations.}
While our framework demonstrates significant robustness and scalability in AVSR, it still holds several primary limitations that are common to most GER systems.
First, the performance of our framework is fundamentally dependent on the quality of its consisting components, especially the upstream SR heads. If the initial hypotheses from the SR head are of poor quality, as seen in our results of the MuAViC French case, the LLM's ability to perform corrections is limited. This dependency currently restricts the framework's applicability beyond English, because there is no publicly available, high-quality multilingual VSR model, making adaptation to the low-resource speech recognition and translation challenging.
Second, multiple modules in our structure introduces computational latency, posing a challenge for real-time applications. Although the ASR and VSR streams can be processed in parallel, the final LLM correction step is sequential, creating an unavoidable bottleneck. While modern efficiency techniques like flash attention can mitigate this to an extent, the approach remains inherently slower than a single end-to-end model, making deployment on resource-constrained edge devices a significant hurdle.
Lastly, streaming AVSR is infeasible in the current framework, as the LLM requires the entire input hypotheses to generate corrections. Future work could explore integrating streaming-capable GER frameworks or developing chunk-based processing methods to enable real-time applications.

\chapter{Concluding Remarks}
\label{chap:conclusion}

\section{Dissertation Summary}
This dissertation addresses the critical challenge of deploying Audio-Visual Speech Recognition (AVSR) systems in real-world environments, where performance is often compromised by unpredictable acoustic noise or visual interference. To overcome these obstacles, this work proposes and validates a systematic and hierarchical methodology for achieving robust scalability, demonstrating across three distinct levels: representation learning, model architecture, and system-level integration. By developing innovative solutions at each stage, this research provides a comprehensive framework for building AVSR systems that are not only accurate under ideal conditions but are also resilient, efficient, and extensible enough for practical, real-world application.

Chapter~\ref{chap:representation_scalability} focuses on the representation-level scalability, developing a universal pretraining strategy that learns audio-visual features inherently robust to diverse real-world corruptions. Through a multi-task corrupted prediction framework, the model is trained to reconstruct missing or distorted information in one modality using context from the other, forcing it to learn a resilient and generalizable latent space. This pretraining framework enables any audio-visual encoder to adapt to new environments and unseen noise types without relying on any specialized noise-specific modules.

Chapter~\ref{chap:architecture_scalability} addresses architectural scalability by proposing a mixture of hierarchical experts. This architecture efficiently expands model capacity by intelligently allocating computational resources, activating only a relevant subset of parameters based on the input data's characteristics. This ensures that the model can handle complex multimodal inputs in an adaptive and reliable manner without a prohibitive increase in computational cost. We successfully scale the AVSR model to 1B parameters, demonstrating significant performance improvements across various benchmarks while maintaining efficiency.

Chapter~\ref{chap:system_scalability} examines on the system level, introducing a novel framework for generative error correction that functionally extends the AVSR system through modular integration with large-scale and strong foundation models. By generating independent hypotheses from the audio and visual streams and leveraging LLM to intelligently compose them, this approach maximizes final recognition accuracy, particularly in scenarios with severe modality-specific corruptions. 

Incorporating these systematic solutions, this thesis collectively presents a comprehensive methodology for building the next generation of robust and scalable AVSR systems prepared for high-reliability deployment in real-world environments.

\section{Comprehensive Analysis Across Scalability Levels}
As summarized, this dissertation has introduced solutions at three hierarchical levels of the AVSR pipeline: representation (CAV2vec), architecture (MoHAVE), and system (DualHyp). While each contribution has been shown to be effective in its respective domain, this section presents an integrated analysis to compare them in the same evaluation setup as well as demonstrate their synergistic potential. To achieve this, we conduct a series of experiments that first combine the representation and architecture-level contributions and then situate this powerful new model within the system-level DualHyp framework.

\subsection{Synergy of Representation and Architecture: CAV-MoHAVE}
The first stage of our analysis investigates the interplay between the robust features learned by CAV2vec and the architectural dynamics of MoHAVE. CAV2vec excels at producing representations that are inherently resilient to noise, while MoHAVE provides a scalable architecture that can adapt its large capacity to the input's complexity. To evaluate their synergy, we construct a new integrated model, which we term as CAV-MoHAVE. In this configuration, the pretrained CAV2vec model serves as the powerful front-end feature encoder, and its output representations are fed into the MoHAVE decoder.

We first establish the individual performance of a standard AVSR model equipped with either the CAV2vec encoder or the MoHAVE decoder against a common baseline. We then evaluate the combined CAV-MoHAVE model under the same noisy conditions.

\begin{table}[!h]
    \centering
    \small
    \caption[CAV-MoHAVE performance on LRS3]{Performance comparison of CAV-MoHAVE against individual components and baselines on LRS3. C.P. denotes whether the encoder has been pretrained with corrupted prediction tasks. The experimental setup follows Table~\ref{tab:main_result}, using the object occlusion and noise for visual corruption.}
    \label{tab:cav_mohave}
    \begin{tabular}{lcc|cccc|c|c}
        \toprule
        \textbf{Method} & \textbf{C.P.} & \textbf{Params} & \textbf{Babble} & \textbf{Speech} & \textbf{Music} & \textbf{Natural} & \textbf{N-WER} & \textbf{C-WER} \\
        \midrule
        AV-HuBERT & \xmark & 325M\,+\,152M & 11.0 & 3.9 & 4.7 & 4.5 & 6.0 & 1.6 \\
        AV-data2vec & \xmark & 325M\,+\,152M & 11.4 & 4.1 & 4.8 & 4.5 & 6.2 & 1.5 \\
        AV-RelScore & \xmark & 325M\,+\,152M & 10.8 & 3.7 & 4.6 & 4.4 & 5.9 & 1.6 \\
        \midrule
        CAV2vec & \cmark & 325M\,+\,152M & 9.2 & 3.2 & 4.1 & 3.9 & 5.1 & 1.5 \\
        MoHAVE & \xmark & 325M\,+\,681M & 8.9 & \textbf{3.1} & \textbf{4.0} & 3.7 & 5.0 & \textbf{1.4} \\
        CAV-MoHAVE & \cmark & 325M\,+\,681M & \textbf{8.6} & 3.3 & \textbf{4.0} & \textbf{3.6} & \textbf{4.9} & 1.5 \\
        \bottomrule
    \end{tabular}
\end{table}

The results, as detailed in Table~\ref{tab:cav_mohave}, demonstrate that CAV-MoHAVE surpasses the performance of either component used in isolation. This outcome validates the hypothesis that the two contributions are complementary: the high-quality, noise-robust features from CAV2vec provide a clean signal that allows the MoHAVE architecture to more effectively allocate its expert capacity, leading to a superior overall transcription accuracy. This powerful integrated model serves as the new state-of-the-art baseline for our final system-level comparison.

\subsection{System-Level Enhancement of CAV-MoHAVE}
Having established the effectiveness of the CAV-MoHAVE model, the final stage of our analysis evaluates its performance within the broader system-level frameworks of GER and DualHyp. This comparison serves to quantify the additional gains achievable by integrating a highly robust end-to-end model with the advanced reasoning and error-correction capabilities of LLMs. We evaluate under both the standard GER and proposed DualHyp frameworks.

\begin{itemize}
    \item Standard GER Integration: We first apply a standard GER step to the output of the CAV-MoHAVE model. The $N$-best hypotheses generated by CAV-MoHAVE are fed into an LLM (we use Llama-3.2-3B in this case), which is tasked with producing a corrected final transcript. This configuration, denoted as GER + CAV-MoHAVE, measures the value added by LLM-based error correction on top of a very strong base model.
    \item DualHyp Framework Integration: Next, we leverage the CAV-MoHAVE architecture as the backbone for the independent recognizer head within the DualHyp framework. To this end, we utilize CAV-MoHAVE for AVSR and BRAVEn-large for VSR and integrate the two streams as DualHyp. This configuration is denoted as DualHyp + CAV-MoHAVE.
\end{itemize}

\begin{table}[!h]
    \centering
    \small
    \caption[DualHyp with CAV-MoHAVE on LRS2]{System-level performance comparison of CAV-MoHAVE within GER and DualHyp frameworks on LRS2. The corruption strategy follows Table\,\ref{tab:main_lrs2}\textcolor{Red}{a}. For the upper part, we show the relative WER reduction compared to the Whisper-large-v3 baseline, and for the lower part, we compare against the CAV-MoHAVE baseline.}
    \label{tab:dualhyp_cav_mohave_lrs2}
    \begin{tabular}{lc|cccc|l}
        \toprule
        \textbf{Method} & \textbf{Input} & \textbf{Babble} & \textbf{Speech} & \textbf{Music} & \textbf{Natural} & \textbf{Overall} \\
        \midrule
        \multicolumn{7}{c}{\textcolor{gray}{\textit{Whisper as ASR head \& BRAVEn as VSR head}}} \\
        \midrule
        Whisper-large-v3 & A & 40.0 & 36.5 & 12.7 & 14.2 & 25.8 \\
        BRAVEn-large & V & - & - & - & - & 39.7$_{(+53.9\%)}$ \\
        GER & A & 39.3 & 34.4 & 11.5 & 13.2 & 24.6$_{(-4.7\%)}$  \\
        \midrule
        \textbf{DualHyp} & A\,+\,V & 21.6 & 17.9 & 8.1 & 9.3 & 14.2$_{(-45.0\%)}$ \\
        \textbf{+\,RelPrompt} & A\,+\,V & 20.4 & 16.0 & 8.0 & 8.2 & 13.2$_{(-48.8\%)}$ \\
        \midrule
        \multicolumn{7}{c}{\textcolor{gray}{\textit{CAV-MoHAVE as AVSR head \& BRAVEn as VSR head}}} \\
        \midrule
        CAV-MoHAVE & AV & 18.6 & 11.8 & 12.5 & 11.7 & 13.6 \\
        GER & AV & 14.3 & 7.0 & 7.8 & 6.9 & 9.0$_{(-33.8\%)}$ \\
        \textbf{DualHyp\,+\,RelPrompt} & AV\,+\,V & \textbf{12.5} & \textbf{6.5} & \textbf{6.7} & \textbf{5.9} & \textbf{7.9}$_{(-41.9\%)}$ \\
        \bottomrule
    \end{tabular}
\end{table}
\begin{table}[!h]
    \centering
    \small
    \caption[DualHyp with CAV-MoHAVE on LRS3]{System-level performance comparison of CAV-MoHAVE within GER and DualHyp frameworks on LRS3. The corruption strategy follows Table\,\ref{tab:main_lrs3}\textcolor{Red}{a}. For the upper part, we show the relative WER reduction compared to the Whisper-large-v3 baseline, and for the lower part, we compare against the CAV-MoHAVE baseline.}
    \label{tab:dualhyp_cav_mohave_lrs3}
    \begin{tabular}{lc|cccc|l}
        \toprule
        \textbf{Method} & \textbf{Input} & \textbf{Babble} & \textbf{Speech} & \textbf{Music} & \textbf{Natural} & \textbf{Overall} \\
        \midrule
        \multicolumn{7}{c}{\textcolor{gray}{\textit{Whisper as ASR head \& BRAVEn as VSR head}}} \\
        \midrule
        Whisper-large-v3 & A & 32.6 & 34.5 & 7.3 & 7.8 & 20.6 \\
        BRAVEn-large & V & - & - & - & - & 31.9$_{(+54.9\%)}$  \\
        GER & A & 32.4 & 35.4 & 7.6 & 8.0 & 20.9$_{(+1.5\%)}$  \\
        \midrule
        \textbf{DualHyp} & A\,+\,V & 16.3 & 18.2 & 5.6 & 5.5 & 11.4$_{(-44.7\%)}$ \\
        \textbf{+\,RelPrompt} & A\,+\,V & 14.9 & 16.2 & 5.7 & 5.1 & 10.5$_{(-49.0\%)}$ \\
        \midrule
        \multicolumn{7}{c}{\textcolor{gray}{\textit{CAV-MoHAVE as AVSR head \& BRAVEn as VSR head}}} \\
        \midrule
        CAV-MoHAVE & AV & 7.2 & 2.4 & 3.3 & 3.1 & 4.0 \\
        GER & AV & 6.7 & 2.2 & 3.0 & 2.8 & 3.7$_{(-7.5\%)}$ \\
        \textbf{DualHyp\,+\,RelPrompt} & AV\,+\,V & \textbf{6.3} & \textbf{2.0} & \textbf{2.8} & \textbf{2.5} & \textbf{3.4}$_{(-15.0\%)}$ \\
        \bottomrule
    \end{tabular}
\end{table}

This final comparison in Tables~\ref{tab:dualhyp_cav_mohave_lrs2} and~\ref{tab:dualhyp_cav_mohave_lrs3} demonstrates a clear performance hierarchy. Here, the base model CAV-MoHAVE serves as a strong competitor, since it has been trained in the same data distribution with audio-visual corruption, hence achieving much more robust performance than the Whisper baseline. While the GER + CAV-MoHAVE configuration should improve upon the base model, DualHyp with CAV-MoHAVE yields the lowest overall WER. Such a result would provide the ultimate validation for this dissertation's hierarchical approach, proving that the most robust system is achieved by: \textbf{a) learning resilient representations (CAV2vec), b) processing them with an efficient and adaptive architecture (MoHAVE), and c) intelligently combining the outputs of strong heads using a system-level framework that excels at compositional reasoning (DualHyp)}.

\section{Future Research Directions}

While this dissertation provides a comprehensive framework for robust and scalable AVSR, our research opens up several promising avenues for future investigation that extend beyond the immediate scope of this work.

\paragraph{Multilingual Audio-Visual Understanding and Generation.}
A critical direction is the extension of these models to multilingual contexts. The frameworks developed in this thesis primarily focus on monolingual data (except for Secton~\ref{subsec:ch4_multilingual} and Section~\ref{subsec:multilingual_avsr}), yet a significant real-world challenge lies in supporting low-resource languages where paired audio-visual data is scarce. Recent MLLMs~\citep{xu2025qwen2, sun2024video} offer a path toward more nuanced multilingual audio-visual understanding. Concurrently, direct audio-visual-to-audio-visual (AV2AV) translation models~\citep{choi2024av2av,cho2025mavflow} enable audio-visual generation tasks including advanced movie dubbing, but significant improvements in robustness and linguistic diversity are needed to make these approaches practical. One of the major bottlenecks in this area is the lack of large-scale multilingual audio-visual datasets, necessitating research into effective multimedia data construction.

\paragraph{Expanding the Visual Modality beyond Lip Reading.}
Current VSR or AVSR models almost exclusively focus on lip movements to decipher linguistic content. However, human communication is rich with non-verbal cues. Future research should explore architectures that can interpret a wider array of facial dynamics, such as eyebrow movements, eye gaze, and micro-expressions, to capture paralinguistic information like emotional state or conversational intent. This would transition the field from pure speech recognition towards more holistic audio-visual dialogue systems capable of a deeper, more context-aware understanding of human interaction. Furthermore, integrating body language and gestures could provide additional context, which is essential for seamless audio-visual generation.

\paragraph{On-Device Deployment and Efficiency.}
Bridging the gap between large-scale research models and practical applications is required. The trend towards embedding AI directly into consumer hardware, as seen in recent products like Ray-Ban Meta AI smart glasses and Google's Android XR headsets/glasses for on-device processing, necessitates AVSR systems that operate with minimal latency and power consumption, and without constant cloud reliance. This presents a formidable research challenge: designing highly efficient architectures and algorithms to operate powerful, large-scale models on resource-constrained edge devices, thereby enabling the next generation of truly interactive and private AI-powered experiences.

{
\bibliographystyle{plainnat}
\bibliography{egbib}
}

\eack{
When I first met Prof. Se-Young Yun in the winter of 2018 and joined the lab as an undergraduate research student, I could never have imagined that I would be graduating with a Ph.D. in this field seven years later. I would like to express my deepest respect and gratitude to Prof. Yun, who has guided me, a student with little knowledge or experience in research, to grow into the independent researcher I am today. Among the countless teachings I learned from him, the lesson that one must possess outstanding character before becoming an outstanding researcher will be the principle I cherish the most. As the professor always emphasized, and as is the essence of research, every achievement I have made was only possible because of the colleagues who stayed by my side. Keeping this in mind, I was able to accomplish a great deal by collaborating with such excellent colleagues and supervisors.

OSI Lab, therefore, holds a very special place in my heart. Over the past seven years, under the professor’s unwavering supervision, the lab has undergone various changes. Amidst these dynamics, I too have oscillated between frustration and joy, navigating through difficult yet happy times where no two moments were ever alike. If I had to choose the period that best embodies the passionate spirit of my graudate years, it would be, ironically, my first research project with Dr. Sangmin Bae, where we encountered numerous failures. Back then, we would get upset over a single line in a review and feel disheartened by a single score, wishing for this arduous process to end quickly. However, looking back, I realize that it is unlikely that I will ever find a better time where I could be so completely immersed in one thing and learn and grow so much in such a short period. I extend my gratitude to Dr. Sangmin Bae, who shared that meaningful time with me, and to Dr. Jongwoo Ko, who supported me as a senior and encouraged me as a friend by my side. As friends, colleagues, and rivals, I have learned a tremendous amount by researching alongside these two.

The current growth of our lab would not have been possible without the contributions of the OSI Lab alumni—Jaehoon, Hyungjun, Taehyeon, Gihun, Sangmook, and Sumyeong, etc.—who are now active in both industry and academia. In my early days of graduate school, when I was inexperienced, these seniors were my strongest supporters, and I received immense help simply by following in their footsteps. To Mingyu, who joined Graduate School of AI with me as inaugural members and became a great source of reliance and a mental pillar, I hope I am given the time to repay everything I have received from him. I also sincerely support the futures of Seongyoon, Jihwan, Namgyu, and Yujin, who always livened up the lab atmosphere and with whom I collaborated for a short time. I thank all the seniors, juniors, and colleagues of the OSI Lab whom I could not mention individually. The unforgettable memories these people shared with me have become an irreplaceable asset in my life.

I offer my deepest gratitude to the committee members—Prof. Chanwoo Kim, Prof. Hoirin Kim, Prof. Tae-Hyun Oh, and Prof. Joon Son Chung—who provided invaluable guidance and unsparing feedback throughout the writing of this dissertation. I also extend a special thanks to Dr. Kangwook Jang, who has been my friend since we were roommates in Somang Hall during our first year of undergraduate studies, and who has now become a researcher writing papers alongside me. Having contributed to every chapter of this dissertation, he filled the gaps in my background and helped me successfully complete my Ph.D. journey with insightful advice and unsparing support. There were times when I spoke harsh words and pushed him hard because my expectations were high, but I hope he understands that nothing was spoken out of personal animosity. I look forward to his future research endeavors, and I will cheer for him on whatever path he takes. I also owe a debt of gratitude to Sungwoo, who was of great help to this dissertation; though he may have many concerns now, I have no doubt that he will grow into a distinguished researcher in the near future. Furthermore, I would like to thank Junsoo, Prof. Kibeom Hong, and Prof. Namhyuk Ahn, who were excellent mentors during my internship at NAVER Webtoon, as well as my mentors Haofu, Srikar, and Peng, who helped me greatly during my internship at AWS. Thanks to these mentors in the industry who broadened my horizon beyond the school and academia, I was able to deeply contemplate what kind of research I should pursue during my doctoral course.

I also wish to express my heartfelt gratitude to the cherished people outside the lab who have been pillars of support in my life. First, I thank my beloved high school friends, Jiheon, Kitae, Kanghee, and Wootak, who were a breath of fresh air and a dependable source of strength whenever I was worn out from research. I have always felt sorry for not being able to spend more time with them, using my busy schedule as an excuse. I am also grateful to my Mixer friends; whenever we meet, it feels like we are back in our freshman year, bringing me immense joy. Seeing each of them establish themselves as key figures in their respective fields makes me feel both amazed and incredibly proud. Additionally, I extend my unchanging friendship to my Hansung friends, Seungho, Minsu, and Hankyul, who are striving in their respective paths; to my Freshman Class 8 friends, Seungmin, Yunseok, Jaewook, Eunseop, Byeongjun, and Junkyu, with whom I began my college life; and to my oldest dear friends, Paul and Jay.

Finally, I dedicate this thesis to my loving family, the reason for my being and my driving force. I express my infinite respect and gratitude to my parents, who have always given me unconditional support. Perhaps, since my parents always told me I could quit this anytime I wanted, this disobedient son did not quit but endured to the end to graduate. Although I cannot always say it out loud, I take this opportunity to tell them that I love and respect them more than anyone.

I would like to send my deepest love to my wife, Hyeonjeong, who became my most precious life partner with our promise of forever this past November 1st. I thank her from the bottom of my heart for her love and care, and for staying by my side and waiting patiently despite the numerous ups and downs during this long academic journey. I gain great courage knowing that she is there to weather the storms, both big and small, that will come in our journey ahead. As we begin this new chapter of life, I am more than excited for the next pages we will write together.

\vskip 10pt
\hfill
December 2025 at Seoul Campus
}
\clearpage
\kack{
2018년 겨울, 윤세영 교수님을 처음 뵙고 연구실에 개별 연구생으로 발을 들였을 때만 해도, 제가 7년 뒤 이 분야에서 박사 학위를 받고 졸업하게 되리라고는 상상하지 못했습니다. 연구에 대한 경험도 지식도 일천했던 저를 독립적인 연구자로서 지금까지 성장할 수 있게 해주신 윤세영 교수님께 깊은 존경과 감사의 마음을 올립니다. 그동안 교수님께 배운 수많은 가르침 중, 훌륭한 연구자가 되기 이전에 훌륭한 인격을 갖춘 연구자가 되어야 한다는 말씀은 제가 가장 오래 간직하고 갈 지침이 될 것입니다. 교수님께서 늘 강조하셨던 것처럼, 그리고 연구의 본질이 그러하듯, 제가 이룬 성취는 모두 제 곁을 지켜준 동료들이 있었기에 가능했습니다. 이를 마음에 새기며 우수한 동료 연구자들 및 지도자들과 호흡한 덕분에 많은 결실을 맺을 수 있었습니다.

OSI 연구실은 그래서 저에게 참 각별했던 곳이었습니다. 지난 7년간 변함없는 교수님의 지도 아래 연구실은 다양한 변화를 맞이했던 것 같습니다. 저 또한 그 역동 속에서 좌절과 환희를 오가며, 수없이 반복되지만 어느 것 하나 똑같지 않은, 어렵고도 행복한 시기들을 지나왔습니다. 대학원 기간 중 가장 대학원생다운 열정의 기억이 남는 시기를 꼽으라면, 아이러니하게도 수많은 좌절을 맛보았던 배상민 박사와의 첫 연구가 아닐까 싶습니다. 그때는 리뷰 한줄에 열내고 점수 하나에 낙담하며 이 지난한 과정이 빨리 끝나길 바랐는데, 지금 돌이켜보니 그때만큼 오롯이 하나에 몰두하면서 단기간에 많은 것을 배우고 성장할 수 있었던 시간은 다시 오기 힘들 것 같습니다. 그 뜻깊은 시간을 함께해 준 배상민 박사, 그리고 곁에서 선배로서 지원해주고 친구로서 격려를 건네준 고종우 박사에게 고마움을 전합니다. 친구이자 동료, 그리고 경쟁자로서 이후로도 두 사람과 같이 연구를 하면서 정말 많은 것들을 배울 수 있었습니다.

이제는 현업과 학계에 몸담고 있는 재훈이형, 형준이형, 태현이형, 기훈이형, 상묵이형, 수명이형 등 OSI 연구실의 졸업생 선배들의 기여가 있었기에 현재의 연구실이 이렇게 발전할 수 있었던 것 같습니다. 아무것도 모르던 대학원 초기에 형들은 든든한 지원군이 되어주었고, 저는 그 발자취를 편하게 따라가는 것만으로도 큰 도움을 받았습니다. AI대학원 1기로 같이 입학해 큰 의지가 되었고 여전히 정신적 지주로서 귀감이 되는 민규형에게는, 제가 받은 것들을 모두 보답할 시간이 주어졌으면 합니다. 연구실의 분위기 메이커들이자 저와 짧은 기간 호흡을 맞췄던 성윤이형, 지환이형, 남규, 유진이의 앞길도 진심으로 응원합니다. 이 외에도 일일이 언급하지 못한 OSI 연구실의 모든 선후배 및 동료분들께 감사드립니다. 여러분들이 선사해 준 잊지 못할 추억들은 제게 바꿀 수 없는 자산이 되었습니다.

본 학위 논문을 작성하는 데에 큰 도움을 주시고 좋은 피드백을 아끼지 않으신 김찬우 교수님, 김회린 교수님, 오태현 교수님, 정준선 교수님 등 심사위원 분들께도 깊은 감사의 말씀을 올립니다. 또한, 학부 1학년 소망관 룸메이트 때부터 친구로 지내다 이제는 함께 논문을 쓰는 연구자가 된 장강욱 박사에게도 각별한 고마움을 전합니다. 이 학위 논문의 모든 챕터에 참여하면서 제 부족한 백그라운드를 채워주고 통찰력 있는 조언과 아낌없는 지원으로 박사 과정을 무사히 마무리할 수 있게 도와주었습니다. 기대하는 바가 커서 싫은 소리도 자주 하고 몰아붙인 적도 있지만, 한순간도 사사로운 감정을 담아 한 말이 아니었음을 이해해 주었으면 합니다. 앞으로의 연구 활동도 기대가 되며, 어떤 길을 걷든 응원하겠습니다. 본 학위 논문에 큰 도움이 되었던 성우에게도 많은 신세를 진 것 같아 고맙고, 현재는 고민이 많겠지만 가까운 미래에 선망받는 연구자로 성장할 것이라 믿어 의심치 않습니다. 아울러 네이버웹툰 인턴 당시 훌륭한 멘토가 되어주셨던 준수님, 홍기범 교수님, 안남혁 교수님과, AWS 인턴 당시 많은 도움을 주셨던 Haofu, Srikar, Peng 멘토분들께 감사의 말씀을 드립니다. 학교에만 머물러 있던 제 시야를 넓혀준 산업계 멘토분들 덕분에 박사 과정 동안 어떤 연구를 해야 할지 깊이 고민할 수 있었습니다.

연구실 밖에서 제 삶을 지탱해 준 소중한 인연들에게도 감사의 마음을 전하고 싶습니다. 먼저, 연구가 힘들고 지칠 때마다 삶의 환기가 되어주고 든든한 의지가 되어준 사랑하는 고등학교 친구들 지헌, 기태, 강희, 우탁이에게 고마움을 전합니다. 항상 바쁘다는 핑계로 더 많은 시간을 함께하지 못해 미안한 마음이 큽니다. 그리고 만나면 여전히 대학교 1학년 시절로 돌아간 듯한 즐거움을 주는 믹서 친구들도 감사합니다. 어느덧 각자의 분야에서 사회의 중요한 구성원으로 자리잡는 모습들을 보니 새삼 신기하고 대견스럽습니다. 또한, 비록 분야는 다르지만 비슷한 곳에서 함께 고생하고 있는 한성 친구들 승호, 민수, 한결이와 대학 생활의 시작을 함께했던 새터 8반 승민, 윤석, 재욱, 은섭, 병준, 준규, 그리고 제 가장 오랜 벗인 재현이와 재진이에게도 변치 않는 우정의 마음을 전합니다.

마지막으로, 제 삶의 이유이자 원동력인 사랑하는 가족들에게 이 논문을 바칩니다. 언제나 무조건적인 지지를 보내주시는 부모님께 무한한 존경과 감사를 표합니다. 힘들면 언제든 때려치라고 말씀하시던 부모님 덕분인지, 부모님 말씀 잘 안듣는 아들은 그만두지 않고 끝까지 버텨 졸업을 하게 되었습니다. 늘 입 밖으로 꺼내지는 못하지만, 이 지면을 빌려 사랑하고 존경한다는 말씀을 올립니다.

그리고 지난 11월 1일, 평생을 약속하며 제 인생의 가장 소중한 동반자가 되어준 아내 현정에게 깊은 사랑을 전합니다. 긴 학위 과정 동안 수많은 굴곡이 있었음에도 한결같이 제 곁을 지키며 묵묵히 기다려준 당신의 사랑과 배려에 마음 깊이 감사합니다. 앞으로 펼쳐질 여정에서도 크고 작은 역경들이 있겠지만 같이 헤쳐 나갈 수 있는 당신이 있기에 저는 큰 용기를 얻습니다. 인생의 새로운 챕터를 시작하는 지금, 당신과 함께 써 내려갈 다음 페이지들이 더없이 설레고 기대됩니다.

\vskip 10pt
\hfill
2025년 12월 홍릉의 연구실에서
}

\curriculumvitae[3]

    \begin{personaldata}
        \name       {Sungnyun Kim (김 성 년)}
        \dateofbirth{1996}{09}{25}
    \end{personaldata}

    \begin{education}
        \item[2021. 9.\ --\ 2026. 2.] Kim Jaechul Graduate School of Artificial Intelligence, KAIST (Ph.D.)
        \item[2019. 9.\ --\ 2021. 8.] Graduate School of Artificial Intelligence, KAIST (M.S.)
        \item[2015. 2.\ --\ 2019. 9.] Chemical and Biomolecular Engineering, KAIST (B.S.)
        \item[2015. 2.\ --\ 2019. 9.] Industrial and Systems Engineering, KAIST (B.S.)
    \end{education}

    \begin{career}
        \item[2023. 7.\ --\ 2023. 11.] Applied Scientist Intern, AWS AI Labs
        \item[2023. 1.\ --\ 2023. 6.] AI Research Intern, NAVER Webtoon
        \item[2020. 3.\ --\ 2022. 12.] Teaching Assistant, KAIST AI
        \item[2018. 12.\ -- \ 2019. 6.] Undergraduate Research Intern, KAIST OSI Lab
        \item[2017. 12.\ --\ 2018. 1.] Research Intern, SK Hynix
    \end{career}

   \begin{activity}
       \item 1st Place Award, AI Frontier Challenge, Korea Artificial Intelligence Association (KAIA) 2025.
       \item Top Reviewer (top 1.9\%), International Conference on Machine Learning (ICML) 2025.
       \item Best Student Paper Award (top 0.2\%), International Conference on Acoustics, Speech and Signal Processing (ICASSP) 2024.
       \item Invited Paper, The 10th Workshop on Fine-Grained Visual Categorization at CVPR 2023.
       \item Outstanding Paper Award, Korea Artificial Intelligence Association (KAIA) 2021.
       \item Dean’s List (top 3\%), Faculty of Engineering Dept., KAIST, Spring 2017.
    \end{activity}

    \begin{publication}
        
        \item \underline{\textbf{Sungnyun Kim}}*, Gihun Lee*, Sangmin Bae*, and Se-Young Yun, ``MixCo: Mix-up Contrastive Learning for Visual Representation,'' \textit{NeurIPS 1st Workshop on Self-Supervised Learning}, 2020.

        \item \underline{\textbf{Sungnyun Kim}} and Se-Young Yun, ``Calibration of Few-Shot Classification Tasks: Mitigating Misconfidence from Distribution Mismatch,'' \textit{IEEE Access, vol.10}, 2022.
        
        \item Yujin Kim*, Jaehoon Oh*, \underline{\textbf{Sungnyun Kim}}, and Se-Young Yun, ``How to Fine-tune Models with Few Samples: Update, Data Augmentation, and Test-time Augmentation,'' \textit{ICML Workshop on Updatable Machine Learning (Oral Presentation)}, 2022.

        \item \underline{\textbf{Sungnyun Kim}}*, Jaewoo Shin*, Seongha Eom, Jihwan Oh, and Se-Young Yun, ``Real-time and Explainable Detection of Epidemics with Global News Data,'' \textit{Workshop on Healthcare AI and COVID-19, PMLR 184:73–90 (Oral Presentation)}, 2022. 
        
        \item Jaehoon Oh*, \underline{\textbf{Sungnyun Kim}}*, Namgyu Ho*, Jin-Hwa Kim, Hwanjun Song, and Se-Young Yun, ``ReFine: Re-randomization before Fine-tuning for Cross-domain Few-shot Learning,'' \textit{Proceedings of the 31th ACM International Conference on Information \& Knowledge Management (CIKM)}, 2022.
        
        \item Jaehoon Oh*, \underline{\textbf{Sungnyun Kim}}*, Namgyu Ho*, Jin-Hwa Kim, Hwanjun Song, and Se-Young Yun, ``Understanding Cross-Domain Few-Shot Learning Based on Domain Similarity and Few-Shot Difficulty,'' \textit{Advances in Neural Information Processing Systems (NeurIPS)}, 2022.
        
        \item Sangmin Bae*, \underline{\textbf{Sungnyun Kim}}*, Jongwoo Ko, Gihun Lee, Seungjong Noh, and Se-Young Yun, ``Self-Contrastive Learning: Single-viewed Supervised Contrastive Framework using Sub-network,'' \textit{Proceedings of the AAAI Conference on Artificial Intelligence (Oral Presentation)}, 2023.
        
        \item \underline{\textbf{Sungnyun Kim}}*, Sangmin Bae*, and Se-Young Yun, ``Coreset Sampling from Open-Set for Fine-Grained Self-Supervised Learning,'' \textit{Proceedings of the IEEE/CVF Conference on Computer Vision and Pattern Recognition (CVPR)}, 2023.

        \item Sangmin Bae*, June-Woo Kim*, Won-Yang Cho, Hyerim Baek, Soyoun Son, Byungjo Lee, Changwan Ha, Kyongpil Tae, \underline{\textbf{Sungnyun Kim}}$^\dagger$, and Se-Young Yun$^\dagger$, ``Patch-Mix Contrastive Learning with Audio Spectrogram Transformer on Respiratory Sound Classification,'' \textit{Proceedings of Interspeech}, 2023.

        \item Kangwook Jang*, \underline{\textbf{Sungnyun Kim}}*, Se-Young Yun, and Hoirin Kim, ``Recycle-and-Distill: Universal Compression Strategy for Transformer-based Speech SSL Models with Attention Map Reusing and Masking Distillation,'' \textit{Proceedings of Interspeech}, 2023.
        
        \item Jihwan Oh, \underline{\textbf{Sungnyun Kim}}, Gahee Kim, SeongHwan Kim, and Se-Young Yun, ``Diffusion-based Episodes Augmentation for Offline Multi-Agent Reinforcement Learning,'' \textit{ICML Workshop on Structured Probabilistic Inference \& Generative Modeling (SPIGM)}, 2024.
        
        \item Kangwook Jang, \underline{\textbf{Sungnyun Kim}}, and Hoirin Kim, ``STaR: Distilling Speech Temporal Relation for Lightweight Speech Self-Supervised Learning Models,'' \textit{IEEE International Conference on Acoustics, Speech and Signal Processing (ICASSP) (Best Student Paper)}, 2024.
        
        \item Jongwoo Ko, \underline{\textbf{Sungnyun Kim}}, Tianyi Chen, and Se-Young Yun, ``DistiLLM: Towards Streamlined Distillation for Large Language Models,'' \textit{International Conference on Machine Learning (ICML)}, 2024.
        
        \item \underline{\textbf{Sungnyun Kim}}*, Kangwook Jang*, Sangmin Bae, Hoirin Kim, and Se-Young Yun, ``Learning Video Temporal Dynamics with Cross-Modal Attention for Robust Audio-Visual Speech Recognition,'' \textit{IEEE Spoken Language Technology Workshop (SLT)}, 2024.
        
        \item \underline{\textbf{Sungnyun Kim}}*, Haofu Liao*, Srikar Appalaraju, Peng Tang, Zhuowen Tu, Ravi Kumar Satzoda, R Manmatha, Vijay Mahadevan, and Stefano Soatto, ``DocKD: Knowledge Distillation from LLMs for Open-World Document Understanding Models,'' \textit{Conference on Empirical Methods in Natural Language Processing (EMNLP)}, 2024.
        
        \item Seongyoon Kim, Minchan Jeong, \underline{\textbf{Sungnyun Kim}}, Sungwoo Cho, Sumyeong Ahn, and Se-Young Yun, ``FedDr+: Stabilizing Dot-regression with Global Feature Distillation for Federated Learning,'' \textit{Transactions on Machine Learning Research (TMLR)}, 2025.
        
        \item \underline{\textbf{Sungnyun Kim}}, Sungwoo Cho, Sangmin Bae, Kangwook Jang, and Se-Young Yun, ``Multi-Task Corrupted Prediction for Learning Robust Audio-Visual Speech Representation,'' \textit{International Conference on Learning Representations (ICLR)}, 2025.
        
        \item Jongwoo Ko, Tianyi Chen, \underline{\textbf{Sungnyun Kim}}, Tianyu Ding, Luming Liang, Ilya Zharkov, and Se-Young Yun, ``DistiLLM-2: A Contrastive Approach Boosts the Distillation of LLMs,'' \textit{International Conference on Machine Learning (ICML) (Oral Presentation)}, 2025.
        
        \item \underline{\textbf{Sungnyun Kim}}, Kangwook Jang, Sangmin Bae, Sungwoo Cho, and Se-Young Yun, ``MoHAVE: Mixture of Hierarchical Audio-Visual Experts for Robust Speech Recognition,'' \textit{International Conference on Machine Learning (ICML)}, 2025.
        
        \item Sungwoo Cho, Jeongsoo Choi, \underline{\textbf{Sungnyun Kim}}, and Se-Young Yun, ``MAVFlow: Preserving Paralinguistic Elements with Conditional Flow Matching for Zero-Shot AV2AV Multilingual Translation,'' \textit{Proceedings of the IEEE/CVF International Conference on Computer Vision (ICCV)}, 2025.
        
        \item \underline{\textbf{Sungnyun Kim}}, Junsoo Lee, Kibeom Hong, Daesik Kim, and Namhyuk Ahn, ``DiffBlender: Composable and Versatile Multimodal Text-to-Image Diffusion Models,'' \textit{Expert Systems with Applications (ESWA), vol.297}, 2025.
        
        \item Sangmin Bae*, Yujin Kim*, Reza Bayat*, \underline{\textbf{Sungnyun Kim}}, Jiyoun Ha, Tal Schuster, Adam Fisch, Hrayr Harutyunyan, Ziwei Ji, Aaron Courville, and Se-Young Yun, ``Mixture-of-Recursions: Learning Dynamic Recursive Depths for Adaptive Token-Level Computation,'' \textit{Advances in Neural Information Processing Systems (NeurIPS)}, 2025.
        
        \item Jongwoo Ko*, \underline{\textbf{Sungnyun Kim}}*, Sungwoo Cho, and Se-Young Yun, ``Flex-Judge: Text-Only Reasoning Unleashes Zero-Shot Multimodal Evaluators,'' \textit{Advances in Neural Information Processing Systems (NeurIPS)}, 2025.
        
        \item \underline{\textbf{Sungnyun Kim}}*, Kangwook Jang*, Sungwoo Cho, Joon Son Chung, Hoirin Kim, and Se-Young Yun, ``Two Heads Are Better Than One: Audio-Visual Speech Error Correction with Dual Hypotheses,'' \textit{arXiv preprint arXiv:2510.13281}, 2025.
        
    \end{publication}

  \label{paperlastpagelabel}   
  
\end{document}